\documentclass[11pt, letterpaper]{article}
\usepackage[margin=1in, centering]{geometry}
\usepackage{mathtools}
\usepackage{amsthm}
\usepackage{thmtools}
\usepackage{dsfont}
\usepackage{amssymb}
\usepackage{amsfonts}
\usepackage{mathrsfs} 
\usepackage{biolinum}
\usepackage{mathpazo, tgpagella}
\DeclareMathAlphabet{\mathsf}{\encodingdefault}{\sfdefault}{m}{n}
\SetMathAlphabet{\mathsf}{bold}{\encodingdefault}{\sfdefault}{b}{n}
\usepackage[table,dvipsnames]{xcolor}
\usepackage{subcaption}
\usepackage[normalem]{ulem}
\definecolor{HANADA}{RGB}{0, 98, 132}
\definecolor{KURENAI}{RGB}{203, 27, 69}
\usepackage{setspace}
\usepackage[
    pdfstartview=FitH,
    pdfpagemode=UseNone,
    bookmarksdepth=2,
    colorlinks=true,
    citecolor=KURENAI,
    linkcolor=HANADA,
    urlcolor=HANADA,
    backref=page,
    linktocpage=true
]{hyperref}
\usepackage{xurl}
\usepackage[capitalise, nameinlink]{cleveref}
\usepackage[nottoc]{tocbibind}
\usepackage{appendix}
\usepackage{array}
\usepackage{braket}
\usepackage{graphicx}
\usepackage{enumitem}
\usepackage{thm-restate}
\usepackage{etoolbox}
\usepackage{authblk}
\usepackage{tikz}
\usetikzlibrary{arrows.meta,positioning,calc,cd}
\usetikzlibrary{decorations.pathreplacing}

\renewcommand\Affilfont{\fontsize{10pt}{11pt}\selectfont}
\makeatletter
\renewcommand\AB@affilsepx{\protect\\[0.8em]\protect\Affilfont}
\makeatother

\usepackage{inconsolata}

\title{\bf $\uqip = \upspace$\\[0.3em]}

\author[1]{Entong He}
\author[2]{Fengning Ou}
\author[2]{Mingnan Zhao}

\makeatletter
\@namedef{@sep6}{\\[0.8em]}
\makeatother

\affil[1]{QICI Quantum Information and Computation Initiative, School of Computing and Data Science,\protect\\[0.3em]
The University of Hong Kong, Pokfulam Road, Hong Kong SAR, China}
\affil[ ]{\texttt{ethe@cs.hku.hk}}
\affil[2]{State Key Laboratory for Novel Software Technology, Nanjing University, Nanjing 210023, China}
\affil[ ]{\texttt{reverymoon@gmail.com, mingnanzh@gmail.com}}

\date{}

\newcommand{\uqip}{\mathsf{unitaryQIP}}
\newcommand{\upspace}{\mathsf{unitaryPSPACE}}

\newcommand{\sqip}{\mathsf{stateQIP}}
\newcommand{\spspace}{\mathsf{statePSPACE}}

\newcommand{\pspace}{\mathsf{PSPACE}}

\newcommand{\U}{\mathsf{U}}

\newcommand{\hdm}{\mathsf{H}}
\newcommand{\tgate}{\mathsf{T}}
\newcommand{\cnot}{\mathsf{CNOT}}

\makeatletter
\renewcommand\paragraph{%
  \@startsection{paragraph}{4}{\z@}%
    {1.5ex plus 0.5ex minus 0.2ex}
    {-1em}
    {\normalfont\normalsize\bfseries}%
}
\renewcommand\subparagraph{%
  \@startsection{subparagraph}{5}{\parindent}%
    {1.3ex plus 0.4ex minus 0.2ex}
    {-1em}
    {\normalfont\normalsize\bfseries}%
}
\makeatother

\newcommand{\norm}[1]{\left\lVert #1 \right\rVert}

\newcommand{\td}{\mathsf{TD}}
\newcommand{\dd}{\mathsf{D}_{\diamond}}
\newcommand{\N}{\mathbb{N}}
\newcommand{\C}{\mathbb{C}}
\newcommand{\Span}{\mathrm{span}}
\newcommand{\bfx}{\mathbf{x}}
\newcommand{\id}{\mathbf{I}}
\newcommand{\HH}{\mathcal{H}}
\newcommand{\I}{\mathrm{i}}
\newcommand{\ad}{\mathrm{Ad}}
\newcommand{\R}{\mathbb{R}}
\newcommand{\ps}[1]{\ket{#1} \!\! \bra{#1}}
\newcommand{\mbf}[1]{\mathbf{#1}}

\newcommand{\linermap}[1]{\mathsf{L}(#1)}

\newcommand{\one}{\mathbf{1}}
\newcommand{\op}{\mathrm{op}}
\newcommand{\interact}{\mathord{\leftrightarrows}}

\def\eqref#1{\textup{(\ref{#1})}}

\newcommand{\TP}{\mathsf{T}}

\newcommand{\F}{\mathbb{F}}

\newcommand{\bfh}{\mathbf{h}}
\newcommand{\rank}{\mathrm{rank}}
\newcommand{\im}{\mathrm{im}}
\newcommand{\Z}{\mathbb{Z}}
\newcommand{\Fou}{\mathscr{F}}
\newcommand{\SU}{\mathsf{SU}}

\newcommand{\ketbra}[1]{| #1 \rangle\! \langle #1 |}

\renewcommand{\Pr}{\mathop{\bf Pr\/}}
\newcommand{\pr}{\mathop{\mathrm{Pr}\/}}
\newcommand{\E}{\mathop{\bf E\/}}

\newcommand{\tr}{\mathrm{tr}} \newcommand{\Tr}{\tr} 
\newcommand{\poly}{\mathrm{poly}}

\newcommand{\diag}{\mathrm{diag}}

\newcommand{\eps}{\varepsilon}
\newcommand{\bigo}{\mathcal{O}}

\newtheorem{theorem}{Theorem}[section]

\newcommand{\theoremname}{testing}
\newtheorem*{namedtheorem}{\theoremname}

\newtheorem{lemma}[theorem]{Lemma}

\newtheorem{proposition}[theorem]{Proposition}

\newtheorem{corollary}[theorem]{Corollary}

\newtheorem{problem}[theorem]{Problem}

\theoremstyle{definition}
\newtheorem{definition}[theorem]{Definition}

\newtheorem{remark}[theorem]{Remark}

\newtheorem{protocol}[theorem]{Protocol}
\newtheorem{fact}[theorem]{Fact}

\begin{document}

\hypersetup{pageanchor=false}
\pagestyle{empty}

\maketitle

\thispagestyle{empty}

\allowdisplaybreaks

\begingroup
\setstretch{1.2}
\begingroup
\setlength{\leftmargini}{2em}
\begin{abstract}
    \normalsize
    \setlength{\emergencystretch}{1em}
    Unitary synthesis complexity studies
    the computational cost of applying a target unitary
    to an unknown quantum state,
    extending complexity theory beyond
    its traditional focus on problems with classical inputs and outputs.
    Rosenthal and Yuen (\hyperlink{cite.RY22}{ITCS 2022}) introduced two complexity classes for unitary synthesis:
    $\mathsf{unitaryQIP}$ and $\mathsf{unitaryPSPACE}$
    for the unitary families implementable
    via interactive proofs and in polynomial space, respectively.
    In this work, we prove the equality
    $\mathsf{unitaryQIP}=\mathsf{unitaryPSPACE}$
    and thereby establish a unitary-complexity
    analogue of $\mathsf{QIP} = \mathsf{PSPACE}$
    (\hyperlink{cite.JJUW11}{Jain, Ji, Upadhyay, and Watrous, J. ACM 2011})
    and $\mathsf{stateQIP}=\mathsf{statePSPACE}$
    (\hyperlink{cite.RY22}{Rosenthal and Yuen, ITCS 2022};
    \hyperlink{cite.MY23}{Metger and Yuen, FOCS 2023}).

    To prove $\mathsf{unitaryQIP} \subseteq \mathsf{unitaryPSPACE}$,
    we use an acceptance rigidity lemma derived from the soundness condition
    to show that the polynomial-space prover constructed for a fixed input
    in state synthesis (\hyperlink{cite.MY23}{Metger and Yuen, FOCS 2023})
    implements the target unitary on every input,
    leading to a polynomial-space simulation of the protocol.
    For the reverse inclusion,
    we build on OpenAI's recently announced solution
    to the unitary synthesis problem
    (\hyperlink{cite.4262757}{Aaronson and Kuperberg, Theory Comput. 2007}),
    and show that $\mathsf{unitaryPSPACE}$ families can be synthesized
    in polynomial time with a $\mathsf{PSPACE}$ oracle.
    We then implement the oracle queries
    using auxiliary states prepared via
    the $\mathsf{statePSPACE} \subseteq \mathsf{stateQIP}$ protocol
    (\hyperlink{cite.RY22}{Rosenthal and Yuen, ITCS 2022}).

    Beyond the equality,
    we show that our protocol for $\mathsf{unitaryPSPACE}$
    can be parallelized to constant rounds
    when the verifier receives polynomially many copies of the input,
    and we also obtain a complexity-theoretic barrier
    to proving the existence of $\mathsf{EFI}$ pairs.
\end{abstract}
\endgroup

\newpage
\tableofcontents

\newpage
\endgroup

\pagenumbering{arabic}
\pagestyle{plain}
\hypersetup{pageanchor=true}


\begingroup
\setstretch{1.1}

\section{Introduction}
\label{sec:intro}
Computational complexity theory traditionally studies the resources required to solve problems on classical strings~\cite{AB09}.
Much of quantum complexity theory adopts the same perspective,
analyzing the resources quantum computers need to solve \emph{classical} decision problems~\cite{BV97,watrous_quantum_2009}.
Quantum computation, however, also encompasses tasks whose desired outputs are \emph{inherently quantum},
including preparing entangled states,
implementing Hamiltonian time evolution \cite{Llo96,LC19}, programming unitary gates \cite{YRC20, He2026resource},
and decoding quantum information transmitted through noisy channels \cite{BDSW96}.
In these tasks, the goal is to physically prepare a quantum state or to apply a transformation to a given quantum system.
Standard quantum complexity classes,
which deal with problems that have classical inputs and outputs,
therefore do not directly capture these quantum tasks.

Moreover, recent developments suggest that traditional computational complexity theory
may not fully align with quantum cryptography.
Indeed, oracle separations~\cite{kretschmer:LIPIcs.TQC.2021.2,KQST23,LMW24,KQT25,LQSY+26}
show that breaking certain quantum cryptographic primitives,
such as pseudorandom state generators ($\mathsf{PRS}$s) \cite{JLS18}
and pseudorandom unitaries ($\mathsf{PRU}$s) \cite{JLS18,MH25,LQSY+25},
may remain hard even when decision problems in $\mathsf{NP}$ or $\mathsf{QMA}$ are easy.
For example, relative to suitable oracles,
$\mathsf{PRS}$s can exist even if $\mathsf{P}=\mathsf{NP}$~\cite{KQST23},
and $\mathsf{PRU}$s can exist even if $\mathsf{BQP}=\mathsf{QMA}$~\cite{kretschmer:LIPIcs.TQC.2021.2}.
As the security of these primitives rests on the hardness of processing quantum states or operations rather than classical strings,
such findings further underscore the need to develop a complexity-theoretic framework for \emph{inherently quantum problems}.

The study of \emph{state synthesis} and \emph{unitary synthesis} addresses this need by treating the preparation of quantum states and the implementation of unitary transformations as computational problems in their own right.
This perspective goes back to the work on quantum proofs, advice, and oracle simulation~\cite{4262757},
and was later developed systematically by Aaronson~\cite{aaronson2016complexityquantumstatestransformations}.
To treat synthesis problems within a similar framework,
Rosenthal and Yuen introduced several complexity classes modeled on those of classical complexity theory~\cite{RY22},
including $\mathsf{statePSPACE}$ and $\mathsf{unitaryPSPACE}$ for states and unitaries synthesizable in polynomial space,
and $\mathsf{stateQIP}$ and $\mathsf{unitaryQIP}$ for those synthesizable via interactive proofs.
They showed that $\mathsf{statePSPACE}\subseteq\mathsf{stateQIP}$,
and Metger and Yuen later proved the converse inclusion~\cite{MY23}, thus establishing $\mathsf{stateQIP}=\mathsf{statePSPACE}$.
In other words, a family of quantum states can be prepared in polynomial space if and only if a polynomial-time quantum verifier can synthesize it with the help of an untrusted prover.
This is a state-synthesis analogue of the celebrated equality \(\textsf{QIP}=\textsf{PSPACE}\) in traditional complexity theory~\cite{JJUW11},
extending it from classical decision problems to quantum state synthesis problems.

This state-synthesis characterization leaves a natural problem of whether the analogous equality extends to unitary synthesis:
\emph{does $\mathsf{unitaryQIP}=\mathsf{unitaryPSPACE}$ hold?}
Answering this question seems to require overcoming additional obstacles.
Whereas state synthesis asks only for a \emph{single} target state for each classical instance, unitary synthesis requires applying a target unitary $U$ coherently to a quantum system in an arbitrary unknown state $\ket{\psi}$. 
Since correctness is required for \emph{every} possible input, preparing the single state $U\!\ket{0^n}$ does not suffice. Moreover, by the no-cloning theorem, the verifier cannot copy $\ket{\psi}$ to run repeated tests against an untrusted prover.
This gap also manifests itself formally in the Boolean-oracle model.
Rosenthal gave a polynomial-time algorithm that approximately prepares
any target state using one query to a suitably chosen Boolean oracle~\cite{Ros24},
while Lombardi, Ma, and Wright proved that one query does not suffice to efficiently synthesize arbitrary unitaries \cite{LMW24}.

Prior work explored the relationship between
$\mathsf{unitaryQIP}$ and $\mathsf{unitaryPSPACE}$ in relaxed settings.
Rosenthal and Yuen gave interactive protocols
for unitary families in $\mathsf{unitaryPSPACE}$
with \emph{polynomial action},
i.e., each unitary acts nontrivially only on a subspace of polynomial dimension,
as well as for general families in $\mathsf{unitaryPSPACE}$
provided that the verifier also knows a polynomial-dimensional subspace containing its input state \cite{RY22}.
For \emph{distributional} unitary synthesis classes,
where correctness is required only on average over a fixed input distribution,
Bostanci, Efron, Metger, Poremba, Qian, and Yuen~\cite{BEM+26}
proved $\mathsf{avgUnitaryQIP}=\mathsf{avgUnitaryPSPACE}$
using the Uhlmann transformation problem as a bridge.
The question in the general worst-case setting,
which asks for correctness on every input without restricting the unitary,
has remained open.

In this work, we resolve this question in the affirmative, proving that
\[
    \uqip=\upspace\enspace.
\]
Together with $\mathsf{QIP}=\pspace$ and $\mathsf{stateQIP}=\mathsf{statePSPACE}$,
this shows that interactive proofs characterize polynomial-space computation
for decision problems, state synthesis, and unitary synthesis.
To show $\uqip\subseteq\upspace$, we simulate interactive synthesis
in polynomial space,
building on the polynomial-space prover construction of Metger and Yuen~\cite{MY23}.
For the converse, we synthesize polynomial-space unitaries interactively,
drawing on a recent solution to the \emph{unitary synthesis problem} (USP)~\cite{4262757,aaronson2016complexityquantumstatestransformations} announced by OpenAI~\cite{OpenAI2026UnitarySynthesis}.
We also provide a self-contained exposition of this solution in the appendix.
Beyond the equality, our techniques also lead to
a constant-round protocol for $\upspace$
when the verifier receives polynomially many copies of the input,
as well as a complexity-theoretic barrier to proving
the existence of $\mathsf{EFI}$ pairs \cite{BCQ23}.

\subsection{Main results}
\label{sec:intro-results}

In this work, we establish inclusions between $\uqip$ and $\upspace$ in both directions.
To state our results, we first introduce these classes informally:
\begin{itemize}[topsep=2pt,itemsep=1pt,leftmargin=1.6em]
    \item $\upspace_\delta$ (see \cref{def:upspace_ubqp}) consists of
        unitary families $\{U_n\}_{n\in\N}$ such that there exists
        a space-uniform polynomial-space quantum algorithm that, on every
        input state $\ket{\psi}$, produces a state within \emph{synthesis error}
        $\delta(n)$ of $U_n\ket{\psi}$.
    \item $\uqip_{c,s,\delta}$ (see \cref{def:para-uqip}) consists of
        unitary families $\{U_n\}_{n\in\N}$ such that there exists
        a polynomial-time quantum verifier that interacts
        with an untrusted prover on an input state $\ket{\psi}$
        with \emph{completeness error} $c$,
        \emph{soundness threshold} $s$, and \emph{synthesis error} $\delta$:
        an honest prover that makes the verifier accept with
        probability at least $1-c(n)$ on every input,
        and for any prover and any input $\ket{\psi}$,
        whenever the verifier accepts with probability greater than $s(n)$,
        its output state conditioned on acceptance is within $\delta(n)$ of $U_n\!\ket{\psi}$.
\end{itemize}

\paragraph{Simulating interactive synthesis in polynomial space.}

We first show that every unitary family in $\uqip$ can be implemented in polynomial space,
i.e., $\uqip\subseteq\upspace$.

\begin{restatable}{theorem}{uqipinupspace}
    \label{thm:uqip-in-upspace}
    Let $c,s,\eps,\delta:\N\to[0,1]$ be such that
    $\eps(n) = 1/q(n)$ for some polynomial $q(n)$,
    and
    $\eps(n)\leq(1-c(n)-s(n))/4$ and
    $\delta(n)\leq(1-c(n)-s(n))^2/8$
    for all $n\in\N$.
    Then
    \[
        \uqip_{c,s,\delta}
        \subseteq
        \upspace_{c+\eps+7\delta} \enspace.
    \]
\end{restatable}

We prove this result by showing that an interactive synthesis protocol can be
approximately simulated in polynomial space.
We start from a result of Metger and Yuen~\cite{MY23}:
when the verifier's input is fixed to the all-zero state,
a near-optimal prover can be implemented in polynomial space.
In our setting, however,
the verifier receives an arbitrary quantum input.
Our key observation is that the soundness condition ensures that
a prover achieving high acceptance probability on one fixed input state
is also accepted with high probability on every input.
Soundness then guarantees that the output is close to the target state on every input.

\paragraph{Synthesizing polynomial-space unitaries interactively.}

Conversely, we construct an interactive protocol in which an efficient verifier, aided by an untrusted prover, synthesizes any polynomial-space unitary with perfect completeness and any prescribed inverse-polynomial soundness, thus proving $\upspace\subseteq\uqip$.

\begin{restatable}{theorem}{upspaceinuqip}
\label{thm:upspace-in-uqip} 
    Let $\delta,s:\N\to[0,1]$ satisfy
    $1/q(n)\leq s(n)<1$ and $\delta(n)\geq 1/r(n)$
    for some polynomials $q,r$ with $q(n)\geq2$ and $r(n)\geq4$.
    Then, it holds that
    \[
        \upspace_{\delta^2/2^{14}} \subseteq \uqip_{0,s,\delta} \enspace.
    \]
\end{restatable}

One crucial component of our protocol is a reduction from
a polynomial-space unitary synthesis circuit to
a polynomial-time quantum circuit with access to a $\mathsf{PSPACE}$ oracle.
This reduction builds on OpenAI's recently announced solution~\cite{OpenAI2026UnitarySynthesis} to
the unitary synthesis problem \cite{4262757,aaronson2016complexityquantumstatestransformations}.
With this reduction in hand, we simulate the oracle queries by consuming polynomially many
auxiliary states in $\mathsf{statePSPACE}$, which can be prepared
through interaction with the prover using Rosenthal and Yuen's
protocol~\cite{RY22}, yielding the desired synthesis
protocol.

Following~\cite{RY22,MY23}, we define the following subscript-free classes by requiring arbitrarily inverse-polynomial synthesis error, perfect completeness and soundness threshold $1/2$\footnote{
    Here we assume perfect completeness, because
    by \cref{thm:uqip-in-upspace,thm:upspace-in-uqip},
    protocols with inverse-polynomial completeness error and synthesis errors
    can be converted into protocols with
    perfect completeness while retaining inverse-polynomial synthesis error.
    The same conversions allow the soundness threshold $1/2$ to be replaced by any $s$ satisfying $1/q(n)\le s(n)\le 1-1/q(n)$ for some polynomial $q$, without affecting the equality below.}:
\[
    \uqip \coloneqq \bigcap_{p(n) \in \poly(n)}\uqip_{0,\frac{1}{2},\frac{1}{p}},
    \quad
    \upspace \coloneqq \bigcap_{p(n) \in \poly(n)}\upspace_{\frac{1}{p}} \enspace,
\]
where both intersections range over all polynomials.
Combining \cref{thm:uqip-in-upspace,thm:upspace-in-uqip}, we have

\begin{theorem}
    \label{thm:uqip-equals-upspace}
    $\uqip=\upspace$.
\end{theorem}

\paragraph{Beyond the equality.}
We also obtain two further results building on our protocol and the underlying synthesis technique with a $\pspace$ oracle.

First, we investigate the round complexity of interactive unitary synthesis.
It is known that both $\pspace$ and $\mathsf{statePSPACE}$ admit
constant-round quantum interactive protocols \cite{KW00,JJUW11,Ros24}, that is, $\mathsf{PSPACE} \subseteq \mathsf{QIP}[3]$ and $\mathsf{statePSPACE} \subseteq \mathsf{stateQIP}[6]$.
We show that constant-round protocols are also possible for $\upspace$
when the verifier is provided with polynomially many copies of the input state.
Combined with \cref{thm:uqip-in-upspace},
this also implies that every $\uqip$ protocol can be compressed to three rounds in this multi-copy setting.

\begin{restatable}{theorem}{constantround}
\label{thm:constant-round-synthesis}
    Let $\eps:\N\to(0,1/4]$ be polynomial-time computable with $1/\eps$ polynomially bounded.
    Then,
    \[
        \upspace_{\eps^4/2^{24}}
        \subseteq \uqip_{0,1/2,\eps}^{(\poly)}[3],
    \]
    where in the notation $\uqip^{(m)}[\ell]$, the verifier receives $m(n)$ copies of the input state and acts in at most $\ell \in \N$ rounds during the interaction.
\end{restatable}

The final result concerns the minimal assumptions of quantum cryptography.
Using our synthesis technique to efficiently implement a polynomial-space distinguisher from \cite{LLW26},
we establish a barrier to proving the existence of $\mathsf{EFI}$ pairs,
answering a question of Brakerski, Canetti, and Qian~\cite{BCQ23}:
any such proof would also separate $\mathsf{BQP}$ from $\pspace$.

\begin{restatable}{theorem}{EFI}
\label{cor:efi-separates-bqp-pspace}
    If $\mathsf{EFI}$ pairs exist, then $\mathsf{BQP}\neq\pspace$.
\end{restatable}

\subsection{Proof overview}
We now outline the proofs of our main results and highlight our technical contribution.

\paragraph{\textbf{\textsf{unitaryQIP}} $\subseteq$ \textbf{\textsf{unitaryPSPACE}}.}
To simulate an interactive synthesis protocol in polynomial space,
our starting point is the result of Metger and Yuen~\cite{MY23} (see \cref{thm:metger-yuen}).
It applies to protocols in which the verifier receives no quantum input,
that is, all of its registers start in the all-zero state, as in state synthesis.
For such protocols, it provides a polynomial-space prover whose acceptance probability is close to the maximum over all provers.
To apply this result to a protocol for unitary synthesis,
we fix the verifier's input register to the all-zero state, turning it into a protocol with no quantum input.
The resulting protocol has no quantum input,
so \cref{thm:metger-yuen} gives a polynomial-space prover that is accepted on the all-zero input with high probability.
By the soundness condition, the interaction with this prover therefore approximately applies the target unitary to the all-zero state.

It remains to examine whether the same prover also works well on arbitrary inputs.
A key observation, formalized in our \emph{acceptance rigidity lemma} (see \cref{lem:soundness-limits-acceptance-variation}),
is that soundness restricts how much the probability that a fixed prover is accepted can vary with the input (see \cref{lem:soundness-limits-acceptance-variation}).
Intuitively, consider an input state in superposition.
If the protocol accepted its components with very different probabilities, conditioning on acceptance would skew the amplitudes of the output state toward the components with higher acceptance probability.
This contradicts soundness, which guarantees that the output state is essentially the target unitary applied to the input state and hence preserves the relative amplitudes of the components.
Applying this to the above prover, which is accepted with high probability on the all-zero input,
we conclude that it is accepted with high probability on every input,
and hence, the interaction with it approximately applies the target unitary to every input.

A polynomial-space simulator can now run the verifier together with this polynomial-space prover.
It returns the verifier's output on acceptance and a fixed state on rejection.
Since both computations use polynomial space, this establishes $\uqip\subseteq\upspace$.

\paragraph{\textbf{\textsf{unitaryPSPACE}} $\subseteq$ \textbf{\textsf{unitaryQIP}}.}
Here a polynomial-time verifier must implement a polynomial-space unitary,
so it has to delegate the space-intensive computation to the prover.
A natural strategy is therefore to reduce the synthesis of a unitary in $\upspace$ to
an efficient quantum computation with queries to a powerful oracle, and then implement these queries with the prover's help.
To obtain this reduction, OpenAI's USP construction \cite{OpenAI2026UnitarySynthesis} offers a natural starting point. 
While the construction shows that some oracle lets the circuit approximate the unitary, it places no bound on the complexity of the language that this oracle encodes. Hence, the construction cannot be used off the shelf in our proof, since the oracle queries may go beyond what the verifier can verify. We overcome this obstacle by showing that every $\upspace$ family can be approximately implemented by polynomial-size quantum oracle circuits \emph{with queries to a $\pspace$ language}. We sketch this construction at the end of this subsection.

The original USP algorithm works for unitaries. Membership in $\upspace$, however, only guarantees that a space-uniform channel approximating the target unitary exists. To connect this implementation to USP, we round the channel to a nearby unitary in the standard way \cite{PhysRevA.90.012110,HKOT23} (see \cref{lemma:exact_unitary_recovery}) and show that the $\pspace$-computability of the entries carries over from the channel to the resulting unitary (see \cref{prop:pspace_computable_of_projected_unitary}).

With this unitary in hand, our $\pspace$-version USP algorithm grants us an oracle and an oracle circuit of polynomial size that approximately implements it (see \cref{lemma:usynthesis_with_pspace_oracle}). The oracle circuit naturally serves as the verifier's computation, so it remains only to implement its oracle queries through interaction with the prover. The standard phase-kickback trick converts each of these queries into the corresponding Boolean phase gate. Using the phase-state injection technique of \cite{11369061}, we reduce applying these phase gates to synthesizing their phase states (see \cref{lemma:exact_injection_of_diagonal_sign_state}). Specifically, for a Boolean function $f:\{0,1\}^m\to\{0,1\}$, the desired gate maps $\ket{x}$ to $(-1)^{f(x)}\ket{x}$, and the corresponding phase state is
\[
    \ket{\varphi_f}
    =\frac{1}{\sqrt{2^m}}
      \sum_{x\in\{0,1\}^m}(-1)^{f(x)}\ket{x}.
\]
A single query to a Boolean function can be implemented by injecting a sequence of phase states via measurements. Each subsequent phase state encodes a Boolean correction function chosen adaptively based on the previous measurement outcomes, and the overall action eventually equals the target query (see \cref{lemma:exact_injection_of_diagonal_sign_state,fig:injection_of_signal_state}).

Since the oracle is a $\pspace$ language, every correction function is $\pspace$-computable as well, so the phase states we need form a $\mathsf{statePSPACE}$ family. The verifier can then synthesize them with the prover's help, using the $\mathsf{statePSPACE} \subseteq \mathsf{stateQIP}$ protocol of Rosenthal and Yuen \cite{RY22}. Our protocol (see \cref{protocol:verifier_for_upspace}) invokes the Rosenthal-Yuen protocol as a subroutine. In every invocation, the honest prover is accepted with certainty, so it yields perfect completeness. To track the soundness, we compare each accepted preparation with an ideal phase state independent of the prover's memory (see \cref{lemma:accepted_map_via_phase_state}). Individual replacement errors add across the computation. Every branch of the ideal computation yields the oracle circuit's output, so conditioning on acceptance leaves it unchanged. With sufficiently accurate preparations, the replacement bound then guarantees that, for every prover accepted with probability above the soundness threshold, the actual output conditioned on acceptance is close to this ideal output (see \cref{lemma:soundness_bound_of_composed_protocol}). Together with the oracle circuit's approximation error, this gives the required closeness to the target unitary output, as per the final proof of \cref{thm:upspace-in-uqip}.

\paragraph{Constructing the $\pspace$ oracle for $\upspace$ families.} OpenAI's algorithm behind \cref{thm:usp} (see \cref{appendix:answer_of_USP}) synthesizes a unitary to within precision $2^{-\beta - 1}$ by first lifting it to a controlled family and then synthesizing this family through a recursion that trades dimension for control labels. At each level, the target is a family of unitaries $\{U_{\lambda}\}_{\lambda}$ living on $\HH_d = \Span\{\ket{j}\}_{j=0}^{d-1}$, indexed by labels $\lambda$ accumulated so far as the algorithm proceeds. We are then to implement $\mbf{U} = \sum_{\lambda} \ps{\lambda} \otimes U_{\lambda}$ with small error. Initially $d = 2^n$ and the target is the $n$-qubit unitary to be synthesized without labels. 

Suppressing $\lambda$, at each level with current dimension $d$, the recursion goes as follows: two preconditioners $P_{\mathrm{in}}, P_{\mathrm{out}}$ from sign gates, Hadamard gates and permutations turn $U$ to $V = P_{\mathrm{out}} U P_{\mathrm{in}}$. In block matrix form, $V = \left(\begin{smallmatrix} A & B\\ C & D \end{smallmatrix}\right)$, where the first $k = d - s$ coordinates serve as ports and the internal block $D$, acting on the last $s$ coordinates, is contractive. The target is passed to the child task by closing the internal coordinates of $V$ in a feedback loop \cite{261572}: for each $\mathbf{z}$ in a finite abelian group $G$, a diagonal unitary $Z(\mathbf{z})$ on the internal coordinates yields the unitary $F(\mathbf{z})=A+BZ(\mathbf{z})(\id -DZ(\mathbf{z}))^{-1}C$ on the ports, and the child's target is the controlled family $\mathbf F=\sum_{\mathbf{z}\in G}\ket{\mathbf{z}}\!\!\bra{\mathbf{z}}\otimes F(\mathbf{z})$. With the labels restored, $\mathbf{z}$ is appended to $\lambda$, and the label thus grows. What the child implements is thus the family $\{F_\lambda(\mathbf{z})\}_{(\lambda, \mbf{z})}$ indexed by $(\lambda,\mathbf{z})$. 

The circuit of the level is then $C_d=P_{\mathrm{out}}^\dagger U_R^\dagger C_kU_EP_{\mathrm{in}}^\dagger$, where an encoder $U_E$ spreads the input over a phase register and a port register, the recursive call $C_k$ applies $F(\mathbf{z})$ in each branch $\mathbf{z}$, and the decoder $U_R^\dagger$ decodes, which nearly recovers $V$ as the two encodings $E_L$ and $R_L$ are nearly isometric and nearly intertwine $V$ with $\mbf{F}$: $\mbf{F}E_L\approx R_LV$. These encodings are constructed by truncating the convergent Neumann series \cite{Conway2007} for the feedback input and output maps at length $L$. After $\bigo(n^3)$ levels, each with a single recursive call, it only suffices to implement controlled phase gates. We obtain the encoders $U_E$ and $U_R$ from block encodings of $E_L$ and $R_L$, respectively, in the Fourier basis of $G$ \cite{Terras1999}, where both maps are row-sparse, via oblivious amplitude amplification \cite{Brassard2002,QSVT} followed by a Fourier transform over $G$. Here the block encoding uses column preparations, preparations of uniform superpositions over candidate index sets, and a flag rotation. The oracle circuit is compiled from a skeleton of programmable slots, also known as uniformly controlled one-qubit gates \cite{PhysRevA.71.052330,PhysRevLett.93.130502,2011670.2011675}. Both are determined by $n$ and $\beta$ alone and have size polynomial in $n$ and $\beta$. The target enters only through the slot tables, which the Boolean oracle encodes as gate words approximating each table entry within $2^{-\kappa}$ by a phase-aware version of the Solovay-Kitaev theorem~\cite{KitaevShenVyalyi2002}.

This construction only shows that suitable tables, and hence a suitable Boolean oracle, exist. However, nothing guarantees their $\pspace$-computability. In fact, the oracles for all $n$ and $\beta$, collected into one language, can be undecidable, as they must be for a family of targets with uncomputable entries. Fortunately, for a family $W=\{W_n\}_n$ with $\pspace$-computable entries, this language can be chosen in $\pspace$ when the algorithm is applied to $W_n$. The proof follows the flow of data through the recursive construction: we first show, by a top-down induction over the levels, that the entries of every target can be computed on demand in polynomial space, then compute the gate tables of all quantum operations from these entries, and finally compute the code bits that the oracle returns.

The induction starts from the $\pspace$-computable entries of $W_n$. The inductive step begins by fixing the preconditioners of the current level from approximations of the entries of its target. The existence proof for good preconditioners (\cref{prop:contracting_internal_block}) uses up to $2^n$ independent random signs, which cannot be stored in polynomial space; we therefore replace them with $\bigo(n)$-wise independent signs generated from $\bigo(n^2)$-bit seeds \cite{ALON1986567,doi:10.1137/0215074}. This suffices to preserve the moments that we use to argue $D$ is contractive, and an exhaustive search over pairs of seeds plus testing the trace moment of $D$ finds us suitable preconditioners in polynomial space (see \cref{lemma:canonical_space_uniform_preconditioners}). This fixes the target $F(\mathbf{z})$ for the next level, and still the truncated Neumann series approximates blocks in $F(\mathbf{z})$ properly. Specifically, a perturbation bound shows that each level costs only $\mathcal O(n)$ extra bits of precision, so a depth-first evaluation that recomputes entries on demand, instead of storing them, computes the entries of the targets at all $\mathcal O(n^3)$ levels (including the phases at the bottom level) in polynomial space (see \cref{lemma:approximating_targets_at_all_levels}).

With these target entries approximated, we start tracking how the gate tables are eventually computed. The signs and permutations in the preconditioners follow directly from the seeds (and rows if $s = 1$) we select. In the encoders, the entries of the near-isometries are sums of products of entries of $V$ over paths of length at most the truncation length $L$ and are thus $\pspace$-computable as well. Tracking the candidate index sets that the block encodings use to uncompute the input follows from exact integer arithmetic alone (see \cref{lemma:computing_entries_and_candidate_sets}). The candidate-set uniform superposition and the flag rotations are implemented easily when the candidate index set is found. For the column preparations, we follow the polynomial-space state preparation of \cite{RY22} and round the preparation vectors to a dyadic grid, so that the prefix weights are exactly computable, zero weights can be detected exactly, and nonzero weights have a known lower bound. This yields the $\pspace$-computable preparation tables with an error $2^{-g_{\mathrm{prep}}}$ per preparation that accumulates at most linearly over the polynomially many occurrences of the preparations and their inverses in the quantum circuit (see \cref{lemma:pspace_computability_of_column_prep,lemma:filled_skeleton_error_bound}). All remaining slot tables are $\pspace$-computable from their explicit formulas.

Finally, for each table entry, an exhaustive search over pairs of polynomially long words with a finite-precision test finds a phase-preserving approximation in polynomial space, and the phase-aware Solovay-Kitaev theorem with sufficiently high precision guarantees that some pair would pass (see \cref{lemma:selecting_gate_words_in_polynomial_space}). 
The language $A_W$ encodes these words for all $n$ and $\beta$, and its decider recomputes any requested code bit along the aforementioned top-down chain in polynomial space, so $A_W \in \pspace$. With suitable choices of the truncation length $L$, the rounding precision $g_{\mathrm{prep}}$, and the compilation precision $\kappa$, the construction of \cref{thm:usp} yields an oracle circuit that has total error below $2^{-\beta-1}$ when instantiated with the language $A_W$ defined above. All three parameters, as well as the space needed to decide $A_W$, are guaranteed to be polynomial in $n$ and $\beta$.

\subsection{Discussion and open problems}
Our work pins down the precise relationship
between interactive proofs and polynomial-space computation
for unitary synthesis,
complementing the well-known characterizations
for decision problems \cite{JJUW11} and state synthesis \cite{RY22,MY23}.
For $\uqip\subseteq\upspace$,
our acceptance rigidity lemma reveals
a global consistency constraint that soundness imposes
on a prover's acceptance behavior across inputs.
We believe this structural consequence of soundness may be
of independent interest in the setting of verifying unknown
quantum computations.
In the reverse direction,
our method also connects unitary synthesis
with classical complexity on decision problems.
Specifically, we show that unitary families with $\pspace$-computable entries
admit efficient synthesis relative to a fixed $\pspace$ language.
This provides a $\pspace$-effective form of unitary synthesis,
addressing a question raised by Kretschmer and Tang
\cite{kretschmer2026unitarycomplexitypolynomialspace}
of whether, in their notation, $
    \mathsf{unitaryPSPACE}
$ equals $
    \mathsf{unitaryP}^{\pspace}$.
We list several open problems for future work.
\begin{itemize}[
    leftmargin=1.6em,
    labelsep=0.5em,
    topsep=4pt,
    itemsep=4pt,
    parsep=0pt
]
    \item Decisional quantum interactive proofs admit completeness–soundness gap amplification by repetition \cite{KW00}. In the unitary synthesis framework, however, this approach seems to break down: the verifier receives only a single copy of an unknown input state, and the no-cloning theorem prevents it from preparing the additional copies that independent repetitions would require. 
    Such a discrepancy motivates us to ask: does $\mathsf{unitaryQIP}$ admit completeness–soundness gap amplification in the single-copy setting?
    \item In traditional quantum interactive proofs, any protocol with polynomially many messages can be parallelized into one with only three messages, i.e., $\mathsf{QIP} = \mathsf{QIP}[3]$ \cite{KW00}. 
    For state synthesis, Rosenthal \cite{Ros24} proved an analogous result: $\mathsf{stateQIP} = \mathsf{stateQIP}[6]$.
    It is therefore natural to ask whether these results extend to unitary synthesis: is $\mathsf{unitaryQIP} = \mathsf{unitaryQIP}[\mathcal{O}(1)]$, or, more ambitiously, $\mathsf{unitaryQIP} = \mathsf{unitaryQIP}[3]$?
    \item In the average case, Bostanci, Efron, Metger, Poremba, Qian, and Yuen
    \cite{BEM+26} have shown that $\textsc{DistSuccinctUhlmann}_1$, the problem of transforming between succinct purifications of the same mixed state, is complete for both $\mathsf{avgUnitaryQIP}$ and $\mathsf{avgUnitaryPSPACE}$.
    By contrast, no complete problem is known for either of the worst-case classes $\mathsf{unitaryQIP}$ and $\mathsf{unitaryPSPACE}$. Finding complete problems for these complexity classes is therefore a natural open question.
\end{itemize}

\paragraph{Organization.}
The rest of the paper is organized as follows.
\Cref{sec:prelim} sets up notation and reviews the necessary background. \Cref{sec:uqip_in_upspace} proves $\uqip\subseteq\upspace$, and \cref{sec:upspace_in_uqip} proves the reverse inclusion $\upspace\subseteq\uqip$. Finally, \cref{sec:extensions} presents constant-round protocols for $\upspace$ given multiple copies of the input state, and shows that the existence of EFI pairs implies $\mathsf{BQP}\neq\pspace$.
The appendices contain the deferred proofs and an exposition of OpenAI's recently announced solution to the unitary synthesis problem.

\paragraph{Acknowledgments.}
The authors thank Penghui Yao for bringing this problem to their attention
and for his invaluable guidance and discussions throughout this work.
Part of this work was done when E.H. was visiting Nanjing University.
F.O. and M.Z. were supported by the National Natural Science Foundation of China (Grant Nos. 62332009 and 12347104), the Quantum Science and Technology-National Science and Technology Major Project (Grant No. 2021ZD0302901), the NSFC/RGC Joint Research Scheme (Grant No. 12461160276), the Natural Science Foundation of Jiangsu Province (No. BK20243060), the Fundamental and Interdisciplinary Disciplines Breakthrough Plan of the Ministry of Education of China (No. JYB2025XDXM118), the ``111 Center'' (No. B26023) and the Fundamental Research Funds for the Central Universities (Grant No. 2026300376).

\paragraph{AI disclosure.} OpenAI GPT-6 was used extensively throughout the preparation of this manuscript, including for mathematical formulation, checking proof arguments, typesetting, and language polishing. The idea behind our proof of $\upspace \subseteq \uqip$ originated in an early discussion among the authors that assumed a positive answer to the unitary synthesis problem. Prompted by the authors, GPT-6 helped sketch the argument that, for $\upspace$ families, the oracles in the unitary synthesis problem can be replaced by a $\pspace$ language. The authors have reviewed and verified the manuscript and take full responsibility for its content.

\section{Preliminaries}
\label{sec:prelim}
\paragraph{Notation.}
We use standard asymptotic notations $\Theta(\cdot), \mathcal{O}(\cdot), \Omega(\cdot)$, and $o(\cdot)$.
We write $\F_2$ for the field with two elements. For $x \in \C$, we write $\Re(x)$ and $\Im(x)$ for its real and imaginary parts, respectively.
We also write
$\poly(x_1, \dots, x_k) \coloneqq \bigcup_{n_1, \dots, n_k \geq 0} \bigo(x_1^{n_1} \cdots x_{k}^{n_k})$.
A function $\nu:\N\to[0,1]$ is \emph{negligible} if, for every positive polynomial $p$, we have $\nu(n)<1/p(n)$ for all sufficiently large $n$.
For $n\in\N$, we write $[n] \coloneqq \{1, 2, \dots, n\}$, and for a subset $S \subseteq [n]$, write $\mathbf{1}_S = (\delta_{j \in S})_{j=1}^n \in \Z^n$ as its indicator vector.
For a vector $\bfx$, $\|\bfx\| = \sqrt{\braket{\bfx , \bfx}}$ is the Euclidean norm.
For a Hilbert space $\HH$,
let $\linermap{\HH}$ denote the set of linear operators on $\HH$.
We write $\id_{\HH}$ for the identity operator on $\HH$,
or simply $\id_d$ when $\dim \HH = d$.
For $X \in \mathsf{L}(\HH)$,
we use either $X_{i, j}$ or $[X]_{i, j}$ to indicate its $(i, j)$ entry, use $\im(X)$ and $\ker(X)$ to denote its image and kernel, respectively,
and use $\|X\|_{\op} = \sqrt{\lambda_{\max}(X^\dagger X)}$ and $\|X\|_1 = \tr[\sqrt{X^\dagger X}]$ to denote its operator norm and trace norm, respectively.
The map $\ad_X: \rho \mapsto X \rho X^\dagger$ is abbreviated by calligraphic $\mathcal{X}(\cdot)$ if $X$ is unitary.

Normally, we assume that Turing machines approximately depict quantum objects in their dyadic rational forms (see, e.g., the proof of \cite[Theorem 2.4.4]{aaronson2016complexityquantumstatestransformations}). To index the entries of operators in a family, we use tuples $(1^n, i, j)$, encoded as strings in a fixed self-delimiting way, where the unary input $1^n$ specifies the dimension and $(i, j)$ the position of the entry. For a Boolean function $f\colon \{0, 1\}^{\ell} \to \{0, 1\}$, we write $O_f$ for the unitary on $\ell + 1$ qubits that maps $\ket{x, b}$ to $\ket{x, b \oplus f(x)}$. We use $\iota\colon \ket{x} \mapsto \ket{x} \ket{0}$ to denote the canonical isometry that appends the workspace qubits with ancillae initialized to $\ket{0}$, with their sizes clear from the text.

\paragraph{Quantum information theory.}

We follow the notation in \cite{TQI}.
For a pure state $\ket{\psi}$, we write $\psi\coloneqq\ketbra{\psi}$
for its density operator. We write $\ket{\mathsf{EPR}_n} = \frac{1}{\sqrt{2^n}} \sum_{x \in \{0, 1\}^n} \ket{x, x}$ as the $n$-qubit maximally entangled state in the computational basis.
The \emph{trace distance} between two quantum states $\rho$ and $\sigma$ is defined as
$$
    \td(\rho, \sigma) \coloneqq \frac{1}{2} \norm{\rho - \sigma}_1 \enspace.
$$
For density operators $\rho$ and $\tau$ on the same Hilbert space,
their \emph{fidelity} is defined as $F(\rho,\tau) \coloneqq \norm{\sqrt{\rho}\sqrt{\tau}}_1$.
The \emph{Fuchs--van de Graaf inequalities} relate the fidelity and trace distance:
$$
    1-F(\rho,\tau)
    \leq \td(\rho,\tau)
    \leq \sqrt{1-F(\rho,\tau)^2} \enspace.
$$
A \emph{quantum channel} is a completely positive, trace-preserving
linear map from operators on an input Hilbert space to operators
on an output Hilbert space.
The \emph{diamond distance} between two quantum channels $\mathcal{E}, \mathcal{F}: \mathsf{L}(\HH_\mathsf{A}) \to \mathsf{L}(\HH_\mathsf{B})$ is defined as
$$
\dd(\mathcal{E}, \mathcal{F}) = \sup_{\HH_\mathsf{R}}
\max_{\ket{\psi} \in \HH_{\mathsf{A}}\otimes \HH_{\mathsf{R}}} \td\left( (\mathcal{E} \otimes \mathcal{I}_\mathsf{R})(\psi), (\mathcal{F} \otimes \mathcal{I}_\mathsf{R})(\psi) \right) \enspace.
$$
Every quantum channel admits a Stinespring isometry.
\begin{definition}[Stinespring isometry]\label{def:stinespring}
    Let $\mathcal{C}$ be a quantum channel acting on a
    finite-dimensional Hilbert space $\mathcal{H}$.
    A \emph{Stinespring isometry} (or \emph{isometric extension})
    of $\mathcal{C}$ is an isometry
    $T:\mathcal{H}\to\mathcal{H}\otimes\mathcal{H}_A$
    satisfying
    $$
        \mathcal{C}(\rho)
        = \Tr_{\mathcal{H}_A}\!\left(T\rho T^\dagger\right)
    $$
    for every density operator $\rho$ on $\mathcal{H}$.
    Here, $\mathcal{H}_A$ is an ancillary Hilbert space.
\end{definition}

\paragraph{Quantum circuits.}
Without loss of generality,
we fix a finite universal gate set $\mathcal{G}$,
for example, $\{\hdm,\tgate,\cnot\}$, where
$$
\hdm = \frac{1}{\sqrt{2}} \begin{pmatrix}
    1 & 1 \\ 1 & -1
\end{pmatrix}, \quad \tgate = \begin{pmatrix}
    1 &0 \\ 0& \mathrm{e}^{\I \pi / 4}
\end{pmatrix}, \quad \cnot = \ps{0} \otimes \id + \ps{1} \otimes \begin{pmatrix}
    0& 1 \\ 1 &0
\end{pmatrix}.
$$
A \emph{quantum circuit} is a finite sequence of operations,
each of which is either the initialization of a new qubit in the state $\ket{0}$,
the partial trace over a qubit,
a unitary gate from $\mathcal{G}$,
or a measurement of a qubit in the computational basis.
A \emph{unitary circuit} is a quantum circuit
that contains only unitary gates.
The \emph{description} of a quantum circuit is
the sequence of its operations
together with a specification of the qubits
on which each operation acts.
The size of a quantum circuit is the number of operations in its description.
We say that a quantum circuit uses space $s$
if the number of qubits involved
at every step of the computation is at most $s$.
We identify a quantum circuit $\mathcal{C}$ with the quantum channel it implements
and use the same symbol $\mathcal{C}$ for both. An \emph{oracle circuit} $\mathcal{C}^{(\cdot)}$ is a quantum circuit that interleaves its own quantum operations with calls to an unspecified Boolean oracle. When the circuit is unitary, we denote it by an operator $C^{(\cdot)}$.

\begin{definition}[Time-uniform circuits]
    A family of quantum circuits $\{\mathcal{C}_n\}_{n \in \N}$
    is \emph{polynomial-size}
    if there exists a polynomial $p$ such that,
    for every $n \in \N$,
    the circuit $\mathcal{C}_n$ contains at most $p(n)$ operations.
    We say that $\{\mathcal{C}_n\}_{n \in \N}$ is \emph{time-uniform}
    if it is polynomial-size
    and there exists a deterministic Turing machine
    running in time polynomial in $n$ that,
    on input $1^n$,
    outputs the description of $\mathcal{C}_n$.
\end{definition}

\begin{definition}[Space-uniform circuits]
    A family of quantum circuits $\{\mathcal{C}_n\}_{n \in \N}$
    is \emph{polynomial-space}
    if there exists a polynomial $p$ such that
    for every $n \in \N$,
    the circuit $\mathcal{C}_n$ uses at most $p(n)$ space.
    We say that $\{\mathcal{C}_n\}_{n \in \N}$ is \emph{space-uniform}
    if it is polynomial-space
    and there exists a deterministic Turing machine using space polynomial in $n$ that, on input $(1^n,i)$,
    outputs the $i$-th operation of $\mathcal{C}_n$.
\end{definition}

We also consider the computational complexity for numbers.
\begin{definition}[$\pspace$-computable complex numbers]
\label{def:pspace_computable_complex_num}
    Let $I \subseteq \{0, 1\}^*$. A family $\{\alpha_x\}_{x \in I}$ of complex numbers is $\pspace$-computable if there exists a deterministic Turing machine, on input $x \in I$ and unary precision $1^h$, outputs a pair of dyadic rationals $(a, b)$ subject to $|a + \I b  - \alpha_x| \leq 2^{-h}$ 
    using $\poly(|x| + h)$ space. 
    We say that a unitary family $\{W_n\}_{n \in \N}$, where each $W_n$ is a $d_n \times d_n$ matrix, has $\pspace$-computable entries if the family $\{\braket{i | W_n | j}\}_{(1^n,i,j)}$ with $i,j \in [d_n]$
    is $\pspace$-computable\footnote{
    Our definition is compatible with \cite[Definition~2.3]{MY23}, which calls a family of matrices $\pspace$-computable if, for every polynomial $p$, a space-uniform machine approximates the $n$-th member to within $2^{-p(n)}$ in trace norm. By the triangle inequality, a $2^n \times 2^n$ matrix whose entries have modulus at most $2^{-h-2n}$ has trace norm at most $2^{2n} \cdot 2^{-h-2n} = 2^{-h}$. Hence, a family of unitaries with $\pspace$-computable entries in our sense is indeed $\pspace$-computable in the sense of \cite{MY23}. A similar equivalence argument is given in \cite[Theorem 6.17]{kretschmer2026unitarycomplexitypolynomialspace}.
    }.    
\end{definition}

\paragraph{Unitary synthesis problem.} Aaronson and Kuperberg \cite{4262757} proposed a fundamental question about unitary complexity, later formalized as the \textit{unitary synthesis problem} (USP) by Aaronson \cite{aaronson2016complexityquantumstatestransformations}. Formally, it asks the following:

\begin{problem}[Unitary synthesis problem, informal]
    Is there a universal $\poly(n)$-size oracle quantum circuit $\mathcal{C}^{(\cdot)}$ such that for every $n$-qubit unitary $U \in \U(2^n)$, there exists a corresponding Boolean function $f_U$ on $\poly(n)$ bits for which $\mathcal{C}^{f_U}$ implements $U$ to a constant error?
\end{problem}

Recently, OpenAI announced a positive answer to USP as part of its release of solutions to numerous open problems \cite{OpenAI2026UnitarySynthesis}. Their result serves as a key ingredient in our proof, and we demonstrate its proof with reinterpretation later in \cref{appendix:answer_of_USP}.

\paragraph{Quantum interactive proof systems}
The model of quantum interactive proof systems considered
in this work is formalized as follows.

A \emph{$k$-round quantum verifier} is a sequence of
quantum circuits $\mathcal{V}=(\mathcal{V}_i)_{i=1}^k$.
Each $\mathcal{V}_i$ acts on a pair of registers $(\mathsf{Z}, \mathsf{M})$,
and represents the action of the verifier
in the $i$-th round of interaction with the prover.
The register $\mathsf{Z}$ is the private register of the verifier
and consists of three sub-registers:
the \emph{work register} $\mathsf{Z}_\mathsf{work}$ for intermediate computations,
the \emph{flag register} $\mathsf{Z}_\mathsf{flag}$ recording the  accept/reject decision,
and the \emph{output register} $\mathsf{Z}_\mathsf{out}$ storing the output state, if any.
The register $\mathsf{M}$ is the \emph{message register} for communication with the prover.
The size of $\mathcal{V}$ is the sum of the sizes of the circuits $\mathcal{V}_i$.

A \emph{quantum prover} is modeled as a unitary channel $\mathcal{P}$ acting on a pair of registers $(\mathsf{M}, \mathsf{Y})$,
where $\mathsf{Y}$ is the private register of the prover.
We note that this unitary model of the prover is without loss of generality,
since the prover may use an arbitrarily large private register to purify its operations and to keep track of the current round.

Let $\ket{\psi}$ be a quantum state on at most as many qubits as $\mathsf{Z}_{\mathsf{work}}$.
The \emph{interaction} between a $k$-round verifier $\mathcal{V}$ and a prover $\mathcal{P}$ on input $\ket{\psi}$,
denoted by $\mathcal{V}(\ket{\psi}) \interact \mathcal{P}$, is defined as follows.
The system is initialized in the state
$$
  \rho_0 = \ketbra{\psi, 0\cdots 0}_{\mathsf{Z}_{\mathsf{work}}}
  \otimes \ketbra{0\cdots 0}_{\mathsf{Z}_{\mathsf{flag}}\mathsf{Z}_{\mathsf{out}}\mathsf{M}\mathsf{Y}} \enspace,
$$
after which the prover and the verifier act alternately, starting with the prover:
for $i = 1, \dots, 2k$,
$$
  \rho_i =
  \begin{cases}
    (\mathcal{I}_{\mathsf{Z}} \otimes \mathcal{P})(\rho_{i-1})      & i \in 2\N - 1, \\
    (\mathcal{V}_{i/2} \otimes \mathcal{I}_{\mathsf{Y}})(\rho_{i-1}) & i \in 2 \N.
  \end{cases}
$$
The interaction $\mathcal{V}(\ket{\psi}) \interact \mathcal{P}$ \emph{accepts} (resp.\ \emph{rejects})
if measuring the register $\mathsf{Z}_{\mathsf{flag}}$ of $\rho_{2k}$
in the computational basis yields the outcome $1$ (resp.\ $0$).
The verifier thus accepts with probability
$\tr\bigl[ \ketbra{1}_{\mathsf{Z}_{\mathsf{flag}}} \cdot \rho_{2k} \bigr]$.
The \emph{output of $\mathcal{V}(\ket{\psi}) \interact \mathcal{P}$ conditioned on accepting} is the state
\begin{equation}
\label{eqn:conditioned_acceptance_state}
  \frac{\Tr_{\mathsf{Z}\mathsf{M}\mathsf{Y} \setminus \mathsf{Z}_{\mathsf{out}}}\!
        \bigl[ \ketbra{1}_{\mathsf{Z}_{\mathsf{flag}}} \cdot \rho_{2k} \bigr]}
       {\Tr \bigl[ \ketbra{1}_{\mathsf{Z}_{\mathsf{flag}}} \cdot \rho_{2k} \bigr]} \enspace,
\end{equation}
which is the reduced state of $\rho_{2k}$ on $\mathsf{Z}_{\mathsf{out}}$
after post-selecting on acceptance in the flag register $\mathsf{Z}_{\mathsf{flag}}$.
If the acceptance probability is $0$, the output is left undefined.

\begin{definition}[Time-uniform quantum verifiers]\label{def:time-uniform-verifier}
    A family of quantum verifiers $\{\mathcal{V}_n\}_{n \in \N}$,
    where $\mathcal{V}_n = (\mathcal{V}_{n,i})_{i=1}^{k(n)}$ is a $k(n)$-round quantum verifier,
    is \emph{polynomial-size}
    if there exists a polynomial $p$ such that,
    for every $n \in \N$,
    both the number of rounds $k(n)$ and the size of $\mathcal{V}_n$ are at most $p(n)$.
    We say that $\{\mathcal{V}_n\}_{n \in \N}$ is \emph{time-uniform}
    if it is polynomial-size
    and there exists a deterministic Turing machine
    running in time polynomial in $n$ that,
    on input $1^n$,
    outputs the descriptions of $\mathcal{V}_{n,1}, \dots, \mathcal{V}_{n,k(n)}$.
\end{definition}

\paragraph{Complexity classes.}
We use the following definitions of complexity classes,
adapted from \cite{RY22,MY23,BEM+26}.

\begin{definition}[$\spspace$]\label{def:statepspace}
    Let $\delta: \N \to [0,1]$.
    The class $\spspace_\delta$ consists of all sequences
    $\{\rho_n\}_{n \in \N}$ of $n$-qubit density operators
    for which there exists a space-uniform family of quantum circuits
    $\{\mathcal{C}_n\}_{n \in \N}$ such that,
    for all sufficiently large $n$,
    $\mathcal{C}_n$ takes no input and outputs a state $\sigma_n$ satisfying
    $$
        \td\!\left(\sigma_n,\rho_n\right)
        \leq \delta(n) \enspace.
    $$
\end{definition}

\begin{definition}[$\upspace$]\label{def:upspace_ubqp}
    Let $\delta: \N \to [0, 1]$.
    The class $\upspace_\delta$ is the set of
    all sequences of unitary operators $\{U_n\}_{n \in \N}$
    such that each $U_n$ acts on $n$ qubits,
    and there exists a space-uniform family of quantum circuits
    $\{\mathcal{C}_n\}_{n \in \N}$ satisfying, for all sufficiently large $n$,
    $$
     \sup_{\ket{\psi} \in \mathbb{C}^{2^n}\, :\, \norm{\ket{\psi}}=1 }  \td\!\left(\mathcal{C}_n(\ketbra{\psi}), U_n \ketbra{\psi} U_n^{\dagger} \right) \leq \delta(n) \enspace.
    $$
\end{definition}

\begin{definition}[$\sqip$]\label{def:stateqip}
    Let $c,s,\delta: \N \to [0,1]$ satisfy
    $1-c(n)>s(n)$ for all $n \in \N$.
    The class $\sqip_{c,s,\delta}$ consists of all sequences
    $\{\rho_n\}_{n \in \N}$ of $n$-qubit density operators
    for which there exists a time-uniform family of quantum verifiers
    $\{\mathcal{V}_n\}_{n \in \N}$ with $n$-qubit output registers
    such that, for all sufficiently large $n$,
    \begin{itemize}
        \item \emph{Completeness.}
        There exists a prover $\mathcal{P}_n^*$ such that
        $$
            \pr\!\left[
                \mathcal{V}_n \interact \mathcal{P}_n^*
                \ \text{accepts}
            \right]
            \geq 1-c(n) \enspace.
        $$

        \item \emph{Soundness.}
        For every prover $\mathcal{P}_n$,
        $$
            \pr\!\left[
                \mathcal{V}_n \interact \mathcal{P}_n
                \ \text{accepts}
            \right] > s(n)
            \quad \Longrightarrow \quad
            \td\!\left(\sigma_{\mathcal{P}_n},\rho_n\right)
            \leq \delta(n) \enspace,
        $$
        where $\sigma_{\mathcal{P}_n}$ is the output of
        $\mathcal{V}_n \interact \mathcal{P}_n$
        conditioned on accepting.
    \end{itemize}
\end{definition}

\begin{definition}[$\uqip$]\label{def:para-uqip}
    Let $c,s: \N \to [0, 1]$ be the \emph{completeness error} and \emph{soundness threshold}
    and suppose that $1-c(n) > s(n)$ for~all $n \in \N$.
    For $\delta: \N \to [0, 1]$ and $\ell:\N \to \N$, the class $\uqip_{c,s,\delta}[\ell]$ is the set of
    all sequences of unitary operators $\{U_n\}_{n \in \N}$
    such that each $U_n$ acts on $n$ qubits,
    and there exists a time-uniform family
    of $\ell(n)$-round quantum verifiers $\{\mathcal{V}_n\}_{n \in \N}$
    such that for sufficiently large $n$,
    \begin{itemize}
        \item \emph{Completeness.}
        There exists a prover $\mathcal{P}_n^*$ such that
        for all $n$-qubit states $\ket{\psi}$,
        $$
            \pr\left[\mathcal{V}_n(\ket{\psi}) \interact \mathcal{P}_n^*\ \text{accepts}\right] \geq 1-c(n) \enspace.
        $$
        \item \emph{Soundness.} 
        For every prover $\mathcal{P}_n$ and every $n$-qubit state $\ket{\psi}$,
        \begin{equation*}
            \pr\left[\mathcal{V}_n(\ket{\psi}) \interact \mathcal{P}_n\ \text{accepts}\right] > s(n)
            \quad \Longrightarrow \quad
            \td\!\left(\rho_{\mathcal{P}_n,\psi}, U_n \ketbra{\psi} U_n^{\dagger} \right) \leq \delta(n) \enspace,
        \end{equation*}
        where $\rho_{\mathcal{P}_n,\psi}$ is the output of
        $\mathcal{V}_n(\ket{\psi}) \interact \mathcal{P}_n$ conditioned on accepting.
    \end{itemize}
    Unless otherwise stated, $\uqip_{c,s,\delta}$ denotes the class with polynomially many rounds.
\end{definition}

\begin{definition}[Multi-copy $\uqip$]
    For $r: \N \to \N$,
    the class $\uqip_{c,s,\delta}^{(r)}$ is defined in the same way as
    $\uqip_{c,s,\delta}$
    except that the verifier $\mathcal{V}_n$ is run on input $\ket{\psi}^{\otimes r(n)}$. In particular, $\uqip_{c,s,\delta}^{(1)} = \uqip_{c,s,\delta}$.
\end{definition}

\section{A polynomial-space simulation of interactive unitary synthesis}
\label{sec:uqip_in_upspace}
In this section, we prove \cref{thm:uqip-in-upspace},
which states that, to within inverse-polynomial error, any unitary that can be synthesized via an interactive protocol can also be implemented by a polynomial-space quantum algorithm. This is the unitary analogue of the state-synthesis inclusion $\mathsf{stateQIP} \subseteq \mathsf{statePSPACE}$~\cite[Theorem 1.2]{MY23}.

\uqipinupspace*

To prove this theorem,
we need to turn the interactive protocol into a polynomial-space algorithm.
Since the verifier is already a polynomial-space quantum circuit,
it therefore suffices to give a polynomial-space implementation of the \emph{honest} prover.
Our starting point is the following result of Metger and Yuen~\cite{MY23} showing that,
for verifiers that receive no input state,
a near-optimal prover can be implemented in polynomial space.\footnote{The original theorem is stated in a slightly different form, but the formulation here follows from its proof.}

\begin{theorem}[{\cite[Theorem 7.1]{MY23}}, reformulated]
    \label{thm:metger-yuen}
    Let $\{\mathcal{V}_n\}_{n \in \N}$ be a time-uniform
    family of quantum verifiers that receive no input state.
    Let $\omega_n^*$ be the maximum acceptance probability of $\mathcal{V}_n$ over all provers that interact with $\mathcal{V}_n$.
    Then, for every polynomial $q(n)$,
    there exists a space-uniform family of unitary circuits
    $\{\mathcal{P}_n\}_{n \in \N}$
    such that the prover $\mathcal{P}_n$ makes $\mathcal{V}_n$ accept
    with probability at least $\omega_n^* - 1/q(n)$.
\end{theorem}

However, we are not able to use this result directly.
The reason is that the verifier in a $\uqip$ protocol receives
an arbitrary quantum state $\ket{\psi}$ as input,
whereas \cref{thm:metger-yuen} only applies to verifiers that receive no input state (or start with the all zero state).

A natural workaround is to fix the input state.
More specifically,
let $\mathcal{V}_n$ be the verifier in a $\uqip_{c,s,\delta}$ protocol for synthesizing a unitary $U_n$.
We define $\mathcal{V}_n^{(0)}$ to be the verifier
that runs $\mathcal{V}_n$ on the all-zero input state $\ket{0}^{\otimes n}$.
Then $\mathcal{V}_n^{(0)}$
is a verifier that receive no input state,
and, by the completeness of $\mathcal{V}_n$,
the honest prover $\mathcal{P}_n^*$
makes $\mathcal{V}_n^{(0)}$ accept with probability at least $1-c(n)$.
Applying \cref{thm:metger-yuen} with $q(n)=1/\eps(n)$,
we obtain a polynomial-space prover
$\widehat{\mathcal{P}}_n$ that makes $\mathcal{V}_n^{(0)}$ accept with probability at least $1-c(n)-\eps(n)>s(n)$. 
By soundness, the output of $\mathcal{V}_n^{(0)}\interact\widehat{\mathcal{P}}_n$
(or equivalently, $\mathcal{V}_n(\ket{0}^{\otimes n})\interact\widehat{\mathcal{P}}_n$)
conditioned on acceptance is close to $U_n\ket{0}^{\otimes n}$.
In other words,
we can synthesize $U_n$ on the fixed input state $\ket{0}^{\otimes n}$
in polynomial space.
The remaining question is how to extend this polynomial-space synthesis of
$U_n$ from $\ket{0}^{\otimes n}$ to arbitrary input states.
The main obstacle is that $\widehat{\mathcal{P}}_n$
is constructed for a specific input,
and \cref{thm:metger-yuen} provides no guarantee on how it behaves on other inputs.

Our solution is simply to
use the same prover on every input.
That is, for every $\ket{\psi}$,
we run $\mathcal{V}_n(\ket{\psi})\interact\widehat{\mathcal{P}}_n$.
The key observation
(formalized in \cref{lem:soundness-limits-acceptance-variation})
is that, by the soundness of $\mathcal{V}_n$,
the probability that $\mathcal{V}_n(\ket{\psi})\interact\widehat{\mathcal{P}}_n$
accepts cannot be much smaller than that of
$\mathcal{V}_n(\ket{0}^{\otimes n})\interact\widehat{\mathcal{P}}_n$.
Consequently, $\widehat{\mathcal{P}}_n$ is accepted with probability greater
than $s(n)$ on every input, and soundness, applied again,
implies that the output conditioned on acceptance is close to $U_n\ket{\psi}$.

In what follows,
we formalize the above key observation
as \cref{lem:soundness-limits-acceptance-variation}
and use it to prove \cref{thm:uqip-in-upspace}
in \cref{sec:proof-of-uqip-in-upspace}.
The proof of \cref{lem:soundness-limits-acceptance-variation}
is deferred to \cref{sec:proof-of-soundness-limits-acceptance-variation}.

\subsection{One prover for all inputs}
\label{sec:proof-of-uqip-in-upspace}

Let $\HH=\C^{2^n}$.
Since $\widehat{\mathcal{P}}_n$ does not depend on the input state,
the accepting branch of
$\mathcal{V}_n(\cdot)\interact\widehat{\mathcal{P}}_n$,
namely the map that sends an input state
to the (unnormalized) output state
in $\mathsf{Z}_\mathsf{out}$
conditioned on the event that $\mathcal{V}_n$ accepts,
is a completely positive and trace-nonincreasing linear map
$\Phi:\linermap{\HH}\to\linermap{\HH}$.
For any input state $\ket{\psi}$, the associated acceptance probability is
$$
    \tr[\Phi(\ketbra{\psi})] = \bra{\psi}\Phi^*(\id_{\HH})\ket{\psi} \enspace,
$$
where $\Phi^*$ is the adjoint of $\Phi$.
Hence, the largest and smallest eigenvalues of $\Phi^*(\id_{\HH})$
are exactly the maximum and minimum acceptance probabilities of
$\mathcal{V}_n(\cdot)\interact\widehat{\mathcal{P}}_n$ over all input states.
Moreover, since the soundness of $\mathcal{V}_n$ holds for every prover,
including $\widehat{\mathcal{P}}_n$,
the map $\Phi$ inherits it:
whenever an input is accepted with probability greater than $s(n)$,
the normalized accepted output is $\delta(n)$-close to the ideal output
(see \cref{eq:soundness-condition} below).
We show that this condition forces the largest and smallest eigenvalues
of $\Phi^*(\id_{\HH})$ to be close,
that is, the acceptance probability of
$\mathcal{V}_n(\cdot)\interact\widehat{\mathcal{P}}_n$
cannot vary much across inputs.

\begin{lemma}
\label{lem:soundness-limits-acceptance-variation}
    Let $s,\delta\in[0,1]$,
    $\Phi:\linermap{\HH}\to\linermap{\HH}$
    be a completely positive and trace-nonincreasing linear map,
    and $U$ be a unitary on $\HH$.
    Suppose that, for every pure state $\ket{\psi}$ in $\HH$,
    \begin{equation}\label{eq:soundness-condition}
            \tr [ \Phi({\psi}) ] > s
            \quad \Longrightarrow \quad
            \td\!\left( \frac{\Phi({\psi})}{\tr [ \Phi({\psi}) ]}, U {\psi} U^{\dagger} \right) \leq \delta \enspace.
    \end{equation}
    Denote by $\lambda_{\mathrm{max}}$ and $\lambda_{\mathrm{min}}$
    the largest and smallest eigenvalues of $\Phi^*(\id_{\HH})$, respectively.
    Then, for every $t\in(0,1)$ satisfying $t\cdot\lambda_{\mathrm  {max}}>s$,
    it holds that
    $$
    	\lambda_{\mathrm{max}} - \lambda_{\mathrm{min}}
        \leq
        \frac{\delta(1+t)}{t(1-t)} \cdot\lambda_{\mathrm{max}} \enspace.
    $$
\end{lemma}

Intuitively, consider the superpositions
$\ket{\psi_\pm} = \sqrt{t}\ket{\psi_1} \pm \sqrt{1-t}\ket{\psi_2}$,
where $\ket{\psi_1}$ and $\ket{\psi_2}$ are the inputs
accepted with probabilities $\lambda_{\mathrm{max}}$ and $\lambda_{\mathrm{min}}$, respectively.
If $\lambda_{\mathrm{max}}$ is much larger than $\lambda_{\mathrm{min}}$,
then the $\ket{\psi_1}$ component is far more likely to be accepted,
so the accepted output would have weight noticeably larger than $t$ on $U\ket{\psi_1}$.
However, soundness forces this output to be close to $U\ket{\psi_\pm}$,
whose weight on $U\ket{\psi_1}$ is exactly $t$.
Hence $\lambda_{\mathrm{min}}$ must be close to $\lambda_{\mathrm{max}}$.
We defer the formal proof to
\cref{sec:proof-of-soundness-limits-acceptance-variation}
and now use it to prove \cref{thm:uqip-in-upspace}.

\begin{proof}[Proof of \cref{thm:uqip-in-upspace}]
    Let $\{U_n\}_{n \in \N} \in \uqip_{c,s,\delta}$
    and let $\{\mathcal{V}_n\}_{n \in \N}$ be the corresponding time-uniform family of verifiers.
    The completeness of $\mathcal{V}_n$ implies that
    there exists an honest prover $\mathcal{P}_n^*$ such that
    for all $n$-qubit states $\ket{\psi}$,
    $$
        \pr\left[\mathcal{V}_n(\ket{\psi}) \interact \mathcal{P}_n^*\ \text{accepts}\right] \geq 1-c(n)\enspace.
    $$
    Now we construct a fixed-input verifier $\mathcal{V}_n^{(0)}$ that runs $\mathcal{V}_n$ on state $\ket{0}^{\otimes n}$.
    Then $\{\mathcal{V}_n^{(0)}\}_{n \in \N}$ is a time-uniform family of verifiers that receive no input state,
    and the honest prover $\mathcal{P}_n^*$ makes $\mathcal{V}_n^{(0)}$ accept with probability at least $1-c(n)$.
    By \cref{thm:metger-yuen} with $q(n) = 1/\eps(n)$,
    there exists a space-uniform family of provers $\{\widehat{\mathcal{P}}_n\}_{n \in \N}$ such that
    \begin{equation}\label{eq:uqip-in-upspace-1}
        \pr\left[\mathcal{V}_n(\ket{0}^{\otimes n}) \interact \widehat{\mathcal{P}}_n\ \text{accepts}\right] \geq 1-c(n)-\eps(n)\enspace.
    \end{equation}
    Our synthesis algorithm $\mathcal{C}_n$ for $U_n$ then works as follows:
    on input $\ket{\psi}$,
    it simulates the interaction $\mathcal{V}_n(\ket{\psi})\interact\widehat{\mathcal{P}}_n$;
    if $\mathcal{V}_n$ accepts, it outputs the state in $\mathsf{Z}_{\mathsf{out}}$,
    and otherwise, it outputs $\ket{0}^{\otimes n}$.
    Since $\{\mathcal{V}_n\}_{n\in\N}$ is time-uniform and
    $\{\widehat{\mathcal{P}}_n\}_{n\in\N}$ is space-uniform,
    $\{\mathcal{C}_n\}_{n\in\N}$ is space-uniform.
    It remains to show that $\mathcal{C}_n$ synthesizes $U_n$ with error at most
    $c(n)+\eps(n)+7\delta(n)$.

    Let $\HH = \mathbb{C}^{2^n}$
    and $\Phi_n : \linermap{\HH} \to \linermap{\HH}$ be the completely positive and trace-nonincreasing map
    corresponding to the accepting branch of $\mathcal{V}_n(\cdot)\interact\widehat{\mathcal{P}}_n$.
    Denote by $\lambda_{\mathrm{max}}$ and $\lambda_{\mathrm{min}}$ the largest and smallest eigenvalues of $\Phi_n^*(\id_{\HH})$, respectively.
    Since $\lambda_{\mathrm{max}}$ is the maximum acceptance probability over all input states,
    the lower bound in \cref{eq:uqip-in-upspace-1} for the input $\ket{0}^{\otimes n}$ implies that
    \begin{equation}\label{eq:uqip-in-upspace-2}
        \lambda_{\mathrm{max}}
        = \sup_{\ket{\psi}: \|\ket{\psi}\| = 1} 
        \bra{\psi} \Phi_n^*(\id_{\HH}) \ket{\psi} 
        = \sup_{\ket{\psi}: \|\ket{\psi}\| = 1} \tr[\Phi_n({\psi})]
        \geq 1-c(n)-\eps(n) \enspace.
    \end{equation}
    Let $t = (1+c(n)+s(n))/2$.
    Since $\lambda_{\mathrm{max}}= \sup_{\ket{\psi}: \|\ket{\psi}\| = 1} \tr[\Phi_n({\psi})]\leq 1$, by \cref{eq:uqip-in-upspace-2}, we have
    \begin{align*}
        t\cdot\lambda_{\mathrm{max}}
        \geq \lambda_{\mathrm{max}}-(1-t)
        &\geq 1-c(n)-\eps(n)-\frac{1-c(n)-s(n)}{2} \\
        &\geq 1-c(n) - \frac{3(1-c(n)-s(n))}{4} \\
        & = s(n) + \frac{1-c(n)-s(n)}{4} > s(n) \enspace,
    \end{align*}
    where the third inequality follows from $\eps(n)\leq(1-c(n)-s(n))/4$.
    Moreover, as discussed before \cref{lem:soundness-limits-acceptance-variation},
    $\Phi_n$ satisfies \cref{eq:soundness-condition}
    Therefore, we can apply \cref{lem:soundness-limits-acceptance-variation} with this $t$
    and obtain
    $$
    	\lambda_{\mathrm{max}} - \lambda_{\mathrm{min}}
        \leq
        \frac{\delta(n)(1+t)}{t(1-t)} \cdot\lambda_{\mathrm{max}} 
        \leq \frac{6\delta(n)}{1-c(n)-s(n)}\cdot\lambda_{\mathrm{max}}
        \leq \frac{3(1-c(n)-s(n))}{4}\cdot\lambda_{\mathrm{max}} \enspace,
    $$
    where the second inequality uses
    $(1+t)/t\leq 3$ since $t\geq 1/2$,
    and the last inequality follows from $\delta(n)\leq(1-c(n)-s(n))^2/8$.
    As a result, we have
    \begin{align*}
        \lambda_{\mathrm{min}}
        &\geq
        \left( 1 - \frac{3(1-c(n)-s(n))}{4} \right) \cdot\lambda_{\mathrm{max}}
        \geq
        \left( 1 - \frac{3(1-c(n)-s(n))}{4} \right) \cdot (1-c(n)-\eps(n))\\
        &\geq 
        \left( 1 - \frac{3(1-c(n)-s(n))}{4} \right) \cdot
        \left( s(n) +  \frac{3(1-c(n)-s(n))}{4}  \right) \\
        & = s(n) + \frac{3c(n)(1-c(n)-s(n))}{4}
        + \frac{3(1-c(n)-s(n))^2}{16}
        > s(n) \enspace,
    \end{align*}
    where the second inequality uses \cref{eq:uqip-in-upspace-2},
    the third inequality uses $\eps(n)\leq(1-c(n)-s(n))/4$.
    Therefore, for every state $\ket{\psi}\in\HH$,
    we have $\tr[\Phi_n(\psi)]=\bra{\psi}\Phi_n^*(\id_{\HH})\ket{\psi}\geq\lambda_{\mathrm{min}}>s(n)$,
    and the soundness of $\mathcal{V}_n$ hence implies
    \begin{equation}\label{eq:uqip-in-upspace-3}
        \td\!\left(\frac{\Phi_n(\psi)}{\tr[\Phi_n(\psi)]},U_n\psi U_n^{\dagger}\right)\leq\delta(n) \enspace.
    \end{equation}
    Furthermore,
    \cref{eq:uqip-in-upspace-3} implies that
    the soundness condition in \cref{eq:soundness-condition}
    is also satisfied with $s = 0$.
    Since $\lambda_{\mathrm{max}}>0$,
    applying \cref{lem:soundness-limits-acceptance-variation}
    with $s = 0$ and $t=1/2$ gives
    $\lambda_{\mathrm{max}}-\lambda_{\mathrm{min}}\leq 6\delta(n)$.
    Combining this with \cref{eq:uqip-in-upspace-2},
    we have that for every state $\ket{\psi}\in\HH$,
    \begin{equation}\label{eq:uqip-in-upspace-4}
        \tr[\Phi_n(\psi)]
        \geq \lambda_{\mathrm{min}} 
        \geq 1-c(n)-\eps(n)-6\delta(n) \enspace.
    \end{equation}

    Finally, by construction,
    the quantum channel implemented by $\mathcal{C}_n$ is
    $$
        \mathcal{C}_n(\rho)
        = \Phi_n(\rho)
        + (\tr[\rho] - \tr[ \Phi_n(\rho)  ]) \cdot\ps{0^n} \enspace.
    $$
    For every state $\ket{\psi}\in\HH$, 
    by the triangle inequality and
    \cref{eq:uqip-in-upspace-3,eq:uqip-in-upspace-4},
    we have
    \begin{align*}
        \td\!\left(\mathcal{C}_n(\psi),U_n\psi U_n^{\dagger}\right) &\leq
        \tr[\Phi_n(\psi)]\cdot\td\!\left(\frac{\Phi_n(\psi)}{\tr[\Phi_n(\psi)]},U_n\psi U_n^{\dagger}\right)
        \\
        &\qquad + (1-\tr[\Phi_n(\psi)])\cdot \td\!\left(\ps{0^n},U_n\psi U_n^{\dagger}\right) \\
        &\quad \leq
        c(n)+\eps(n)+7\delta(n) \enspace.
    \end{align*}
    Therefore, we conclude that $\{U_n\}_{n\in\N}\in\upspace_{c+\eps+7\delta}$.
\end{proof}

\subsection{Soundness limits acceptance variation}
\label{sec:proof-of-soundness-limits-acceptance-variation}

We now give the proof of \cref{lem:soundness-limits-acceptance-variation}.

\begin{proof}[Proof of \cref{lem:soundness-limits-acceptance-variation}]
    We assume $ \lambda_{\mathrm{max}} \neq \lambda_{\mathrm{min}} $,
    since otherwise the lemma holds trivially.
    Fix $t\in(0,1)$ such that $t\cdot\lambda_{\mathrm{max}}>s$
    and let $\ket{\psi_1}$ and $\ket{\psi_2}$ be the eigenvectors of $\Phi^*(\id_{\HH})$ corresponding to $\lambda_{\mathrm{max}}$ and $\lambda_{\mathrm{min}}$, respectively.
    Since $\tr[\Phi({\psi_1})] = \tr[ \Phi^*(\id_{\HH})\cdot{\psi_1} ] = \lambda_{\mathrm{max}} > s$,
    we have by \cref{eq:soundness-condition},
    $$
        \td\!\left( \frac{\Phi({\psi_1})}{\tr[\Phi({\psi_1})]}, U {\psi_1} U^{\dagger} \right) \leq \delta \enspace.
    $$
    Therefore,
    \begin{align*}
        \tr\!\left[ U {\psi_1} U^{\dagger} \cdot \frac{\Phi({\psi_1})}{\tr[\Phi({\psi_1})]} \right] 
        & = 
        1 - \tr\!\left[ U {\psi_1} U^{\dagger}\cdot \left( U {\psi_1} U^{\dagger} - \frac{\Phi({\psi_1})}{\tr[\Phi({\psi_1})]} \right) \right] \\
        &\geq
        1 - \td\!\left( \frac{\Phi({\psi_1})}{\tr[\Phi({\psi_1})]}, U {\psi_1} U^{\dagger} \right)  \geq 1 - \delta \enspace.
    \end{align*}
    Rescaling by $\tr[\Phi({\psi_1})]$, we have
    \begin{equation} \label{eq:soundness-limits-acceptance-variation-1}
        \tr\!\left[ U {\psi_1} U^{\dagger} \cdot \Phi({\psi_1}) \right] 
        \geq (1 - \delta) \cdot \tr[\Phi({\psi_1})] = (1 - \delta) \cdot \lambda_{\mathrm{max}} \enspace.
    \end{equation}
    Now define two pure states $\ket{\psi_+}$ and $\ket{\psi_-}$ as
    $$
        \ket{\psi_\pm} \coloneqq \sqrt{t}\ket{\psi_1} \pm \sqrt{1-t}\ket{\psi_2} \enspace.
    $$
    Since $\tr[\Phi({\psi_+})] = \tr[ \Phi^*(\id_{\HH}) \cdot {\psi_+} ] = t\cdot\lambda_{\mathrm{max}} + (1-t)\cdot\lambda_{\mathrm{min}} > s$,
    by \cref{eq:soundness-condition},
    $$
        \td\!\left( \frac{\Phi({\psi_+})}{\tr[\Phi({\psi_+})]}, U {\psi_+} U^{\dagger} \right) \leq \delta \enspace. 
    $$
    Therefore, we have
    \begin{align*}
        \tr\!\left[ U {\psi_1} U^{\dagger} \cdot \frac{\Phi({\psi_+})}{\tr[\Phi({\psi_+})]} \right]  
        &  =
        \left| \braket{ \psi_1 | \psi_+} \right|^2
        + \tr\!\left[ U {\psi_1} U^{\dagger}\cdot \left(\frac{\Phi({\psi_+})}{\tr[\Phi({\psi_+})]} - U {\psi_+} U^{\dagger} \right) \right] \\
        & \leq
        t + \td\!\left( \frac{\Phi({\psi_+})}{\tr[\Phi({\psi_+})]}, U {\psi_+} U^{\dagger} \right)  \leq t + \delta \enspace.
    \end{align*}
    Rescaling by $\tr[\Phi({\psi_+})]$, we have
    \begin{equation} \label{eq:soundness-limits-acceptance-variation-2}
        \tr\!\left[ U {\psi_1} U^{\dagger} \cdot \Phi({\psi_+}) \right]
        \leq (t + \delta) \cdot \left( t\cdot\lambda_{\mathrm{max}} + (1-t)\cdot\lambda_{\mathrm{min}} \right) \enspace.
    \end{equation}
    Similarly, we have
    \begin{equation} \label{eq:soundness-limits-acceptance-variation-3}
        \tr\!\left[ U {\psi_1} U^{\dagger} \cdot \Phi({\psi_-}) \right]
        \leq (t + \delta) \cdot \left( t\cdot\lambda_{\mathrm{max}} + (1-t)\cdot\lambda_{\mathrm{min}} \right) \enspace.
    \end{equation}
    Note that $\Phi({\psi_+} + {\psi_-}) = 2t\cdot\Phi({\psi_1}) + 2(1-t)\cdot\Phi({\psi_2})$.
    \cref{eq:soundness-limits-acceptance-variation-1} then gives
    \begin{equation} \label{eq:soundness-limits-acceptance-variation-4}
        \tr\left[ U {\psi_1} U^{\dagger} \cdot \Phi({\psi_+} + {\psi_-}) \right]
        \geq 2t \cdot (1 - \delta) \cdot \lambda_{\mathrm{max}} \enspace.
    \end{equation}
    On the other hand, combining \cref{eq:soundness-limits-acceptance-variation-2,eq:soundness-limits-acceptance-variation-3},
    we have
    \begin{equation} \label{eq:soundness-limits-acceptance-variation-5}
        \tr\left[ U {\psi_1} U^{\dagger} \cdot \Phi({\psi_+} + {\psi_-}) \right]
        \leq 2(t + \delta) \cdot \left( t\cdot\lambda_{\mathrm{max}} + (1-t)\cdot\lambda_{\mathrm{min}} \right) \enspace.
    \end{equation}
    Combining \cref{eq:soundness-limits-acceptance-variation-4,eq:soundness-limits-acceptance-variation-5}, we have
    $$
        2t \cdot (1 - \delta) \cdot \lambda_{\mathrm{max}} \leq 2(t + \delta) \cdot \left( t\cdot\lambda_{\mathrm{max}} + (1-t)\cdot\lambda_{\mathrm{min}} \right) \enspace.
    $$
    Rearranging the terms, we obtain
    \[
        \lambda_{\mathrm{max}} - \lambda_{\mathrm{min}}
        \leq
        \frac{\delta(1+t)}{(t+\delta)(1-t)} \cdot \lambda_{\mathrm{max}}
        \leq
        \frac{\delta(1+t)}{t(1-t)} \cdot\lambda_{\mathrm{max}} \enspace.  \qedhere
    \]
\end{proof}

\section{An interactive protocol for polynomial-space unitary synthesis}
\label{sec:upspace_in_uqip}
In this section, we establish the equality $\uqip = \upspace$ by proving the reverse inclusion $\upspace \subseteq \uqip$ stated in \cref{thm:upspace-in-uqip}, which is the unitary analogue of the state-synthesis inclusion $\mathsf{statePSPACE} \subseteq \mathsf{stateQIP}$~\cite[Theorem~1.1]{RY22}.

To this end, we have to show that
any unitary family $\{U_n\}_{n\in\N}\in\upspace$
can be synthesized interactively.
Our high-level idea is to use
techniques developed for the unitary synthesis problem
to turn a polynomial-space implementation of $U_n$
into a polynomial-size quantum oracle circuit
with access to a $\pspace$ oracle.
We then replace each query to the $\pspace$ oracle
with an interactive protocol.
The following subsections develop this approach in five steps:
\begin{itemize}[
    label=$\circ$,
    leftmargin=1.6em,
    labelsep=0.5em,
    topsep=4pt,
    itemsep=4pt,
    parsep=0pt
]
  \item The definition of $\upspace$ only guarantees a polynomial-space quantum channel $\mathcal{C}_n$
  that approximates the target unitary $U_n$.
  To fit the setting of the unitary synthesis problem,
  we first show that this channel can be rounded to an exact unitary $W_n$ that remains close to the target unitary $U_n$
  in \cref{subsec:rounding_UPSPACE}.

  \item \cref{subsec:UPSPACE_to_UBQP_PSPACE} then
  borrows techniques from the unitary synthesis problem
  to synthesize $W_n$
  by a polynomial-size quantum oracle circuit
  with access to a fixed $\pspace$ oracle $A_W$.

  \item We next show
  in \cref{subsec:Boolean_queries_with_phase_states}
  that the query to the oracle $A_W$
  can be simulated by an efficient quantum circuit
  supplied with polynomially many resource states from
  $\mathsf{statePSPACE}$.

  \item To obtain these resource states,
  \cref{subsec:from_state_synthesis_to_accepted_maps}
  uses the interactive state-synthesis protocol of Rosenthal
  and Yuen \cite{RY22} to construct polynomial-time verifiers
  that prepare the states required at each stage of the
  oracle simulation.

  \item Finally, \cref{subsec:the_interactive_proof_protocol}
  combines these components into an interactive protocol for $U_n$,
  establishes its completeness and soundness,
  and proves the main theorem.
\end{itemize}

\subsection{Rounding \texorpdfstring{$\upspace$}{unitaryPSPACE}
implementations to exact unitaries}
\label{subsec:rounding_UPSPACE}
The membership in $\upspace_{\gamma}$ provides only a channel family $\{\mathcal{C}_n\}_{n \in \N}$ that is ``almost unitary'', i.e., close to $\{\mathcal{U}_n\}_{n \in \N}$.
Fortunately, the definition is robust enough that we can extract from each $\mathcal{C}_n$ an exact unitary close to $U_n$.
A similar robustness argument appears in the proof of the robustness of polynomial-space state synthesis under purification \cite[Theorem~6.1]{MY23}.
Here, we carry out its unitary analogue
in \cref{sec:rounding}:
an approximately unitary channel can be rounded to an exact unitary at a square-root cost.
Moreover, we show in \cref{sec:pspace_computability}
that the resulting unitary has entries computable in $\pspace$,
which is crucial for the subsequent unitary synthesis step.

\subsubsection{Constructing the rounded unitary}
\label{sec:rounding}
We first show how to round an approximately unitary channel,
not necessarily arising from a polynomial-space implementation,
to an exact unitary.
The idea is to construct a linear operator by setting
its image of a fixed pure input state
to be the leading eigenvector of the channel output,
then complete the construction using the channel's action on off-diagonal operators.
Finally, we apply polar decomposition to obtain an exact unitary. 

A similar procedure for the channel-to-unitary rounding appears in \cite{PhysRevA.90.012110,HKOT23}. We nevertheless give a full proof below, since \cref{prop:pspace_computable_of_projected_unitary} relies on the details of the construction to show that the rounded unitary has $\pspace$-computable entries whenever the channel does (see \cref{lemma:entries_of_space_uniform_channel}).

\begin{lemma}[Recovering the exact unitary]
\label{lemma:exact_unitary_recovery}
    Let $\mathcal{C}$ be a quantum channel and $U$ a unitary, both acting on $\HH$, such that $\td(\mathcal{C}(\psi),\mathcal{U}(\psi))\leq\gamma$ for every $\ket{\psi}\in\HH$, and assume that $0\leq\gamma\leq 2^{-10}$. Fix $\ket{\phi} \in \HH$ and set $\sigma \coloneqq  \mathcal{C}(\ket{\phi} \!\! \bra{\phi})$, then
    \begin{enumerate}
        \item The largest eigenvalue $\lambda_{\max}(\sigma)$ has multiplicity $1$, and satisfies $\lambda_{\max}(\sigma) \geq 1 - \gamma$.
        \item Let $\ket{v_{\max}}$ be the eigenstate of $\sigma$ for $\lambda_{\max}(\sigma)$, and $Q_{v_{\max}} = \id - \ket{v_{\max}} \!\! \bra{v_{\max}} $. Define a linear operator $B \in \mathsf{L}(\HH)$ via the following actions:
        \begin{equation}
        \label{eqn:def_of_aux_operator}
        B \ket{\phi} = \ket{v_{\max}}, \quad \forall \ket{\phi_{\perp}}: \braket{\phi | \phi_{\perp}} = 0, \quad B \ket{\phi_{\perp}} = Q_{v_{\max}} \mathcal{C}\left( \ket{\phi_{\perp}} \!\! \bra{\phi} \right) \ket{v_{\max}}.
        \end{equation}
        Then there exists a complex number $z \in \C$ with $|z| = 1$ such that $\| B - z U\|_{\op} \leq 8 \sqrt{\gamma}$.

        \item The singular values of $B$ lie in the interval $[1 - 8 \sqrt{\gamma}, 1 + 8 \sqrt{\gamma}] \subseteq [3/4, 5/4]$, and the unitary projection $W \coloneqq B(B^\dagger B)^{-1/2}$ satisfies
        $$
        \left\| W - z U \right\|_{\op} \leq 16 \sqrt{\gamma}.
        $$
    \end{enumerate}
\end{lemma}

To prove \cref{lemma:exact_unitary_recovery},
we first establish the following estimate,
which will allow us to controls the action of the channel
on vectors orthogonal to the fixed input.

\begin{lemma}[Tangent estimate]
\label{lemma:tangent_estimate}
    Let $\gamma\in[0,1]$, and let $\mathcal{C}$ be a quantum channel and $U$ a unitary, both acting on Hilbert space $\HH$ such that $\td(\mathcal{C}(\psi), \mathcal{U}(\psi)) \leq \gamma$ for every $\ket{\psi} \in \HH$. Let the isometry $T: \HH \to \HH \otimes \HH_{\mathsf{A}}$ be the isometric extension of $\mathcal{C}$.
    Fix a state $\ket{\phi} \in \HH$,
    we put
    \begin{equation}
    \label{eqn:var_substitution}
    \ket{u} = U \ket{\phi}, \quad Q_u \coloneqq  \id_{\HH} - \ket{u} \! \bra{u}, \quad \ket{\tilde{v}} \coloneqq \left( \bra{u} \otimes \id_{\HH_{\mathsf{A}}} \right) T \ket{\phi} \in \HH_{\mathsf{A}}.
    \end{equation}
    Then it holds that $1 - \gamma \leq \|\ket{\tilde{v}} \|^2 \leq 1$, and for every state $\ket{\phi_{\perp}}$ such that $\braket{\phi | \phi_{\perp}} = 0$, it holds that
    $$
    \left\| (Q_u \otimes \id_{\HH_{\mathsf{A}}}) T \ket{\phi_{\perp}} - U \ket{\phi_{\perp}} \otimes \ket{\tilde{v}} \right\| \leq \frac{16}{\sqrt{15}}\sqrt{\gamma}.
    $$
\end{lemma}

\begin{proof}
    Since $T$ is an isometry and $\| \bra{u} \otimes \id_{\HH_{\mathsf{A}}} \|_{\op} \leq 1$, $\|\ket{\tilde{v}}\| \leq 1 $ is immediate. Moreover, by definition we have $1 - \|\ket{\tilde{v}}\|^2 = 1 - \braket{\phi | T^\dagger (\ket{u} \!\! \bra{u} \otimes \id_{\HH_{\mathsf{A}}}) T  |\phi} = 1 -\braket{u | \mathcal{C}(\phi) |u} = \Tr[  \ket{u} \!\! \bra{u}(\ket{u} \!\! \bra{u} - \mathcal{C}(\phi)) ] \leq \gamma$.

    Fix a unit vector $\ket{\phi_{\perp}} \perp \ket{\phi}$ and set $\ket{u_{\perp}} = U \ket{\phi_{\perp}}$. For a complex $z \in \C$ with $|z| = 1$, and numbers $a, b \in \R$ such that $a^2+ b^2 = 1$, we set
    $$
    \ket{\phi_z} \coloneqq a \ket{\phi} + bz \ket{\phi_{\perp}}, \quad L_z = Q_u - \frac{b}{a} z \ket{u_{\perp}} \!\! \bra{u} = I_{\HH} - \left(\ket{u} - \frac{b}{a}z\ket{u_{\perp}}\right) \bra{u}.
    $$
    $L_z$ annihilates $U \ket{\phi_z}$, since $L_z U\ket{\phi_z}
    = \left(Q_u-bz/a\ket{u_{\perp}}\!\!\bra{u}\right)
      \left(a\ket{u}+bz\ket{u_{\perp}}\right)
    = bz\ket{u_{\perp}}-bz\ket{u_{\perp}}$ $=0$.
    Moreover, 
    $$
    \begin{aligned}
    \left\| L_z \right\|_{\op} = \max_{|\alpha|^2 + |\beta|^2 = 1} \left\| L_z(\alpha \ket{u} + \beta \ket{u_{\perp}}) \right\| &= \max_{|\alpha|^2 + |\beta|^2 = 1}\left\| \left(\beta - \frac{b}{a} z \alpha \right) \ket{u_{\perp}} \right\| \\
    &= \max_{|\alpha|^2 + |\beta|^2 = 1} \frac{\sqrt{|\alpha|^2 + |\beta|^2}}{|a|} = \frac{1}{|a|},
    \end{aligned}
    $$
    where the first equality holds since $L_z$ acts non-trivially only on $\text{span}\{\ket{u}, \ket{u_{\perp}}\}$.
    If we set $Q_z = \id_{\HH} - U \ket{\phi_z} \!\! \bra{\phi_z} U^\dagger$, then $L_z = L_z Q_z$, and 
    $$
    \left\| (Q_z \otimes \id_{\HH_{\mathsf{A}}}) T \ket{\phi_z} \right\|^2 = \Tr\left[ Q_z \mathcal{C}(\phi_z) \right] = \Tr\left[ Q_z \left( \mathcal{C}(\phi_z) - U \ket{\phi_z} \!\! \bra{\phi_z} U^\dagger \right) \right] \leq \gamma.
    $$
    If we define $\ket{\xi(z)} = (L_z \otimes \id_{\HH_{\mathsf{A}}}) T \ket{\phi_z} = (L_z Q_z \otimes \id_{\HH_{\mathsf{A}}}) T \ket{\phi_z}$, then we have
    \begin{equation}
    \label{eqn:upper_bound_of_xi_z}
    \left\| \ket{\xi(z)} \right\| = \left\| (L_z Q_z \otimes \id_{\HH_{\mathsf{A}}}) T \ket{\phi_z} \right\| \leq \| L_z \|_{\op} \| (Q_z \otimes \id_{\HH_{\mathsf{A}}}) T \ket{\phi_z} \| \leq \frac{\sqrt{\gamma}}{|a|}.
    \end{equation}
    Observe that if we expand $T \ket{\phi_z} = a T \ket{\phi} + b z T \ket{\phi_{\perp}}$, it holds that
    $$
    \begin{aligned}
        \ket{\xi(z)} &= (L_z \otimes \id_{\HH_{\mathsf{A}}}) T \ket{\phi_z} \\
        &= \left( Q_u \otimes \id_{\HH_{\mathsf{A}}} - \frac{b}{a} z \left(\ket{u_{\perp}} \!\!\bra{u} \otimes \id_{\HH_{\mathsf{A}}}  \right) \right) \left( a T \ket{\phi} + bz T \ket{\phi_{\perp}} \right) \\
        &= a(Q_u \otimes \id_{\HH_{\mathsf{A}}}) T \ket{\phi} + z \cdot \underbrace{b \left(  (Q_u \otimes \id_{\HH_{\mathsf{A}}}) T \ket{\phi_{\perp}} -\left( \ket{u_{\perp}}\!\! \bra{u} \otimes \id_{\HH_{\mathsf{A}}} \right) T \ket{\phi} \right)}_{\ket{\xi_1}} \\
        &\quad - z^2 \cdot \frac{b^2}{a} \left( \ket{u_{\perp}} \!\! \bra{u} \otimes \id_{\HH_{\mathsf{A}}} \right) T \ket{\phi_{\perp}}.
    \end{aligned}
    $$
    Note that by \cref{eqn:var_substitution} we can express $\ket{\xi_1} = b ((Q_u \otimes \id_{\HH_{\mathsf{A}}})T \ket{\phi_{\perp}} - \ket{u_{\perp}} \otimes \ket{\tilde{v}})$, which up to a scalar is exactly what we want to bound. Observe that we can express $\ket{\xi_1}$ as a weighted sum 
    $$
    \ket{\xi_1} = \frac{1}{3} \sum_{k=0}^{2} \mathrm{e}^{ -\I\frac{2\pi k}{3}  } \ket{\xi(\mathrm{e}^{\I \frac{2\pi k}{3} })}.
    $$
    Therefore, by the triangle inequality and \cref{eqn:upper_bound_of_xi_z}, and setting $a = \sqrt{15}/4$, $b =1/4$,
    $$
    \begin{aligned}
        \left\| (Q_u \otimes \id_{\HH_{\mathsf{A}}}) T \ket{\phi_{\perp}} - U \ket{\phi_{\perp}} \otimes \ket{\tilde{v}} \right\| &= \frac{1}{|b|} \|\ket{\xi_1}\| \leq \frac{ \| \ket{\xi(1)}\| + \|\ket{\xi(\mathrm{e}^{\I \frac{2\pi}{3} })} \| + \|\ket{\xi(\mathrm{e}^{\I \frac{4\pi}{3} })}\| }{3|b|}
        \\
        &\leq \frac{1}{|b|} \max_{z \in \C: |z| =1 } \left\| \xi(z)  \right\| \leq \frac{\sqrt{\gamma}}{|a| |b|} = \frac{16}{\sqrt{15}} \sqrt{\gamma},
    \end{aligned}
    $$
    concluding the proof.
\end{proof}

Now, we are ready to prove \cref{lemma:exact_unitary_recovery}.

\begin{proof}[Proof of \cref{lemma:exact_unitary_recovery}]
    We first set $\ket{u} = U \ket{\phi}$ and $Q_u \coloneqq  \id - \ket{u} \!\! \bra{u}$, $\Delta = (\mathcal{C} - \mathcal{U})(\ps{u}) = \sigma - \ps{u}$. Note that $\Tr[\Delta] = 0$, then by assumption 
    $\|\Delta\|_{\op}  \leq \frac{1}{2}\|\Delta\|_1 \leq  \gamma$.
    We have $\lambda_{\max}(\sigma)
    \geq \braket{u | \sigma | u} \geq 1-\gamma > 1/2$.
    Since $\sigma$ is positive semidefinite and has trace $1$,
    this eigenvalue has multiplicity $1$.
    This proves the first statement.

    Secondly, $\lambda_{\max} Q_u \ket{v_{\max}} = Q_u \sigma \ket{v_{\max}} = Q_u \Delta \ket{v_{\max}}$, then by the above argument,
    $$
    x \coloneqq  \left\| Q_u \ket{v_{\max}}  \right\| = \frac{\| Q_u \Delta \ket{v_{\max}} \|}{\lambda_{\max}} \leq \frac{ \| Q_u \|_{\op} \|\Delta\|_{\op} }{\lambda_{\max}} \leq \frac{\gamma}{\lambda_{\max}} \leq \frac{\gamma}{1 - \gamma}.
    $$
    As $|\braket{u | v_{\max}}|^2 = 1 - x^2 > 0$ we set the scalar $z = \braket{u | v_{\max}} / |\braket{u | v_{\max}}|$ and $\ket{v_0} \coloneqq  \overline{z} \ket{v_{\max}}$, so that we have $\braket{u | v_{0}} = \overline{z} \braket{u|v_{\max}} = |\braket{u | v_{\max}}|= \sqrt{1 - x^2}$, so that
    $$
    \|\ket{v_0} - \ket{u} \|^2  = 2 - 2 \sqrt{1 - x^2} \leq 2 x^2, \quad \| Q_{v_{\max}} - Q_u \|_{\op} = \| \ps{u} - \ps{v_0} \|_{\op} = x.
    $$
    Since $\gamma \leq 2^{-10}$ we have $x \leq \frac{\gamma}{1 - \gamma} \leq \frac{{2^{10}}}{2^{10} - 1} \gamma$, so $ \|\ket{v_0} - \ket{u} \|, \| Q_{v_{\max}} - Q_u \|_{\op} \leq 2 \gamma$. Recall \cref{eqn:def_of_aux_operator}, let $B_0$ defined identically as $B$ with $\ket{v_{\max}}$ replaced by $\ket{v_0}$. By linearity, we have $B = z B_0$, and to prove statement 2 it suffices to show $\|B_0 - U \|_{\op} \leq 8 \sqrt{\gamma}$.

    We track the actions of $B_0$ on the complete bases. 
    For any $\ket{\phi_{\perp}} \perp \ket{\phi}$, we take $T : \HH \to \HH \otimes \HH_A$ as the isometric extension of $\mathcal{C}$. We would like to bound $\|(B_0 - U) \ket{\phi_{\perp}} \|$. The operator $\mathcal{C}(\ket{\phi_{\perp}} \!\! \bra{\phi}) = \Tr_{\HH_{\mathsf{A}}}(T \ket{\phi_{\perp}} \!\! \bra{\phi} T^\dagger)$, then choose an orthonormal basis $\{ \ket{j} \}_j$ of $\HH_{\mathsf{A}}$ gives $T \ket{\phi_{\perp}} = \sum_j \ket{t_j}\!\! \ket{j}$ and $T \ket{\phi} = \sum_j \ket{s_j}\!\! \ket{j}$.
    Then $\mathcal{C}(\ket{\phi_{\perp}} \!\! \bra{\phi}) = \sum_j \ket{t_j} \!\! \bra{s_j}$ and we set $\ket{g} = \sum_j \braket{u | s_j} \!\ket{j}$, hence
    $$
        \mathcal{C}(\ket{\phi_{\perp}} \!\! \bra{\phi}) \ket{u} = \sum_j \ket{t_j} \braket{s_j | u} = (\id \otimes \bra{g}) T \ket{\phi_{\perp}}.
    $$
    We have $1 \geq \|\ket{g}\|^2 = \sum_{j} | \braket{u | s_j} |^2 = \braket{u | \sigma | u} \geq 1 - \gamma$, and by Cauchy-Schwarz,
    $$
    \begin{aligned}
    \| \mathcal{C}(\ket{\phi_{\perp}} \!\! \bra{\phi}) \|_{\op}^2  &= \max_{\ket{\psi}, \ket{\varphi}} \left| \braket{\psi |\mathcal{C}(\ket{\phi_{\perp}} \!\! \bra{\phi}) |\varphi } \right|^2 \\
    &= \max_{\ket{\psi}, \ket{\varphi}} \left|\sum_{j} \braket{\psi | t_j}  \braket{s_j | \varphi} \right|^2 \leq \max_{\ket{\psi}, \ket{\varphi}} \left( \sum_j |\braket{\psi | t_j}|^2 \right) \left( \sum_j |\braket{s_j | \varphi}|^2 \right) \leq 1.
    \end{aligned}
    $$
    Then using the result in \cref{lemma:tangent_estimate}, we have
    $$
    \begin{aligned}
    &\left\| Q_u\mathcal{C}(\ket{\phi_{\perp}} \!\! \bra{\phi}) \ket{u} - (\id \otimes \bra{g})(U \ket{\phi_{\perp}} \otimes \ket{g})  \right\| \\
    &\qquad = \left\| Q_u (\id \otimes \bra{g}) T \ket{\phi_{\perp}} - (\id \otimes \bra{g})(U \ket{\phi_{\perp}} \otimes \ket{g})  \right\| \\
    &\qquad \leq \| \id \otimes \bra{g} \|_{\op} \left\|(Q_u \otimes \id) T \ket{\phi_{\perp}} - U \ket{\phi_{\perp}} \otimes \ket{g} \right\| \leq \frac{16}{\sqrt{15}} \sqrt{\gamma}.
    \end{aligned}
    $$
    And hence
    $$
    \begin{aligned}
    &\left\| Q_u \mathcal{C}(\ket{\phi_{\perp}} \!\! \bra{\phi}) \ket{u} - U \ket{\phi_{\perp}} \right\| \\
    &\qquad \leq \left\| Q_u\mathcal{C}(\ket{\phi_{\perp}} \!\! \bra{\phi}) \ket{u} - \|\ket{g}\|^2 U \ket{\phi_{\perp}}  \right\| + \left\| \|\ket{g}\|^2 U \ket{\phi_{\perp}} - U \ket{\phi_{\perp}} \right\| \\
    &\qquad \leq \frac{16}{\sqrt{15}} \sqrt{\gamma} + (1 - \|\ket{g}\|^2) \|U \ket{\phi_{\perp}}\| \leq \gamma + \frac{16}{\sqrt{15}} \sqrt{\gamma}.
    \end{aligned}
    $$
    Therefore, recall in \cref{eqn:def_of_aux_operator} that $B_0 \ket{\phi_{\perp}} = Q_{v_0} \mathcal{C}(\ket{\phi_{\perp}} \!\! \bra{\phi}) \ket{v_0}$, we have
    $$
    \begin{aligned}
    \left\| (B_0 - U) \ket{\phi_{\perp}} \right\| &\leq \|Q_{v_0} \mathcal{C}(\ket{\phi_{\perp}} \!\! \bra{\phi}) \ket{v_0} - Q_u \mathcal{C}(\ket{\phi_{\perp}} \!\! \bra{\phi}) \ket{u} \| + \left\| Q_u \mathcal{C}(\ket{\phi_{\perp}} \!\! \bra{\phi}) \ket{u} - U \ket{\phi_{\perp}} \right\| \\
    &\leq \left\| Q_{v_0} - Q_u  \right\|_{\op} \|\mathcal{C}(\ket{\phi_{\perp}} \!\! \bra{\phi})\|_{\op} + \|Q_u \|_{\op} \| \mathcal{C}(\ket{\phi_{\perp}} \!\! \bra{\phi}) \|_{\op} \|\ket{v_0} - \ket{u}\| \\
    &\quad + \left\| Q_u \mathcal{C}(\ket{\phi_{\perp}} \! \bra{\phi}) \ket{u} - U \ket{\phi_{\perp}} \right\| \\
    &\leq 4 \gamma + \gamma + \frac{16}{\sqrt{15}} \sqrt{\gamma} = 5 \gamma + \frac{16}{\sqrt{15}} \sqrt{\gamma}.
    \end{aligned}
    $$
    On the input state $\ket{\phi}$, we have $\|(B_0 - U) \ket{\phi} \| = \| \ket{v_0} - \ket{u} \| \leq 2 \gamma$. Then for any legitimate input $\ket{\psi} = \alpha \ket{\phi} + \beta \ket{\phi_{\perp}}$, we have
    \begin{equation}
    \label{eqn:near_unitary_condition-2}
    \begin{aligned}
   \| B - z U\|_{\op} &= \max_{\phi_{\perp}}\max_{|\alpha|^2 + |\beta|^2 = 1} \|(B_0 - U) \ket{\psi} \| \\&\leq \max_{\phi_{\perp}} \max_{|\alpha|^2 + |\beta|^2 = 1} |\alpha| \| (B_0 - U) \ket{\phi} \| + |\beta| \| (B_0 - U) \ket{\phi_{\perp}}\| \\
    &\leq \max_{|\alpha|^2 + |\beta|^2 = 1} 2 |\alpha| \gamma + \left( 5 \gamma + \frac{16}{\sqrt{15}} \sqrt{\gamma} \right) |\beta| \leq 7 \gamma + \frac{16}{\sqrt{15}} \sqrt{\gamma} \leq 8 \sqrt{\gamma},
    \end{aligned}
    \end{equation}
    where we have used $\sqrt{\gamma} \leq 1/32$ and thus $7 \gamma \leq 7\sqrt{\gamma}/32$ in the last inequality.

    To complete the unitary roundup, we set $\eps_B \coloneqq 8 \sqrt{\gamma} \leq 1/4$. The near-unitary condition in \cref{eqn:near_unitary_condition-2} ensures that the eigenvalues of $B^\dagger B$ lies within $[(1 - \eps_B)^2, (1 + \eps_B)^2] \subseteq [(3/4)^2, (5/4)^2]$. Since $B$ is well-defined on the complete orthonormal basis of $\HH$, it is invertible, so is $|B| = (B^\dagger B)^{1/2}$. The operator $W = B|B|^{-1}$ is unitary, which is precisely the unitary factor in the polar decomposition $B = W|B|$ \cite[Theorem 7.3.1]{Horn2012}. Finally, we have
    $$
    \left\| W - z U\right\|_{\op} \leq \| W - B\|_{\op} + \|B - z U\|_{\op} = \|\id - |B|\|_{\op} + \|B - zU\|_{\op} \leq 2 \eps_B \leq 16 \sqrt{\gamma},
    $$
    this completes the proof.
\end{proof}

\subsubsection{$\mathsf{PSPACE}$ computability}
\label{sec:pspace_computability}
We now apply the rounding construction to a channel family
that implements a $\upspace$ unitary family,
and show that the resulting unitary family has
$\pspace$-computable entries.
We will use the following two facts,
whose proofs are given in \cref{sec:pspace_arthm}.

\begin{restatable}[Arithmetic of $\pspace$-computable numbers]{lemma}{pspaceArithmeticLemma}
    \label{lemma:arthm_of_PSPACE_num}
    Let $k, p$ be polynomials. Then we have the following closure properties:
    \begin{enumerate}
        \item \textbf{Finite sums.} If a sequence $\{\alpha_{y, c}\}_{y, c}$ indexed by $y$ and $c \in \{0, 1\}^{k(|y|)}$ is $\pspace$-computable, so is $\{ \sum_{c} \alpha_{y, c} \}_{y}$.
        \item \textbf{Products, quotients and roots.} Let both $\{\alpha_y\}_y, \{\beta_y\}_y$ be $\pspace$-computable with $|\alpha_y|, |\beta_y| \leq 2^{p(|y|)}$, then $\{\alpha_y \beta_y\}_y$ is $\pspace$-computable. If $|\beta_y| \geq 2^{-p(|y|)}$, then $\{\alpha_y/\beta_y\}_y$ is $\pspace$-computable. If each $\beta_y$ is real and nonnegative, then $\{\sqrt{\beta_y}\}_y$ is $\pspace$-computable.
        \item \textbf{(Exponentially) long products.} For each $y$, let $\{M_{y, t}\}_{t=1}^{N(y)}$ be $2^{k(|y|)} \times 2^{k(|y|)}$ matrices subject to $N(y) \leq 2^{p(|y|)}$ is polynomial-space computable from $y$. If the entries $\{\braket{i | M_{y, t} | j}\}_{(y, t, i, j)}$ are $\pspace$-computable and every product of consecutive factors $M_{y, t_2} \cdots M_{y, t_1}$ with $t_1 \leq t_2 \in [N(y)]$ has all entries of modulus bounded by $1$, the entries of the matrix product $M_{y, N(y)} \cdots M_{y, 1}$ are then $\pspace$-computable.
    \end{enumerate}
\end{restatable}

\begin{restatable}[Computability of entries of a space-uniform channel, {\cite[Lemma~6.2]{MY23}}, extended]{lemma}{channelEntriesLemma}
    \label{lemma:entries_of_space_uniform_channel}
    Let $\{\mathcal{C}_n\}_{n\in\N}$ be space-uniform, each with $n$ input qubits
    and $n$ output qubits.
    Then, we have that the family $\{ \braket{i | \mathcal{C}_n(\ket{x}\!\! \bra{y}) | j} \}_{(1^n, i, j, x, y)}$ consists of numbers of modulus at most $1$ and is $\pspace$-computable.
\end{restatable}

We can now apply these facts to the construction in \cref{lemma:exact_unitary_recovery},
and show that we can compute the leading eigenspace of the output on a fixed input, recover the auxiliary operator, and approximate its polar factor in polynomial space.

\begin{proposition}[$\pspace$-computability of the rounded unitary]
\label{prop:pspace_computable_of_projected_unitary}
    Let the unitary family $\{U_n \}_{n \in \N} \in \upspace_{\gamma}$ with $\gamma:\N\to[0,2^{-10}]$, witnessed by a space-uniform circuit family $\{\mathcal{C}_n\}_{n \in \N}$.
    Suppose that there exists $n_0$ such that, for every $n\geq n_0$ $\td(\mathcal{C}_n(\psi),\mathcal{U}_n(\psi))\leq\gamma$ for every $\ket{\psi}$.
    For each $n \geq n_0$, let $W_n$ be the unitary obtained from \cref{lemma:exact_unitary_recovery} with $\mathcal{C}=\mathcal{C}_n$ and $\ket{\phi}=\ket{0^n}$, and set $W_n=\id$ for $n<n_0$.
    Then $\{W_n\}_{n \in \N}$ is a family of unitaries with $\pspace$-computable entries, and for each $n\geq n_0$ there exists $z_n \in \C$ with $|z_n| = 1$ such that
    $$
    \left\| W_n - z_n U_n \right\|_{\op} \leq 16 \sqrt{\gamma(n)}.
    $$
\end{proposition} 

\begin{proof}
    The bound is directly due to \cref{lemma:exact_unitary_recovery} with the phase of $\ket{v_{\max}}$ arbitrarily chosen. We hardwire the dummy $n < n_0$ unitaries and consider $n \geq n_0$. Write $\sigma_n = \mathcal{C}_n(\ps{0^n})$ and sort its eigenvalues as $\lambda_{\max}(\sigma_n) = \lambda_1 \geq \lambda_2 \geq \cdots$ with eigenstates $\ket{e_1}, \ket{e_2}, \dots$, where $\ket{e_1}$ is the vector $\ket{v_{\max}}$ in \cref{lemma:exact_unitary_recovery}.
    Hence, $\lambda_1 \geq 1- \gamma$. Let $\Pi_1 = \ps{e_1}$ be the projection onto the eigenspace of the largest eigenvalue. By \cref{lemma:entries_of_space_uniform_channel} entries of $\sigma_n$ and $M_x \coloneqq  \mathcal{C}_n(\ket{x} \!\! \bra{0^n})$ have modulus bounded by $1$ and are $\pspace$-computable. We take the precision parameter $h$, drop the explicit dependence on $n$ and take steps to show that the following are $\pspace$-computable:

    \paragraph{The projector $\Pi_1$.} We borrow the main idea from \cite{Seif2023} to approximate $\Pi_1$ via $\sigma$. For $k \geq 1$, we take $\Pi^{(k)} = \sigma^k / \Tr[\sigma^k]$, then in the eigenbasis, the diagonal operator
    $$
    \Pi^{(k)} - \Pi_1 = \frac{1}{\Tr[\sigma^k]} \left( \sum_{j \geq 2} \lambda_j^k \ps{e_j} -  \left( \sum_{j \geq 2} \lambda_j^k \right) \Pi_1\right)
    $$
    has entries with modulus at most $\sum_{j \geq 2} \lambda_j^k / \Tr[\sigma^k]$. Then it holds that
    $$
    \begin{aligned}
    \left\| \Pi^{(k)} - \Pi_1 \right\|_{\op} &= \frac{\left\| \sum_{j \geq 2} \lambda_j^k \ps{e_j} -  \left( \sum_{j \geq 2} \lambda_j^k \right) \Pi_1 \right\|_{\op}}{\Tr[\sigma^k]} \\
    &= \frac{\sum_{j \geq 2} \lambda_j^k }{\Tr[\sigma^k]} \leq \frac{\left( \sum_{j\geq 2}\lambda_j \right)^k}{\Tr[\sigma^k]} \leq \left( \frac{\gamma}{1 - \gamma} \right)^k \leq \left( 2 \gamma \right)^k \leq 2^{-9k},
    \end{aligned}
    $$
    where we have used the fact that $\sum_j a_j^k \leq (\sum_j a_j)^k$ for nonnegative $a_j$. Therefore, to approximate $\braket{i | \Pi_1 | j}$ within $2^{-h}$ it suffices to take $k = h + 1$ to approximate $\braket{i | \Pi^{(k)} |j}$ within $2^{-h-1}$. By \cref{lemma:arthm_of_PSPACE_num} and note that $\Tr[\sigma^k] \geq (1 - \gamma)^k \geq 2^{-k}$, the entries of $\sigma^k$, $\Tr[\sigma^k]$, and thus the quotient $\Pi^{(k)} = \sigma^k / \Tr[\sigma^k]$ are $\pspace$-computable, so $\Pi_1$ can be approximated in polynomial space.

    \paragraph{The canonical relative phase representative.} To recover a canonical and stably computable representative $\ket v$ of $\Pi_1 = \ps{e_1}$, we choose a reliably nonzero coordinate to fix its global phase. 
    For each $j \in \{0, 1\}^n$, we set $\widehat{\Pi}_{j, j}$ as a real number approximation to $\braket{j|\Pi_1|j}$ within $2^{-n-6}$, obtained by taking the real part of the algorithm's output.
    And let $j_0$ be the lexicographically first index maximizing $\Re \widehat{\Pi}_{j, j}$, which depends only on $n$\footnote{
    As the index-finding algorithm is deterministic and the precision is only $n$-dependent.
    }. Note that $\sum_{j \in \{0, 1\}^n } \braket{j | \Pi_1 | j} = \Tr[\Pi_1] = 1$, then there must exist some $j_1 \in \{0, 1\}^n$ such that $\braket{j_1 |\Pi_1 | j_1} \geq 2^{-n}$, and therefore by the assumption,
    $$ 
    \begin{aligned}
    \braket{j_0 | \Pi_1 | j_0} &\geq\widehat{\Pi}_{j_0,j_0} - 2^{-n - 6} \geq \widehat{\Pi}_{j_1, j_1} - 2^{-n - 6} \geq 2^{-n} - 2 \cdot 2^{-n - 6} > 2^{-n - 1}.
    \end{aligned}
    $$
    Then we fix the representative state for $\ket{e_1} = \ket{v_{\max}}$:
    $$
    \ket{v} = \frac{\Pi_1 \ket{j_0}}{\sqrt{ \braket{j_0 | \Pi_1 | j_0} }}.
    $$
    For each coordinate $i \in \{0, 1\}^n$, $\braket{i | v} = \braket{i | \Pi_1 | j_0} / \sqrt{\braket{j_0 | \Pi_1 | j_0}}$ is $\pspace$-computable due to \cref{lemma:arthm_of_PSPACE_num} since $|\braket{i | \Pi_1 | j_0}| \leq 1$ and the denominator is at least $2^{-n}$.

    \paragraph{The operator $B$.} Recall the defining relation in \cref{eqn:def_of_aux_operator} with $\ket{\phi}$ set to $\ket{0^n}$, and every $x \in \{0, 1\}^n \setminus \{0^n\}$ gives rise an orthogonal basis to $\ket{0^n}$. By the identities
    $$
    \begin{aligned}
    \braket{i | B |0^n} = \braket{i | v}; \quad \forall x \neq 0^n,~ \braket{i | B | x} &= \braket{i | (\id - \ps{v}) M_x |v} \\
    &= \sum_{c \in \{0, 1\}^n } \braket{i | M_x | c} \braket{c | v} - \braket{i | v} \sum_{a, c \in \{0, 1\}^n } \braket{v | a} \braket{a | M_x |c} \braket{c | v}.
    \end{aligned}
    $$
    Then all factors have modulus at most $1$ and the $\pspace$-computability follows from \cref{lemma:arthm_of_PSPACE_num}.
    \paragraph{The inverse square root $S^{-1/2}$.} Set $S = B^\dagger B$ with entries $\braket{i | S | j} = \sum_{a \in \{0, 1\}^n } \overline{\braket{a | B |i}} \braket{a | B |j}$. We define $Z \coloneqq  \id - \frac{1}{2} S$. Recall \cref{lemma:exact_unitary_recovery}, the eigenvalues of $S$ lies in $[(3/4)^2, (5/4)^2]$, so $Z$ is directly Hermitian with spectrum within $[7/32, 23/32]$. The entries of $Z^k$ for any $k \in \N$ satisfies $|\braket{i | Z^k | j}| \leq \|Z^k\|_{\op} \leq \|Z\|_{\op}^k < 1$. In the same vein as \cite[Corollary 67]{QSVT}, for $|x| < 1$ we have the power series $(1 - x)^{-1/2} = \sum_{k \geq 0} 4^{-k} \binom{2k}{k} x^k = \sum_{k \geq 0} \prod_{j=1}^{k} \frac{2j-1}{2j} x^k$, so by the spectral theorem,
    $$
    S^{-1/2} = \left( 2 (\id - Z) \right)^{-1/2} = \frac{1}{\sqrt{2}} \sum_{k \geq 0} \frac{1}{4^k} \binom{2k}{k} Z^k.
    $$
    And we use the truncated approximation at level $L$: take $Y_L = \frac{1}{\sqrt{2}} \sum_{0 \leq k \leq L} 4^{-k} \binom{2k}{k} Z^k$, then
    \begin{equation}
    \label{eqn:truncated_series_error}
    \begin{aligned}
    \left\| S^{-1/2} - Y_L \right\|_{\op} &= \left\| \frac{1}{\sqrt{2}} \sum_{k > L} \frac{1}{4^k} \binom{2k}{k} Z^k \right\|_{\op} \leq \frac{1}{\sqrt{2}} \sum_{k > L} \frac{1}{4^k} \binom{2k}{k} \left\|Z \right\|_{\op}^k \\
    &< \frac{1}{\sqrt{2}} \sum_{k > L} \left\|Z \right\|_{\op}^k \leq \frac{1}{\sqrt{2}} \sum_{k > L} \left( \frac{23}{32} \right)^k = \frac{32}{9\sqrt{2}} \left( \frac{23}{32}\right)^{L+1}.
    \end{aligned}
    \end{equation}
    The coefficients $\prod_{j=1}^k \frac{2j-1}{2j}$ are exact rationals, and entries of each $Z^k$ are $\pspace$-computable by \cref{lemma:arthm_of_PSPACE_num}. Therefore, the entries of $Y_L$ are $\pspace$-computable uniformly in $L$ given in unary.

    \paragraph{The projected unitary $W$.} Finally, recall that $W = B S^{-1/2}$. It suffices to control the approximation properly via \cref{eqn:truncated_series_error}. Specifically, we take $L = 3h + 8$, and since $(23/32)^3 < 1/2$, and $\|B\|_{\op} \leq 5/4$, for all $i, j \in \{0, 1\}^n$, the entry-wise approximation error gives
    $$
    \left| \braket{i | W | j} - \braket{i | BY_L | j} \right| \leq \left\| W - BY_L \right\|_{\op} \leq \|B\|_{\op} \left\| S^{-1/2} - Y_L \right\|_{\op} \leq \frac{5}{4} \cdot \frac{32}{9 \sqrt{2}} \cdot 2^{-h - 3} < 2^{-h - 1}.
    $$
    With the guarantee that the $\pspace$-computable number $\braket{i | BY_L | j}$ (since both $B$ and $Y_L$ are bounded and proved $\pspace$-computable, we apply \cref{lemma:arthm_of_PSPACE_num}) is approximated within $2^{-h-1}$, the entries $\braket{i | W | j}$ is guaranteed to yield an error at most $2^{-h - 1} \cdot 2 = 2^{-h}$. This concludes the $\pspace$-computability of each building block of \cref{lemma:exact_unitary_recovery} and thus the recovered unitary.
\end{proof}

\subsection{Synthesizing unitaries in
\texorpdfstring{$\upspace$}{unitaryPSPACE} efficiently
with a \texorpdfstring{$\pspace$}{PSPACE} oracle}
\label{subsec:UPSPACE_to_UBQP_PSPACE}

This subsection shows that the unitary family obtained in
\cref{prop:pspace_computable_of_projected_unitary}
can be synthesized efficiently
with access to a fixed $\pspace$ oracle.
Our starting point is the following positive answer to the unitary synthesis problem recently announced by OpenAI, which we restate in a form aligned with our setting.

\begin{theorem}[{\cite[Theorem~1.1]{OpenAI2026UnitarySynthesis}}]
\label{thm:usp}
    There exists an algorithm such that,
    on input $(1^n,1^t,1^\beta,d)$ with integers
    $n,\beta\geq1$, $t\geq0$, and $d\in[1,2^n]$ encoded in binary,
    outputs an oracle quantum circuit $\mathcal{C}_{n,d,t,\beta}^{(\cdot)}$
    of size $\poly(n+t+\beta)$ such that
    \begin{itemize}
        \item The circuit consists only of unitary gates and oracle queries.
        \item The workspace consists of a $t$-qubit register $\mathsf{L}$,
        an $n$-qubit register $\mathsf{Q}$, and an ancillary register
        $\mathsf{A}$.
    \end{itemize}
    Write $\HH_d\coloneqq\Span\{\ket{0},\ldots,\ket{d-1}\}
    \subseteq(\C^2)^{\otimes n}$.
    For every unitary family $\{U_\lambda\}_{\lambda\in\{0,1\}^t}$
    acting on $\HH_d$, put
    $\mathbf{U}\coloneqq\sum_{\lambda\in\{0,1\}^t}
    \ps{\lambda}\otimes U_\lambda$.
    Let $\iota$ be the isometry
    $\ket{\lambda}_{\mathsf{L}}\ket{\psi}_{\mathsf{Q}}
    \mapsto\ket{\lambda}_{\mathsf{L}}\ket{\psi}_{\mathsf{Q}}
    \ket{0}_{\mathsf{E}}$ for $\ket{\psi}\in\HH_d$.
    Then there exists a Boolean function $f_{\mathbf{U},\beta}$ such that
    \[
        \left\|
            \mathcal{C}_{n,d,t,\beta}^{f_{\mathbf{U},\beta}}
            \cdot\iota-\iota\cdot\mathbf{U}
        \right\|_{\op}
        \leq 2^{-\beta-1}.
    \]
\end{theorem}

The theorem above places no restriction on the complexity of the oracle.
For unitary families with $\pspace$-computable entries,
the following lemma shows that a fixed $\pspace$ oracle
suffices for efficient synthesis.
For the sake of simplicity, we defer its proof to \cref{appendix:proof_of_pspace_USP}.

\begin{lemma}[Effective synthesis from $\pspace$-computable entries]
\label{lemma:usynthesis_with_pspace_oracle}
    Let $W=\{W_n\}_{n\in\N}$ be a unitary family
    with $\pspace$-computable entries,
    where each $W_n$ acts on $n$ qubits.
    Then there exists a language $A_W\in\pspace$
    and a deterministic polynomial-time algorithm that,
    on input $(1^n,1^\beta)$ with integers $n,\beta\geq1$,
    outputs an oracle quantum circuit $C_{n,\beta}^{(\cdot)}$
    of size $\poly(n, \beta)$ such that
    \begin{itemize}
      \item The circuit $C_{n,\beta}^{(\cdot)}$ consists only of unitary gates and oracle queries.
      \item The workspace of $C_{n,\beta}^{(\cdot)}$ consists of an $n$-qubit register $\mathsf{Q}$
    and an ancillary register $\mathsf{E}$.
    \end{itemize}
    Let $\iota$ be the isometry $\ket{\psi}_{\mathsf{Q}}\mapsto\ket{\psi}_{\mathsf{Q}}\ket{0}_{\mathsf{E}}$.
    Then
    $$
    \left\| C_{n, \beta}^{A_W} \cdot \iota - \iota \cdot W_n \right\|_{\op} \leq 2^{-\beta - 1}.
    $$
\end{lemma}

Combining \cref{lemma:usynthesis_with_pspace_oracle}
with \cref{prop:pspace_computable_of_projected_unitary}
shows that any unitary family in $\upspace_{\gamma}$
can be approximated by efficient quantum oracle circuits
with access to a fixed $\pspace$ oracle.

\begin{corollary}
\label{corollary:unitary_synthesis_in_BQP_with_PSPACE_oracle} 
    Let $\{ U_n\}_{n \in \N} \in \upspace_\gamma$
    with $\gamma(n) \leq 2^{-10}$ for all $n$.
    Take $W=\{W_n\}_{n\in\N}$ and $n_0$ as in
    \cref{prop:pspace_computable_of_projected_unitary},
    and $A_W,C_{n,\beta}^{(\cdot)},\mathsf{E},\iota$ as in
    \cref{lemma:usynthesis_with_pspace_oracle}.
    Let $\beta: \N \to \N$ be such that
    $1^{\beta(n)}$ is polynomial-time computable from $1^n$.
    Define
    $\Phi_n(\rho)\coloneqq
      \Tr_{\mathsf{E}}\!\left[
        C_{n,\beta(n)}^{A_W}\cdot \iota\,\rho\,\iota^\dagger\cdot
        \bigl(C_{n,\beta(n)}^{A_W}\bigr)^\dagger
    \right].
    $
    Then $C_{n,\beta(n)}^{(\cdot)}$ can be generated from $1^n$
    in polynomial time,
    and for every $n \geq n_0$,
    \begin{equation}
    \label{eqn:closeness_of_oracle_circuit}
    \dd\!\left( \Phi_n, \mathcal{U}_n \right) \leq 16 \sqrt{\gamma(n)} + 2^{-\beta(n) - 1}.
    \end{equation}
\end{corollary}

\begin{proof}
    By \cref{prop:pspace_computable_of_projected_unitary}, members of $\{W_n\}_{n \in \N}$ has $\pspace$-computable entries, so \cref{lemma:usynthesis_with_pspace_oracle} applies directly. Since $1^{\beta(n)}$ is polynomial-time computable, $\beta(n) \in \poly(n)$ and $C_{n,\beta(n)}^{(\cdot)}$ has polynomial size and polynomial-time generated.
    For each $n\geq n_0$, let $z_n$ be the phase supplied by
    \cref{prop:pspace_computable_of_projected_unitary}.
    Combining \cref{lemma:usynthesis_with_pspace_oracle} and \cref{prop:pspace_computable_of_projected_unitary}, it follows directly that
    $$
    \begin{aligned}
    \dd\!\left( \Phi_n, \mathcal{U}_n \right) \leq  \left\| C_{n, \beta(n)}^{A_{W}} \cdot \iota - z_n \iota \cdot U_n \right\|_{\op} &\leq \left\| C_{n, \beta(n)}^{A_{W}} \cdot \iota - \iota \cdot W_n \right\|_{\op} + \left\| \iota \left( W_n - z_n U_n \right)\right\|_{\op} \\
   &\leq 2^{-\beta(n) - 1} + 16 \sqrt{\gamma(n)}.
   \end{aligned}
    $$
    This completes the proof.
\end{proof}

\subsection{Implementing \texorpdfstring{$\pspace$}{PSPACE} oracles with \texorpdfstring{$\mathsf{statePSPACE}$}{statePSPACE} resources}
\label{subsec:Boolean_queries_with_phase_states}

The previous subsection shows how to efficiently synthesize a unitary family in $\upspace$ using a $\pspace$ oracle. We now show how to implement each $\pspace$ oracle query using resource states in $\mathsf{statePSPACE}$.
This will allow us to invoke $\mathsf{statePSPACE}=\mathsf{stateQIP}$ in the next subsection to obtain an interactive proof protocol for unitary synthesis.

The argument in this subsection consists of two parts:
first, we express Boolean oracle queries as Boolean phase gates
and implement these gates by recursively consuming phase states
(\cref{lemma:exact_injection_of_diagonal_sign_state});
second, we show that, for a $\pspace$ oracle, all phase states
required by this procedure belong to $\mathsf{statePSPACE}$
(\cref{fact:correction_predicate_in_PSPACE}).
Combining these two parts, we show the existence of a polynomial-size circuit that implements the unitary family in $\upspace$ using only resource states in $\mathsf{statePSPACE}$
(\cref{lemma:unitary_synthesis_with_state_resources}).

We first consider how to implement a single query to the oracle
$O_f$ for a Boolean function $f$.
By the standard fact in \cite{10.1098/rspa.1998.0164},
such a query can be implemented using Hadamard gates and
a Boolean phase gate, defined as follows.

\begin{definition}[Boolean phase gate, Boolean phase state]
\label{def:basic_identities_in_phase_state_to_gate}
    For a Boolean function $g: \{0, 1\}^m \to \{0, 1\}$,
    its associated \emph{Boolean phase gate} $D_g$ and \emph{phase state} $\ket{\varphi_g}$ are defined as
    $$
    \begin{aligned}
    &D_g \coloneqq  \sum_{x \in \{0, 1\}^m } (-1)^{g(x)} \ps{x}, \quad \ket{\varphi_g} \coloneqq  D_g \ket{+}^{\otimes m} = \frac{1}{\sqrt{2^m}} \sum_{x \in \{0, 1\}^m } (-1)^{g(x)} \ket{x}.
    \end{aligned}
    $$
\end{definition}

\begin{fact}[{\cite{10.1098/rspa.1998.0164}}]
\label{lemma:phase_kickback}
    For every Boolean function $f: \{0, 1\}^{\ell} \to \{0, 1\}$, let $h(y, t) \coloneqq  t \cdot f(y)$ for $y \in \{0, 1\}^{\ell}, t \in \{0, 1\}$. Then it holds that $O_f = (\id \otimes \hdm) D_h (\id \otimes \hdm)$.
\end{fact}

Thus, implementing $O_f$ reduces to implementing the Boolean phase gate $D_h$.
We next show how to implement such gates using phase states.
We adapt the recursive technique used to teleport the QRAM gate in \cite[Section~2.3]{11369061}.
This method requires only phase states as auxiliary resources. We describe the procedure in detail below.

Suppose we want to implement $D_g$ for a Boolean function $g: \{0, 1\}^m \to \{0, 1\}$.
Given an arbitrary input state
$\ket{\psi}=\sum_x\alpha_x\ket{x}$, our goal is to obtain
\[
    D_g\ket{\psi}
    =\sum_x\alpha_x(-1)^{g(x)}\ket{x}.
\]
A single step of the procedure uses an auxiliary register
prepared in the phase state $\ket{\varphi_g}$
and applies $\mathsf{CNOT}$ gates from each input qubit to the corresponding auxiliary qubit,
obtaining
\[
    \frac{1}{\sqrt{2^m}}
    \sum_{x,y}\alpha_x(-1)^{g(y)}
    \ket{x}\ket{y\oplus x}.
\]
The auxiliary register is then measured in the computational basis.
Each outcome $a\in\{0,1\}^m$ occurs with probability $2^{-m}$,
and the resulting normalized output state is
\[
    \sum_x\alpha_x(-1)^{g(x\oplus a)}\ket{x}
    =D_{\Delta_a g} \cdot D_g\ket{\psi},
    \quad
    \text{where}
    \quad
    \Delta_a g(x)\coloneqq g(x)\oplus g(x\oplus a).
\]
In other words, the output state differs from the target state
$D_g\ket{\psi}$ by a Boolean phase gate $D_{\Delta_a g}$.
Hence, the goal of the next step is to implement $D_{\Delta_a g}$
as a \emph{correction} since $D_{\Delta_a g}^2 = \id$,
using the same procedure with the phase state
$\ket{\varphi_{\Delta_a g}}$.

This recursion terminates because, by the following lemma,
the Boolean function defining the correction becomes constant
after at most $m$ steps.
The corresponding phase gate is then $\pm\id$, which requires no further correction.

\begin{lemma}[Iterated differences, cf. {\cite[Section 4.6]{11369061}}]
\label{lemma:iterated_differences}
    Let $m\ge1$ and $j\ge0$ be integers,
    $g:\{0,1\}^m\to\{0,1\}$ be a Boolean function, and
    $a_1,\ldots,a_j\in\{0,1\}^m$ be bitstrings.
    Define $g_0 \coloneqq g$ and $g_i \coloneqq  \Delta_{a_i} g_{i-1}$ for $i\in[j]$.
    For all subsets $S \subseteq [j]$, define $a_S \coloneqq  \bigoplus_{i \in S} a_i$ with the convention that $a_{\varnothing} \coloneqq  0^m$. Then, we have
    \begin{equation}
    \label{eqn:iterated_differences}
    g_j(x) = \bigoplus_{S \subseteq [j] } g(x \oplus a_S),
    \end{equation}
    and for $j=m$, the function $g_m$ is a constant.
\end{lemma}

\begin{proof}
    We prove this by induction. \cref{eqn:iterated_differences} holds trivially for $j = 0$. Assume that it holds for $j - 1$, then by assumption we can write $g_j(x)$ as
    $$
    \begin{aligned}
       g_j(x) = g_{j-1}(x \oplus a_j) \oplus g_{j-1}(x) &= \left( \bigoplus_{S \subseteq [j - 1] } g(x \oplus a_j \oplus a_S) \right) \oplus \left( \bigoplus_{S \subseteq [j - 1] } g(x \oplus a_S) \right) \\
        &= \bigoplus_{T \subseteq [j]: j \in T} g(x \oplus a_T) \oplus \bigoplus_{S \subseteq [j-1]} g(x \oplus a_S) \\
        &= \bigoplus_{T \subseteq [j]: j \in T} g(x \oplus a_T) \oplus \bigoplus_{T \subseteq [j]: j \not \in T} g(x \oplus a_T) = \bigoplus_{S \subseteq [j]} g(x \oplus a_S).
    \end{aligned}
    $$
    As for $j = m$, we identify $S \subseteq [m]$ with its indicator $\mathbf{1}_S \in \F_2^m$, then we can write $a_S = A \mathbf{1}_S$ where $A = (a_1, a_2, \dots, a_m)$. If $A$ is invertible, since $S \mapsto \mathbf{1}_S$ is a bijection, it holds that $g_m(x) = \bigoplus_{S \subseteq [m] } g(x \oplus a_S) = \bigoplus_{y \in \{0, 1\}^m } g(x \oplus y) = \bigoplus_{y \in \{0, 1\}^m} g(y)$ is independent of $x$. If $A$ is not invertible, every vector in the image of the map $S \mapsto A \mathbf{1}_S$ has $|\ker(A)| = 2^{m - \rank(A)}$ preimages:
    $$
    g_m(x) = \bigoplus_{s \in \F_2^m } g(x \oplus As) = \bigoplus_{y \in \im(A)} \bigoplus_{ s: As = y } g(x \oplus y) = \bigoplus_{y \in \im(A)} 0 = 0,
    $$
    $g_m$ is therefore still a constant. This concludes the proof.
\end{proof}

Thus, the full procedure for implementing $D_g$
is given below and illustrated in
\cref{fig:injection_of_signal_state}.

\paragraph{Phase-state injection procedure.}
Let $g:\{0,1\}^m\to\{0,1\}$ be a Boolean function, and let
$\mathsf{Q}$ be an $m$-qubit register containing an arbitrary
input state $\ket{\psi}$.
Let $g_0\coloneqq g$.
For each $j=1,\ldots,m$, perform the following steps,
as illustrated in \hyperref[fig:injection_of_signal_state]{\cref*{fig:injection_of_signal_state}(a)}:
\begin{enumerate}
    \item Prepare a fresh $m$-qubit auxiliary register
    $\mathsf{B}_j$ in the state $\ket{\varphi_{g_{j-1}}}$.

    \item Apply $\cnot$ from the $k$-th qubit of $\mathsf{Q}$
    to the $k$-th qubit of $\mathsf{B}_j$ for every $k\in[m]$.

    \item Measure $\mathsf{B}_j$ in the computational basis,
    record the outcome $a_j\in\{0,1\}^m$, and discard
    $\mathsf{B}_j$. Define
    $g_j\coloneqq\Delta_{a_j}g_{j-1}$.
\end{enumerate}
After $m$ rounds, output register $\mathsf{Q}$.
The composition of these rounds is illustrated in
\hyperref[fig:injection_of_signal_state]{\cref*{fig:injection_of_signal_state}(b)}.

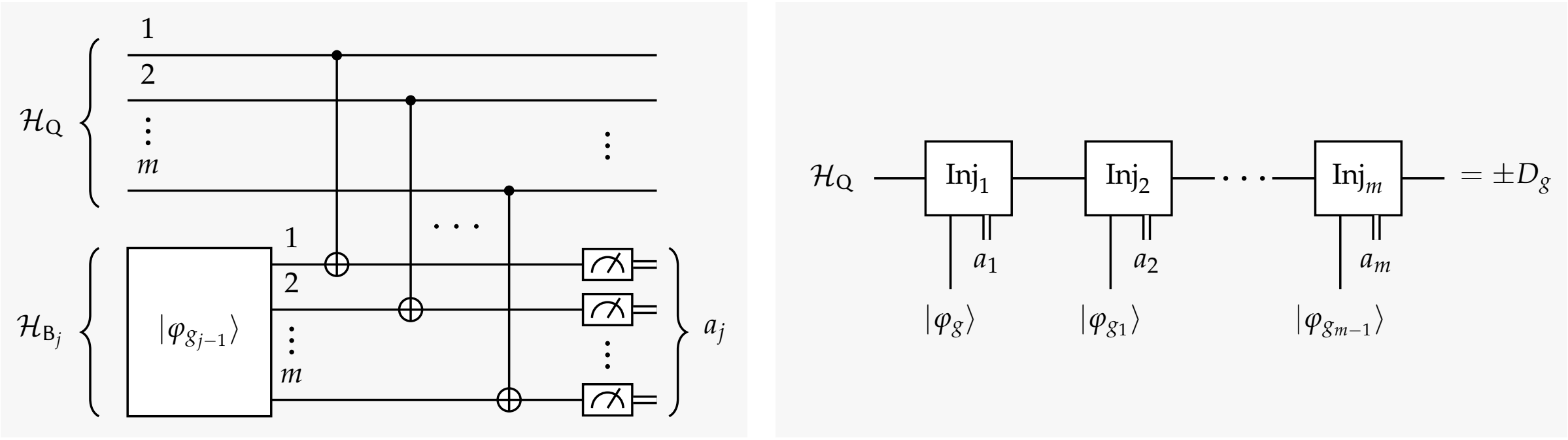
\begin{figure}[!t]
  \centering
  \begin{tikzpicture}[
    scale=0.83, transform shape,
    font=\small,
    wire/.style={line width=0.65pt},
    classical/.style={
      wire, double, double distance=1.2pt
    },
    meter/.style={
      draw, wire, fill=white,
      minimum width=0.60cm, minimum height=0.38cm,
      inner sep=1pt
    },
    grouping/.style={
      wire, decorate, decoration={brace,amplitude=5pt}
    },
    injection/.style={
      draw, wire, fill=white,
      minimum width=1.05cm, minimum height=0.90cm,
      inner sep=3pt
    }
  ]

    \fill[black!3] (-1.20,-4.65) rectangle (7.90,0.65);
    \fill[black!3] (8.25,-4.65) rectangle (17.90,0.65);

    \draw[grouping] (0,-1.85) -- (0,0.20)
      node[midway,left=8pt] {$\mathcal H_{\mathsf{Q}}$};
    \draw[grouping] (0,-4.40) -- (0,-2.35)
      node[midway,left=8pt] {$\mathcal H_{\mathsf{B}_j}$};

    \draw[wire,fill=white]
      (0.35,-4.40) rectangle (2.1,-2.35);
    \node at (1.225,-3.375)
      {$\ket{\varphi_{g_{j-1}}}$};

    \foreach \k/\yq/\yb in {
      1/0/-2.55,
      2/-0.55/-3.10,
      m/-1.65/-4.20
    } {
      \draw[wire] (0.35,\yq) -- (6.8,\yq);
      \node[above=2pt] at (0.60,\yq) {$\k$};

      \node[meter] (M\k) at (6.2,\yb) {};
      \begin{scope}[shift={(M\k.center)}]
        \draw[wire] (-0.19,-0.10)
          arc[start angle=180,end angle=0,radius=0.19cm];
        \draw[wire] (0,-0.10) -- (0.15,0.14);
      \end{scope}

      \draw[wire] (2.1,\yb) -- (M\k.west);
      \node[above=2pt] at (2.35,\yb) {$\k$};
      \draw[classical] (M\k.east) -- (6.8,\yb);
    }

    \draw[grouping] (6.95,-2.35) -- (6.95,-4.40)
      node[midway,right=8pt] {$a_j$};

    \foreach \x/\yq/\yb in {
      2.9/0/-2.55,
      3.8/-0.55/-3.10,
      5.0/-1.65/-4.20
    } {
      \draw[wire] (\x,\yq) -- (\x,\yb);
      \fill (\x,\yq) circle[radius=1.8pt];
      \draw[wire,fill=white]
        (\x,\yb) circle[radius=0.14cm];
      \draw[wire]
        ({\x-0.14},\yb) -- ({\x+0.14},\yb);
      \draw[wire]
        (\x,{\yb-0.14}) -- (\x,{\yb+0.14});
    }

    \foreach \x/\y in {
      0.60/-1.10,
      2.35/-3.65
    } {
      \node[font=\Large,inner sep=0pt,yshift=8pt]
        at (\x,\y) {$\vdots$};
    }

    \foreach \x/\y in {
      6.2/-1.10,
      6.2/-3.65
    } {
      \node[font=\Large,inner sep=0pt,yshift=3pt]
        at (\x,\y) {$\vdots$};
    }

    \node[font=\Large,inner sep=0pt]
      at (4.4,-2.10) {$\cdots$};
    \node at (3.3,-5.1) {\large (a) Single-round injection $\mathrm{Inj}_j$.};

    \begin{scope}[xshift=9.50cm]
      \node[anchor=east] at (-0.15,-1.50)
        {$\mathcal H_{\mathsf{Q}}$};

      \draw[wire] (-0.05,-1.50) -- (4.10,-1.50);
      \node[font=\Large,inner sep=0pt]
        at (4.5,-1.50) {$\dots$};
      \draw[wire] (4.80,-1.50) -- (6.90,-1.50);
      \node[anchor=west] at (6.95,-1.50)
        {$= \pm D_g$};

      \foreach \x/\j/\f in {
        1.10/1/g,
        3.05/2/{g_1},
        5.85/m/{g_{m-1}}
      } {
        \node[injection] (I\j) at (\x,-1.50)
          {$\mathrm{Inj}_{\j}$};

        \draw[wire]
          ({\x-0.22},-2.85) -- ({\x-0.22},-1.95);
        \node[anchor=north] at ({\x-0.22},-2.90)
          {$\ket{\varphi_{\f}}$};

        \draw[classical]
          ({\x+0.22},-1.95) -- ({\x+0.22},-2.25);
        \node[anchor=north] at ({\x+0.22},-2.28)
          {$a_{\j}$};
      }

      \node at (3.45,-5.1) {\large (b) Multiple-round injection.};
    \end{scope}

  \end{tikzpicture}
  \caption{The phase-state injection for the Boolean phase oracle $D_g$.}
  \label{fig:injection_of_signal_state}
\end{figure}

\begin{lemma}[Boolean queries from phase state,
    cf. {\cite[Sections 2.3, 4.5, 4.6]{11369061}}]
\label{lemma:exact_injection_of_diagonal_sign_state}
    For every sequence of measurement outcomes $\mathbf{a} = (a_1, a_2, \dots, a_m)$,
    the above phase-state injection procedure acts only nontrivially on $\HH_{\mathsf{Q}}$ by the operator
    \[
        M_{\mathbf{a}}
        \coloneqq \frac{\chi_{\mathbf{a}}}{\sqrt{2^{m^2}}}D_g
        \quad \text{for some }\ \chi_{\mathbf{a}}\in\{+1,-1\}.
    \]
    Consequently, discarding the outcomes yields $\mathcal{D}_g(\cdot) = D_g (\cdot) D_g^\dagger$ as desired.
\end{lemma}

\begin{proof}
    We consider a single-step injection with resource state $\ket{\varphi_f} \in \HH_{\mathsf{B}_j}$ and outcome $a \in \{0, 1\}^m$. For every $x \in \{0, 1\}^m$, 
    $$
    \begin{aligned}
    \cnot^{\otimes m} \left( \ket{x}_{\mathsf{Q}} \otimes \ket{\varphi_f}_{\mathsf{B}_j} \right) &= \cnot^{\otimes m} \left( \frac{1}{\sqrt{2^m}} \sum_{y \in \{0, 1\}^m } (-1)^{f(y)} \ket{x}_{\mathsf{Q}} \otimes \ket{y}_{\mathsf{B}_j} \right) \\
    &= \frac{1}{\sqrt{2^m}} \sum_{y \in \{0, 1\}^m} (-1)^{f(y)} \ket{x}_{\mathsf{Q}} \otimes \ket{y \oplus x}_{\mathsf{B}_j}.
    \end{aligned}
    $$
    Then, for outcome $a$, we contract it with $\id_{\mathsf{Q}} \otimes \bra{a}_{\mathsf{B}_j}$, which keeps only the terms with $y = x \oplus a$.
    By linearity, the single-step injection with outcome $a$
    is described by the operator that introduces the phase flip $(-1)^{f(x \oplus a)}$ for each basis $x \in \{0, 1\}^m$, that is,
    $$
    M_{a}^{(f)} = \frac{1}{\sqrt{2^m}} D_{f(\cdot \oplus a)} = \frac{1}{\sqrt{2^m}} D_{f} D_{f \oplus f(\cdot \oplus a)} = \frac{1}{\sqrt{2^m}} D_{f} D_{\Delta_a f},
    $$
    here we identify the action of $M_{a}^{(f)}$ via $M_{a}^{(f)} \ket{x} = \frac{1}{\sqrt{2^m}} D_{f(\cdot \oplus a)} \ket{x} = \frac{1}{\sqrt{2^m}} (-1)^{f(x \oplus a)} \ket{x}$.
    Since diagonal operators commute and $D_{g_j}^2 = \id$,
    conditioned on $\mathbf{a}$, the whole procedure can be described by the composite action
    $$
    \begin{aligned}
    M_{\mathbf{a}} \coloneqq  M_{a_m}^{(g_{m-1})} M_{a_{m-1}}^{(g_{m-2})} \cdots M_{a_1}^{(g_0)} &= \prod_{j=1}^{m} \left( \frac{1}{\sqrt{2^m}}  D_{g_{j-1}} D_{\Delta_{a_j} g_{j-1}}  \right) \\
    &= \frac{1}{\sqrt{2^{m^2}}} \left( D_{g_0} D_{g_1} \right) \left(D_{g_1} D_{g_2} \right) \cdots \left( D_{g_{m-1}} D_{g_m} \right) = \frac{1}{\sqrt{2^{m^2}}} D_{g} D_{g_m}.
    \end{aligned}
    $$
    By \cref{lemma:iterated_differences}, $g_m$ is a constant in $\{0, 1\}$ that may depend on $\mathbf{a}$. Thus $D_{g_m} = (-1)^{g_m} \id =: \chi_{\mathbf{a}} \id$. It suffices to collect all the branches to obtain the channel expression:
    $$
    \sum_{\substack{\mathbf{a} = (a_1, \dots, a_m): \\ a_j \in  \{0, 1\}^{m} }} M_{\mathbf{a}}(\cdot) M_{\mathbf{a}}^\dagger = \frac{1}{2^{m^2}} \sum_{\substack{\mathbf{a} = (a_1, \dots, a_m): \\ a_j \in  \{0, 1\}^{m} }} D_g(\cdot) D_g^\dagger = D_g(\cdot) D_g^\dagger = \mathcal{D}_g(\cdot).
    $$
    This concludes the proof.
\end{proof}

It remains to show that, when the procedure in
\cref{lemma:exact_injection_of_diagonal_sign_state} is used
to implement a $\pspace$ oracle, the required phase states
belong to $\mathsf{statePSPACE}$.
To this end, it suffices to describe their phases using
a fixed $\pspace$ predicate.
The subset expansion formula in
\cref{lemma:iterated_differences} gives such a description:

\begin{lemma}[Correction predicates are in $\pspace$]
\label{fact:correction_predicate_in_PSPACE}
    Let a language $A \in \pspace$. For a bitstring $x$ with $|x| = \ell + 1$, we split $x = (y, t)$ with $y \in \{0, 1\}^{\ell}$ and $t \in \{0, 1\}$, and define $h_A(y, t) = t \cdot A(y)$. For $j \geq 0$ and a list $\mathbf{a}_{\leq j} = (a_1, \dots, a_j)$ of $j$ strings of length $|x|$ (empty if $j = 0$), implicitly encoded as a binary string, write $a_S\coloneqq \bigoplus_{i\in S}a_i$ for $S\subseteq[j]$, with $a_\varnothing=0^{|x|}$. We define the language
    $$
    \Gamma_A\!\left( \mathbf{a}_{\leq j}, x \right) \coloneqq \bigoplus_{S \subseteq [j] } h_A\!\left( x \oplus a_S \right),
    $$
    and we set $\Gamma_A = 0$ on illegitimate input. Then $\Gamma_A \in \pspace$.
\end{lemma}

\begin{proof}
    For a machine on input $(\mathbf{a}_{\leq j}, x)$, it checks the format of the list and runs a $j$-bit counter through all the subsets $S \subseteq [j]$. For each $S$, it computes $x \oplus a_S$ by scanning $\mathbf{a}_{\leq j}$, evaluates $h_A$ with a $\pspace$ decider for $A$, and takes the XOR. Since the space is reused across evaluations, the machine uses only polynomial space in the input length at any time. 
\end{proof}

We are now ready to replace the $\pspace$ oracle used for
unitary synthesis in
\cref{corollary:unitary_synthesis_in_BQP_with_PSPACE_oracle}
by resource states in $\mathsf{statePSPACE}$.

\begin{lemma}[Unitary synthesis from $\mathsf{statePSPACE}$ resources]
\label{lemma:unitary_synthesis_with_state_resources} 
    Let $\Phi_n$ and $A_W$ be as in
    \cref{corollary:unitary_synthesis_in_BQP_with_PSPACE_oracle}.
    Then $\Phi_n$ can be implemented by a
    time-uniform family of quantum circuits
    supplied with polynomially many resource states in $n$,
    selected adaptively according to the preceding measurement
    outcomes. These states are phase states of the fixed
    predicate $\Gamma_{A_W}\in\pspace$ defined in \cref{fact:correction_predicate_in_PSPACE},
    and form a family in $\mathsf{statePSPACE}$.
\end{lemma}

\begin{proof}
    Let $C_n\coloneqq C_{n,\beta(n)}^{A_W}$ be the oracle circuit from
    \cref{corollary:unitary_synthesis_in_BQP_with_PSPACE_oracle}.
    Apply \cref{lemma:exact_injection_of_diagonal_sign_state} to each oracle query of $C_n$. Suppose $C_n$ makes $T$ oracle queries, with the $i$-th query acting on $\ell_i+1$ qubits. The circuit in \cref{lemma:exact_injection_of_diagonal_sign_state} then uses $K=\sum_{i=1}^T(\ell_i+1)$ resource states. Each query consumes $\ell_i+1$ states on $\ell_i+1$ qubits. The resource states are the phase states of $\Gamma_{A_W}$. By \cref{fact:correction_predicate_in_PSPACE}, these states can be prepared in polynomial space by preparing a uniform superposition, computing the predicate into an auxiliary qubit, applying its phase, and uncomputing the auxiliary register.  Thus they form a family in $\mathsf{statePSPACE}$.
    
    Since $C_n$ has polynomial size, the number of resource states is polynomially bounded in $n$.
    Each query is therefore implemented exactly.
    As a result, initializing the ancilla to the all-zero state and tracing it out at the end yields
    the channel $\Phi_n$.
\end{proof}

\subsection{Preparing
\texorpdfstring{$\mathsf{statePSPACE}$}{statePSPACE}
resource states interactively}
\label{subsec:from_state_synthesis_to_accepted_maps}

As illustrated in \cref{subsec:Boolean_queries_with_phase_states}, to implement $\pspace$ oracle queries it suffices to consume phase states of a $\pspace$ predicate, which are polynomial-space synthesizable. In a quantum interactive proof, however, the verifier running in polynomial time may not be able to prepare these states on its own. Fortunately, prior work by Rosenthal and Yuen \cite{RY22} constructed an interactive protocol in which the verifier can synthesize any space-uniform family of states with the help of an untrusted prover.
We state their result as follows:

\begin{theorem}[$\mathsf{statePSPACE} \subseteq \mathsf{stateQIP}$ {\cite[Theorem 4.5]{RY22}}]
\label{lemma:statePSPACE_in_stateQIP}
    Let $\{ \ket{\psi_n} \}_{n \in \N}$ be a state family in $\mathsf{statePSPACE}$\footnote{
We also use the extension of this theorem to state families
$\{\ket{\psi_u}\}_{u\in\{0,1\}^*}$ indexed by classical strings.  The proof extends by including $u$ in the classical inputs to the $\pspace$ languages used in the state-synthesis protocol of \cite[Section 6.1]{RY22}.
},
    and let $Q\geq 1$ be a polynomial.
    There exists a time-uniform family of quantum verifiers $\{\mathcal{V}_n\}_{n\in\N}$,
    each taking no quantum input,
    such that, for sufficiently large $n$, the following statements hold for $\mathcal{V}_n$: 
    \begin{enumerate}
        \item 
        There exists a prover such that $\mathcal{V}_n$
        accepts with probability $1$, and its output state
        $\rho_n$ conditioned on acceptance
        (cf. \cref{eqn:conditioned_acceptance_state}) satisfies
        $\td(\rho_n,\psi_n)\leq \mathrm{e}^{-Q(n)}$.
        \item For every prover, if the verifier accepts with probability $p>0$
          and outputs $\rho_n$ conditioned on acceptance, then
        \begin{equation}
        \label{eqn:acceptance_state_closeness_as_soundness}
        p \leq \exp\left( \mathrm{e}^{-Q(n)} - Q(n) \cdot \left( \td(\rho_n, \psi_n) \right)^4 \right).
        \end{equation}
    \end{enumerate}
\end{theorem}

Building on this theorem,
we establish the main technical lemma in this subsection,
which intuitively states that the verifier's accepted map
is close to one that prepares the desired phase state
in tensor product with the prover's remaining memory.

\begin{lemma}[Accepted map via verified phase states]
\label{lemma:accepted_map_via_phase_state}
    Let $\Gamma(z, x)$ be a $\pspace$ predicate.
    There exists a deterministic polynomial-time algorithm that,
    on input $(z, 1^m, 1^R)$ with $z\in\{0,1\}^*$ and integers $m, R \geq 1$,
    outputs a $\poly(|z| + m + R)$-size quantum verifier
    with no quantum input and an $m$-qubit output register $\mathsf{B}$,
    such that
    \begin{enumerate}
        \item There exists an honest prover such that the verifier accepts with probability $1$. 
        \item For every fixed prover,
        regard its initial private state on $\mathsf{P}$ as an input,  
        and let $\mathcal{A}: \mathsf{L}(\HH_{\mathsf{P}}) \to \mathsf{L}(\HH_{\mathsf{B}} \otimes \HH_{\mathsf{P}})$ be the unnormalized accepted map.
        Then there exists a completely positive and trace-nonincreasing map $\Lambda: \mathsf{L}(\HH_{\mathsf{P}}) \to \mathsf{L}(\HH_{\mathsf{P}})$ such that
        $$
       \dd\!\left( \mathcal{A}, \ps{\varphi_{\Gamma_{z,m}}} \otimes \Lambda \right) \leq \frac{1}{R},
        $$
        where $\Gamma_{z,m}\colon\{0,1\}^m\to\{0,1\}$ is defined by
        $\Gamma_{z,m}(x)=\Gamma(z,x)$, and
        $\ket{\varphi_{\Gamma_{z,m}}}$ is its associated phase state
        as defined in \cref{def:basic_identities_in_phase_state_to_gate}.
    \end{enumerate}
    Notably, with $\Gamma$ fixed, the verifier is generated uniformly from $(z,1^m,1^R)$.
\end{lemma}

\begin{proof}
    Set $N=|z|+m+R$.
    By the same construction as in the proof of
    \cref{lemma:unitary_synthesis_with_state_resources},
    the phase-state family
    $\{\ket{\varphi_{\Gamma_{z,m}}}\}_{(z,1^m,1^R)}$
    is in $\mathsf{statePSPACE}$;
    Applying \cref{lemma:statePSPACE_in_stateQIP}
    with $Q(N)=N^8$ yields the required uniformly generated
    verifier and proves statement~1.

    Since the proof of \cref{lemma:statePSPACE_in_stateQIP} (cf. \cite[Section 4.1]{RY22}) allows arbitrary prover actions, the bound in \cref{eqn:acceptance_state_closeness_as_soundness} with $n = N$ holds for every initial private state of any dimension. If we write $\ket{\varphi}$ shorthand for $\ket{\varphi_{\Gamma_{z,m}}}$ and set $\Pi = \id - \ps{\varphi}$. For any density operator $\rho \in \mathsf{L}(\HH_{\mathsf{P}})$, and the accepted map turns it to $\mathcal{A}(\rho)$, the acceptance probability reads $p = \Tr[\mathcal{A}(\rho)]$. 

    We start by showing that $\Tr[(\Pi \otimes \id_{\mathsf{P}}) \mathcal{A}(\rho)] \leq 1/Q^{1/4}$, and without loss of generality we assume $p > 0$. The prover starts with $\rho$ and is accepted with probability $p$, then the output conditioned on acceptance reads $\rho_{\mathsf{B}} = \Tr_{\HH_{\mathsf{P}}}[\mathcal{A}(\rho)] / p$. By the reformulated soundness bound \cref{eqn:acceptance_state_closeness_as_soundness}, we have
    \begin{equation}
    \label{eqn:overlap_upper_bound}
    \begin{aligned}
    \Tr\left[ (\Pi \otimes \id_{\mathsf{P}}) \mathcal{A}(\rho) \right] &= \Tr\left[ (\id - \ps{\varphi} \otimes \id_{\HH_{\mathsf{P}}}) \mathcal{A}(\rho) \right] = p \cdot \Tr\left[ (\id - \ps{\varphi}) (\rho_{\mathsf{B}} - \ps{\varphi}) \right] \\
    &\leq p \cdot \td\left( \rho_{\mathsf{B}}, \varphi \right) \leq p  \left( \frac{\mathrm{e}^{-Q} + \log(1/p)}{Q} \right)^{1/4} \leq \frac{1}{Q^{1/4}},
    \end{aligned}
    \end{equation}
    where the last inequality is due to $Q \geq 1$ and $$\sup_{p \in (0, 1] } p^4 \cdot (\mathrm{e}^{-Q} + \log(1/p)) \leq 1/\mathrm{e} + \sup_{p \in (0, 1] } p^4 \log(1/p) \leq 1/\mathrm{e} + 1/(4 \mathrm{e}) < 1. $$
    Let $V: \HH_{\mathsf{P}} \to \HH_{\mathsf{B}} \otimes \HH_{\mathsf{P}} \otimes \HH_{\mathsf{F}}$ be the isometric extension of $\mathcal{A}$. Since the inequality \cref{eqn:overlap_upper_bound} holds for all input $\rho$ and $\Pi^2 = \Pi = \Pi^\dagger$, it holds that
    $$
    \begin{aligned}
    \left\| (\Pi \otimes \id_{\HH_{\mathsf{P}} \otimes \HH_{\mathsf{F}}}) V \right\|_{\op}^2 &= \left\| V^\dagger \left( \Pi \otimes \id_{\HH_{\mathsf{P}} \otimes \HH_{\mathsf{F}}} \right) V \right\|_{\op} \\
    &= \left\| \mathcal{A}^*\left( \Pi \otimes \id_{\HH_{\mathsf{P}}}\right) \right\|_{\op} = \max_{\rho } \Tr\left[ (\Pi \otimes \id_{\mathsf{P}}) \mathcal{A}(\rho) \right] \leq \frac{1}{Q^{1/4}}.
    \end{aligned}
    $$
    Then we put the operator $L = (\bra{\varphi} \otimes \id_{\HH_{\mathsf{P}} \otimes \HH_{\mathsf{F}}}) V$ and $V_0 \coloneqq  \left( \ps{\varphi} \otimes \id_{\HH_{\mathsf{P}} \otimes \HH_{\mathsf{F}}} \right) V = \ket{\varphi} \otimes L$. The map $\Lambda$ is constructed via $\Lambda(\cdot) \coloneqq  \Tr_{\HH_{\mathsf{F}}}(L (\cdot) L^\dagger)$ is trace non-increasing since $\|L\|_{\op} \leq 1$. Here $V - V_0 = (\Pi \otimes \id_{\HH_{\mathsf{P}} \otimes \HH_{\mathsf{F}}}) V$, so that $\Tr_{\HH_{\mathsf{F}}}(V_0 (\cdot) V_0^\dagger) = \ps{\varphi} \otimes \Lambda(\cdot)$. Using \cite[Proposition I.6]{HKOT23} and the above inequality, we have
    $$
    \begin{aligned}
    \dd\!\left( \mathcal{A},  \ps{\varphi_{\Gamma_{z,m}}} \otimes \Lambda  \right) &\leq \dd\!\left( V(\cdot) V^\dagger,  V_0(\cdot) V_0^\dagger  \right) \\
    &\leq \left\| V - V_0 \right\|_{\op} \leq \left\| (\Pi \otimes \id_{\HH_{\mathsf{P}} \otimes \HH_{\mathsf{F}}}) V\right\|_{\op} \leq \frac{1}{Q^{1/8}} = \frac{1}{N} \leq \frac{1}{R},
    \end{aligned}
    $$
    completing the proof. Specifically, the construction of $\Lambda$ should be independent of the prover's initial state $\rho$.
\end{proof}

\subsection{Putting things together}
\label{subsec:the_interactive_proof_protocol}

We now combine the ingredients
from the preceding subsections
into an interactive protocol
for any unitary family in $\upspace$.
We first describe the protocol in \cref{subsec:interactive_protocol},
then bound the accumulated error in
\cref{subsec:composed_protocol_error}.
Finally, \cref{subsec:completeness_and_soundness}
establishes completeness and soundness,
completing the proof of \cref{thm:upspace-in-uqip}.

\subsubsection{The interactive synthesis protocol}
\label{subsec:interactive_protocol}

\begin{figure}[!t]
\begingroup
\small
\newcommand{\kw}[1]{\textbf{#1}}

\setlength{\fboxrule}{0.4pt} 
\setlength{\fboxsep}{10pt} 

\setlength{\parindent}{0pt}
\setlength{\parskip}{0pt}
\setlength{\emergencystretch}{1em}

\setlist[enumerate,1]{
  label=\arabic*.,
  leftmargin=1.8em,
  labelsep=.5em,
  topsep=6pt,
  itemsep=7pt,
  parsep=2pt
}
\setlist[enumerate,2]{
  label=(\roman*),
  leftmargin=2.6em,
  labelsep=.55em,
  topsep=4pt,
  itemsep=5pt,
  parsep=2pt
}
\setlist[enumerate,3]{
  label=(\alph*),
  leftmargin=2.1em,
  labelsep=.5em,
  topsep=4pt,
  itemsep=4pt,
  parsep=1pt
}
\setlist[itemize]{
  leftmargin=1.8em,
  labelsep=.5em,
  topsep=6pt,
  itemsep=4pt,
  parsep=0pt
}

\noindent\fbox{%
\begin{minipage}{
  \dimexpr\linewidth-2\fboxsep-2\fboxrule\relax
}
\begin{protocol}[An interactive protocol for
$\{U_n\}_{n}\in\upspace_{\delta^2/2^{14}}$
with soundness $s$]
\label{protocol:verifier_for_upspace}
\normalfont
\leavevmode\par\medskip

\underline{\textbf{Parameters.}}
\quad Fix polynomials $q,r$ such that
$q(n)\geq 2$, $r(n)\geq 4$,
$s(n)\geq 1/q(n)$ and $\delta(n)\geq 1/r(n)$.
\par\medskip

\underline{\textbf{Input state.}}
\quad An $n$-qubit pure state $\ket{\psi}$.

\par\medskip
\underline{\textbf{Verifier $\mathcal{V}_n$.}}
\begin{enumerate}
\item Compute $q(n)$, $r(n)$ and $\beta(n) = \left\lceil \log_2\!\left( 8 r(n) \right) \right\rceil$.

\item Obtain the description of the oracle circuit
  $C_{n,\beta(n)}^{(\cdot)}$,
  which acts on an $n$-qubit register $\mathsf{Q}$
  and an ancillary register $\mathsf{E}$,
  and has query access to the fixed oracle $A_W\in\pspace$,
  as in \cref{corollary:unitary_synthesis_in_BQP_with_PSPACE_oracle}.
  Let $T$ be its number of oracle queries,
  ordered by their occurrence in the circuit,
  with the $i$-th query acting on $\ell_i+1$ qubits.
  Set
  \begin{equation}
    \label{eqn:aux_params_in_protocol}
    K \coloneqq \sum_{i=1}^{T}(\ell_i+1),
    \qquad
    R_n \coloneqq
    \left\lceil 8K\cdot q(n)r(n)\right\rceil.
  \end{equation} 

\item Identify $\mathsf{Q}$ with the verifier's register
  holding the input $\ket{\psi}$,
  and initialize $\mathsf{E}$ to the all-zero state.

\item \kw{For each} operation of $C_{n,\beta(n)}^{(\cdot)}$ in circuit order:
  \begin{enumerate}
  \item \kw{If} it is an elementary unitary gate,
    apply it and \kw{continue} to the next operation.

  \item \kw{Else}, it is the $i$-th oracle query,
  acting on an $\ell_i$-qubit register $\mathsf{Y}_i$
  and a single-qubit target register $\mathsf{t}_i$.
  Apply $\hdm$ to $\mathsf{t}_i$. \\ ~ \\
  \kw{For} $j=1,\ldots,\ell_i+1$:
    \begin{enumerate}
    \item Let $z_j\gets (a_1,\ldots,a_{j-1})$, with $z_1\coloneqq \varnothing$,
      be the sequence of measurement outcomes from the preceding iterations.
      Send $(i,j,z_j)$ to the prover.

    \item Run the verifier specified in
      \cref{lemma:accepted_map_via_phase_state}
      with parameters $(z_j,1^{\ell_i+1},1^{R_n})$
      and the predicate $\Gamma_{A_W}$ defined in
      \cref{fact:correction_predicate_in_PSPACE},
      using fresh verifier and message registers.

    \item \kw{If} the verifier in step (b) rejects, \kw{REJECT AND HALT}.
      Otherwise, denote its output register by $\mathsf{B}_{i,j}$,
      retain it, and discard its remaining verifier and message registers.

    \item 
      Apply $\cnot$ from the $w$-th qubit of
      $(\mathsf{Y}_i,\mathsf{t}_i)$ to the $w$-th qubit of
      $\mathsf{B}_{i,j}$ for each $w\in[\ell_i+1]$.
      Measure $\mathsf{B}_{i,j}$ in the computational basis,
      record the outcome as $a_j\in\{0,1\}^{\ell_i+1}$,
      and discard $\mathsf{B}_{i,j}$.
    
    \end{enumerate}

    Apply $\hdm$ to $\mathsf{t}_i$.
  \end{enumerate}

\item Discard the ancillary register $\mathsf{E}$,
  output the register $\mathsf{Q}$ and \kw{ACCEPT}.
\end{enumerate}

\medskip
\underline{\textbf{Honest prover $\mathcal{P}_n^*$.}} \quad 
For each execution of step 4(ii)(b), follow the corresponding
honest prover strategy in \cref{lemma:accepted_map_via_phase_state}
using fresh private registers.

\end{protocol}
\end{minipage}%
}
\par
\endgroup
\end{figure}

Fix $\delta,s\colon\N\to[0,1]$ as in
\cref{thm:upspace-in-uqip}.
Let $\{U_n\}_{n\in\N}\in\upspace_{\delta^2/2^{14}}$
be the unitary family to be synthesized interactively.
The protocol in \cref{protocol:verifier_for_upspace} specifies a verifier with perfect completeness
and soundness $s$ for $\{U_n\}_{n\in\N}$,
together with a corresponding honest prover.

At a high level, the protocol combines three components,
which we explain in turn below.
First, the verifier reduces the synthesis of $U_n$
to running an efficient circuit with queries to a fixed $\pspace$ oracle,
as obtained in \cref{subsec:UPSPACE_to_UBQP_PSPACE} (steps 1--3).
Second, while running this circuit on its input,
the verifier applies elementary gates directly
and implements each oracle query by phase-state injection procedure in
\cref{subsec:Boolean_queries_with_phase_states},
which consumes a few phase states chosen adaptively
by earlier measurement outcomes (step 4).
Third, since these phase states are $\mathsf{statePSPACE}$ states,
the verifier obtains each of them from the prover via
the state-synthesis subprotocol in
\cref{subsec:from_state_synthesis_to_accepted_maps},
and rejects as soon as any subprotocol rejects
(steps 4(ii)(a)--(c)).
We now describe each component in more detail.

\paragraph{Reducing to oracle circuit implementation.}
The first two steps of the verifier $\mathcal{V}_n$
aim to reduce the synthesis of $U_n$ to the implementation of an efficient oracle circuit,
as disscussed in \cref{subsec:UPSPACE_to_UBQP_PSPACE}.
By \cref{corollary:unitary_synthesis_in_BQP_with_PSPACE_oracle},
the verifier can compute the description of an oracle circuit $C_{n,\beta(n)}^{(\cdot)}$
with query access to a fixed $\pspace$ oracle $A_W$. 
When instantiated with $A_W$, this circuit, with its ancillary
register $\mathsf{E}$ initialized to the all-zero state
and discarded at the end, induces a channel approximating
$\mathcal{U}_n(\cdot)=U_n(\cdot)U_n^\dagger$ on its another register $\mathsf{Q}$.

Therefore, to achieve the goal of applying the target unitary $U_n$ to the input state $\ket{\psi}$,
the verifier identifies $\mathsf{Q}$ with its register holding $\ket{\psi}$
and initializes $\mathsf{E}$ to the all-zero state in step 3.
It then starts to simulate $C_{n,\beta(n)}^{A_W}$ using the methods
developed in \cref{subsec:Boolean_queries_with_phase_states,subsec:from_state_synthesis_to_accepted_maps} in step 4.

\paragraph{Implementing oracle queries via phase-state injection.}
Specifically, the verifier reads the circuit $C_{n,\beta(n)}^{(\cdot)}$,
and by \cref{lemma:usynthesis_with_pspace_oracle},
$C_{n,\beta(n)}^{(\cdot)}$ consists of two types of operations:
elementary unitary gates and oracle queries.
The verifier processes these operations in circuit order.
For elementary unitary gates, it applies them directly to the corresponding qubits.
For oracle queries, the verifier simulates them using the phase-state injection procedure
introduced in \cref{subsec:Boolean_queries_with_phase_states}.
By \cref{lemma:phase_kickback}, the $i$-th query reduces to
realizing the Boolean phase gate $D_{h_{A_W}}$ on $\ell_i+1$ qubits
by conjugating with a Hadamard gate on the target qubit,
where $h_{A_W}(y,t)=t\cdot A_W(y)$ is defined in
\cref{fact:correction_predicate_in_PSPACE}.
By \cref{lemma:exact_injection_of_diagonal_sign_state},
the phase-state injection procedure then 
implements this Boolean phase gate exactly using $\ell_i+1$ ideal phase states as resources,
selected adaptively according to the preceding measurement outcomes.

\paragraph{Preparing phase states interactively.}
For the $i$-th query, we apply the injection procedure in
\cref{lemma:exact_injection_of_diagonal_sign_state}
with $m=\ell_i+1$ and $g=h_{A_W}$.
In round $j$, this procedure requires the phase state
$\ket{\varphi_{(h_{A_W})_{j-1}}}$, where $(h_{A_W})_{j-1}$
is obtained from the preceding outcomes $z_j=(a_1,\ldots,a_{j-1})$
by the iterated-difference recursion in \cref{lemma:iterated_differences}.
To prepare this state, the verifier invokes
\cref{lemma:accepted_map_via_phase_state}
with the fixed predicate $\Gamma=\Gamma_{A_W}$ from
\cref{fact:correction_predicate_in_PSPACE}
and parameters $(z_j,1^{\ell_i+1},1^{R_n})$.
The resulting subprotocol has
$\ket{\varphi_{\Gamma_{z_j,\ell_i+1}}}$ as its target state.
By the subset expansion in \cref{lemma:iterated_differences}
and the definition of $\Gamma_{A_W}$ in \cref{fact:correction_predicate_in_PSPACE},
for every $x\in\{0,1\}^{\ell_i+1}$ we have
\[
  \Gamma_{z_j,\ell_i+1}(x)
  =\Gamma_{A_W}(z_j,x)
  =\bigoplus_{S\subseteq[j-1]}h_{A_W}(x\oplus a_S)
  =(h_{A_W})_{j-1}(x).
\]
Hence, this target state is exactly the resource state required for injection.
If the subprotocol rejects, the verifier rejects the whole computation;
otherwise, it uses the output register $\mathsf{B}_{i,j}$
in the subsequent phase-state injection.
After completing the $\ell_i+1$ injection rounds,
it applies the final Hadamard to $\mathsf{t}_i$ to complete the phase kickback in \cref{lemma:phase_kickback} and then proceeds to the next circuit operation.

\paragraph{Completing the protocol.}
Once all operations have been processed, the verifier
discards $\mathsf{E}$, outputs $\mathsf{Q}$, and accepts.
The honest prover follows the corresponding honest strategy
in each invocation of the state-synthesis subprotocol,
using fresh private registers.

\subsubsection{Bounding the oracle circuit simulation error}
\label{subsec:composed_protocol_error}

Perfect completeness follows readily from the honest prover strategy
described in \cref{protocol:verifier_for_upspace}.
To analyze soundness, a crucial step is to bound the accumulated error
in simulating $C_{n,\beta(n)}^{A_W}$
introduced by the state-synthesis subprotocols of the composed protocol.

To this end, we fix an arbitrary prover $\mathcal{P}_n$
with private register $\mathsf{P}$ and let
$\mathcal{F}\colon
\mathsf{L}(\HH_{\mathsf{Q}}\otimes\HH_{\mathsf{P}})
\to \mathsf{L}(\HH_{\mathsf{Q}})$
denote the unnormalized accepted map of the whole protocol.

We refer to each phase-state injection in \cref{protocol:verifier_for_upspace},
together with its preceding state-synthesis subprotocol, as a \emph{stage}.
There are $K$ stages in total, as defined in
\cref{eqn:aux_params_in_protocol}.
For the $i$-th oracle query, the $j$-th stage has global \emph{stage index}
\[
    k=k(i,j)\coloneqq\kappa_i+j,
    \quad\text{where}\quad
    \kappa_i\coloneqq\sum_{u=1}^{i-1}(\ell_u+1).
\]
We use $\mathsf{B}_k$ to denote the resource register
$\mathsf{B}_{i,j}$ at this stage.
To describe the state-synthesis step at each stage,
let us assume the verifier record all preceding phase-injection outcomes
in a classical auxiliary register $\mathsf{C}$.
Before stage $k=\kappa_i+j$,
the contents of $\mathsf{C}$
form the \emph{history} $\mathbf{h}_k$, given by
\begin{equation}
\label{eqn:history_in_terms_of_outcomes}
\mathbf{h}_k
=
\left(
    \mathbf{a}^{(1)}_{\leq m_1},
    \ldots,
    \mathbf{a}^{(i-1)}_{\leq m_{i-1}},
    z_j
\right),
\qquad
z_j=\mathbf{a}^{(i)}_{\leq j-1},
\end{equation}
where $m_r=\ell_r+1$ and
$\mathbf{a}^{(r)}_{\leq s}$ denotes the first $s$
measurement outcomes in the simulation of the $r$-th oracle query.
Thus, the history $\mathbf{h}$ stored in $\mathsf{C}$
before stage $k$ determines the classical parameters
of the $k$-th state-synthesis subprotocol invoked
in \cref{protocol:verifier_for_upspace}.
For each stage $k$ and each possible history $\mathbf{h}_k$,
let
\[
    \mathcal{A}_{k,\mathbf{h}_k}\colon
    \mathsf{L}(\HH_{\mathsf{P}})
    \to
    \mathsf{L}(\HH_{\mathsf{B}_k}\otimes\HH_{\mathsf{P}})
\]
denote the unnormalized accepted map of the corresponding
state-synthesis subprotocol under the fixed prover strategy.

Using the above notation,
the overall accepted map
$\mathcal{F}: \mathsf{L}(\HH_{\mathsf{Q}} \otimes \HH_{\mathsf{P}}) \to \mathsf{L}(\HH_{\mathsf{Q}})$
for the whole protocol factors into the verifier's actions,
the $\mathsf{statePSPACE} \subseteq \mathsf{stateQIP}$ subprotocol operations,
and the phase state injections conditioned on histories stored classically in $\HH_{\mathsf{C}}$:
\begin{equation}
\label{eqn:overall_protocol_accepted_map}
\mathcal{F} = \mathcal{E} \circ \bigcirc_{k = 1}^{K}
\mathcal{J}_k \circ \mathcal{S}_k \circ \mathcal{G}_k, \quad \text{where} \quad \forall\, k \in [K],
~\mathcal{S}_k(\cdot) = \sum_{\mathbf{h}_k}
\ps{\mathbf{h}_k}_{\mathsf{C}}
\otimes
\mathcal{A}_{k,\mathbf{h}_k}\!
\left(\braket{\mathbf{h}_k|(\cdot)|\mathbf{h}_k}\right).
\end{equation}
More specifically, in the above decomposition:
\begin{itemize}
    \item
    $\mathcal{G}_k \colon \mathsf{L}( \HH_{\mathsf{Q}} \otimes \HH_{\mathsf{E}} ) \to \mathsf{L}( \HH_{\mathsf{Q}} \otimes \HH_{\mathsf{E}} )$
    consists of the verifier's circuit gates
    applied before the state-synthesis step at stage $k$
    and after the preceding phase-state injection, if any.

    \item 
    $\mathcal{S}_k \colon
    \mathsf{L}(\HH_{\mathsf{C}}\otimes\HH_{\mathsf{P}})
    \to
    \mathsf{L}(\HH_{\mathsf{C}}\otimes\HH_{\mathsf{B}_k}
    \otimes\HH_{\mathsf{P}})$
    is the completely positive and trace-nonincreasing map
    describing the accepted branch of
    the $k$-th state-synthesis subprotocol,
    controlled by the history stored in $\mathsf{C}$,
    in steps 4(ii)(a)--(c)
    of \cref{protocol:verifier_for_upspace}.
    
    \item 
    $\mathcal{J}_k \colon
    \mathsf{L}(\HH_{\mathsf{Q}}\otimes\HH_{\mathsf{E}}
    \otimes\HH_{\mathsf{C}}\otimes\HH_{\mathsf{B}_k})
    \to
    \mathsf{L}(\HH_{\mathsf{Q}}\otimes\HH_{\mathsf{E}}
    \otimes\HH_{\mathsf{C}})$
    is the channel describing step 4(ii)(d)
    of \cref{protocol:verifier_for_upspace}.
    It performs the phase-state injection, appends the
    measurement outcome to the history in $\mathsf{C}$,
    and discards $\mathsf{B}_k$.

    \item 
    $\mathcal{E} \colon
    \mathsf{L}(\HH_{\mathsf{Q}}\otimes\HH_{\mathsf{E}}
    \otimes\HH_{\mathsf{C}}\otimes\HH_{\mathsf{P}})
    \to
    \mathsf{L}(\HH_{\mathsf{Q}})$
    is the channel that applies the remaining verifier
    circuit gates and discards $\mathsf{E}$ and $\mathsf{C}$,
    retaining only the output register $\mathsf{Q}$.
\end{itemize}
With this decomposition in hand, we can now track the implementation errors stage by stage:

\begin{lemma}[Soundness of the composed protocol]
\label{lemma:soundness_bound_of_composed_protocol}
    Fix $n$ and a prover $\mathcal{P}_n$ with initial private register $\mathsf{P}$. Let $\mathcal{F}$ be the accepted map of \cref{protocol:verifier_for_upspace}, and let $\Phi_n$ be the channel on $\mathsf{Q}$ defined by
    \[
        \Phi_n(\rho)\coloneqq
        \Tr_{{\mathsf{E}}}\!\left[
            C_{n,\beta(n)}^{A_W}
            \bigl(\rho\otimes\ps{0}_{\mathsf{E}}\bigr)
            \bigl(C_{n,\beta(n)}^{A_W}\bigr)^\dagger
        \right].
    \]
    For every state $\sigma \in \mathsf{L}(\HH_{\mathsf{P}})$, there exists a constant $\zeta(\sigma) \in [0, 1]$ independent of $\ket{\psi}$ such that
    $$
    \dd\left( \mathcal{F}_{\sigma}, \zeta(\sigma) \cdot \Phi_{n} \right) \leq \frac{K}{R_n},
    \quad \text{where} \quad \mathcal{F}_{\sigma}(\cdot) \coloneqq \mathcal{F}((\cdot) \otimes \sigma):\mathsf{L}(\HH_{\mathsf{Q}}) \to \mathsf{L}(\HH_{\mathsf{Q}}).
    $$
\end{lemma}

\begin{proof}
    For $k=\kappa_i+j$, write
    $\ket{\varphi_{k,\mathbf{h}_k}}
    \coloneqq\ket{\varphi_{\Gamma_{z_j,m_i}}}$.
    Let $\Lambda_{k,\mathbf{h}_k}$ denote the map obtained by
    \cref{lemma:accepted_map_via_phase_state}
    for the accepted map $\mathcal{A}_{k,\mathbf{h}_k}$.
    Specifically, we have $\dd(\mathcal{A}_{k, \bfh_k}, \varphi_{k, \bfh_k} \otimes \Lambda_{k, \bfh_k}) \leq 1/R_n$. We define the map $\mathcal{F}
    '$ to $\mathcal{F}$ in \cref{eqn:overall_protocol_accepted_map} with $\mathcal{S}_k$ replaced by
    \begin{equation}
    \label{eqn:state_preparation_with_history}
    \mathcal{S}_k'(\cdot) = \sum_{\bfh_k} \ps{\bfh_k}_{\mathsf{C}} \otimes \ps{\varphi_{k, \bfh_k}}_{\mathsf{B}_k} \otimes \Lambda_{k, \bfh_k} \left( \braket{\bfh_k | (\cdot) | \bfh_k} \right).
    \end{equation}
    Then it holds that
    $$
    \begin{aligned}
    \dd\left( \mathcal{S}_k, \mathcal{S}_k' \right)  &= \frac{1}{2} \sup_{\HH_{\mathsf{Z}'}} \max_{\ket{\rho} \in \HH_{\mathsf{C}} \otimes \HH_{\mathsf{P}} \otimes \HH_{\mathsf{Z}'} } \sum_{\bfh_k}\left\| \left( \mathcal{A}_{k, \bfh_k} - \varphi_{k, \bfh_k} \otimes \Lambda_{k, \bfh_k} \right) \otimes \id_{\HH_{\mathsf{Z}'}} \left( \braket{\bfh_k | \rho | \bfh_k} \right) \right\|_1 \\
    &\leq \sup_{\HH_{\mathsf{Z}'}} \max_{\ket{\rho} \in \HH_{\mathsf{C}} \otimes \HH_{\mathsf{P}} \otimes \HH_{\mathsf{Z}'} } \sum_{\bfh_k} \Tr[\braket{\bfh_k | \rho | \bfh_k}] \cdot \dd\!\left( \mathcal{A}_{k, \bfh_k}, \varphi_{k, \bfh_k} \otimes \Lambda_{k, \bfh_k} \right) \\
    &\leq \frac{1}{R_n} \sup_{\HH_{\mathsf{Z}'}} \max_{\ket{\rho} \in \HH_{\mathsf{C}} \otimes \HH_{\mathsf{P}} \otimes \HH_{\mathsf{Z}'} } \sum_{\bfh_k} \Tr[\braket{\bfh_k | \rho | \bfh_k}] = \frac{1}{R_n}.
    \end{aligned}
    $$
    Therefore a simple telescoping with the triangle inequality (cf.~\cite[Lemma~13]{10.1145/276698.276708}) 
    yields
    \begin{equation}
    \label{eqn:hybrid_upper_bound}
    \dd\left( \mathcal{F}, \mathcal{F}' \right) \leq \sum_{k=1}^{K} \dd\!\left( \mathcal{S}_k, \mathcal{S}_k' \right) \leq \frac{K}{R_n}.
    \end{equation}
    We likewise put $\mathcal{F}_{\sigma}'(\cdot) = \mathcal{F}'((\cdot) \otimes \sigma)$.
    After all rounds, by \cref{eqn:history_in_terms_of_outcomes}, the history $\mathbf{h}$ reads
    $$
    \mathbf{h}= \left( \mathbf{a}^{(1)}_{\leq m_1}, \mathbf{a}^{(2)}_{\leq m_2}, \dots, \mathbf{a}^{(T)}_{\leq m_T} \right).
    $$
    For a fixed full history $\bfh$,
    let $C_{n,\bfh}$ denote the unnormalized branch operator
    on $\mathsf{Q}\mathsf{E}$ obtained from step 4 of
    \cref{protocol:verifier_for_upspace}
    using ideal resource phase states and selecting
    the measurement outcomes specified by $\bfh$.
    By \cref{lemma:exact_injection_of_diagonal_sign_state},
the injection branch for the $i$-th oracle query
implements that query up to the scalar
$2^{-m_i^2/2}\chi_{\mathbf a_{\leq m_i}^{(i)}}$,
where $\chi_{\mathbf a_{\leq m_i}^{(i)}}\in\{+1,-1\}$.
Consequently,
\[
C_{n,\bfh}
=
\left[
    \prod_{i=1}^{T}
    \left(
        2^{-m_i^2/2}
        \chi_{\mathbf a_{\leq m_i}^{(i)}}
    \right)
\right]
C_{n,\beta(n)}^{A_W}
=
2^{-\frac12\sum_{i=1}^{T}m_i^2}
\left[
    \prod_{i=1}^{T}
    \chi_{\mathbf a_{\leq m_i}^{(i)}}
\right]
C_{n,\beta(n)}^{A_W}.
\]
By \cref{eqn:state_preparation_with_history},
each $\mathcal{S}_k'$ prepares the ideal phase state
and applies $\Lambda_{k,\mathbf h_k}$ to the prover's
private register.
For any density operator $\rho$ on $\mathsf Q$,
expanding $\mathcal{F}'_\sigma$ over all complete
measurement histories therefore gives
    \[
    	\begin{aligned}
\mathcal{F}'_\sigma(\rho)
&=
\left(
    \mathcal{E}\circ
    \bigcirc_{k=1}^{K}
    \bigl(\mathcal{J}_k\circ\mathcal{S}'_k\circ\mathcal{G}_k\bigr)
\right)
\left(
    \rho\otimes\ps{0}_{\mathsf E}
    \otimes\ps{\varnothing}_{\mathsf C}\otimes\sigma
\right)
\\
&=
\Tr_{\mathsf C\mathsf E\mathsf P}\!\left[
    \sum_{\bfh}
    \ps{\bfh}_{\mathsf C}
    \otimes
    C_{n,\bfh}
    \bigl(\rho\otimes\ps{0}_{\mathsf E}\bigr)
    C_{n,\bfh}^\dagger
    \otimes
    \left(
        \Lambda_{K,\bfh_K}\circ\cdots\circ
        \Lambda_{1,\bfh_1}
    \right)(\sigma)
\right]
\\
&=
\underbrace{
    2^{-\sum_{i=1}^{T}m_i^2}
    \sum_{\bfh}
    \Tr\!\left[
        \left(
            \Lambda_{K,\bfh_K}\circ\cdots\circ
            \Lambda_{1,\bfh_1}
        \right)(\sigma)
    \right]
}_{\displaystyle =:\zeta(\sigma)} \cdot~
\Tr_{\mathsf E}\!\left[
    C_{n,\beta(n)}^{A_W}
    \bigl(\rho\otimes\ps{0}_{\mathsf E}\bigr)
    \bigl(C_{n,\beta(n)}^{A_W}\bigr)^\dagger
\right]
\\
&=\zeta(\sigma)\,\Phi_n(\rho),
\end{aligned}
    \]
    where the input-independent scalar $\zeta(\sigma) \in [0, 1]$ follows from the trace-non-increasing property of the maps $\{\Lambda_{j, \bfh_{j}}\}_{j=1}^K$, so is their composition. Combining this identity with \cref{eqn:hybrid_upper_bound} and the data processing inequality, we can readily conclude that
    $$
    \dd\!\left( \mathcal{F}_{\sigma}, \zeta(\sigma) \cdot \Phi_{n} \right) = \dd\!\left( \mathcal{F}_{\sigma}, \mathcal{F}_{\sigma}' \right) \leq \dd\!\left( \mathcal{F}, \mathcal{F}' \right) \leq \frac{K}{R_n}.
    $$
    This completes the proof.
\end{proof}

\subsubsection{Establishing completeness and soundness}
\label{subsec:completeness_and_soundness}

Now we prove \cref{thm:upspace-in-uqip}, which is restated here for convenience.

\upspaceinuqip*

\begin{proof}
  Let $\{U_n\}_{n\in\N}\in\upspace_{\delta^2/2^{14}}$.
Fix polynomials $q,r$ satisfying the hypotheses, and consider
the verifier $\mathcal V_n$ in
\cref{protocol:verifier_for_upspace}.
Write $\mathcal U_n(\cdot)=U_n(\cdot)U_n^\dagger$,
and let $\Phi_n$ be the channel defined in
\cref{lemma:soundness_bound_of_composed_protocol}.

\paragraph{Uniformity.}
By \cref{corollary:unitary_synthesis_in_BQP_with_PSPACE_oracle},
the oracle circuit and the protocol parameters can be computed
in polynomial time from $1^n$.
By \cref{lemma:accepted_map_via_phase_state},
each state-synthesis verifier can be generated uniformly
in polynomial time from the corresponding classical history
and parameters, whose lengths are polynomially bounded.
Since the protocol invokes polynomially many such subprotocols
and all remaining operations take polynomial time,
$\{\mathcal V_n\}_{n\in\N}$ is time-uniform.

\paragraph{Perfect completeness.}
The honest prover follows the strategy specified in
\cref{protocol:verifier_for_upspace}, using fresh private
registers for each state-synthesis subprotocol.
By \cref{lemma:accepted_map_via_phase_state},
each subprotocol accepts with probability $1$.
Thus the same honest prover makes the verifier accept
with probability $1$ on every input, establishing
perfect completeness.

\paragraph{Soundness.}
Fix a sufficiently large $n$ so that
\cref{corollary:unitary_synthesis_in_BQP_with_PSPACE_oracle}
applies. We have
\begin{equation}\label{eq:approximation_bound_for_Phi_n}
  \dd(\Phi_n,\mathcal U_n)
    \leq \frac{\delta(n)}{8}+\frac{1}{16r(n)}
     \leq \frac{3\delta(n)}{16}.
\end{equation}
Fix an arbitrary prover $\mathcal P_n$ and an arbitrary
$n$-qubit input $\ket{\psi}$.
Let $\sigma$ be the prover's initial private state,
which is independent of the input, and let
$\mathcal F_\sigma$ be the unnormalized accepted map
of the protocol with this prover and initial state,
as defined in \cref{lemma:soundness_bound_of_composed_protocol}.
By \cref{lemma:soundness_bound_of_composed_protocol},
there exists $\zeta=\zeta(\sigma)\in[0,1]$ such that
\[
    \dd(\mathcal F_\sigma,\zeta\cdot\Phi_n)
    \leq \frac{K}{R_n}.
\]
Consequently,
\[
    \|\mathcal F_\sigma(\psi)
      -\zeta(\sigma)\Phi_n(\psi)\|_1
    \leq \frac{2K}{R_n}.
\]
Since $\Tr[\Phi_n(\psi)]=1$, taking traces yields
\[
    \left|\Tr[\mathcal F_\sigma(\psi)]-\zeta(\sigma)\right|
    \leq
    \|\mathcal F_\sigma(\psi)
      -\zeta(\sigma)\Phi_n(\psi)\|_1
    \leq \frac{2K}{R_n}.
\]
Suppose that the acceptance probability satisfies
$\Tr[\mathcal F_\sigma(\psi)]>s(n)$.
The output conditioned on acceptance is
$\mathcal F_\sigma(\psi)/\Tr[\mathcal F_\sigma(\psi)]$.
By the triangle inequality,
\[
\begin{aligned}
    \td\!\left(
        \frac{\mathcal F_\sigma(\psi)}
             {\Tr[\mathcal F_\sigma(\psi)]},
        \Phi_n(\psi)
    \right)
    &\leq
    \frac{
        \|\mathcal F_\sigma(\psi)
          -\zeta(\sigma)\Phi_n(\psi)\|_1
        +\left|\zeta(\sigma)-\Tr[\mathcal F_\sigma(\psi)]\right|
    }{2\cdot \Tr[\mathcal F_\sigma(\psi)]}
    \leq
    \frac{2K}{R_n\cdot \Tr[\mathcal F_\sigma(\psi)]}.
\end{aligned}
\]
Combining this with the approximation bound for $\Phi_n$ in \cref{eq:approximation_bound_for_Phi_n} gives
\[
\begin{aligned}
    \td\!\left(
        \frac{\mathcal F_\sigma(\psi)}
             {\Tr[\mathcal F_\sigma(\psi)]},
        \mathcal U_n(\psi)
    \right)
    &\leq
    \frac{2K}{R_n \cdot \Tr[\mathcal F_\sigma(\psi)]}
    +\dd(\Phi_n,\mathcal U_n)
    \leq
    \frac{1}{4q(n)r(n)\cdot\Tr[\mathcal F_\sigma(\psi)]}
    +\frac{3\delta(n)}{16}\\
    &\leq
    \frac{\delta(n)}{4}+\frac{3\delta(n)}{16}
    =\frac{7\delta(n)}{16}<\delta(n),
\end{aligned}
\]
where we used $R_n\geq 8 K \cdot q(n)r(n)$,
$\Tr[\mathcal F_\sigma(\psi)]>s(n)\geq1/q(n)$,
and $1/r(n)\leq\delta(n)$.
This establishes the required soundness condition.

Thus, $\{\mathcal V_n\}_{n\in\N}$ is a time-uniform verifier
family witnessing
$\{U_n\}_{n\in\N}\in\uqip_{0,s,\delta}$,
so we conclude that
\[
    \upspace_{\delta^2/2^{14}}
    \subseteq \uqip_{0,s,\delta}. \qedhere
\]
\end{proof}

\section{Beyond the equality}
\label{sec:extensions}
In this section,
we present two further results
built on the techniques developed for $\uqip = \upspace$.
In \cref{sec:round-compression},
we show that every $\upspace$ unitary admits a constant-round protocol
when multiple copies of the input are available.
\Cref{sec:applications} then turns to cryptography
and proves a complexity-theoretic barrier for $\mathsf{EFI}$ pairs by showing that the existence of $\mathsf{EFI}$ pairs implies $\mathsf{BQP}\neq\pspace$.

\subsection{Constant-round protocols with multiple input copies}
\label{sec:round-compression}

In this subsection, we obtain constant-round protocols for synthesizing $\upspace$ unitaries by combining round compression (\cref{lem:round-compression-quantum-input}) with parallel repetition (\cref{lem:parallel-repetition-synthesis}) on multiple input copies.
The following theorem establishes that the resulting protocol has perfect completeness and a constant soundness threshold.
We defer its proof to the end of this section.

\constantround*

Following Kitaev and Watrous~\cite{KW00} and Chia et al.~\cite{CCHLS26}, we encode the execution history of the original interactive protocol in a history state, a superposition of time labels paired with the corresponding interaction states, and check consistency between consecutive times. As formalized in \cref{lem:round-compression-quantum-input}, this yields a protocol with three verifier rounds, but its soundness guarantee requires an acceptance probability close to one. We then apply \cref{lem:parallel-repetition-synthesis} to lower this threshold to $1/2$ through parallel repetition with multiple input copies.

\begin{lemma}[Round compression]
\label{lem:round-compression-quantum-input}
    Let $1/q_0(n) \leq\eps_0 \leq 1/4$ for some polynomial $q_0$.
    Starting from \cref{protocol:verifier_for_upspace} and using the same soundness argument as in \cref{thm:upspace-in-uqip}, we obtain
    \[
        \upspace_{\eps_0^2/2^{20}}
        \subseteq
        \uqip_{0,1-p_0,\eps_0}^{(1)}[3]
    \] for some $p_0:\N\to(0,1)$ with polynomially bounded $1/p_0$.
    Moreover, for an original protocol with $k$ rounds at precision $\eps_0/8$, the construction achieves soundness threshold $1-\Omega(\eps_0^3/k^2)$.
\end{lemma}

To prove \cref{lem:round-compression-quantum-input}, we use the following form of Uhlmann's theorem to relate the soundness of the compressed protocol to that of the original protocol.

\begin{lemma}[Uhlmann's theorem, {\cite[Theorem~3.22 and Corollary~3.23]{TQI}}]
\label{lem:compression-uhlmann}
    Let $\ket u,\ket v$ be arbitrary vectors in a finite-dimensional space $\HH_{\mathsf A}\otimes\HH_{\mathsf B}$, and let $\rho=\tr_{\mathsf B}\ketbra u$ and $\tau=\tr_{\mathsf B}\ketbra v$. Write $F(\rho,\tau)=\norm{\sqrt\rho\sqrt\tau}_1$. For every operator $Q$ on $\HH_{\mathsf B}$ with $\norm{Q}_{\op}\leq1$,
    \[
        \left|\bra u(\id_{\mathsf A}\otimes Q)\ket v\right|
        \leq F(\rho,\tau).
    \]
    Moreover, equality is attained by a unitary $Q$.
\end{lemma}

\begin{proof}[Proof of \cref{lem:round-compression-quantum-input}]
    Choose $N$ to be the least power of two at least $64(k+1)/\eps_0$, and set $p_0=\eps_0^2/(2^{20}kN)$.
    Run \cref{protocol:verifier_for_upspace} with the parameters of \cref{thm:upspace-in-uqip}, taking precision $\eps_0/8$ and soundness threshold $1/2$. By \cref{thm:upspace-in-uqip}, it has perfect completeness and output error at most $\eps_0/8$ for every family in $\upspace_{\eps_0^2/2^{20}}$.
    Fix the input length and write $U$ for the target unitary. By deferring intermediate measurements, recording their outcomes coherently in private ancillas, and retaining discarded registers, we represent the original verifier's actions by unitaries $V_1,\ldots,V_k$. 

    The compressed verifier uses private registers $\mathsf Z=\mathsf{Z_0Z_1C}$ and message registers $\mathsf M=\mathsf{M_0M_1}$, with $\mathsf C$ the clock. Write $\mathsf E$ for the prover's private memory, initially independent of the input. 
    For honest prover operations $P_1,\ldots,P_k$ and input state $\ket{\psi}$, define
    \[
        \ket{h_0}=\iota\ket\psi\otimes\ket0_{\mathsf E},
        \qquad
        \ket{h_j}=V_jP_j\ket{h_{j-1}}\quad(1\leq j\leq k),
    \]
    where the isometry is defined as $\iota: \ket{\psi} \longmapsto \ket{\psi} \ket{0^*}$.
    We construct the compressed verifier in three rounds, discarding the initial prover message required by our convention.

    \paragraph{First round: requesting the history.}
    To fix the initial state while allowing the prover to prepare later states, the verifier prepares
    \[
        \frac1{\sqrt N}\left(\ket0_{\mathsf C}\bigl(\iota\ket\psi\bigr)_{\mathsf Z_0}\ket0_{\mathsf M_0}\ket0_{\mathsf M_1}+\sum_{t=1}^{N-1}\ket t_{\mathsf C}\ket0_{\mathsf Z_0}\bigl(\iota\ket\psi\bigr)_{\mathsf M_0}\ket{\min\{t,k\}}_{\mathsf M_1}\right).
    \]
    It sends $\mathsf{M_0M_1}$ to the prover and retains $\mathsf{Z_0C}$, with $\mathsf Z_1$ still zero. The honest prover applies the corresponding prefix of the original interaction and returns $\mathsf{M_0M_1}$. The joint state is then
    \[
        \frac1{\sqrt N}\left(\ket0_{\mathsf C}\bigl(\iota\ket\psi\bigr)_{\mathsf Z_0}\ket0_{\mathsf{M_0E}}\ket0_{\mathsf M_1}+\sum_{t=1}^{N-1}\ket t_{\mathsf C}\ket0_{\mathsf Z_0}\ket{h_{\min\{t,k\}}}_{\mathsf{M_0E}}\ket{\min\{t,k\}}_{\mathsf M_1}\right).
    \]

    \paragraph{Second round: comparing adjacent times.}
    Upon receiving $\mathsf{M_0M_1}$, the verifier checks the label in $\mathsf M_1$ with $\min\{t,k\}$ and uncomputes it using $\mathsf C$. The verifier then rearranges the registers and projects $\mathsf Z_0$ onto zero, rejecting on failure. The resulting unnormalized state is
    \[
        \frac1{\sqrt N}\left(\ket0_{\mathsf C}\ket0_{\mathsf Z_0}\bigl(\iota\ket\psi\otimes\ket{e_0}\bigr)_{\mathsf{Z_1M_0E}}+\sum_{t=1}^{N-1}\ket t_{\mathsf C}\ket0_{\mathsf Z_0}\ket{v_t}_{\mathsf{Z_1M_0E}}\right),
    \]
    where $\ket{v_0}=\iota\ket\psi\otimes\ket{e_0}$, $\ket{v_t}$ is on $\mathsf{Z_1M_0E}$, and $\ket{e_0}$ is independent of $\psi$.

    To compare consecutive states, the verifier chooses $b\in\{0,1\}$ uniformly and pairs $j-1$ with $j$ whenever $1\leq j\leq k$ and $j\equiv b\;(\mathrm{mod}\,2)$. All other times are unpaired. It undoes $V_j$ at the later endpoint of each pair and prepares
\[
    \resizebox{\linewidth}{!}{$\displaystyle
    \frac1{\sqrt N}\!\left[
    \sum_{\substack{1\leq j\leq k\\j\equiv b \bmod{2}}}\!
    \left(
    \ket{j-1}_{\mathsf C}\!\ket{b,j-1}_{\mathsf M_1}\!\ket{v_{j-1}}_{\mathsf{Z_1M_0E}}
    +\!
    \ket j_{\mathsf C}\!\ket{b,j}_{\mathsf M_1}\!V_j^\dagger\!\ket{v_j}_{\mathsf{Z_1M_0E}}
    \right)
    +\!
    \sum_{t\text{ unpaired}}\!
    \ket t_{\mathsf C}\!\ket{b,t}_{\mathsf M_1}\!\ket{v_t}_{\mathsf{Z_1M_0E}}
    \right].
    $}
\]
    It sends $\mathsf{M_0M_1}$ to the prover, retaining $\mathsf{Z_1C}$ and the bit $b$ in $\mathsf Z_0$. The honest prover applies $P_j^\dagger$ to register $\mathsf{M_0E}$ at the later endpoint of each pair, acts as the identity at all other times, and returns $\mathsf{M_0M_1}$. With honest replies in both exchanges, the joint state becomes
\[
    \resizebox{\linewidth}{!}{$\displaystyle
    \frac1{\sqrt N}\!\left[
    \sum_{\substack{1\leq j\leq k\\j\equiv b \bmod{2}}}\!
    \left(
    \ket{j-1}_{\mathsf C}\!\ket{b,j-1}_{\mathsf M_1}
    +\!
    \ket j_{\mathsf C}\!\ket{b,j}_{\mathsf M_1}
    \right)\!\ket{h_{j-1}}_{\mathsf{Z_1M_0E}}
    +\!
    \sum_{t\text{ unpaired}}\!
    \ket t_{\mathsf C}\!\ket{b,t}_{\mathsf M_1}\!
    \ket{h_{\min\{t,k\}}}_{\mathsf{Z_1M_0E}}
    \right].
    $}
\]

    \paragraph{Third round: checking and returning the output.}
    The verifier checks the returned label against $(b,t)$ and erases it, rejecting on failure. It then measures the clock in the orthonormal basis
    \[
        \left\{\frac{\ket{j-1}+\ket j}{\sqrt2},
        \frac{\ket{j-1}-\ket j}{\sqrt2}:
        1\leq j\leq k,\ j\equiv b\;(\mathrm{mod}\,2)\right\}
        \ \cup\ \{\ket t:t\text{ unpaired}\}.
    \]
    At a paired edge $(j-1,j)$, a minus outcome rejects; a plus outcome accepts and returns $\ket{0^n}$. At an unpaired time $t\geq k$, the verifier checks the original acceptance flag and returns the original output on acceptance. At any other unpaired time, it accepts and returns $\ket{0^n}$. This completes the third verifier round.

    \paragraph{Perfect completeness.}
    For honest replies, $\ket{v_t}=\ket{h_{\min\{t,k\}}}$ after the initial checks. At the later endpoint of each pair, the two inverse operations give $P_j^\dagger V_j^\dagger\ket{h_j}=\ket{h_{j-1}}$. Thus the endpoint states agree, and every consistency test accepts. The original protocol's perfect completeness gives acceptance at all final times as well. State preparation and clock checks take polynomial time, as do the original verifier operations, so the compressed verifier has polynomial size.

    \paragraph{Soundness.}
    Fix an arbitrary prover and let $p$ be its acceptance probability. The initial checks give $\norm{v_t}\leq1$ and $\ket{v_0}=\iota\ket\psi\otimes\ket{e_0}$, where $\ket{e_0}$ is independent of $\psi$. Also, $\ket{v_t}=\ket{v_k}$ for all $t\geq k$, because the prover receives the same input and label at these times.

    We first use the consistency tests to recover operations for the original prover. Choose unitaries $Q_j$ on $\mathsf{M_0E}$ minimizing $\norm{Q_jv_{j-1}-V_j^\dagger v_j}$. Projecting the returned label onto $\ket{b,t}$ induces an operator on $\mathsf{M_0E}$ with operator norm at most one. By \cref{lem:compression-uhlmann}, the rejection probability on a tested pair is at least $\norm{Q_jv_{j-1}-V_j^\dagger v_j}^2/(2N)$. Averaging over the two choices of parity and including the $\mathsf{Z_0}$ checks in the second round gives
    \begin{equation}
    \label{eq:compression-rejection}
        1-p\geq
        \frac1N\sum_{t=0}^{N-1}\bigl(1-\norm{v_t}^2\bigr)
        +\frac1{4N}\sum_{j=1}^k\norm{Q_jv_{j-1}-V_j^\dagger v_j}^2.
    \end{equation}

    Suppose $p>1-p_0$. Run the original protocol with these operations, starting from $\ket{\phi_0}=\ket{v_0}/\norm{v_0}$ and setting $\ket{\phi_j}=V_jQ_j\ket{\phi_{j-1}}$. This is a valid execution on the prescribed input, since the initial prover state is independent of $\psi$. By \cref{eq:compression-rejection} and Cauchy-Schwarz,
    \[
        \norm{\phi_j-v_j}
        \leq 1-\norm{v_0}+\sum_{i=1}^j\norm{Q_iv_{i-1}-V_i^\dagger v_i}
        \leq Np_0+2\sqrt{kNp_0}< \frac{\eps_0}{256}.
    \]
    Thus the supplied state at every final time is close to the final state of this execution.

    We now apply the original soundness guarantee to the recovered execution. Let $p_{\mathrm{rej}}$ be its rejection probability. Since $\norm{\phi_k-v_k}<\eps_0/256$, applying the original acceptance test to $v_k$ gives rejection weight at least $p_{\mathrm{rej}}-\eps_0/128$. At an unpaired final time, the prover's second reply cannot change the original acceptance flag's statistics before the label check. That check commutes with the acceptance test, so any rejection by the latter also causes the compressed protocol to reject. For either parity, there are at least $N-(k+1)$ such clock values, each carrying $v_k$ with coefficient $1/\sqrt N$. Hence
    \[
        1-p\geq\left(1-\frac{k+1}{N}\right)\left(p_{\mathrm{rej}}-\frac{\eps_0}{128}\right)
        \quad\Rightarrow\quad
        p_{\mathrm{rej}}\leq\frac{1-p}{1-(k+1)/N}+\frac{\eps_0}{128}<\frac12.
    \]
    The original soundness guarantee therefore bounds the recovered execution's accepting output error by $\eps_0/8$.

    Let $\rho^{\mathrm{acc}}$ denote the compressed protocol's output state conditioned on acceptance. We now compare it with the recovered execution's accepting output. At unpaired final times, replacing $v_k$ by $\phi_k$ changes both the accepting output and its weight, contributing at most $\eps_0/128$. The label check discards weight at most $1-p$, and all other clock values contribute accepted weight at most $(k+1)/N$. Dividing by the total acceptance probability $p$ gives
    \[
        \td(\rho^{\mathrm{acc}},U\psi U^\dagger)
        \leq\frac{\eps_0/128+\eps_0/8+(1-p)+(k+1)/N}{p}
        <\eps_0.
    \]
    Finally, the choices of $N$ and $p_0$ give $p_0=\Theta(\eps_0^3/k^2)$. 
\end{proof}

We next reduce the soundness threshold of the compressed protocol to $1/2$ by running several instances in parallel, accepting only if all of them accept, and returning one of their outputs uniformly at random.

\begin{lemma}[Parallel repetition]
\label{lem:parallel-repetition-synthesis}
    Let $m,r,\ell:\N\to\N$ be polynomially bounded and computable in polynomial time. For $0<p_0\leq1/2$ and $0\leq\eps_0,\eps\leq1$ satisfying $2\eps_0+\log2/(rp_0)\leq\eps^2$, we have
    \[
        \uqip_{0,1-p_0,\eps_0}^{(m)}[\ell]
        \subseteq
        \uqip_{0,1/2,\eps}^{(2mr)}[\ell].
    \]
\end{lemma}

The proof of \cref{lem:parallel-repetition-synthesis} relies on the following product rule.

\begin{lemma}[Product rule, consequence of {\cite[Theorem~9]{GW07}}]
\label{lem:interactive-payoff-product}
    Run $r$ quantum verifier-prover interactions in parallel, with the same number of rounds, product initial verifier states, and separate verifier operations. For nonnegative outcome payoffs $w_i(x_i)$, let $\omega_i$ be the maximum expected payoff in instance $i$. Then
    \[
        \sup_{\mathcal P}\E\!\left[\prod_{i=1}^r w_i(x_i)\right]
        =\prod_{i=1}^r \omega_i,
    \]
    where $\mathcal P$ ranges over all joint prover strategies.
\end{lemma}

\begin{proof}[Proof of \cref{lem:parallel-repetition-synthesis}]
    Let $r'$ be the least power of two at least $r$. Run $r'$ instances of the original protocol in parallel on independent groups of $m$ input copies, accept if all instances accept, and return a uniformly chosen output. Since $r'\leq2r$, the verifier uses at most $2mr$ input copies.

    Fix $\ket\psi$ and put $\sigma=U\psi U^\dagger$. Let $\mathcal A$ be the event that all instances accept and suppose $p_3=\pr[\mathcal A]>1/2$. Let $\rho^{\mathrm{acc}}$ be the uniformly chosen output conditioned on $\mathcal A$. In hypothetical measurements $\{\sigma,\id-\sigma\}$ of the $r'$ outputs, let $N_{\mathrm{err}}$ count the outcomes $\id-\sigma$. Then
    \[
        1-\tr(\sigma\rho^{\mathrm{acc}})
        =\frac{\E[N_{\mathrm{err}}\mid\mathcal A]}{r'}.
    \]
    As $\sigma$ is pure, the Fuchs-van de Graaf inequality shows that it suffices to bound this conditional average by $\eps^2$.

    We bound this average against joint prover strategies using nonnegative payoffs. Assign payoff $0$ on rejection, $1$ on acceptance with outcome $\sigma$, and $1+\lambda$ on acceptance with outcome $\id-\sigma$, where $\lambda=p_0/(1-p_0)$. For one instance, let $p_1$ be its acceptance probability and $p_2$ the joint probability of acceptance and outcome $\id-\sigma$. Its expected payoff is $p_1+\lambda p_2$. Soundness gives $p_2\leq p_1\eps_0$ when $p_1>1-p_0$, while $p_2\leq p_1$ always. Thus
    \[
        p_1+\lambda p_2\leq
        \max\{1+\lambda\eps_0,(1+\lambda)(1-p_0)\}
        =1+\lambda\eps_0.
    \]

    The product of the payoffs is $\one_{\mathcal A}(1+\lambda)^{N_{\mathrm{err}}}$. Applying \cref{lem:interactive-payoff-product} and Jensen's inequality conditioned on $\mathcal A$ gives
    \[
        p_3(1+\lambda)^{\E[N_{\mathrm{err}}\mid\mathcal A]}
        \leq
        \E\!\left[\one_{\mathcal A}(1+\lambda)^{N_{\mathrm{err}}}\right]
        \leq(1+\lambda\eps_0)^{r'}.
    \]
    Taking logarithms yields
    \[
        \frac{\E[N_{\mathrm{err}}\mid\mathcal A]}{r'}
        \leq\frac{\log(1/p_3)}{r'\log(1+\lambda)}
          +\frac{\log(1+\lambda\eps_0)}{\log(1+\lambda)}
        \leq\frac{\log2}{rp_0}+2\eps_0\leq\eps^2,
    \]
    where $\log(1+\lambda)=-\log(1-p_0)\geq p_0$ and $\lambda\leq2p_0$.

    Independent honest prover strategies give perfect completeness, and sending the parallel messages together preserves the number of rounds.
\end{proof}

We now combine \cref{lem:round-compression-quantum-input,lem:parallel-repetition-synthesis}
to prove our main result in this subsection.

\begin{proof}[Proof of \cref{thm:constant-round-synthesis}]
    Fix a family in $\upspace_{\eps^4/2^{24}}$, and let $k$ be the number of rounds of its original protocol at precision $\eps^2/32$. Applying \cref{lem:round-compression-quantum-input} with $\eps_0=\eps^2/4$ gives membership in $\uqip_{0,1-p_0,\eps^2/4}^{(1)}[3]$ for $p_0=\Theta(\eps^6/k^2)$.
    Choose $r=\lceil2/(p_0\eps^2)\rceil$ and apply \cref{lem:parallel-repetition-synthesis} with $m=1$ and $\ell=3$. The family therefore belongs to $\uqip_{0,1/2,\eps}^{(2r)}[3]$. Since $k$ and $1/\eps$ are polynomially bounded, so is $r$. All parameters and verifier circuits are computable in polynomial time, and we conclude the proof.
\end{proof}

\subsection{A complexity-theoretic barrier for \texorpdfstring{$\mathsf{EFI}$}{EFI} pairs}
\label{sec:applications}

This subsection uses the unitary synthesis lemma (\cref{lemma:usynthesis_with_pspace_oracle}) to show that the existence of $\mathsf{EFI}$ pairs separates $\mathsf{BQP}$ from $\pspace$. Intuitively, $\mathsf{EFI}$ pairs are two families of quantum states that are statistically far but cannot be efficiently distinguished. We now recall the formal definition.

\begin{definition}[$\mathsf{EFI}$ pairs, {\cite[Definition~7]{BCQ23}}]
\label{def:efi}
    An $\mathsf{EFI}$ pair is a family $\{\rho_{0,n},\rho_{1,n}\}_{n\in\N}$ of pairs of mixed states satisfying:
    \begin{enumerate}
        \item \textbf{Efficient (E) generation.} A uniform quantum polynomial-time algorithm $G$, on input $(1^n,b)$, outputs $\rho_{b,n}$ for each $b\in\{0,1\}$.
        \item \textbf{Statistical farness (F).} There is a polynomial $q\geq1$ such that
        $\td(\rho_{0,n},\rho_{1,n})\geq1/q(n)$ for all sufficiently large $n$.
        \item \textbf{Computational indistinguishability (I).} For every quantum polynomial-time distinguisher $D$ with polynomial-size quantum advice $\{\sigma_n\}_n$, the advantage
        $$
            \left|\pr[D(1^n,\rho_{0,n}\otimes\sigma_n)=0]
            -\pr[D(1^n,\rho_{1,n}\otimes\sigma_n)=0]\right|
        $$
        is negligible in $n$.
    \end{enumerate}
\end{definition}

We have the following theorem, which establishes a complexity-theoretic barrier to proving the existence of $\mathsf{EFI}$ pairs, answering a question posed by Brakerski, Canetti, and Qian~\cite{BCQ23}.

\EFI*

We use the following lemma, which gives a polynomial-space measurement distinguishing two efficiently generated state families with advantage close to their trace distance. By \cref{lemma:usynthesis_with_pspace_oracle}, this attack can be implemented efficiently with a $\pspace$ oracle. If $\mathsf{BQP}=\pspace$, simulating the oracle queries yields an efficient quantum attack on any $\mathsf{EFI}$ pair.

\begin{lemma}[Algorithmic Holevo-Helstrom measurement, {\cite[Theorem~16]{LLW26}}]
\label{lem:algorithmic-helstrom-measurement}
     Consider efficiently generated state families $\{\rho_{0,n},\rho_{1,n}\}_{n\in\N}$ and a polynomial-time computable function $\eps:\N\to(0,1/4]$ with polynomially bounded $1/\eps$.
    There is a space-uniform unitary circuit family $W_n$ with ancillas initialized to zero. Measuring a designated output qubit on inputs $\rho_{0,n},\rho_{1,n}$ gives outcome-zero probabilities $p_0(n),p_1(n)$ satisfying
    $$
        p_0(n)-p_1(n)
        \geq\td(\rho_{0,n},\rho_{1,n})-\eps(n).
    $$
\end{lemma}

\begin{proof}[Proof of \cref{cor:efi-separates-bqp-pspace}]
    Assume $\mathsf{BQP}=\pspace$, and fix an $\mathsf{EFI}$ pair $\{\rho_{0,n},\rho_{1,n}\}_{n\in\N}$.
    Choose an polynomial $q\geq1$ such that $\td(\rho_{0,n},\rho_{1,n})\geq1/q(n)$ for all sufficiently large $n$, and set $\eps(n)=1/(16q(n))$. Applying \cref{lem:algorithmic-helstrom-measurement} at this precision gives circuits $W_n$ whose outcome probabilities satisfy $p_0(n)-p_1(n)\geq1/q(n)-\eps(n)$.

    The entries of $W_n$ are $\pspace$-computable
    by \cref{lemma:arthm_of_PSPACE_num}.
    By padding with unused qubits, we can reindex the family by its qubit count, taking the identity at all remaining sizes.
    By \cref{lemma:usynthesis_with_pspace_oracle}, there is a fixed language $A\in\pspace$ and a time-uniform family of polynomial-size oracle circuits $C_n^A$ satisfying
    $$
        \left\|C_n^A\iota-\iota W_n\right\|_{\op}\leq\eps(n),
    $$
    where $\iota$ appends the synthesis workspace in the zero state.

    Since $\mathsf{PSPACE} = \mathsf{BQP}$, we have $A\in\mathsf{BQP}$ and thus each query can be simulated by an efficient quantum circuit. Let $T(n)\geq1$ be a polynomial bound on the number of queries of $C_n^A$, and amplify the decision algorithm for $A$ to error at most $\eps(n)^2/(4T(n)^2)$. To simulate the oracle transformation $\ket{x,b}\mapsto\ket{x,b\oplus A(x)}$ coherently, we use the standard procedure: preserve the query register $x$, run the amplified decision circuit without measurements, XOR its output into the response qubit $b$, and apply the inverse circuit to uncompute the workspace. The resulting query simulation has operator-norm error at most $\eps(n)/T(n)$ on inputs with zero workspace. Replacing all queries with fresh workspace for each adds at most $\eps(n)$ to the synthesis error, giving a circuit within $2\eps(n)$ of $W_n$.

    Measuring the output qubit gives a uniform quantum polynomial-time distinguisher $D$. Each outcome probability changes by at most $2\eps(n)$, so
    $$
    \begin{aligned}
        &\pr[D(1^n,\rho_{0,n})=0]-\pr[D(1^n,\rho_{1,n})=0] \geq p_0(n)-p_1(n)-4\eps(n)
        \geq\frac{1}{2q(n)}.
    \end{aligned}
    $$
    This contradicts the computational indistinguishability of the $\mathsf{EFI}$ pair.
\end{proof}

\endgroup

\newpage
\bibliographystyle{alphaurl}
\bibliography{ref}

@InProceedings{RY22,
  author =	{Rosenthal, Gregory and Yuen, Henry},
  title =	{Interactive Proofs for Synthesizing Quantum States and Unitaries},
  booktitle =	{13th Innovations in Theoretical Computer Science Conference (ITCS 2022)},
  pages =	{112:1--112:4},
  series =	{Leibniz International Proceedings in Informatics (LIPIcs)},
  ISBN =	{978-3-95977-217-4},
  ISSN =	{1868-8969},
  year =	{2022},
  volume =	{215},
  doi =		{10.4230/LIPIcs.ITCS.2022.112},
  note = {Full version available at \href{https://arxiv.org/abs/2108.07192}{\path{arXiv:2108.07192}}}
}

@INPROCEEDINGS{MY23,
  author={Metger, Tony and Yuen, Henry},
  booktitle={2023 IEEE 64th Annual Symposium on Foundations of Computer Science (FOCS)}, 
  title={{$\mathsf{stateQIP} = \mathsf{statePSPACE}$}},
  year={2023},
  volume={},
  number={},
  pages={1349-1356},
  note = {Full version available at \href{https://arxiv.org/abs/2301.07730}{\path{arXiv:2301.07730}}},
  doi={10.1109/FOCS57990.2023.00082}}

@InProceedings{BEM+26,
  author =	{Bostanci, John and Efron, Yuval and Metger, Tony and Poremba, Alexander and Qian, Luowen and Yuen, Henry},
  title =	{Unitary Complexity and the {Uhlmann} Transformation Problem},
  booktitle =	{17th Innovations in Theoretical Computer Science Conference (ITCS 2026)},
  pages =	{24:1--24:17},
  series =	{Leibniz International Proceedings in Informatics (LIPIcs)},
  ISBN =	{978-3-95977-410-9},
  ISSN =	{1868-8969},
  year =	{2026},
  volume =	{362},
  doi =		{10.4230/LIPIcs.ITCS.2026.24}
}

@INPROCEEDINGS{4262757,
  author={Aaronson, Scott and Kuperberg, Greg},
  booktitle={Twenty-Second Annual IEEE Conference on Computational Complexity (CCC'07)}, 
  title={Quantum versus Classical Proofs and Advice}, 
  year={2007},
  volume={},
  number={},
  pages={115-128},
  doi={10.1109/CCC.2007.27}}

@misc{aaronson2016complexityquantumstatestransformations,
      title={The complexity of quantum states and transformations: from quantum money to black holes},
      author={Scott Aaronson},
      year={2016},
      eprint        = {1607.05256},
      archivePrefix = {arXiv}
}

@misc{OpenAI2026UnitarySynthesis,
  author = {{OpenAI}},
  title  = {Polynomial-Time Unitary Synthesis from a {Boolean} Oracle},
  year   = {2026},
  note = {\url{https://github.com/openai/math/blob/main/preprints/Polynomial-Time-Unitary-Synthesis-from-a-Boolean-Oracle-October-5-2026/paper.pdf}}
}

@book{Nielsen2012,
  title = {Quantum computation and quantum information: 10th anniversary edition},
  ISBN = {9780511976667},
  DOI = {10.1017/cbo9780511976667},
  publisher = {Cambridge University Press},
  author = {Nielsen, Michael A. and Chuang, Isaac L.},
  year = {2012}
}

@INPROCEEDINGS{HKOT23,
  author={Haah, Jeongwan and Kothari, Robin and O’Donnell, Ryan and Tang, Ewin},
  booktitle={2023 IEEE 64th Annual Symposium on Foundations of Computer Science (FOCS)}, 
  title={Query-optimal estimation of unitary channels in diamond distance}, 
  year={2023},
  volume={},
  number={},
  pages={363-390},
  doi={10.1109/FOCS57990.2023.00028}}

@article{PhysRevLett.93.130502,
  title = {Quantum Circuits for General Multiqubit Gates},
  author = {M{\"o}tt{\"o}nen, Mikko and Vartiainen, Juha J. and Bergholm, Ville and Salomaa, Martti M.},
  journal = {Phys. Rev. Lett.},
  volume = {93},
  issue = {13},
  pages = {130502},
  numpages = {4},
  year = {2004},
  publisher = {American Physical Society},
  doi = {10.1103/PhysRevLett.93.130502}
}

@article{PhysRevA.71.052330,
  title = {Quantum circuits with uniformly controlled one-qubit gates},
  author = {Bergholm, Ville and Vartiainen, Juha J. and M{\"o}tt{\"o}nen, Mikko and Salomaa, Martti M.},
  journal = {Phys. Rev. A},
  volume = {71},
  issue = {5},
  pages = {052330},
  numpages = {7},
  year = {2005},
  publisher = {American Physical Society},
  doi = {10.1103/PhysRevA.71.052330}
}

@ARTICLE{1629135,
  author={Shende, V.V. and Bullock, S.S. and Markov, I.L.},
  journal={IEEE Transactions on Computer-Aided Design of Integrated Circuits and Systems}, 
  title={Synthesis of quantum-logic circuits}, 
  year={2006},
  volume={25},
  number={6},
  pages={1000-1010},
  doi={10.1109/TCAD.2005.855930}}

@book{Saad2003,
  title = {Iterative methods for sparse linear systems},
  DOI = {10.1137/1.9780898718003},
  publisher = {Society for Industrial and Applied Mathematics},
  author = {Saad,  Yousef},
  year = {2003}
}

@misc{kretschmer2026unitarycomplexitypolynomialspace,
      title={Unitary complexity in polynomial space}, 
      author={William Kretschmer and Ewin Tang},
      year={2026},
      eprint={2610.03705},
      archivePrefix={arXiv},
      primaryClass={quant-ph}
}

@book{Horn2012,
  title = {Matrix analysis},
  ISBN = {9780521548236},
  DOI = {10.1017/cbo9781139020411},
  publisher = {Cambridge University Press},
  author = {Horn, Roger A. and Johnson, Charles R.},
  year = {2012},
  month = Oct 
}

@article{WATROUS1999281,
title = {Space-Bounded Quantum Complexity},
journal = {Journal of Computer and System Sciences},
volume = {59},
number = {2},
pages = {281-326},
year = {1999},
issn = {0022-0000},
doi = {https://doi.org/10.1006/jcss.1999.1655},
author = {John Watrous}
}

@Article{Watrous2003,
author={Watrous, John},
title={On the complexity of simulating space-bounded quantum computations},
journal={computational complexity},
year={2003},
month={Jun},
day={01},
volume={12},
number={1},
pages={48-84},
issn={1420-8954},
doi={10.1007/s00037-003-0177-8}
}

@article{SAVITCH1970177,
title = {Relationships between nondeterministic and deterministic tape complexities},
journal = {Journal of Computer and System Sciences},
volume = {4},
number = {2},
pages = {177-192},
year = {1970},
issn = {0022-0000},
doi = {https://doi.org/10.1016/S0022-0000(70)80006-X},
author = {Walter J. Savitch}
}

@book{Montalban2021,
  author    = {Montalb{\'a}n, Antonio},
  title     = {Computable structure theory: within the arithmetic},
  series    = {Perspectives in Logic},
  publisher = {Cambridge University Press},
  year      = {2021},
  doi       = {10.1017/9781108525749},
  isbn      = {9781108423298}
}

@article{Seif2023,
  title = {Shadow distillation: quantum error mitigation with classical shadows for near-term quantum processors},
  volume = {4},
  DOI = {10.1103/prxquantum.4.010303},
  number = {1},
  journal = {PRX Quantum},
  publisher = {American Physical Society (APS)},
  author = {Seif,  Alireza and Cian,  Ze-Pei and Zhou,  Sisi and Chen,  Senrui and Jiang,  Liang},
  year = {2023}
}

@inproceedings{QSVT,
author = {Gily{\'e}n, Andr{\'a}s and Su, Yuan and Low, Guang Hao and Wiebe, Nathan},
title = {Quantum singular value transformation and beyond: exponential improvements for quantum matrix arithmetics},
year = {2019},
isbn = {9781450367059},
doi = {10.1145/3313276.3316366},
booktitle = {Proceedings of the 51st Annual ACM SIGACT Symposium on Theory of Computing},
pages = {193–204},
numpages = {12},
note = {Full version available at \href{https://arxiv.org/abs/1806.01838}{\path{arXiv:1806.01838}}}
}

@INPROCEEDINGS{11369061,
  author={Dalzell, Alexander M. and Gilyén, András and Hann, Connor T. and McArdle, Sam and Salton, Grant and Nguyen, Quynh T. and Kubica, Aleksander and Brandão, Fernando G.S.L.},
  booktitle={2025 IEEE 66th Annual Symposium on Foundations of Computer Science (FOCS)}, 
  title={A distillation-teleportation protocol for fault-tolerant {QRAM}},
  year={2025},
  volume={},
  number={},
  pages={38-74},
  note={ Full version available at \href{https://arxiv.org/abs/2505.20265}{\path{arXiv:2505.20265}}},
  doi={10.1109/FOCS63196.2025.00008}}

@article{10.1098/rspa.1998.0164,
    author = {Cleve, R. and Ekert, A. and Macchiavello, C. and Mosca, M.},
    title = {Quantum algorithms revisited},
    journal = {Proceedings of the Royal Society A: Mathematical, Physical and Engineering Sciences},
    volume = {454},
    number = {1969},
    pages = {339-354},
    year = {1998},
    doi = {10.1098/rspa.1998.0164},
}

@inproceedings{10.1145/276698.276708,
author = {Aharonov, Dorit and Kitaev, Alexei and Nisan, Noam},
title = {Quantum circuits with mixed states},
year = {1998},
isbn = {0897919629},
doi = {10.1145/276698.276708},
booktitle = {Proceedings of the Thirtieth Annual ACM Symposium on Theory of Computing},
pages = {20–30},
numpages = {11}
}

@Article{Gough2009,
author={Gough, J.
and James, M. R.},
title={Quantum feedback networks: {Hamiltonian} formulation},
journal={Communications in Mathematical Physics},
year={2009},
day={01},
volume={287},
number={3},
pages={1109-1132},
doi={10.1007/s00220-008-0698-8}
}

@book{Conway2007,
  title = {A course in functional analysis},
  ISBN = {9781475743838},
  ISSN = {0072-5285},
  DOI = {10.1007/978-1-4757-4383-8},
  journal = {Graduate Texts in Mathematics},
  publisher = {Springer New York},
  author = {Conway,  John B.},
  year = {2007}
}

@INPROCEEDINGS{261572,
  author={Doyle, J. and Packard, A. and Zhou, K.},
  booktitle={[1991] Proceedings of the 30th IEEE Conference on Decision and Control}, 
  title={Review of {LFTs}, {LMIs}, and {$\mu$}},
  year={1991},
  volume={},
  number={},
  pages={1227-1232 vol.2},
  doi={10.1109/CDC.1991.261572}}

@article{doi:10.1142/S1793536911000787,
author = {Tropp, Joel A.},
title = {Improved analysis of the subsampled randomized {Hadamard} transform},
journal = {Advances in Adaptive Data Analysis},
volume = {03},
number = {01n02},
pages = {115-126},
year = {2011},
doi = {10.1142/S1793536911000787}
}

@article{JJUW11,
  author        = {Jain, Rahul and Ji, Zhengfeng and Upadhyay, Sarvagya and Watrous, John},
  title         = {{$\mathsf{QIP} = \mathsf{PSPACE}$}},
  journal       = {Journal of the ACM},
  volume        = {58},
  number        = {6},
  pages         = {30:1--30:27},
  year          = {2011},
  doi           = {10.1145/2049697.2049704}
}

@inproceedings{Ros24,
author = {Gregory Rosenthal},
title = {Efficient Quantum State Synthesis with One Query},
booktitle = {Proceedings of the 2024 Annual ACM-SIAM Symposium on Discrete Algorithms (SODA)},
pages = {2508-2534},
year          = {2024},
doi = {10.1137/1.9781611977912.89}
}

@inproceedings{LMW24,
author = {Lombardi, Alex and Ma, Fermi and Wright, John},
title = {A One-Query Lower Bound for Unitary Synthesis and Breaking Quantum Cryptography},
year = {2024},
isbn = {9798400703836},
doi = {10.1145/3618260.3649650},
booktitle = {Proceedings of the 56th Annual ACM Symposium on Theory of Computing},
pages = {979–990},
numpages = {12}
}

@misc{yuen2022unitarysynthesis,
  author       = {Yuen, Henry},
  title        = {Lecture 6: unitary synthesis},
  year         = {2022},
  howpublished = {Lecture Notes for COMS E6998:
                  Frontiers of Quantum Complexity
                  and Cryptography},
  institution  = {Columbia University},
  note         = {\url{https://www.henryyuen.net/classes/spring2022/lec6-unitarysynthesis.pdf}}
}

@misc{tang2026amplitudeamplificationestimationrequire,
      title={Amplitude amplification and estimation require inverses}, 
      author={Ewin Tang and John Wright},
      year={2026},
      eprint={2507.23787},
      archivePrefix={arXiv},
      primaryClass={quant-ph}
}

@book{symm_func_1995,
    author = {Macdonald, I G},
    title = {Symmetric functions and {Hall} polynomials},
    publisher = {Oxford University Press},
    year = {1995},
    month = {03},
    isbn = {9780198534891},
    doi = {10.1093/oso/9780198534891.001.0001},
    url = {https://doi.org/10.1093/oso/9780198534891.001.0001},
}

@inproceedings{10.1145/2591796.2591854,
author = {Berry, Dominic W. and Childs, Andrew M. and Cleve, Richard and Kothari, Robin and Somma, Rolando D.},
title = {Exponential improvement in precision for simulating sparse {Hamiltonians}},
year = {2014},
isbn = {9781450327107},
doi = {10.1145/2591796.2591854},
booktitle = {Proceedings of the Forty-Sixth Annual ACM Symposium on Theory of Computing},
pages = {283–292},
numpages = {10}
}

@article{Brassard2002,
  title = {Quantum amplitude amplification and estimation},
  ISBN = {9780821878958},
  ISSN = {0271-4132},
  DOI = {10.1090/conm/305/05215},
  journal = {Quantum Computation and Information},
  publisher = {American Mathematical Society},
  author = {Brassard,  Gilles and Høyer,  Peter and Mosca,  Michele and Tapp,  Alain},
  year = {2002},
  pages = {53–74}
}

@Article{Tropp2012,
author={Tropp, Joel A.},
title={User-Friendly Tail Bounds for Sums of Random Matrices},
journal={Foundations of Computational Mathematics},
year={2012},
day={01},
volume={12},
number={4},
pages={389-434},
doi={10.1007/s10208-011-9099-z}
}

@book{KitaevShenVyalyi2002,
  author    = {Kitaev, A. Yu. and Shen, A. H. and Vyalyi, M. N.},
  title     = {Classical and quantum computation},
  series    = {Graduate Studies in Mathematics},
  volume    = {47},
  publisher = {American Mathematical Society},
  year      = {2002},
  doi       = {10.1090/gsm/047}
}

@misc{grover2002creatingsuperpositionscorrespondefficiently,
      title={Creating superpositions that correspond to efficiently integrable probability distributions}, 
      author={Lov Grover and Terry Rudolph},
      year={2002},
      eprint={quant-ph/0208112},
      archivePrefix={arXiv},
      primaryClass={quant-ph}
}

@article{2011670.2011675,
author = {M{\"o}tt{\"o}nen, Mikko and Vartiainen, Juha J. and Bergholm, Ville and Salomaa, Martti M.},
title = {Transformation of quantum states using uniformly controlled rotations},
year = {2005},
issue_date = {September 2005},
publisher = {Rinton Press, Incorporated},
volume = {5},
number = {6},
issn = {1533-7146},
journal = {Quantum Info. Comput.},
month = sep,
pages = {467–473},
numpages = {7},
doi = {10.26421/QIC5.6-5}
}

@article{PhysRevA.54.147,
  title = {Quantum networks for elementary arithmetic operations},
  author = {Vedral, Vlatko and Barenco, Adriano and Ekert, Artur},
  journal = {Phys. Rev. A},
  volume = {54},
  issue = {1},
  pages = {147--153},
  numpages = {0},
  year = {1996},
  month = {Jul},
  publisher = {American Physical Society},
  doi = {10.1103/PhysRevA.54.147}
}

@book{Terras1999,
  title = {{Fourier} analysis on finite groups and applications},
  ISBN = {9780511626265},
  DOI = {10.1017/cbo9780511626265},
  publisher = {Cambridge University Press},
  author = {Terras,  Audrey},
  year = {1999},
  month = Mar 
}

@article{10.1063/1.4954983,
    author = {De las Cuevas, G. and Cubitt, T. S. and Cirac, J. I. and Wolf, M. M. and Pérez-García, D.},
    title = {Fundamental limitations in the purifications of tensor networks},
    journal = {Journal of Mathematical Physics},
    volume = {57},
    number = {7},
    pages = {071902},
    year = {2016},
    month = {07},
    issn = {0022-2488},
    doi = {10.1063/1.4954983}
}

@misc{grier2026mathsfqac0containsmathsftc0with,
      title={$\mathsf{QAC}^0$ Contains $\mathsf{TC}^0$ (with Many Copies of the Input)}, 
      author={Daniel Grier and Jackson Morris and Kewen Wu},
      year={2026},
      eprint={2601.03243},
      archivePrefix={arXiv},
      primaryClass={cs.CC}
}

@article{ntvs-c48s,
  title = {Quantum State Preparation without Coherent Arithmetic},
  author = {McArdle, Sam and Gily{\'e}n, Andr{\'a}s and Berta, Mario},
  journal = {Phys. Rev. Lett.},
  volume = {136},
  issue = {24},
  pages = {240603},
  numpages = {8},
  year = {2026},
  publisher = {American Physical Society},
  doi = {10.1103/ntvs-c48s}
}

@book{TQI,
  author    = {Watrous, John},
  title     = {The theory of quantum information},
  publisher = {Cambridge University Press},
  year      = {2018},
  doi = {10.1017/9781316848142}
}

@inproceedings{KW00,
author = {Kitaev, Alexei and Watrous, John},
title = {Parallelization, amplification, and exponential time simulation of quantum interactive proof systems},
year = {2000},
doi = {10.1145/335305.335387},
booktitle = {Proceedings of the Thirty-Second Annual ACM Symposium on Theory of Computing},
pages = {608–617},
numpages = {10}
}

@inproceedings{GW07,
author = {Gutoski, Gus and Watrous, John},
title = {Toward a general theory of quantum games},
year = {2007},
isbn = {9781595936318},
doi = {10.1145/1250790.1250873},
booktitle = {Proceedings of the Thirty-Ninth Annual ACM Symposium on Theory of Computing},
pages = {565–574},
numpages = {10}
}

@misc{CCHLS26,
  author        = {Chia, Nai-Hui and Chung, Kai-Min and Huang, Tzu-Hsiang and Lu, Chuhan and Shih, Jhih-Wei},
  title         = {Complexity Theory for Quantum Promise Problems},
  year          = {2026},
  eprint        = {2411.03716},
  archivePrefix = {arXiv}
}

@InProceedings{BCQ23,
  author        = {Brakerski, Zvika and Canetti, Ran and Qian, Luowen},
  title         = {On the Computational Hardness Needed for Quantum Cryptography},
  booktitle     = {14th Innovations in Theoretical Computer Science Conference (ITCS 2023)},
  series        = {Leibniz International Proceedings in Informatics (LIPIcs)},
  volume        = {251},
  year          = {2023},
  pages =	{24:1--24:21},
  doi           = {10.4230/LIPIcs.ITCS.2023.24}
}

@InProceedings{LLW26,
  author =	{Le Gall, Fran\c{c}ois and Liu, Yupan and Wang, Qisheng},
  title =	{A Slightly Improved Upper Bound for Quantum Statistical Zero-Knowledge},
  booktitle =	{51st International Symposium on Mathematical Foundations of Computer Science (MFCS 2026)},
  pages =	{12:1--12:20},
  series =	{Leibniz International Proceedings in Informatics (LIPIcs)},
  year =	{2026},
  volume =	{386},
  doi =		{10.4230/LIPIcs.MFCS.2026.12}
}

@misc{shen2026quantumalgorithmsmultivariablepolynomial,
      title={Quantum algorithms for multivariable polynomial transformations: from efficient synthesis to quantum channel transformations},
      author={Zheyu Shen and Yusen Wu and Xiao Yuan and Xiao-Ming Zhang and Yukun Zhang},
      year={2026},
      eprint={2610.08714},
      archivePrefix={arXiv},
      primaryClass={quant-ph}
}

@book{DummitFoote2004,
  author    = {Dummit, David S. and Foote, Richard M.},
  title     = {Abstract algebra},
  edition   = {3},
  publisher = {John Wiley \& Sons},
  year      = {2004},
  isbn      = {9780471433347}
}

@article{PhysRevA.87.032332,
  title = {Exact synthesis of multiqubit {$\mathrm{Clifford}+T$} circuits},
  author = {Giles, Brett and Selinger, Peter},
  journal = {Phys. Rev. A},
  volume = {87},
  issue = {3},
  pages = {032332},
  numpages = {7},
  year = {2013},
  month = {Mar},
  publisher = {American Physical Society},
  doi = {10.1103/PhysRevA.87.032332}
}

@misc{BrentZimmermann2010,
  author        = {Brent, Richard P. and Zimmermann, Paul},
  title         = {Modern Computer Arithmetic},
  year          = {2010},
  note          = {Version 0.5.1, 5 March 2010},
  eprint        = {1004.4710},
  archivePrefix = {arXiv}
}

@article{ALON1986567,
title = {A fast and simple randomized parallel algorithm for the maximal independent set problem},
journal = {Journal of Algorithms},
volume = {7},
number = {4},
pages = {567-583},
year = {1986},
issn = {0196-6774},
doi = {https://doi.org/10.1016/0196-6774(86)90019-2},
author = {Noga Alon and László Babai and Alon Itai}
}

@article{doi:10.1137/0215074,
author = {Luby, Michael},
title = {A Simple Parallel Algorithm for the Maximal Independent Set Problem},
journal = {SIAM Journal on Computing},
volume = {15},
number = {4},
pages = {1036-1053},
year = {1986},
doi = {10.1137/0215074}
}

@book{Durrett2019,
    author    = {Rick Durrett},
    title     = {Probability: Theory and Examples},
    edition   = {5},
    year      = {2019},
    publisher = {Cambridge University Press},
    series    = {Cambridge Series in Statistical and Probabilistic Mathematics},
    volume    = {49},
    doi       = {10.1017/9781108591034},
    isbn      = {9781108473682}
}

@article{Horner1819,
    author  = {Horner, William George},
    title   = {XXI. A New Method of Solving Numerical Equations
               of All Orders, by Continuous Approximation},
    journal = {Philosophical Transactions of the Royal Society
               of London},
    volume  = {109},
    pages   = {308--335},
    year    = {1819},
    doi     = {10.1098/rstl.1819.0023},
    url     = {https://doi.org/10.1098/rstl.1819.0023}
}

@misc{cook2025structureinplacespaceboundedcomputation,
      title={The Structure of In-Place Space-Bounded Computation}, 
      author={James Cook and Surendra Ghentiyala and Ian Mertz and Edward Pyne and Nathan S. Sheffield},
      year={2025},
      eprint={2510.12005},
      archivePrefix={arXiv},
      primaryClass={cs.CC}
}

@InProceedings{kretschmer:LIPIcs.TQC.2021.2,
  author =	{Kretschmer, William},
  title =	{Quantum Pseudorandomness and Classical Complexity},
  booktitle =	{16th Conference on the Theory of Quantum Computation, Communication and Cryptography (TQC 2021)},
  pages =	{2:1--2:20},
  series =	{Leibniz International Proceedings in Informatics (LIPIcs)},
  year =	{2021},
  volume =	{197},
  doi =		{10.4230/LIPIcs.TQC.2021.2}
}

@book{Hislop1996,
  title = {Introduction to Spectral Theory: With Applications to Schr\"{o}dinger Operators},
  ISBN = {9781461207412},
  ISSN = {0066-5452},
  DOI = {10.1007/978-1-4612-0741-2},
  journal = {Applied Mathematical Sciences},
  publisher = {Springer New York},
  author = {Hislop,  P. D. and Sigal,  I. M.},
  year = {1996}
}

@incollection{watrous_quantum_2009,
	title = {Quantum Computational Complexity},
	doi = {10.1007/978-0-387-30440-3_428},
	booktitle = {Encyclopedia of {Complexity} and {Systems} {Science}},
	publisher = {Springer New York},
	author = {Watrous, John},
	year = {2009},
	pages = {7174--7201},
}

@article{BV97,
author = {Bernstein, Ethan and Vazirani, Umesh},
title = {Quantum Complexity Theory},
journal = {SIAM Journal on Computing},
volume = {26},
number = {5},
pages = {1411-1473},
year = {1997},
doi = {10.1137/S0097539796300921}
}

@book{AB09,
  author    = {Arora, Sanjeev and Barak, Boaz},
  title     = {Computational Complexity: A Modern Approach},
  publisher = {Cambridge University Press},
  year      = {2009}
}

@article{Llo96,
author = {Seth Lloyd},
title = {Universal Quantum Simulators},
journal = {Science},
volume = {273},
number = {5278},
pages = {1073-1078},
year = {1996},
doi = {10.1126/science.273.5278.1073}
}

@article{LC19,
  doi = {10.22331/q-2019-07-12-163},
  title = {Hamiltonian Simulation by Qubitization},
  author = {Low, Guang Hao and Chuang, Isaac L.},
  journal = {{Quantum}},
  volume = {3},
  pages = {163},
  year = {2019}
}

@article{BDSW96,
  title = {Mixed-state entanglement and quantum error correction},
  author = {Bennett, Charles H. and DiVincenzo, David P. and Smolin, John A. and Wootters, William K.},
  journal = {Phys. Rev. A},
  volume = {54},
  issue = {5},
  pages = {3824--3851},
  numpages = {0},
  year = {1996},
  publisher = {American Physical Society},
  doi = {10.1103/PhysRevA.54.3824}
}

@inproceedings{KQST23,
author = {Kretschmer, William and Qian, Luowen and Sinha, Makrand and Tal, Avishay},
title = {Quantum Cryptography in Algorithmica},
year = {2023},
doi = {10.1145/3564246.3585225},
booktitle = {Proceedings of the 55th Annual ACM Symposium on Theory of Computing},
pages = {1589–1602},
numpages = {14}
}

@InProceedings{JLS18,
  author={Ji, Zhengfeng and Liu, Yi-Kai and Song, Fang},
  title={Pseudorandom Quantum States},
  booktitle={Advances in Cryptology -- CRYPTO 2018},
  year={2018},
  publisher={Springer},
  pages={126--152},
  doi={10.1007/978-3-319-96878-0_5}
}

@article{LQSY+26,
author = {Lu, Chuhan and Qin, Minglong and Song, Fang and Yao, Penghui and Zhao, Mingnan},
title = {Quantum Pseudorandom Scramblers},
journal = {SIAM Journal on Computing},
volume = {55},
number = {4},
pages = {787-850},
year = {2026},
doi = {10.1137/24M170990X}
}

@inproceedings{MH25,
  title={How to construct random unitaries},
  author={Ma, Fermi and Huang, Hsin-Yuan},
  booktitle={Proceedings of the 57th Annual ACM Symposium on Theory of Computing -- STOC 2025},
  pages={806--809},
  year={2025},
  url={https://doi.org/10.1145/3717823.3718254},
  doi={10.1145/3717823.3718254}
}

@misc{LQSY+25,
      title={Parallel {Kac's} Walk Generates {PRU}}, 
      author={Chuhan Lu and Minglong Qin and Fang Song and Penghui Yao and Mingnan Zhao},
      year={2025},
      eprint={2504.14957},
      archivePrefix={arXiv},
      primaryClass={quant-ph},
      note={To appear in Asiacrypt 2026}
}

@inproceedings{KQT25,
author = {Kretschmer, William and Qian, Luowen and Tal, Avishay},
title = {Quantum-Computable One-Way Functions without One-Way Functions},
year = {2025},
doi = {10.1145/3717823.3718144},
booktitle = {Proceedings of the 57th Annual ACM Symposium on Theory of Computing},
pages = {189–200},
numpages = {12}
}

@article{PhysRevA.90.012110,
  title = {Quantum process tomography of unitary and near-unitary maps},
  author = {Baldwin, Charles H. and Kalev, Amir and Deutsch, Ivan H.},
  journal = {Phys. Rev. A},
  volume = {90},
  issue = {1},
  pages = {012110},
  numpages = {10},
  year = {2014},
  publisher = {American Physical Society},
  doi = {10.1103/PhysRevA.90.012110}
}

@article{YRC20,
  title = {Optimal Universal Programming of Unitary Gates},
  author = {Yang, Yuxiang and Renner, Renato and Chiribella, Giulio},
  journal = {Phys. Rev. Lett.},
  volume = {125},
  issue = {21},
  pages = {210501},
  numpages = {5},
  year = {2020},
  month = {Nov},
  publisher = {American Physical Society},
  doi = {10.1103/PhysRevLett.125.210501}
}

@article{He2026resource,
  doi = {10.22331/q-2026-07-20-2166},
  title = {Resource quantification for programming low-depth quantum circuits},
  author = {He, Entong and Yang, Yuxiang},
  journal = {{Quantum}},
  issn = {2521-327X},
  publisher = {{Verein zur F{\"{o}}rderung des Open Access Publizierens in den Quantenwissenschaften}},
  volume = {10},
  pages = {2166},
  year = {2026}
}

\begingroup
\setstretch{1.1}

\newpage
\appendix
\hypersetup{bookmarksdepth=2}
\section*{Appendices}
\addcontentsline{toc}{section}{Appendices}
\addtocontents{toc}{\protect\setcounter{tocdepth}{2}}
\section{\textbf{\textsf{PSPACE}} arithmetic}\label{sec:pspace_arthm}

This section connects circuits with numbers. We handle exponentially large states (vectors) and matrices through their entries, as they can not be written down in polynomial space, but a single entry can often be approximated to any prescribed number of bits by a polynomial-space (quantum) machine\footnote{
Note that, for approximating a single entry, a polynomial-space classical Turing machine is as powerful as its quantum counterpart, since $\mathsf{BQPSPACE} = \pspace$~\cite{WATROUS1999281}.
}. These entries are therefore $\pspace$-computable, and two basic facts make the notion useful. First, it is closed under the arithmetic of exponentially large objects by recycling the workspace. Second, it contains the numbers that characterize space-bounded quantum computations, such as acceptance probabilities and the entries of output states, since a quantum computation in space $s$ can be simulated classically in space $\bigo(s^2)$~\cite{WATROUS1999281, Watrous2003}. Both facts are standard in state and unitary synthesis, and are used, e.g., in the polynomial-space tomography algorithm of $\mathsf{statePSPACE}$ states \cite[Section 5]{RY22} and in polynomial approximations of matrix functions with $\pspace$-computable coefficients~\cite[Section 2.2 and Lemma 6.2]{MY23}. We record several basic facts required in an entry-wise form in which a universal machine serves all precisions. Notably, up to the encoding of precision, this notion is equivalent to the fully polynomial-space exponential approximation scheme (FPSEAS) recently introduced by Kretschmer and Tang \cite[Definition 4.1]{kretschmer2026unitarycomplexitypolynomialspace} applied to the real and imaginary parts separately.
We now prove the arithmetic facts, starting with closure under sums, products, quotients and roots.

\pspaceArithmeticLemma*

\begin{proof}
    Properties 1 and 2 are standard facts about arithmetic with polynomial-space approximations, and also appear, for instance, in the proofs of \cite[Lemma~3.6]{MY23} and \cite[Corollaries~5.4 and~5.5]{RY22} up to some notational differences. For brevity, we omit their proofs and refer the reader to these works for details. It suffices to prove property 3. We use a divide-and-conquer evaluation of long products, similar to the one in the proof of Savitch's theorem \cite[Theorem 1]{SAVITCH1970177}. Equivalently, we evaluate a balanced tree of matrix products depth-first, keeping one frame per~level.

    Set $k = k(|y|)$ and $N = N(|y|)$, and put the matrix-valued polynomial $P(t_1, t_2) = M_{y, t_2} \cdots M_{y, t_1}$. An entry of $P(t_1,t_1)$ is approximated by invoking the machine for approximating $M_{y,t_1}$. For $t_1 < t_2$ we set the midpoint $t_m \coloneqq  \lfloor \frac{t_1 + t_2}{2} \rfloor$ and we approximate $P(t_1, t_2)$ as follows: To approximate $\braket{i | P(t_1, t_2) | j}$ within $2^{-h}$, we approximate $\braket{i | P(t_m + 1, t_2) | c}$ and $\braket{c | P(t_1, t_m) | j}$ recursively within $2^{-h - k - 3}$ for all $c \in \{0, 1\}^k$ and sum them up by the basic matrix product equality. Since each entry is bounded and an erroneous estimation is bounded by $2$, the error accumulates as $2^k \cdot 3 \cdot 2^{-h - k - 3} + 2^{-h - 1} < 2^{-h}$, so that when rounding up to $h + 1$ fractional bits the approximation is exact. Starting from the interval $[1, N]$, the recursion halves the problem dimension at each level so the depth is $\lceil \log_2 N \rceil \leq p(|y|)$ and requests precision at most $h + \lceil \log_2 N \rceil (k + 3)$. By space recycling, the space thus scales as $\poly(|y| + h)$.
\end{proof}

The next lemma extends the entrywise $\pspace$-computability of space-uniform states to space-uniform channels via the Choi-Jamio\l{}kowski isomorphism \cite{TQI}.

\channelEntriesLemma*

\begin{proof}
    For each $n \in \N$, we take $J_n = (\mathcal{C}_n \otimes \mathcal{I})(\ketbra{\mathsf{EPR}_n})$ to be the Choi state of $\mathcal{C}_n$, and thus $\braket{i | \mathcal{C}_n(\ket{x} \bra{y}) |j} = 2^n \braket{i, x |J_n | j, y }$.
    The modulus claim follows readily via the Cauchy-Schwarz inequality:
    $$| \braket{i, x |J_n | j, y } | = | \braket{ i, x | \sqrt{J_n}^\dagger J_n |j, y } | \leq \sqrt{\braket{i, x | J_n | i, x } \braket{j, y | J_n | j, y } } \leq 2^{-n}.$$

    Note that the states $\{\ket{\mathsf{EPR}_n}\}_{n \in \N}$ is in $\mathsf{statePSPACE}_0$ (more precisely, in $\mathsf{stateBQP}_0$), so that for space-uniform circuit $\{\mathcal{C}_n\}_{n \in \N}$, the family $\{J_n\}_{n \in \N}$ are space-uniform, and we borrow the proof techniques of \cite[Theorem~5.1]{RY22} to show that it has $\pspace$-computable entries over the universal gate set $\{\hdm, \tgate, \cnot\}$ with algebraic entries~\cite{Watrous2003}.

    Take $a, b \in \{0, 1\}^{2n}$ with $a \neq b$ and for a scalar $z \in \{1, \I, -1, -\I\}$, we set $\ket{w_z} = \frac{1}{\sqrt{2}} ( \ket{a} + \overline{z}\ket{b} )$. The pure-state polarization identity \cite[Equation 5.1]{RY22} that extends to all operators via linearity tells us that
    $$
    \begin{aligned}
    \frac{1}{2} \sum_{z \in \{1, \I, -1, -\I\} } z \braket{ w_z | J_n | w_z } &= \frac{1}{4}\sum_{z\in\{1,\I,-1,-\I\}}
\left( z\bra aJ_n\ket a +z\bar z\bra aJ_n\ket b +z^2\bra bJ_n\ket a +z\bra{b}J_n\ket{b}
\right)  \\
&= \frac{1}{4}\sum_{z\in\{1,\I,-1,-\I\}} |z|^2 \braket{a | J_n | b} = \braket{a | J_n | b}.
    \end{aligned}
    $$
    Here the expression $\braket{ w_z | J_n | w_z }$ is the probability that the following circuit outputs $1$ on input $(a, b, z)$: prepare the state $J_n$, apply the inverse of the state preparation circuit of $\ket{w_z}$ in \cite[Proof of Corollary 5.5]{RY22}, and test whether the state is in the all-$0$ state. For $a = b$, the number $\braket{a | J_n | a}$ is simply the outcome via a computational basis measurement. These tests form a space-uniform family of one-output-qubit circuits, so we apply the same argument as \cite[Theorem 5.1]{RY22} indexed by $N = \left<n ,h \right>$ under the Cantor's pairing $\left< \cdot, \cdot\right>$ (see, e.g. \cite[Page xiv]{Montalban2021}), so that $n + h \leq N \leq (n + h + 1)^2$ with inputs padded to length $4N + 2$ and $m(N) = N + 2$ (cf. theorem statement of \cite[Theorem 5.1]{RY22}). This yields a machine running in $\poly(N) = \poly(n + h)$ space and approximates all probability within $2 \cdot 2^{-m(N)} \leq 2^{-n - h - 1}$. Taking $a = (i, x)$ and $b = (j, y)$ then yields a dyadic approximation of the entry $\braket{i | \mathcal{C}_n(\ket{x} \bra{y}) |j}$ within $2^{-n - h - 1} \cdot 2^{n-1} \cdot 4 = 2^{-h}$, finishing the proof.
\end{proof}

Besides the closure properties of $\pspace$-computable numbers from \cref{lemma:arthm_of_PSPACE_num}, our subsequent construction of a $\pspace$ oracle for synthesizing $\upspace$ unitary families relies on the following two facts.

\begin{lemma}[$\pspace$-computability of elementary functions and precision bounds]
\label{lemma:more_computation_within_pspace}
    The following statements hold:
    \begin{enumerate}
        \item The families $\{\mathrm{e}^{2\pi\I\theta}\}_{\theta \in \N \times \N}$ and $\{\sin(\pi\theta)\}_{\theta \in \N \times \N}$, indexed by rational numbers $\theta$ with their numerators and denominators given in binary are $\pspace$-computable.
        \item Let $a\geq0$ and $v\geq1$ be integers, and consider the sum of products $\sum_{t\in\mathcal{T}}c_{t}\prod_{\ell\in\mathcal{L}_{t}}x_{t,\ell}$ with $|\mathcal{T}|\leq2^{a}$, $|\mathcal{L}_{t}|\leq v$, dyadic $|c_{t}|\leq1$ and $|x_{t,\ell}|\leq1$. Replacing every $x_{t,\ell}$ by their approximation $\widetilde{x}_{t,\ell}$ with $|\widetilde{x}_{t,\ell}-x_{t,\ell}|\leq2^{-g}$, where $2^{g}\geq v$ changes this sum by less than $3v\cdot2^{a-g}$.
    \end{enumerate}
\end{lemma}

\begin{proof}
    1. Note that for two complex numbers $x, y \in \C$ satisfies $|x - y| \leq |\Re(x) - \Re(y)| + |\Im(x) - \Im(y)|$, and $\sin(\pi \theta) = \Im(\mathrm{e}^{2 \pi \I (\theta / 2)})$, it suffices to treat $\{\mathrm{e}^{2\pi\I\theta}\}_{\theta}$, and we may assume $0\leq\theta<1$ by replacing $\theta$ with $\theta-\lfloor\theta\rfloor$. The functions $\arctan{x}$, $\sin{x}$ and $\cos{x} - 1$ are holonomic (cf. \cite[Section 4.9.2]{BrentZimmermann2010}), vanish at $x = 0$ and have no singularities on $[-1, 8]$, so by \cite[Theorem 4.9.1]{BrentZimmermann2010} they can be approximated to within $2^{-m}$ at any $m$-bit floating-point number in $(-1, 8)$, in time $\poly(m)$. We thus approximate $\pi = 4\arctan{1}$ in this way and multiply the result by $2\theta$ and round to obtain a dyadic rational $\widetilde{x} \in (-1, 8)$ with $g + \mathcal{O}(1)$ bits, and it satisfies $|\widetilde{x} - 2\pi\theta| \leq 2^{-g-2}$. Since both $\sin$ and $\cos$ are $1$-Lipschitz, approximating $\cos({\widetilde{x}})$ and $\sin({\widetilde{x}})$ to within $2^{-g-2}$ then gives an approximation of $\mathrm{e}^{2\pi\I\theta}$ to within $\sqrt{2} \cdot 2^{-g-1} < 2^{-g}$. All approximations are conducted in time polynomial in $g$ and the bit length of $\theta$ in binary.

    2. It suffices to combine the arguments in \cref{lemma:arthm_of_PSPACE_num}. For a term with $k\le v$ factors, suppressing $t$ and ordering its
factors, telescoping and using the the handy inequality $(1+1/x)^x \leq \mathrm{e}$ give
$$
 \begin{aligned}
 \left|\prod_{\ell=1}^k\widetilde x_\ell-\prod_{\ell=1}^kx_\ell\right|
 &\le\sum_{r=1}^k|\widetilde x_r-x_r|
       \prod_{\ell<r}|\widetilde x_\ell|\prod_{\ell>r}|x_\ell| \\
       &\le v2^{-g}(1+2^{-g})^{v-1} \leq  v 2^{-g}\left(1 + \frac{1}{v} \right)^{v}
 \le \mathrm{e}\cdot v2^{-g},
 \end{aligned}
$$
since $|\widetilde x_\ell|\le|x_\ell|+2^{-g}\le1+2^{-g}$ and $2^g\ge v$.
The change in the sum is at most the sum of the absolute changes in its terms, so the change is at most $\sum_{t\in\mathcal T}|c_t| \mathrm{e}\cdot v2^{-g} \le \mathrm{e}\cdot v2^{a-g}<3v\cdot2^{a-g}$. We then sum over at most $2^{a}$ terms with $|c_t| \leq 1$ and use the triangle inequality to obtain the desired bound. This concludes the proof.
\end{proof}

\begin{remark}
\label{remark:rounding_error_for_sum_of_products}
    Suppose we replace the approximation precision of the sum of products in statement 2 of \cref{lemma:more_computation_within_pspace} from $2^{-g}$ to within $2^{-g-a-\lceil \log_2{v} \rceil - 3}$, and round to $g + 2$ binary bits. The error incurred by approximation is bounded by $3c \cdot 2^{-g - a - \lceil \log_2 v \lceil - 3} \leq \frac{3}{8} \cdot 2^{-g}$, and the rounding error is bounded by $\sqrt{2} \cdot \frac{2^{-g - 2}}{2} = \frac{\sqrt{2}}{8} \cdot2^{-g}$ with $\frac{1}{2} 2^{-g-2}$ on the real and imaginary parts. So the total approximation error reads $(\frac{3}{8} + \frac{\sqrt{2}}{8}) 2^{-g} < 2^{-g}$.
\end{remark}

\section{Polynomial-time unitary synthesis from a Boolean oracle: an exposition}
\label[appendix]{appendix:answer_of_USP}

\paragraph{(Ad hoc) defintions and notational conventions.} Fix $n \in \N$, for any $ 1 \leq d \leq 2^n$, we use the subspace encoding
$$
\HH_d \coloneqq \Span\left\{ \ket{0}, \ket{1}, \dots, \ket{d -1} \right\} \subseteq (\C^2)^{\otimes n}.
$$

\begin{definition}[Controlled synthesis task and clean error {\cite[Section 2]{OpenAI2026UnitarySynthesis}}]
\label{def:controlled_unitary_synthesis_task}
    A controlled unitary synthesis task consists of a $t$-qubit label register and a unitary $U_{\lambda}$ acting on $\HH_d$ for each $\lambda \in \{0, 1\}^t$, which the operation prescribed to be
    $$
    \mathbf{U} \coloneqq \sum_{\lambda \in \{0, 1\}^t} \ket{\lambda} \! \bra{\lambda} \otimes U_{\lambda}.
    $$
    A unitary circuit $C$ acting on the complete space\footnote{
    With the label, input data, and workspace registers included.
    } solves the task with \textit{clean error} $\delta$ if
    $$
    \left\| C \cdot \iota - \iota \cdot \mathbf{U} \right\|_{\op} = \max_{\ket{\psi}} \left\| C\left( \ket{\psi} \ket{0}_{\mathsf{anc}} \right) - \mathbf{U}(\ket{\psi}) \ket{0}_{\mathsf{anc}} \right\|  \leq \delta.
    $$
\end{definition}

\subsection{Problem setup and main result}

For clarity, we formally state Aaronson and Kuperberg's USP:
\begin{problem}[Unitary synthesis problem \cite{4262757, aaronson2016complexityquantumstatestransformations}]
\label{prob:USP}
    For every $n \in \N$, find a deterministic polynomial-time algorithm that, given $1^n$, outputs a polynomial-size oracle circuit $\mathcal{C}_n^{(\cdot)}$ with queries of polynomial length $\ell(n)$, such that
    $$
    \forall \, U \in \U(2^n), \quad \exists f_U: \{0, 1\}^{\ell(n)} \to \{0, 1\}: \quad \dd( \Phi_{n, f_U}, \mathcal{U} ) \leq \frac{1}{2}.
    $$
    Here $\Phi_{n, f_U}$ is the channel of circuit $\mathcal{C}_n^{f_U}$ that keeps $n$ of its output qubits and discards any ancillae.
\end{problem}

\begin{corollary}
    By the data processing inequality \cite{Nielsen2012} and \cite[Proposition I.6]{HKOT23}, setting $d = 2^n$, $t = 0$ and $\beta = \Theta(1)$ in \cref{thm:usp} yields
    $$
    \begin{aligned}
    \dd\!\left( \Phi_{n, f_U} - \mathcal{U} \right) &= \dd\!\left( \Tr_{\mathsf{anc}} \circ \ad_{\mathcal{C}_{n, d, t, \beta}^{f_{\mathbf{U}, \beta}} \cdot \iota},  \Tr_{\mathsf{anc}} \circ \ad_{\iota \cdot \mathbf{U}} \right) \\
    &\leq \dd\!\left( \ad_{\mathcal{C}_{n, d, t, \beta}^{f_{\mathbf{U}, \beta}} \cdot \iota},  \ad_{\iota \cdot \mathbf{U}} \right)
    \leq \left\|\mathcal{C}_{n, d, t, \beta}^{f_{\mathbf{U}, \beta}} \cdot \iota - \iota \cdot \mathbf{U}  \right\|_{\op} \leq 2^{-\beta - 1} = \Theta(1),
    \end{aligned}
    $$
    suggesting a valid construction to \cref{prob:USP}.
\end{corollary}

\subsection{Proof overview}
\label[appendix]{appendix:proof_overview}
As illustrated in \cref{def:controlled_unitary_synthesis_task}, the proof reduces the implementation of a $d$-dimensional unitary to that of a smaller, coherently controlled unitary. Each reduction step consists of three parts: encode the unknown input, apply the smaller controlled unitary, and decode. For the sake of a time- and space-efficient oracle circuit, the central difficulty lies in striking a balance between the encoder's information-preserving capability and its implementation efficiency. 

We have to produce, from the sizes alone, a circuit that implements an arbitrary target unitary once the Boolean oracle is filled in. The construction consists of two stages: (\textit{i}) building the circuit from \emph{programmable slots}, defined below; (\textit{ii}) replacing each slot with gates from the fixed universal gate set and oracle queries.

\paragraph{The programmable slots.} We specify the programmable-slot model, motivated by the study of uniformly controlled one-qubit gates \cite{PhysRevLett.93.130502, PhysRevA.71.052330}: a slot on the control and target space $\HH_{\mathsf{C}} \otimes \HH_{\mathsf{T}}$ is specified by a target qubit and an ordered list of $c$ control qubits, and acts as
\begin{equation}
\label{eqn:programmable_slot}
\sum_{y \in \{0, 1\}^{c}} \ket{y} \!\!\bra{y}_{\mathsf{C}} \otimes W_y, \quad W_y \in \U(2),
\end{equation}
where $y$ runs over the values of the control qubits in $\mathcal{C}$, and the small unitary \emph{table} $\{W_y\}_{y \in \{0, 1\}^{c}}$ is chosen arbitrarily. We will need two properties of this model to shape the proof. First, it separates what a circuit generator (usually a Turing machine in the complexity-theoretic formulation) should produce from what the target determines. The generator prints the circuit \emph{skeleton} consisting of the registers and, for each slot, its target and the control list. These depend solely on the circuit sizes and encode nothing computationally hard. Everything that depends on the large target unitary goes into the tables, with hardness resolved by the power of oracles.

Second, the labels are (almost) free. Suppose we have in hand a circuit skeleton that implements each $U \in \U(d)$ with suitable tables. Then padding a $t$-qubit label to the control list of each slot and setting the $(\lambda, y)$-entry of the enlarged table as the $y$-entry of the table that implements $U_{\lambda}$ solves the controlled unitary synthesis task $\mathbf{U} = \sum_{\lambda \in \{0, 1\}^t} \ket{\lambda} \!\! \bra{\lambda} \otimes U_{\lambda}$ (cf. \cref{def:controlled_unitary_synthesis_task}), with the same number of slots and longer control lists. In terms of operator norm, this does not amplify the approximation error. This is what lets the recursion below trade unitary dimension for labels.

\subsubsection{Polynomial-size recursion with one call per step} 
\label[appendix]{appendix:poly_size_recursion_by_one_call_per_step}
The induction hypothesis is the bounded operator norm distance statement of \cref{thm:usp} in smaller dimensions formulated in the slot model. For each $k < d$ and each number of label qubits $t$, one skeleton implements every controlled family over the space $\HH_k$ with clean error at most $e(k)$ using at most $S(k)$ slots. A step shall implement a family $\{U_{\lambda}\}_{\lambda}$ on the whole space $\HH_d$ by calling such circuits with extra gates of its own. Before asking how, let us see what a step may cost and, for the purpose of solving USP, what the shape of a step should be.

\paragraph{What makes a step good.} Suppose a reduction step lowers the dimension from $d$ to $k = d - s$, runs the circuit for dimension $k$ a total of $c$ times, and adds $p$ slots of its own. Then the slot count would satisfy
$$
S(d) \leq c S(k) + p,
$$
and after $m$ iterative steps the expanded circuit holds $\Theta(c^m)$ copies of the innermost circuit. 

The sub-task count $c = 2$ typically occurs in conventional recursion models. For instance, the construction in \cite{PhysRevA.71.052330} based on the cosine-sine decomposition writes an $n$-qubit unitary in blocks according to its first qubit geometry as
$$
\forall \, U \in \U(2^n),~ U = \begin{pmatrix}
    L_0 & \\ & L_1
\end{pmatrix} \begin{pmatrix}
    \mathfrak{C} & -\mathfrak{S} \\ \mathfrak{S} & \mathfrak{C}
\end{pmatrix} \begin{pmatrix}
    R_0 & \\ & R_1
\end{pmatrix}: ~ L_0, L_1, R_0, R_1 \in \U(2^{n-1}), ~\mathfrak{C}^2 + \mathfrak{S}^2 = \id,
$$
with $\mathfrak{C}$ and $\mathfrak{S}$ being real diagonal matrices. Note that the middle operator corresponds to the rotation on the first qubit controlled by the remaining $n-1$ ones, and is thus a single slot. The outer operators are singly controlled $(n-1)$-qubit unitaries that halve the dimension. Therefore, the recursive relation reads $S(2^n) = 2S(2^{n-1}) + 1$ and $S(2) = 1$, so that $S(2^n) = 2^n - 1$. It is essentially the presence of the second call that pushes the slot count to exponential. A similar idea of trading data qubits for control qubits underlies the quantum Shannon decomposition~\cite{1629135}, where uniformly controlled gates are called quantum multiplexors; without an oracle, it is likewise exponentially expensive.

What if a step could make do with a single call, that is, $c=1$? Then the recurrence becomes $S(d)\leq  S(k)+p$, and unrolling it over $m$ levels with $S(1)=1$ gives $S(d)\leq  1+mp$: the slot count grows only linearly in the recursion depth with $p$ slots per level. It remains to keep $m$ polynomial. Removing a constant number of dimensions per step would take about $d$ levels, so a good step removes at least a $1/\poly(n)$ fraction of the dimension, adds only polynomially many slots of its own, and calls the smaller circuit only once.

Our subsequent construction achieves this cheaply, with $s \geq d/(16384(n+1)^2)$, so after $j$ steps the remaining dimension is at most
$$
2^n \cdot \left( 1 - \frac{1}{16384(n+1)^2} \right)^j \leq 2^n \mathrm{e}^{ -\frac{j}{16384(n+1)^2} },
$$
which goes below $1$ when the recursion level $j \geq J= 32768(n+1)^3 = \bigo(n^3)$, and each level adds $p = \poly(n, L)$ slots.

\paragraph{The shape of a single step.} The good step sandwiches a single call between an \emph{encoder} and the adjoint of a possibly different encoder, which serves as the \emph{decoder}. The encoder writes the $d$-dimensional input into two registers: a \emph{port register} (we will explain later) with valid dimension $k < d$ and a \emph{phase register} with basis states $\{\ket{\mathbf{z}}\}_{\mathbf{z} \in G}$ labeled by a finite abelian group $G$. The smaller circuit acts on the port register and reads the phase register only as controls. Since $k < d$, the port register alone cannot hold the whole input, and the rest of the information is spread over the superposition over $\mathbf{z}$. For each label $\lambda$, which we suppress for simplicity, we proceed with the (almost) commutative diagram:
\begin{equation}\label{eqn:near_commutative_diagram}
 \begin{tikzcd}[column sep=large, row sep=large]
  \C^d \arrow[r, "V"] \arrow[d, "E_L"'] & \C^d \arrow[d, "R_L"]\\
  \C^G\otimes\C^k \arrow[r, "\mathbf F"'] & \C^G\otimes\C^k
 \end{tikzcd}
\end{equation}
in the sense that $\mathbf{F} E_L \approx R_L V$. Here $V$ is the target unitary after an efficient preconditioning detailed subsequently. $E_L$ and $R_L$ are nearly isometric encodings of the input and output with \emph{truncation length $L$}, and $\mathbf{F} = \bigoplus_{\mathbf{z} \in G} F(\mathbf{z})$ is the smaller call. For nearly isometric $R_L$, we have approximate recovery $V \approx R_L^\dagger \mathbf{F} E_L$. For dimension $d$, the circuit is
\begin{equation}
\label{eqn:dim_d_circuit_single_step}
C_d \coloneqq P_{\mathrm{out}}^\dagger U_R^\dagger C_k U_E P_{\mathrm{in}}^\dagger, \quad k = d - s < d.
\end{equation}
At this level, we build every factor other than $C_k$ from $\lambda$-dependent tables:

\begin{itemize}
    \item $P_{\mathrm{in}}$ and $P_{\mathrm{out}}$ are efficient preconditioners made of signs, Hadamard gates, and permutations. They replace the target $U$ with $V = P_{\mathrm{out}} U P_{\mathrm{in}}$, whose last $s$ coordinates correspond to a norm-contracting block; see \cref{appendix:convergence_of_feedback_loop}. Since $U = P_{\mathrm{out}}^\dagger V P_{\mathrm{in}}^\dagger$, it suffices to equivalently implement $V$ between $P_{\mathrm{in}}^\dagger$ and $P_{\mathrm{out}}^\dagger$.
    \item $U_E$ and $U_R$ are the encoders that approximately apply $E_L$ and $R_L$, respectively, in the sense of \cref{eqn:ideal_identities} below; see \cref{appendix:encoder_circuit_with_block_encoding} for detail. $U_R^\dagger$ serves as a decoder instead.
    \item $C_k$ is the recursive call that approximates the controlled family $\mathbf{F} = \sum_{\mathbf{z} \in G} \ket{\mathbf{z}} \!\! \bra{\mathbf{z}} \otimes F(\mathbf{z})$ with $F(\mathbf{z}) \in \U(k)$. Each label $\mathbf{z} \in G$ determines a diagonal unitary $Z(\mathbf{z})$, and $F(\mathbf{z}) = F(Z(\mathbf{z}))$ is obtained from $V$ by closing a feedback loop; see \cref{appendix:unitary_for_child_task}.
\end{itemize}

\paragraph{What should the ideal pieces do?} The map $\iota$ (cf. \cref{sec:prelim}) puts $\ket{\psi} \in \C^d$ in the index register $\mathsf{X}$, and $\gamma$ puts vectors in $\C^G \otimes \C^k$ into the phase and port registers and initializes register $\mathsf{X}$ to $\ket{0}$ and the flag qubits to $\ket{11}$ (as explained later). Both maps put the workspace of $C_k$ into $\ket{0}$. Suppose the following identities hold exactly:
\begin{equation}
\label{eqn:ideal_identities}
\mathbf{F} E_L = R_L V, \quad U_E  \iota = \gamma E_L, \quad U_R \iota = \gamma R_L, \quad C_k \gamma = \gamma \mathbf{F}.
\end{equation}
As illustrated in \cref{fig:registers}, every input state $\ket{\psi} \in \C^d$ evolves as
\begin{equation}
\label{eqn:state_evolution_chain}
\iota \ket{\psi} \xmapsto{ \ U_E \ } \gamma E_L \ket{\psi} \xmapsto{ \ C_k \ } \gamma \mathbf{F} E_L \ket{\psi} = \gamma R_L V \ket{\psi } = U_R \iota V \ket{\psi}  \xmapsto{ \ U_R^\dagger  \ } \iota V \ket{\psi},
\end{equation}
so the middle three factors of \cref{eqn:dim_d_circuit_single_step} implement $V$.

\begin{figure}[!tbp]
\centering
\begin{tikzpicture}[>=Latex, semithick, font=\small,
  gate/.style={draw, rounded corners=1.5pt, fill=gray!8,
               minimum width=9mm, inner sep=2pt},
  ctrl/.style={circle, fill, inner sep=1.3pt},
  note/.style={font=\footnotesize, inner sep=1pt}]
 \foreach \y/\nm in {0/{label $\lambda$}, -0.8/{index $\mathsf{X}$},
                     -1.6/{$\mathsf{flags} = \mathsf{ab}$}, -2.4/{phase $\mathsf{Z}$}, -3.2/{port $\mathsf{P}$},
                     -4.0/{workspace}}
 {
   \draw (0,\y) -- (13.3,\y);
   \node[left] at (0,\y) {\nm};
 }
 \foreach \x in {0.45,2.65,5.25,7.95,10.55,12.85}
   \draw[gray!60, densely dashed, thin] (\x,0.35) -- (\x,-4.45);
 \draw (1.4,0) -- (1.4,-0.5); \node[ctrl] at (1.4,0) {};
 \node[gate, minimum height=6mm] at (1.4,-0.8) {$P_{\mathrm{in}}^{\dagger}$};
 \draw (3.95,0) -- (3.95,-0.5); \node[ctrl] at (3.95,0) {};
 \node[gate, minimum height=30mm, minimum width=11mm] at (3.95,-2.0) {$U_E$};
 \draw (6.6,0) -- (6.6,-2.95); \node[ctrl] at (6.6,0) {};
 \node[ctrl] at (6.6,-2.4) {};
 \node[gate, minimum height=14mm, minimum width=11mm] at (6.6,-3.6) {$C_k$};
 \draw (9.25,0) -- (9.25,-0.5); \node[ctrl] at (9.25,0) {};
 \node[gate, minimum height=30mm, minimum width=11mm] at (9.25,-2.0) {$U_R^{\dagger}$};
 \draw (11.75,0) -- (11.75,-0.5); \node[ctrl] at (11.75,0) {};
 \node[gate, minimum height=6mm] at (11.75,-0.8) {$P_{\mathrm{out}}^{\dagger}$};
 \node[note] at (0.45,-4.85)  {$\ket\psi_{\mathsf{X}}$};
 \node[note] at (2.65,-4.85)  {$\ket{\psi'}_{\mathsf{X}}=P_{\mathrm{in}}^{\dagger}\ket\psi_{\mathsf{X}}$};
 \node[note] at (5.25,-4.85)  {$\approx(E_L\ket{\psi'})_{\mathsf{ZP}}$};
 \node[note] at (7.95,-4.85)  {$\approx(R_LV\ket{\psi'})_{\mathsf{ZP}}$};
 \node[note] at (10.55,-4.85) {$\approx(V\ket{\psi'})_{\mathsf{X}}$};
 \node[note] at (12.85,-4.85) {$\approx(U\ket\psi)_{\mathsf{X}}$};
\end{tikzpicture}
\caption{The circuit for a single step presented in \cref{eqn:dim_d_circuit_single_step}. The label register holds $\lambda$, the index register $\mathsf{X}$ the data, the phase and port registers $\mathsf{Z},\mathsf{P}$ the encoded data, the flags mark the good subspace, and the workspace serves the levels inside $C_k$. The preconditioners act on $\mathsf{X}$; the encoder moves the data from $\mathsf{X}$ into $\mathsf{Z},\mathsf{P}$, the child acts on $\mathsf{P}$ with $(\lambda,\mathbf{z})$ as controls, and the decoder moves the data back into $\mathsf{X}$. To make the direction of data flow clear, we label the state at each stage with its corresponding subspace.}
\label{fig:registers}
\end{figure}
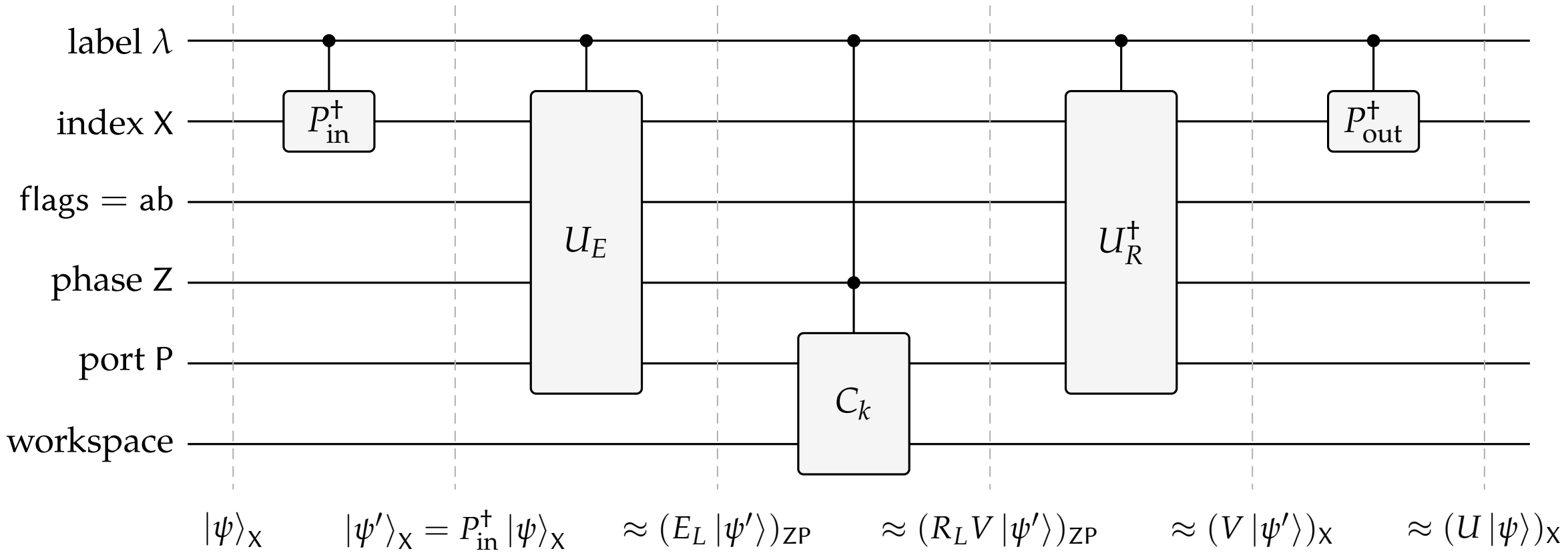

Concretely, with the maps $M_L(\mathbf{z}): \mathsf{L}(\C^k, \C^d)$ defined later, $\gamma E_L \ket{\psi}$ takes the form
$$
\gamma E_L \ket{\psi} = \ket{0}_{\mathsf{X}} \otimes \frac{1}{\sqrt{|G|}} \sum_{\mathbf{z}  \in G } \ket{\mathbf{z}}_{\mathsf{Z}} \otimes M_L(\mathbf{z})^\dagger \ket{\psi}_{\mathsf{P}} \otimes \ket{11}_{\mathsf{flags}}.
$$

Two features of \cref{eqn:ideal_identities} shape the rest of the construction. Since both $U_E$ and $U_R$ are unitary, $E_L$ and $R_L$ must (nearly) preserve norms, although each branch $\mathbf{z}$ holds only a $k$-dimensional compression of the input. And the child must uncompute its workspace to $\ket{0}$ for $U_R^{\dagger}$ to undo the encoding. The induction is stated in terms of clean error to control the overall effect of approximate implementation and imperfect uncomputation. Both identities hold approximately: with $r = 3/4$, the first has truncation error $8r^L$, each encoder has error $5r^{2L}$, and a child error $e(k)$ gives $e(d) \leq e(k) + 18r^L$, as we wrap up in \cref{appendix:wrap_up_the_proof}.

\paragraph{The child task and the recursion unrolled.}  The task solved by the child circuit $C_k$, with the labels written out, is given by
$$
\mathbf{F} = \sum_{\lambda \in \{0, 1\}^t } \sum_{\mathbf{z} \in G} \ps{\lambda, \mathbf{z}} \otimes F_{\lambda}(\mathbf{z}), 
$$
which reduces the data and control sizes of the synthesis task from $(d, t)$ to $(k, t + K)$, where $K = \log_2 |G| = \poly(n, L)$ is the number of qubits of the phase register. The induction hypothesis allows any number of label qubits, so it covers this task, whose data dimension is $k < d$; it would not if the phase register were counted as data, because $|G| \cdot k$ need not be smaller than $d$. Unrolling, the level $j$ of the recursion gives
\begin{equation}
\label{eqn:recursion_of_parameters_setup}
d_{j + 1} = d_j - s_j, \quad \lambda_j = (\lambda, \mathbf{z}^{(1)}, \dots, \mathbf{z}^{(j)}), \quad U_{(\lambda_j, \mathbf{z}^{(j + 1)})}^{(j + 1)} = F_{\lambda_j}^{(j)}(\mathbf{z}^{(j + 1)}), \quad t_j = t + j \cdot K,
\end{equation}
and the boundary condition $d_0 \coloneqq d$ and $U_{\lambda}^{(0)} \coloneqq U_{\lambda}$; here $\mathbf{z}^{(j+1)} \in G$ is the phase label created at level $j$, and $F^{(j)}$ is the feedback family of that level. The recursion stops when $d_m = 1$, where the targets are phase gates $\mathrm{e}^{\I \alpha_{\lambda_m}} \id \in \U(1)$, i.e., relative phases between label branches, and specifically they are implemented at once by one slot $C^{(m)}$ controlled by all labels. The expanded circuit reads
$$
C^{(0)} = \left( P_{\mathrm{out}}^{(0)} \right)^{\dagger} \left( U_{R}^{(0)} \right)^\dagger \cdots \left( P_{\mathrm{out}}^{(m- 1)} \right)^{\dagger} \left( U_{R}^{(m-1)} \right)^\dagger C^{(m)} U_{E}^{(m - 1)} \left( P_{\mathrm{in}}^{(m - 1)} \right)^{\dagger} \cdots U_{E}^{(0)} \left( P_{\mathrm{in}}^{(0)} \right)^{\dagger}.
$$
Since $m < J$, the whole circuit has at most $1 + Jp$ slots and clean error at most $18 J r^L$. From now on we suppress the index $\lambda$ again, and all bounds in the subsequent context hold blockwise with the same universal constants.

\subsubsection{Unitaries for the child task from a feedback loop}
\label[appendix]{appendix:unitary_for_child_task}
Where can a unitary on smaller dimensions that we pass to the child tasks come from? Split $\C^d = \C^{k} \oplus \C^{s}$ into the first $k$ coordinates as \emph{ports} and the last $s$ coordinates as the \emph{internal} ones, and we decompose a unitary $V \in \U(d)$ in the block matrix form
\begin{equation}
\label{eqn:block_decomposition_of_unitary}
V = \begin{pmatrix}
    A & B \\ C & D
\end{pmatrix}, \quad A \in \mathsf{L}(\C^k),~D \in \mathsf{L}(\C^s).
\end{equation}
We borrow these names from systems theory (see, e.g., \cite[Section 5]{Gough2009}), where they call the port \emph{external} coordinates. The signal enters and leaves via the ports, and the $s$ internal coordinates eliminated by the step are closed off by the feedback loop detailed below. The child will act on the ports, and hence the port register we defined in \cref{appendix:poly_size_recursion_by_one_call_per_step}. It seems that in \cref{eqn:block_decomposition_of_unitary} the port block $A$ is the immediate candidate to assign to the child task. $A$ itself is, however, not a unitary in general since $V^\dagger V = \id_{d}$ gives $A^\dagger A + C^\dagger C = \id_{k}$.

Instead of discarding the action of $C$, we feed it back. Specifically, we fix a diagonal unitary $Z$, for any port input $\mathbf{u} \in \C^{k}$, we look for a port input $\mathbf{v} \in \C^s$ and a port output y such that
\begin{equation}
\label{eqn:inner_loop_system}
V\left( \mathbf{u} \oplus Z \mathbf{v} \right) = \mathbf{y} \oplus \mathbf{v}, \quad \mathbf{v} = C \mathbf{u} + D Z \mathbf{v}, ~\mathbf{y} = A \mathbf{u} + BZ \mathbf{v}.
\end{equation}
Whenever $\id_{s} - DZ$ is invertible, solving the system in \cref{eqn:inner_loop_system} gives $\mathbf{y} = (A + BZ (\id_{s} - DZ)^{-1} C ) \mathbf{u}$. By the Neumann series \cite{Conway2007}, when $\|DZ\|_{\op} < 1$, denote
\begin{equation}
\label{eqn:expansion_of_closing_internal_coordinates}
\begin{aligned}
    F(Z) = A + BZ (\id_{s} - DZ)^{-1} C = A + BZ \sum_{\ell = 0}^{\infty}(DZ)^{\ell} C.
\end{aligned}
\end{equation}
We say that $F(Z)$ is obtained by \emph{closing the internal coordinates of $V$ via $Z$}, as illustrated in \cref{fig:closing_internal_coorinate_loop}, by iteratively feeding the internal output back and keeping only the map $\mathbf{u} \mapsto \mathbf{y}$. For every diagonal unitary $Z$, the operator $F(Z)$ is provably unitary and yields a smaller dimension, thus passing a smaller unitary synthesis problem to the child task.

\begin{figure}[!t]
\centering
\begin{tikzpicture}[>=Latex, semithick, font=\small]
 \node[draw, rounded corners=2pt, fill=gray!8, minimum width=32mm,
       minimum height=24mm] (V) at (0,0)
       {$V=\begin{pmatrix}A&B\\C&D\end{pmatrix}$};
 \draw[->] (-4.6,0.7) node[left]{$\ket u \in \C^k$} -- (-1.6,0.7);
 \draw[->] (1.6,0.7) -- (4.6,0.7) node[right]{$\ket y=F(Z) \ket u$};
 \node[above] at (-3.1,0.7) {port input};
 \node[above] at (3.1,0.7) {port output};
 \node[draw, fill=white, minimum size=7mm] (Z) at (0,-2.1) {$Z$};
 \draw[->] (1.6,-0.7) -- (3.0,-0.7) -- (3.0,-2.1) -- (Z.east);
 \draw[->] (Z.west) -- (-3.0,-2.1) -- (-3.0,-0.7) -- (-1.6,-0.7);
 \node[right] at (3.0,-1.4) {$\ket v \in \C^s$};
 \node[left] at (-3.0,-1.4) {$Z\ket v$};
 \node[above] at (2.3,-0.7) {\footnotesize internal};
 \node[above] at (-2.3,-0.7) {\footnotesize internal};
\end{tikzpicture}
\caption{Closing the internal coordinates of $V$ through a fixed diagonal unitary
$Z$. See also the first figure of \cite{261572}.}
\label{fig:closing_internal_coorinate_loop}
\end{figure}
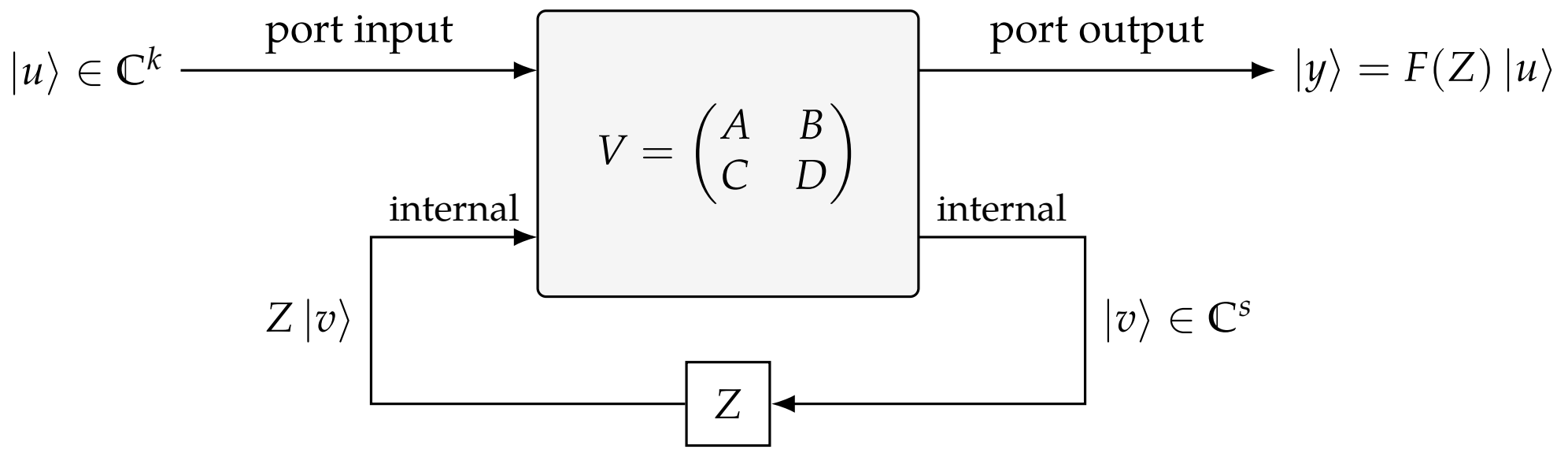

Practically, we will truncate the series with the finite-loop operation $BZ (DZ)^{L} C$ instead. If we can guarantee that $\| DZ \|_{\op} = \|D\|_{\op} < 1$, the truncation error decays exponentially, so choosing a polynomial loop depth would suffice.

\subsubsection{Convergence of the feedback loop}
\label[appendix]{appendix:convergence_of_feedback_loop}
Note that the idealized setting of \cref{appendix:unitary_for_child_task} need not hold in general. First, setting $V = \id_d$ so that $D = \id_{s}$ prevents signal transmission between the port and internal coordinate, so the procedure runs forever. Second, the internal block should not be set too large: since $CC^\dagger = \id_s - DD^\dagger$, when $s > k$, since $\rank(C) \leq k < s$, $DD^\dagger$ has an eigenvalue of $ 1$, so $\|D\|_{\op} = 1$ prohibits the loop from converging. So before obtaining $V$ for \cref{fig:closing_internal_coorinate_loop}, a proper preprocessing of the target unitary should choose the port-internal basis so that the block $D$ is contracting, while keeping the internal space small.

We borrow the preconditioning techniques in numerical linear algebra (see, e.g., \cite[Chapter 9]{Saad2003}). Recall \cref{eqn:dim_d_circuit_single_step}, we invoke the preconditioning on the target unitary $U$ via $V = P_{\mathrm{out}} U P_{\mathrm{in}}$. A na\"ive yet effective approach is to take both $P_{\mathrm{out}}$ and $P_{\mathrm{in}}$ Haar-random, and consequently so is $V$. This ensures that $D$ in the corner has $\mathcal{O}(\sqrt{s/d})$ operator norm with high probability. In general, Haar-random unitaries have no efficient circuits \cite[Section 4.5.4]{Nielsen2012}, but we show that this small-corner property can be achieved by random signs combined with Hadamard transforms, in the same vein as~\cite{doi:10.1142/S1793536911000787}, at the cost of polynomial overhead in the recursion level. To illustrate, let $q = 2^{\lfloor \log_2 d \rfloor}$, $\hdm_q$ is the Hadamard transform on last $q$ coordinates, and let $S_1, S_2$ be diagonal sign matrices in $\mathsf{L}(\C^q)$ embedded into the whole space. The actions of $S_1, S_2$ act locally on the corner of $U$, while $\hdm_q$ then pinches to spread them globally, so the internal block gives
\begin{equation}
\label{eqn:sum_of_rank_one_expression_of_internal_block}
D = \sum_{j=1}^{q} \tau_j \mathbf{h}_j \mathbf{m}_j^\dagger: \quad S_1 = \mathrm{diag}\left( \{\tau_j\}_{j=1}^q \right),
\end{equation}
where the rank-one terms arise from matrix operations involving $\hdm_q, U$, and $S_2$. These signs in $S_1, S_2$ play different roles: by keeping the amplitudes flat, a suitable choice of $S_2$ makes every $\mathbf{c}_j$ small in norm. For such an $S_2$ and a random choice of signs in $S_1$, a tail bound for matrix Rademacher series \cite[Corollary~4.2]{Tropp2012} ensures that whenever $s \leq q/(4096(n + 1)^2)$,
$$
\Pr_{S_1}\left[ \left\| D \right\|_{\op} \geq \frac{1}{4} \right] = \Pr_{\tau_1, \dots, \tau_q}\left[ \left\| D \right\|_{\op} \geq \frac{1}{4} \right] \leq 2s \exp\left( -\frac{q}{512 s(n + 1)} \right) < 1.
$$
So good signs for $S_1$ that make $D$ contractive exist, and we fix one such choice and hardwire it into the circuit. For small $d$ we keep $s = 1$ and permute an entry of modulus at most $1/\sqrt{2}$ to the corner. In all cases we are guaranteed that $s \geq d/(16384(n+1)^2)$ and $\|D\|_{\op} < 3/4 =: r$ (recall \cref{appendix:poly_size_recursion_by_one_call_per_step}), while both $P_{\mathrm{out}}$ and $P_{\mathrm{in}}$ cost $\bigo(n)$ slots, which keeps us within the ideal framework.

\subsubsection{Recovering the preconditioned unitary from the child task}
\label[appendix]{appendix:recovering_preconditioned_unitary}
As remarked earlier, we are to pass $F(\mathbf{z}) = F(Z(\mathbf{z}))$ coherently to the child task for all $\mathbf{z} \in G$. Even after we pass $F(\mathbf{z})$ to the child task and receive it back, we still need to decode to recover $V$ (cf. \cref{fig:closing_internal_coorinate_loop}). Note that the loop grants us more information than $F(Z)$: for a port input $\ket{u}$, the internal signal is fixed, and the loop determines the full input $M(Z) \ket{u}$ and the full output $N(Z) F(Z) \ket{u}$ of $V$, where analytically
\begin{equation}
\label{eqn:blocks_of_applications}
M(Z) = \begin{pmatrix}
    \id_{k} \\ Z(\id_s - DZ)^{-1} C
\end{pmatrix}, \quad N(Z) = \begin{pmatrix}
    \id_k \\ \left(\id_s - Z^\dagger D^\dagger\right)^{-1} Z^\dagger B^\dagger
\end{pmatrix},
\end{equation}
subject to $V M(Z) = N(Z) F(Z)$, or $F(Z) M(Z)^\dagger = N(Z)^\dagger V$ since both $V$ and $F(Z)$ are unitaries. Therefore for any $\ket{\psi} \in \C^d$ two routes would agree: (\textit{i}) compress $\ket{\psi}$ to the ports by $M(Z)^\dagger$ then apply $F(Z)$; (\textit{ii}) apply $V$ then compress via $N(Z)^\dagger$.

A compression in $\mathbb{C}^{k \times d}$ can not be inverted in general. The remedy is to use many diagonal unitaries at once coherently: fix a finite abelian group $G$ and a diagonal unitary $Z(\mathbf{z})$ for each member $\mathbf{z} \in G$. The child instead implements the controlled unitary $\mathbf{F} = \sum_{\mathbf{z} \in G} \ps{\mathbf{z}} \otimes F(\mathbf{z})$. Write $M(\mathbf{z}) = M(Z(\mathbf{z}))$ and $N(\mathbf{z}) = N(Z(\mathbf{z}))$ and setting the encodings
\begin{equation}
\label{eqn:encoder_and_decoder}
E \ket{\psi} \coloneqq \frac{1}{\sqrt{|G|}} \sum_{\mathbf{z} \in G} \ket{\mathbf{z}}_{\mathsf{Z}} \otimes M(\mathbf{z})^\dagger \ket{\psi}_{\mathsf{P}}, \quad R \ket{\psi} = \frac{1}{\sqrt{|G|}} \sum_{\mathbf{z} \in G} \ket{\mathbf{z}}_{\mathsf{Z}} \otimes N(\mathbf{z})^\dagger \ket{\psi}_{\mathsf{P}}.
\end{equation}
Then, extending branchwise by linearity gives us $\mathbf{F} E = RV$. If both $E$ and $R$ are isometries, we would be done since $V = R^\dagger R V = R^\dagger \mathbf{F} E$ can be synthesized by calling the child once. However, they need not be: $E^\dagger E = |G|^{-1} \sum_{\mathbf{z} \in G} M(\mathbf{z}) M(\mathbf{z})^\dagger$ averages rank-$k < d$ matrices, so for a constant family $Z(\mathbf{z}) \equiv Z$ it gives rise to a rank-$k$ matrix $M(Z) M(Z)^\dagger$, similar for $R$. It therefore remains to choose the family so that $E$ and $R$ are nearly isometric.

To obtain near-isometry, we use phase averaging to cancel out cross terms between contributions with different numbers of passes through the loop. A finite cyclic phase distinguishes these counts only modulo its order, so we first bound the number of cycles by truncating the Neumann series after $L$ terms. Still with $Z = Z(\mathbf{z})$ implicitly, we define the truncated version of \cref{eqn:blocks_of_applications}:
$$
M_L(\mathbf{z}) = \begin{pmatrix}
    \id_{k} \\ Z \sum_{\ell \geq 0}(DZ)^{\ell} C
\end{pmatrix}, \quad N_L(\mathbf{z}) = \begin{pmatrix}
    \id_k \\ \sum_{\ell \geq 0} (Z^\dagger D^\dagger)^{\ell} Z^\dagger B^\dagger,
\end{pmatrix}
$$
and define $E_L, E_L$ by replacing $M(\mathbf{z}), N(\mathbf{z}) $ with their truncated version in \cref{eqn:encoder_and_decoder}. As $\|D\|_{\op} \leq r$ this replacement changes each branch by at most $r^L / (1 - r) \leq 4 r^L$ for both $E$ and $R$.

These truncated maps involve only the cycle counts $0, 1, \dots, L$, which a common \emph{clock phase} can distinguish. Specifically, take $E_L^\dagger E_L$ averages $M_L(\mathbf{z}) M_L(\mathbf{z})^\dagger$ over $G$, and the cross terms yields interference between different cycles through $Z$. If each diagonal entry of $Z(\mathbf{z})$ shares a common factor $\omega$, then any term with $\ell + 1$ cycles carries a factor of $\omega^{\ell + 1}$. Averaging over all $m_0$-th roots of unity $\omega = 2\pi \I / m_0$ with $m_0 > L$ thus removes these cross terms. The remaining diagonal terms telescope via $CC^\dagger = \id_s - DD^\dagger$, which tells us that the component kept inside the cycle is at most an $r^{2L}$-fraction of the total. The same holds for $R_L$ via $B^\dagger B = \id_s - D^\dagger D$. If we take the diagonal entries of $Z(\mathbf{z})$ to be characters of the direct product group
\begin{equation}
\label{eqn:phase_group_setup}
G \coloneqq \left(\Z / m_0 \Z \right)^{L + 1}, \quad m_0 \coloneqq 2^b, \quad b = nL + \lceil \log_2(L + 1) \rceil,
\end{equation}
so that for $\mathbf{z} = (z_0, z_1, \dots, z_L) \in G$, the diagonal entries
$$
\forall\, i \in [s], \quad \left[ Z(\mathbf{z}) \right]_{i, i}= \chi_{\mathbf{v}_i}(\mathbf{z}) = \exp\left( \frac{2\pi \I}{m_0} \sum_{j=0}^{L} (\mathbf{v}_i)_j z_j  \right) , \quad \mathbf{v}_i = (1, i, i^2, \dots, i^L) \in G,
$$
where $\mathbf{v}_i$ is the frequency vector, and the first coordinate supplies the clock $\omega = \mathrm{e}^{2\pi \I z_0 / m_0}$ since $(\mathbf{v}_i)_0 = i^0 = 1$. With these constructions, the phase register is on $K = b(L + 1)$ qubits, and we can prove the near-isometric properties of $E_L, R_L$:
\begin{equation}
\label{eqn:near_isometry_condition}
\left( 1 - r^{2L} \right) \id_{d} \preceq E_L^\dagger E_L, R_L^\dagger R_L \preceq \id_d.
\end{equation}
The truncation perturbs the exact identity $\mathbf{F} E = RV$ by at most $8r^L$ branchwise among $\lambda$s.

\subsubsection{Encoder circuits via block encodings}
\label[appendix]{appendix:encoder_circuit_with_block_encoding}
The encoding operators $E_L, R_L$ constructed in \cref{appendix:recovering_preconditioned_unitary} ensures that the recovered unitary stays close to $V$ as we hope:
$$
\left\| R_L^\dagger \mathbf{F} E_L - V \right\|_{\op} \leq \left\| R_L^\dagger \left( \mathbf{F} E_L - R_L V \right) \right\|_{\op} + \left\| \left(R_L^\dagger R_L - \id_d \right) V  \right\|_{\op} \leq 8r^L + r^{2L},
$$
They are, however, now immediately implementable on a quantum circuit.
 To this end, the usual convention is to invoke the \emph{block encoding} technique \cite{QSVT} to build circuits for them. In addition, a key step between column preparation and the construction of the corresponding unitaries is to uncompute the input index coherently, as illustrated below.

 A natural but unsuccessful approach to unitary synthesis is to reduce it to the provably efficient state synthesis \cite[Proposition 3.3.5]{aaronson2016complexityquantumstatestransformations}. To illustrate, for a matrix $T$ on $d$ columns, such as the encoders $E_L$ or $R_L$ in our case, and an unknown input $\ket{\psi} = \sum_i \alpha_i \ket{i} \in \HH_{\mathsf{X}}$. We are to apply $T$ to $\ket{\psi}$, with a circuit that depends only on $T$. Since $T \ket{\psi} = \sum_i \alpha_i T \ket{i}$, one might start preparing $T \ket{i}$ branch by branch coherently. It acts on an output register with basis states $\ket{u}$ indexed by rows of $T$ and on a flag bit for normalization:
 \begin{equation}
 \label{eqn:column_wise_preparation_of_T}
 \ket{i}_{\mathsf{X}} \ket{0}_{\mathsf{ZP}} \ket{0}_{\mathsf{flag}} \mapsto \ket{i}_{\mathsf{X}} \left( \sum_{u} T_{u,i} \ket{u}_{\mathsf{ZP}} \ket{1}_{\mathsf{flag}} + \sqrt{1 - \left\|T \ket{i} \right\|^2 } \ket{0}_{\mathsf{ZP}} \ket{0}_{\mathsf{flag}} \right).
 \end{equation}
Extending by linearity, we readily see that the component marked with flag $1$ is incompatible with the desired clean implementation:
$$
\sum_{i} \alpha_i \ket{i}_{\mathsf{X}} \otimes T \ket{i}_{\mathsf{ZP}} \otimes \ket{1}_{\mathsf{flag}} \neq \sum_{i} \alpha_i \ket{0}_{\mathsf{X}} \otimes T \ket{i}_{\mathsf{ZP}} \otimes \ket{1}_{\mathsf{flag}}.
$$
To obtain $T \ket{\psi}$, one needs to uniformly uncompute the index register to $\ket{0}$, which brings us back to the main obstacle of the USP \cite{yuen2022unitarysynthesis}: for a general dense $T$, this is usually impossible.

Fortunately, the encoders themselves are not ``general matrices''. Uncomputing the index $i$ from $\ket{i} \ket{u}$ only requires a short list of candidates for $i$ given $u$, i.e., few non-zero entries within each row. In the basis $\{\ket{\mathbf{z}}\}_{\mathbf{z} \in G}$, the rows of $E_L$ and $R_L$ are dense, while in the phase-register Fourier basis they are sparse. Let $\Fou_G: \ket{\mathbf{w}} \mapsto |G|^{-1/2} \sum_{\mathbf{z} \in G} \chi_\mathbf{w}(\mathbf{z}) \ket{\mathbf{z}}$ be the Fourier transform of group $G$ and put the Fourier-transformed operators $\widehat{E}_{L} = (\Fou_G^\dagger \otimes \id_{k}) E_L$ and similarly $\widehat{R}_{L} = (\Fou_G^\dagger \otimes \id_{k}) R_L$ with rows indexed by a frequency $w \in G$ and port $a$. As detailed later, expanding $M_L(\mathbf{z})^\dagger \ket{i}$ for an internal input $i$ gives a sum over \emph{(Feynman) paths} of the signal, i.e., over index sequences $(i_1, i_2, \dots, i_m = i)$ with length $1 \leq m \leq L$. Each path picks a diagonal phase per cycle, and these multiply to a single character:
$$
\overline{Z(\mathbf{z})_{i_1, i_1}} \cdot \overline{Z(\mathbf{z})_{i_2, i_2}} \cdot \cdots \cdot \overline{Z(\mathbf{z})_{i_m, i_m}} = \chi_{-(\mathbf{v}_{i_1} + \mathbf{v}_{i_2} + \cdots + \mathbf{v}_{i_m})}(\mathbf{z}).
$$
By orthogonality among distinct characters, the Fourier coefficient at frequency $w$ keeps only the paths with $\mathbf{v}_{i_1} + \mathbf{v}_{i_2} + \cdots + \mathbf{v}_{i_m} = -\mathbf{w}$\footnote{
A similar histogram-tracking formulation also appears in \cite{tang2026amplitudeamplificationestimationrequire}.
}, so the entries of $\widehat{E}_L$ give
$$
\left[\widehat{E}_L \right]_{(\mathbf{w}, a), i} = \sum_{m=1}^{L} \sum_{\substack{(i_1, i_2, \dots, i_m = i): \\ \mathbf{v}_{i_1} + \mathbf{v}_{i_2} + \cdots + \mathbf{v}_{i_m}  = -\mathbf{w}} } \left[ C^\dagger\right]_{a, i_1} \left[ D^\dagger\right]_{i_1, i_2} \cdots \left[ D^\dagger\right]_{i_{m-1}, i_m}.
$$
And similarly for $\widehat{R}_L$ with $B, D$ in the place of $C^\dagger, D^\dagger$, respectively, and with $\mathbf{w}$ in place of $-\mathbf{w}$.

Here the choice of the frequency vectors $\mathbf{v}_i$ pays off. Specifically, the sum $\sum_{j=1}^m v_{i_j}$ gives
$$
\sum_{j=1}^m v_{i_j} = \left( p_0 = m, p_1, \dots, p_L \right): \quad p_j = \sum_{\ell = 1}^{m} i_{\ell}^j,
$$
of the size and the power sums of multiset $\{i_1, i_2, \dots, i_m\}$. Each coordinate is bounded from above by $L \cdot (2^n)^{L} = L \cdot 2^{n L} < 2^b$ since $j \leq L$ and $i \leq s < d = 2^n$, so by the setup of $b$, nothing wraps modulo $m_0$. By Newton’s identities \cite{symm_func_1995}, the
size and the power sums jointly determine the multiset. Hence all paths that contribute to a row $(\mathbf{w}, a)$ visit the same multiset of at most $L$ coordinates, and each of them contains its input coordinate $i$. As port inputs contribute solely at frequency $\mathbf{0}$, which is not the sum of any nonempty multiset, every row of $\widehat{E}_L$ and $\widehat{R}_L$ has at most $L$ nonzero entries, yielding sufficient sparsity.

With at most $L$ candidates per row, the index can then be uncomputed with a known amplitude. Let $T$ be either $\widehat{E}_L$ or $\widehat{R}_L$. For each row $u$, we choose a list $I_u$ on exactly $L$ indices collecting all $i$ with non-zero $T_{u, i}$, padded with dummy indices if necessary, which exist with index register extended by $h = \lceil \log_2(L + 1) \rceil$ qubits. Suppose $W_u$ prepares uniform superposition over $I_u$:
$
W_u \ket{0} = \ket{s_u} = \frac{1}{\sqrt{L}} \sum_{i \in I_u} \ket{i}
$. Then, after the column-wise application in \cref{eqn:column_wise_preparation_of_T}, apply $W_u^\dagger$ controlled on $u$. The branch we want is the one with the joint index-flag register in $\ket{0}_{\mathsf{X}} \ket{1}_{\mathsf{flag}}$: there the state reads
$$
\begin{aligned}
\sum_{u} \left( \sum_{i \in I_u} \alpha_i T_{u, i} \braket{0 | W^\dagger_u | i} \right) \ket{0}_{\mathsf{X}} \ket{u}_{\mathsf{ZP}} \ket{1}_{\mathsf{flag}} &= \frac{1}{\sqrt{L}} \sum_{u} \braket{u | T | \psi} \ket{0}_{\mathsf{X}} \ket{u}_{\mathsf{ZP}} \ket{1}_{\mathsf{flag}} \\
&= \frac{1}{\sqrt{L}} \ket{0}_{\mathsf{X}} \otimes T \ket{\psi}_{\mathsf{ZP}} \ket{1}_{\mathsf{flag}}.
\end{aligned}
$$
Thus, for every input, the component is exactly $T \ket{\psi} / \sqrt{L}$ with the index uncomputed. The circuit is therefore a block encoding of $T / \sqrt{L}$.

It suffices to distill the good subspace by amplifying this component. Recall that $T^\dagger T \approx \id$, the amplitude of the good component is nearly identical for every input, so we invoke the oblivious amplitude amplification \cite{10.1145/2591796.2591854} in its robust version \cite{QSVT}. To make $L$ rounds land exactly, we add a rotation on a second flag to lower the factor $1/\sqrt{L}$ to $t_0 = \sin\left( \pi / (4L + 2) \right)$. Let the resulting unitary be $B_T$, then $\gamma^\dagger \cdot B_T \cdot \iota = t_0 T$, where $\iota$ supplies clean workspace and $\gamma$ embeds the rows into the \emph{good subspace}, in which the index and flags has joint state $\ket{0}_{\mathsf{X}} \ket{11}_{\mathsf{flag}}$. Then we iterate the amplification operator $\mathscr{G}_T = ( 2 B_T \iota \iota^\dagger B_T^\dagger - \id ) ( \id - 2 \gamma \gamma^\dagger )$ \cite{Brassard2002} for $L$ rounds to obtain the good subspace exactly. Take the unitary $U_T \coloneqq \mathscr{G}_T^L B_T$, it can be proved that
\begin{equation}
\label{eqn:near_unitary_condition}
(1 - \eps) \id \preceq T^\dagger T \preceq \id \quad \implies \quad \left\| U_T \cdot \iota - \gamma \cdot T \right\|_{\op} \leq 5 \eps.
\end{equation}
The amplification stays inside the encoders by interleaving $B_T$ and $B_T^\dagger$. Since the Fourier transform is unitary, the near-isometric condition \cref{eqn:near_isometry_condition} holds for $T = \widehat{E}_L, \widehat{R}_L$ as well, and after the amplification we apply $\mathscr{F}_G$ to take everything back to the primitive basis. Therefore, substituting $\eps = r^{2L}$ into \cref{eqn:near_unitary_condition} yields a bound on the error incurred by implementing $E_L$ and $R_L$ via block encodings.

\subsubsection{Error, size, and the Boolean oracle}
\label[appendix]{appendix:wrap_up_the_proof}
With all components in place, we now combine the arguments above to complete the solution to \cref{prob:USP}: we bound the error of a single step and then the error and size of the whole recursion, tune the characterizing parameters, and finally replace the programmable slots with elementary gates and, most significantly, calls to the Boolean oracle.

\paragraph{One-step error.} We begin with analyzing the error. Suppose the child yields clean error $e(k)$, that is, $\|C_k \cdot \gamma - \gamma \cdot \mathbf{F} \|_{\op} \leq e(k)$, then we telescope along \cref{eqn:state_evolution_chain} and combine inequalities established in \cref{appendix:recovering_preconditioned_unitary} and \cref{eqn:near_unitary_condition}, we can get
\begin{equation}
\label{eqn:error_bound_per_step}
\begin{aligned}
e(d) = \left\| U_R^\dagger C_k U_E \cdot \iota -   \iota \cdot V \right\|_{\op} \leq e(k) + 18 r^{L}.
\end{aligned}
\end{equation}

\paragraph{The whole recursion.} At the bottom level we have $d = 1$, all that suffices to implement is the single exact slot $\ket{\lambda} \ket{0^n} \mapsto \mathrm{e}^{\I \alpha_{\lambda}} \ket{\lambda} \ket{0^n}$, and therefore $e(1) = 0$. The recursion level $J = \mathcal{O}(n^3)$. Summing \cref{eqn:error_bound_per_step} over the levels gives clean error $e(d) \leq 18 J r^{L}$. The circuit sizes add up similarly, since each level makes a single call to the child. A level contains two encoders, each requiring $L$ amplification rounds of $\bigo(K + n + h)$ slots and a Fourier transform of $\bigo(K)$ slots, and preconditioners of $\bigo(n)$ slots. Each recursion level also adds $K + n + h + 2$ qubits in total for phase and port registers, index extension, and two flags (cf. \cref{fig:registers}), and $\bigo(n)$ temporary qubits for aligning the $q$-dimensional subspace. With $t$ controlled label qubits, summing the per-level costs over all levels quantifies the total number of slots $S$ and qubits $Q$:
$$
S = \bigo\left( 1 + J \left[ K + n + L(K + n + h) \right] \right), \quad Q = t + \bigo\left( n + J(K + n + h) \right).
$$
For a final operator-norm error $\delta = 2^{-\beta - 1}$ we take $L = 4(\beta + \lceil \log_2(72J) \rceil)$ so $r^L < 2^{-\beta} / 72J = \delta/36J$ since $r^4 < 1/2$. Then $L, h = \bigo(n + t + \beta)$ and thus $K, Q, S = \poly(n, t, \beta)$.

\paragraph{Transforming slots to one Boolean oracle.} Here the separation made at the start of this section pays off: the skeleton depends only on the sizes, while everything that depends on the target sits in the tables, which may be exponentially long and need not be computable. Compilation hands the tables to the oracle. For each slot and each control value $y$ (labels and possibly data or workspace bits), the oracle stores a short word over $\hdm, \tgate, \tgate^\dagger$ that approximates $W_y$, while $\cnot$ gates help implement the controlled operations in the slots (cf. \cref{eqn:programmable_slot}). The controls serve as the query address, so each branch of a superposition fetches, applies, and uncomputes its own entry, and the length of a table costs nothing in the circuit. It remains to keep the words short and phase-exact, to lay out the oracle, and finally to read the words coherently. We take steps to illustrate how they are handled.

\subparagraph{Words.} The phase of $W_y$ is a relative phase between control branches and must be preserved, but every word over $\hdm, \tgate, \tgate^\dagger$ has
determinant a power of $\mathrm{e}^{\I \pi/4}$, so words cannot approximate a general $W_y$ arbitrarily well. We therefore split off the phase: write $W_y=\mathrm{e}^{\I \phi_y}S_y$ with $S_y\in\SU(2)$, apply $S_y$ to the target, and apply $P_y=\diag(\mathrm{e}^{\I \phi_y},\mathrm{e}^{-\I \phi_y})$ to an auxiliary qubit in $\ket{0}$ to get the phase. It suffices to approximate $S_y$ and $P_y$ via the Solovay-Kitaev theorem \cite{KitaevShenVyalyi2002} with word length polylogarithmic in the inverse precision and thus affordable for the oracle circuit. $S_y$ and $P_y$ are approximated to $\kappa$ precision bits by determinant-one words, padded with $\id$ to a common length $\ell_\kappa = \poly(\kappa)$.

\subparagraph{The Boolean oracle.} We insert the words of all slots and control values into the Boolean function at addresses that the circuit can compute from the skeleton alone. We encode each element $g$ in the gate set $\{\id, \hdm, \tgate, \tgate^\dagger\}$ by two bits, that is, $\mathrm{code}(g) \in \{0, 1\}^2$. Let $\widehat y$ be the control string obtained by padding $y$ to the full width of the skeleton on $Q$ bits, and let $g_{j,\widehat y,i}$ and $g'_{j,\widehat y,i}$ be the $i$-th letters of the target word and of the phase word in slot $j$, respectively. We then set the oracle functions
\begin{equation}
\label{eqn:boolean_function_definition}
 f(j,\widehat y,0,i,\cdot)=\mathrm{code} \left(g_{j,\widehat y,i} \right), \quad 
 f(j,\widehat y,1,i,\cdot)=\mathrm{code} \left(g'_{j,\widehat y,i} \right),
\end{equation}
where the last address bit selects one of the two code bits.

\subparagraph{Coherent reading of gate instructions from the oracle.} We extend the preceding discussion branchwise, with each branch reading its own word and then uncomputing. Each letter is therefore queried into clean qubits, applied by an exact controlled circuit, and uncomputed by a second query. By linearity, the slot is reproduced within $2^{-\kappa}$ on superpositions of control values, and the errors of all these $S$ slots add up to at most $S \cdot 2^{-\kappa}$.

With $\kappa=\beta+2+\lceil\log_2(S+1)\rceil$ the compilation error is less than $\delta/2$, so the total clean error is less than $\delta =2^{-\beta-1}$. The Turing machine prints only the skeleton and the address computations, which depend solely on $n,d,t,\beta$. Like all other data that depend on the target unitary $U$, the words come for free with the oracle, since its truth table can depend arbitrarily on $U$. Wrapping up everything, we finally conclude the positive answer to \cref{prob:USP}.

\subsection{The construction in full}
In this section, we give the construction in full. We define precisely the objects that the overview introduced only informally, and prove the claims that it stated without proof in the same order. Some of the context therefore overlaps slightly with \cref{appendix:proof_overview}, and we repeat it so that this section can be read on its own, and refer back to the overview for the intuition.

\subsubsection{The programmable slots} 
\label[appendix]{appendix:programmable_slots}
We will formalize the slot model previously defined and provide several basic slot-based constructions used in the proof. Slot models are what Bergholm, Vartiainen, M\"ott\"onen and Salomaa \cite{PhysRevA.71.052330} call uniformly controlled gates, also known as quantum multiplexors \cite{1629135}.

\begin{definition}[Programmable slots and skeletons {\cite{ PhysRevA.71.052330, 1629135}}]
\label{def:programming_slot_and_skeleton}
    A \textit{slot} specifies a target qubit and an ordered list of qubits as controls. A \emph{table} is a list of unitaries $\{W_y\}_{y \in \{0, 1\}^* } \subseteq \U(2)$. The table is filled into a slot to jointly act as $W = \sum_{y \in \{0, 1\}^* } \ps{y} \otimes W_y$. A \emph{skeleton} is a list of registers and a sequence of slots on them, and its \emph{size} is the number of slots. It suffices to choose the table for each slot to turn a skeleton into a circuit. 
\end{definition}

\begin{lemma}[\cite{grover2002creatingsuperpositionscorrespondefficiently, PhysRevA.71.052330}, extended]
\label{lemma:diagonal_gates_controlled_state_prep_in_slot_formalism}
    A diagonal unitary on $u \geq 1$ qubits is one slot. For a family of unit vectors $\{ \ket{a_{\lambda}} \}_{\lambda} \subseteq \C^{2^u}$, the map $\ket{\lambda} \ket{0^u} \mapsto \ket{\lambda} \ket{a_{\lambda}}$ extends to a unitary given by a skeleton of $u + 1$ slots. Both constructions implement the prescribed maps exactly.
\end{lemma}

\begin{proof}
    We inherit the ideal of Grover-Rudolph type decomposition \cite{grover2002creatingsuperpositionscorrespondefficiently}. For a diagonal unitary $\Lambda = \sum_{x \in \{0, 1\}^u } \mathrm{e}^{\I \phi_x} \ps{x}$, we split $x = yb$ with $y \in \{0, 1\}^{u - 1}$ and $b \in \{0, 1\}$, and choose the last qubit as target and first $u - 1$ qubits as controls. $\Lambda$ is a slot in the sense of
    $$
    \Lambda = \sum_{y \in \{0, 1\}^{u - 1} } \sum_{b \in \{0, 1\} } \mathrm{e}^{\I \phi_{yb}} \ps{y, b} = \sum_{y \in \{0, 1\}^{u - 1} } \ps{y} \otimes W_y, \quad W_y = \begin{pmatrix}
        \mathrm{e}^{\I \phi_{y0}} & 0 \\ 0 & \mathrm{e}^{\I \phi_{y 1}}
    \end{pmatrix} \in \U(2).
    $$

    For the state preparation, we use the standard construction from uniformly controlled rotations by splitting the probability mass one bit at a time \cite{grover2002creatingsuperpositionscorrespondefficiently, 2011670.2011675, aaronson2016complexityquantumstatestransformations, RY22}. Fix a label and write $\ket{a} = \sum_x a_x \ket{x}$, and for a prefix $p$, we set probability $\Pr[p] =\sum_y |a_{py}|^2$. For $j \in [u]$ a real rotation of bit $j$ controlled by first $j - 1$ bits splitting $\sqrt{\Pr[p]}$ into $\sqrt{\Pr[p0]}$ and $\sqrt{\Pr[p1]}$, and for $\Pr[p]= 0$ any rotation would do. A final diagonal gate multiplies the non-zero amplitude of $x$ by $a_x / |a_x|$. In the slot model, each of these $u+1$ uniformly controlled gates is one slot, and adding $\lambda$ to every control list makes all choices label-wise.
\end{proof}

\begin{lemma}[Permutations with clean ancillae \cite{PhysRevA.54.147}]
\label{lemma:perm_with_clean_ancillae}
    For a family of permutations of the $u$-qubit computational basis $\{\pi_{\lambda}\}_{\lambda \in \{0, 1\}^t} \subseteq \mathfrak{S}_{2^u}$, the controlled map $\ket{\lambda} \ket{x} \mapsto \ket{\lambda} \ket{\pi_{\lambda}(x)}$ has a common skeleton of $5u$ slots using $u$ clean ancillae\footnote{
    Specifically, it intertwines with the isometry $\iota$ (cf. \cref{sec:prelim}).
    }.
\end{lemma}

\begin{proof}
    First, $u$ slots map $\ket x\ket0\mapsto\ket x\ket{\pi_\lambda(x)}$: the $j$-th slot targets the $j$-th ancilla, is controlled by $\lambda$ and $x$, and has the table entry $X$ or $\id$ according to the $j$-th bit of $\pi_\lambda(x)$. Next, $u$ slots controlled by $\lambda$ and by the ancilla register storing $y$ XOR $\pi_\lambda^{-1}(y)$ into the first register, which gives $\ket0\ket{\pi_\lambda(x)}$. Finally, three $\cnot$s per bit swap the two registers. Each step permutes basis states, so by linearity this action extends naturally to superpositions. The total number of slots reads $u + u + 3u = 5u$. This completes the proof.
\end{proof}

\begin{lemma}[Skeleton for Fourier transform \cite{Nielsen2012}]
\label{lemma:fourier_transform_skeleton}
    For every $b \in \N$, the transform
    $$
    \mathscr{F}_b \ket{x} = \frac{1}{\sqrt{2^b}} \sum_{z = 0}^{2^b - 1} \exp\left( \frac{2\pi \I x \cdot z}{2^b} \ket{z}\right)
    $$
    has a uniformly generated skeleton with $\bigo(b)$ slots. For tensor products of these transforms, the total number of slots scales linearly with its total width.
\end{lemma}

\begin{proof}
Write $x = \sum_{j=1}^b x_j 2^{b - j}$ in its binary and number the qubits $\mathsf{X}_1, \dots, \mathsf{X}_b$ from most to least significant. We refer to \cite[Figure 5.1]{Nielsen2012} for a detailed illustration, which rests on the product of phased uniform superpositions:
    $$
    \mathscr{F}_b \ket{x} = \bigotimes_{ j = 1 }^{b } \left( \frac{\ket{0} + \mathrm{e}^{ \frac{2 \pi \I}{2^{j}} x } \ket{1} }{\sqrt{2}} \right).
    $$
    To prepare this state, we process $\mathsf{X}_k$ in the order $k = 1, 2, \dots, b$ as follows: apply a Hadamard gate in one slot. For $k < b$, the controlled rotations following are combined into one slot in the form of \cref{def:programming_slot_and_skeleton} with target $\mathsf{X}_k$ controlled by all $\mathsf{X}_j$ with $k < j < b$. The skeleton labels $y$ and unitaries $W_y^{(k)}$ are given by
    $$
    W_y^{(k)} = \diag\left( 1, \exp\left( 2 \pi \I \cdot \sum_{\ell = k + 1}^b \frac{y_{\ell}}{2^{\ell - k + 1}} \right)  \right), \quad y = y_{k+1} y_{k+2} \cdots y_b \in \{0, 1\}^{b - k}.
    $$
    There are $b$ Hadamard slots, $b - 1$ slots for phase gates, and $\lfloor b/2 \rfloor$ final swaps to resume the desired output order, each using three $\cnot$ slots. The total gate count is $b + (b -1) + 3 \lfloor b/2 \rfloor = \bigo(b)$, with the target and control lists generated uniformly from $b$ alone.
\end{proof}

\subsubsection{A contracting internal block}
\label[appendix]{appendix:contracting_internal_block}
In this section, we illustrate how properly chosen preconditioners $P_{\mathrm{out}},P_{\mathrm{in}}$ with short skeletons help make the internal block of $V= P_{\mathrm{out}}UP_{\mathrm{in}}$ (cf. \cref{fig:closing_internal_coorinate_loop}) contractive.

\paragraph{Extracting the internal block.} To start with, we fix $U \in \U(d)$ with $2 \leq d \leq 2^n$, and let $q = 2^{\lfloor \log_2 d \rfloor}$ and take the internal block size $s \in [q]$ with value later chosen. Let $J_q = (0_{q \times (d - q)}, \id_{q})^{\TP} \in \C^{d \times q}$, $E = (0_{s \times (q-s)}, \id_{s})^{\TP} \in \C^{q \times s}$ be coordinate embeddings. Then we extract the bottom-right $q \times q$ corner of $U$ by $M_0 = J_q^\dagger U J_q$.
Define the smaller Hadamard gate $\hdm_q = \hdm^{\otimes \log_2 q}$ and let $W = \hdm_q E \in \C^{q \times s}$ with entries $W_{j, a} = [\hdm_q]_{j, q - s + a}$ for $j \in [q]$ and $a \in [s]$. If follows that $W^\dagger W = \id_s$ and $|W_{j, a}| = 1/\sqrt{q}$ for all $j, a$.

For two diagonal sign unitaries $S_1, S_2 \in \C^{q \times q}$ with entries in $\{-1, 1\}$, we put 
$$
P_{\mathrm{out}} = \id_{d - q} \oplus \hdm_q S_1, \quad P_{\mathrm{in}} = \id_{d - q} \oplus S_2 \hdm_q.
$$
We can therefore extract the bottom-right $s \times s$ block $D$ of the encoded unitary via
\begin{equation}
\label{eqn:internal_block_D_expression}
D = \left( J_q E \right)^\dagger P_{\mathrm{out}}UP_{\mathrm{in}} J_q E = E^\dagger \hdm_q S_1 M_0 S_2 \hdm_q E = \underbrace{E^\dagger \hdm_q^\dagger}_{=: W^\dagger} S_1 \underbrace{M_0 S_2 \hdm_q E}_{=: M_1} = W^\dagger S_1 M_1 \in \C^{s \times s}.
\end{equation}
For $i \in [q]$, let $\mathbf{e}_i$ be the $i$-th standard basis vector, the column vector $\mathbf{m}_i = M_i^\dagger \mathbf{e}_i$. We first choose $S_2$ so that these norms are small. The following lemma bounds the individual entries and hence the row norms, as in the analysis of the subsampled randomized Hadamard transform \cite{doi:10.1142/S1793536911000787} extended to complex-valued matrices.

\begin{lemma}
\label{lemma:random_sign_entry_and_row_bounds}
    Let $T \in \C^{q \times q}$ with $\|T\|_{\op} \leq 1$ and $W \in \C^{q \times s}$ have orthonormal columns with entry modulus $1/\sqrt{q}$. Let $S = \diag(\eps_1,  \dots, \eps_q)$ with i.i.d. uniform signs $\eps_j \in \{-1, 1\}$, and put $M_1 = T S W$. Then $\|M_1\|_{\op} \leq 1$ and 
    $$
   \forall\, i \in [q], \quad  \E_{\eps_1, \dots, \eps_q}\left[ \left|[M_1]_{i, a} \right|^2 \right] = \frac{\left[ T T^\dagger \right]_{i, i}}{q} \leq \frac{1}{q}.
    $$
    And for every $x > 0$, it holds that
    $$
    \Pr_{\eps_1, \dots, \eps_q}\left[ \max_{i \in [q], a \in [s] } \left| [M_1]_{i, a} \right| \geq x\right] \leq 4 qs \mathrm{e}^{ - \frac{qx^2}{4} }.
    $$
    Outside this event, for each $i \in [q]$, take $\mathbf{m}_i = M_1^\dagger \mathbf{e}_i \in \C^s$, then $\| \mathbf{m}_i \|^2 \leq s x^2$.
\end{lemma}

\begin{proof}
    The operator norm bound follows readily by $$\|M_1\|_{\op}=\|TSW\|_{\op} \leq\|T\|_{\op}\|S\|_{\op}\|W\|_{\op} =\|T\|_{\op} \leq 1.$$
    Expanding entries as sums gives
    $$
   \forall \, i \in [q],~a \in [s], \quad [M_1]_{i, a} = \sum_{j \in [q] } \eps_j \alpha_j, \quad \alpha_j = T_{i, j} W_{j, a}, \quad \sum_{j \in [q]} |\alpha_j|^2 = \frac{1}{q} \sum_{j \in [q]} |T_{i, j}|^2 \leq \frac{1}{q}.
    $$
    The second-moment argument follows from the basic identity
    $$
    \E_{\eps_1, \dots, \eps_q} \left[ \left|\sum_{j \in [q] } \eps_j \alpha_j \right|^2 \right] = \sum_{j \in [q]} |\alpha_j|^2 + \sum_{k \neq \ell} \E_{\eps_1, \dots, \eps_q} \left[ \eps_{k} \eps_{\ell} \overline{\alpha_{k}} \alpha_{\ell} \right] = \sum_{j \in [q]} |\alpha_j|^2 \leq \frac{1}{q}.
    $$
    By Hoeffding's inequality, for a real sign sum $\sum_j \eps_j a_j$ with $\sum_j a_j^2 \leq 1/q$ has modulus at least $x$ with probability at most $2 \mathrm{e}^{-q x^2 / 2}$. Suppose $|\sum_{j \in [q]} \eps_j \alpha_j| \geq x$, its real or imaginary part certainly has modulus $\geq x/\sqrt{2}$. Extending this to every entry of $M_1$, by a union bound we have
    $$
    \Pr_{\eps_1, \dots, \eps_q}\left[ \max_{i \in [q], a \in [s] } \left| [M_1]_{i, a} \right| \geq x\right] \leq \sum_{i \in [q] , a \in [s]} \Pr_{\eps_1, \dots, \eps_q} \left[  \left| [M_1]_{i, a} \right| \geq x\right] \leq 4 qs \cdot \mathrm{e}^{ - \frac{qx^2}{4} }.
    $$
    Outside this event $\| \mathbf{m}_i \|^2 = \sum_{a \in [s] } |[M_1]_{i, a}|^2 \leq \sum_{a \in [s]} x^2 \leq sx^2$. This concludes the proof.
\end{proof}

\paragraph{Choosing the signs $S_2$.} We choose the $\varepsilon_j$ independently and uniformly from $\{-1,1\}$. Apply \cref{lemma:random_sign_entry_and_row_bounds} by plugging in $T=M_0$, $S=S_2$, and $x =4\sqrt{(n+1)/q}$. Since $qs\leq 4^n$, the failure probability is bounded by $4\cdot4^n\mathrm{e}^{-4(n+1)}=(4\mathrm{e}^{-4})^{n+1}<1$, so there exists some $S_2$ satisfying
\begin{equation}
\label{eqn:short-rows}
 \max_{i \in [q]} \|\mathbf{m}_i\|^2\leq sx^2 = \frac{16s(n+1)}q, \quad \|M_1\|_{\op} \leq 1.
\end{equation}
We keep this $S_2$ fixed and similarly choose the signs $\tau_i$ of $S_1 = \diag(\tau_1, \dots, \tau_q)$ independently and uniformly from $\{-1,1\}$. Define $\mathbf{h}_i=W^{\dagger}\mathbf{e}_i\in\C^s$, so that the $i$-th rows of $W,M_1$ are
$\mathbf{h}_i^{\dagger},\mathbf{m}_i^{\dagger}$, respectively. Then \cref{eqn:internal_block_D_expression} becomes
\begin{equation}
\label{eqn:internal_block_via_rank-one-sum}
 D=W^\dagger S_1 M_1  = \sum_{i=1}^q W^\dagger \left( \tau_i \mathbf{e}_i \mathbf{e}_i^\dagger \right)   M_1 = \sum_{i=1}^q\tau_i \mathbf{h}_i \mathbf{m}_i^{\dagger},
\end{equation}
as previously formulated in \cref{eqn:sum_of_rank_one_expression_of_internal_block}. We now show that a suitable choice of $S_1$ makes $D$ contractive via the following standard tail inequality.

\begin{lemma}[Matrix Rademacher series {\cite[Corollary 4.2]{Tropp2012}}]
\label{lemma:matrix_rademacher}
    Let $B_1, B_2, \dots, B_q$ be fixed complex $d_1 \times d_2$ matrices, let $\tau_1, \tau_2, \dots, \tau_q$ be independent random signs uniformly distributed on $\{-1, 1\}$, define
    $$
    \sigma^2 = \max\left\{  \left\| \sum_{i=1}^q B_i B_i^\dagger \right\|_{\op}, \left\| \sum_{i=1}^q B_i^\dagger B_i \right\| \right\}.
    $$
    Then for every $x \geq 0$,
    $$
    \Pr_{\tau_1, \dots, \tau_q}\left[ \left\| \sum_{i=1}^q \tau_i B_i \right\| \geq x \right] \leq (d_1 + d_2) \mathrm{e}^{ -\frac{x^2}{2 \sigma^2} }.
    $$
\end{lemma}

\paragraph{The internal block can be contractive.}
Take $B_i = \mathbf{h}_i \mathbf{m}_i^\dagger$ in \cref{lemma:matrix_rademacher} and by \cref{eqn:short-rows},
$$
\begin{aligned}
\left\| \sum_{i=1}^q B_i B_i^\dagger \right\|_{\op} &= \left\| \sum_{i=1}^q \|\mathbf{m}_i\|^2 \mathbf{h}_i \mathbf{h}_i^\dagger \right\|_{\op} \leq  \frac{16s(n + 1)}{q} \left\| W^\dagger W \right\|_{\op} = \frac{16s(n + 1)}{q}, \\
\left\| \sum_{i=1}^q B_i^\dagger B_i \right\|_{\op} &= \left\| \sum_{i=1}^q \|\mathbf{h}_i\|^2 \mathbf{m}_i \mathbf{m}_i^\dagger \right\|_{\op} = \frac{s}{q} \left\| \sum_{i=1}^q \mathbf{m}_i \mathbf{m}_i^\dagger \right\|_{\op} = \frac{s}{q}  \left\| M_1^\dagger M_1 \right\|_{\op} \leq \frac{s}{q}.
\end{aligned}
$$
And therefore $\sigma^2 \leq 16s(n+ 1) / q$. Then \cref{lemma:matrix_rademacher} tells us that $D = \sum_{i=1}^q \tau_i B_i$ with $d_1 = d_2 = s$ and $x = 1/4$ gives
\begin{equation}
\label{eqn:operator_norm_probability_bound}
\Pr_{S_1} \left[ \left\| D \right\|_{\op} \geq \frac{1}{4} \right] \leq 2 s \mathrm{e}^{ -\frac{1}{32} \cdot \frac{q}{16s(n + 1)} } = 2s \mathrm{e}^{- \frac{q}{512 s(n + 1)} }.
\end{equation}
Hence, for every $U$ and every $1 \leq s \leq  q$ for which the right-hand side is less than $1$, there exists a pair of sign matrices $S_1, S_2$ with $\|D\|_{\op} < 1/4$.

\paragraph{Tuning the block size.} With the above arguments, we choose
\begin{equation}
\label{eqn:block_size_parameters}
 s= \max \left\{1,\left\lfloor\frac q{4096(n+1)^2}\right\rfloor\right\},
 \quad k=d-s, \quad r=\frac34.
\end{equation}
Then $1 \leq s < d$, and both the $k$ port coordinates and the $s$ internal coordinates are nonempty.

\begin{proposition}[A contracting internal block]
\label{prop:contracting_internal_block}
Let $s$ and $r$ be as in \cref{eqn:block_size_parameters}. For every $U\in\U(d)$ there exist preconditioners $P_{\mathrm{out}},P_{\mathrm{in}}\in\U(d)$ such that the bottom-right block $D \in \C^{s \times s}$ of $V=P_{\mathrm{out}}UP_{\mathrm{in}}$ satisfies $\|D\|_{\op} \leq r$. The preconditioners and their inverses can be implemented by skeletons with $\bigo(n)$ slots and $\bigo(n)$ clean ancillae. These skeletons depend only on $n$ and $d$, so for a labeled family $\{U_\lambda\}_{\lambda}$
only the tables depend on $\lambda$.
\end{proposition}

\begin{proof}
    If $s>1$, then $s\leq q/[4096(n+1)^2]$ and $2s\le2^{n+1}$, so the right-hand side of \cref{eqn:operator_norm_probability_bound} is at most $(2/\mathrm{e}^{8})^{n+1}<1$, then some choice of $S_1$ gives $\|D\|_{\op}<1/4<r$. If $s=1$, some entry of the last column of $U$ has modulus at most $1/\sqrt {d} \leq 1/\sqrt{2} <r$. We then let $P_{\mathrm{out}}$ be a permutation that moves it to the last row, and let $P_{\mathrm{in}}= \id$, then the modulus bound ensures that $\|D\|_{\op} < r$.

For $s=1$, \cref{lemma:perm_with_clean_ancillae} implements $P_{\mathrm{out}}$ directly. For $s>1$, we first use \cref{lemma:perm_with_clean_ancillae} to move the last $q$ valid indices, in order, to $0, \dots, q-1$. On this block, $\hdm_q$ consists of $\log_2q$ Hadamard slots controlled on the higher bits being zero, and each sign matrix, extended by $1$ outside the block, is one diagonal slot (cf. \cref{lemma:diagonal_gates_controlled_state_prep_in_slot_formalism}). Finally, we undo the permutation. In both cases, the circuits fix the invalid subspace and use $\bigo(n)$ slots and $\bigo(n)$ clean ancillae, and their reverses implement the inverses with the same resources. Only the tables depend on $U$. For a labeled family, adding the label to every control list gives one skeleton for all $\lambda$, and the internal block $\bigoplus_\lambda D_\lambda$ yields the same operator norm upper bound. This completes the proof.
\end{proof}

\paragraph{The number of recursion levels.} Suppose the recursion halts after $J$ levels. By the choice of $s$ in \cref{eqn:block_size_parameters}, using $\max\{1, \lfloor x \rfloor\} \geq x/2$ for $x \geq 0$ and $q \leq d < 2q$, we have
$$
s \geq \max \left\{1,\left\lfloor\frac q{4096(n+1)^2}\right\rfloor\right\} \geq \frac{q}{8192(n + 1)^2} > \frac{d}{16384 (n + 1)^2}.
$$
Therefore, each replacement $d\mapsto d-s$ removes at least a $1/(16384(n+1)^2)$ fraction of the current dimension. We claim that the recursion level is at most $J = 32768(n + 1)^3 = \bigo(n^3)$, since
$$
2^n \cdot \left( 1 - \frac{1}{16384(n + 1)^2} \right)^J \leq 2^n \mathrm{e}^{- \frac{J}{16384(n + 1)^2} } = 2^n \mathrm{e}^{-2(n + 1)} < 1,
$$
so the recursion would have halted.

\subsubsection{Establishing the closed-loop identity}
\label[appendix]{appendix:closed_loop_identity}
Here we prove the identities used in \cref{appendix:unitary_for_child_task}, writing the unitary $V$ from \cref{prop:contracting_internal_block} with internal block $D$ satisfying $\|D\|_{\op}<r$.

\begin{lemma}[Feedback identity]\label{lemma:feedback_identity}
For every diagonal unitary $Z\in\C^{s\times s}$, the feedback relation
\eqref{eqn:inner_loop_system} uniquely determines $\mathbf{v},\mathbf{y}$ for each port input
$\mathbf{u}\in\C^k$. The resulting map $F(Z)\colon \mathbf{u}\mapsto \mathbf{y}$ is unitary.
For these feedback solutions, define the linear maps $M,N \in \mathsf{L}(\C^k,\C^d)$ by
$$
 M\mathbf{u}=\mathbf{u}\oplus Z\mathbf{v},\quad 
 N\mathbf{y}=\mathbf{y}\oplus\mathbf{v}.
$$
Then it holds that
\begin{equation}\label{eqn:feedback-equations}
 F(Z)M^{\dagger}=N^{\dagger}V.
\end{equation}
\end{lemma}

\begin{proof}
For a port input $\mathbf{u}$, start from the desired feedback relation
\eqref{eqn:inner_loop_system}:
$$
 V\left(\mathbf{u}\oplus Z\mathbf{v}\right) = \begin{pmatrix}
     A & B \\ C & D
 \end{pmatrix} \begin{pmatrix}
     \mathbf{u} \\ Z \mathbf{v}
 \end{pmatrix} = \begin{pmatrix}
     A \mathbf{u} + BZ \mathbf{v} \\ C \mathbf{u} + DZ \mathbf{v}
 \end{pmatrix}
 =\mathbf{y}\oplus\mathbf{v}.
$$
Reading its lower and upper blocks gives
$$
 (\id_{s}-DZ)\mathbf{v}=C\mathbf{u},\quad  \mathbf{y}=A\mathbf{u}+BZ\mathbf{v}.
$$
Since $\|DZ\|_{\op}\leq r<1$, the matrix $\id_{s}-DZ$ is invertible by its norm-convergent Neumann series \cite{Conway2007}. Thus the unique solution is $\mathbf{v}=X\mathbf{u}$,
$\mathbf{y}=F(Z)\mathbf{u}$, where
\begin{equation}
\label{eqn:feedback-definitions}
 X=(\id_{s}-DZ)^{-1}C,\quad 
 F(Z)=A+BZX=A+BZ(\id_{s}-DZ)^{-1}C.
\end{equation}
Unitarity of $V$ and $Z$ in the feedback relation ensures that
$$
 \|\mathbf{u}\|^2+\|\mathbf{v}\|^2
 =\left\|\mathbf{u}\oplus Z\mathbf{v}\right\|^2
 =\left\|\mathbf{y}\oplus\mathbf{v}\right\|^2
 =\|\mathbf{y}\|^2+\|\mathbf{v}\|^2.
$$
Hence $\|F(Z)\mathbf{u}\|=\|\mathbf{u}\|$ for every $\mathbf{u}$, and $F(Z)$ is unitary since it is a square matrix. We can therefore write $\mathbf{u}=F(Z)^{\dagger}\mathbf{y}$ and $\mathbf{v}=Y\mathbf{y}$, with
$Y=XF(Z)^{\dagger}$. Applying $V^{\dagger}$ to the feedback relation and taking
the lower block gives
$$
 B^{\dagger}\mathbf{y}+D^{\dagger}\mathbf{v}=Z\mathbf{v},
 \quad \implies \quad 
 (\id_{s}-Z^{\dagger}D^{\dagger})Y =Z^{\dagger}B^{\dagger}.
$$
Here $\mathbf{y}$ ranges over all of $\C^k$ because $F(Z)$ is unitary. Since
$\|Z^{\dagger}D^{\dagger}\|_{\op}\leq r<1$, another Neumann series then gives
$$
 Y=(\id_{s}-Z^{\dagger}D^{\dagger})^{-1}Z^{\dagger}B^{\dagger}.
$$
The feedback input and output maps are consequently
$$
 M=\begin{pmatrix}\id_{k}\\ZX\end{pmatrix},\quad 
 N=\begin{pmatrix}\id_{k}\\Y\end{pmatrix},\quad 
 VM=\begin{pmatrix}F(Z)\\X\end{pmatrix}=NF(Z),
$$
implying that $V^{\dagger}N=MF(Z)^{\dagger}$, and taking adjoints on both sides gives
\cref{eqn:feedback-equations}.
\end{proof}

\subsubsection{Nearly isometric encodings over a finite phase group}
\label[appendix]{appendix:nearly_isometric_encodings_over_phase_group}
In this section, we formally define the phase group and the truncated field maps used in \cref{appendix:recovering_preconditioned_unitary}, and show that the field maps are nearly isometric and nearly intertwine $V$ with the controlled smaller unitary passed to the child task. We start by defining some auxiliary parameters: fix some $L \geq 1$, we set
\begin{equation}
\label{eqn:group_order_setup}
b := n L + \lceil \log_2(L + 1) \rceil, \quad m_0 = 2^b.
\end{equation}

\paragraph{The phase group.} We first recall the finite-group facts used for phase averaging and the Fourier representation used subsequently.

\begin{definition}[Phase group and characters \cite{Terras1999}]
    The \emph{phase group} is the finite abelian group $G=(\Z/m_0\Z)^{L+1}$ of order $m_0^{L+1}$. We represent its elements by $\mathbf{z} = (z_0, z_1, \dots, z_L)$ with $z_j \in \{0, 1, \dots, m_0 - 1\}$. A \emph{character} of $G$ is a homomorphism $G \to \U(1)$. For each $\mathbf{w} \in G$, define
    \begin{equation}
    \label{eqn:character_of_phase_group}
  \forall \, \mathbf{z} \in G, \quad  \chi_{\mathbf{w}}(\mathbf{z}) \coloneqq \exp\left( \frac{2\pi \I}{m_0} \sum_{j=0}^{L} w_j z_j \right) = \exp\left( \frac{2\pi \I}{m_0} \left<\mathbf{w}, \mathbf{z} \right> \right).
    \end{equation}
\end{definition}

\begin{lemma}[Character identities and orthogonality {\cite[Chapter 10]{Terras1999}}]
\label{lemma:character_basic_properties}
    The functions $\chi_{\mathbf{w}}$ defined in \cref{eqn:character_of_phase_group} are well-defined\footnote{
    That is, the value of the character is independent of the choice of integer representatives of $G$.
    } characters with $|\chi_{\mathbf{w}}(\mathbf{z})| \equiv 1$. For any $\mathbf{u}, \mbf{w}, \mbf{z}, \mbf{z}' \in G$, 
    \begin{equation} \label{eqn:basic_properties_of_characters}
    \begin{aligned}
        \chi_{\mbf{w}}(\mbf(z) + \mbf{z}') &= \chi_{\mbf{w}}(\mbf{z}) \chi_{\mbf{w}}(\mbf{z}'), ~ \chi_{\mbf{u} + \mbf{w}}(\mbf{z}) = \chi_{\mbf{u}}(\mbf{z}) \chi_{\mbf{w}}(\mbf{z}), ~ \chi_{-\mbf{w}}(\mbf{z}) = \overline{\chi_{\mbf{w}}(\mbf{z})}, ~ \chi_{\mbf{0}}(\mbf{z}) = 1.
        \end{aligned}
    \end{equation}
    And the characters satisfy the orthogonality relation
    $$
    \frac{1}{|G|} \sum_{\mbf{z} \in G} \overline{\chi_{\mbf{u}}(\mbf{z})} \cdot \chi_{\mbf{w}}(\mbf{z}) = \delta_{\mbf{u}, \mbf{w}}.
    $$
    Let $z_0$ be uniformly distributed over $\Z/m_0 \Z$, and fix $z_1, z_2, \dots, z_L \in \Z / m_0 \Z$, then the clock phase $\omega = \mathrm{e}^{ 2 \pi \I z_0 / m_0 }$ satisfies, for every integer $a \in \Z$, 
    $$
    \E_{z_0}\left[ \omega^a \right] = \delta_{0, a \bmod{m}}.
    $$
\end{lemma}

\begin{proof}
    The characters are well-defined since changing an integer representative changes the exponent by an integer multiple of $2 \pi \I$. The modulus claim and \cref{eqn:basic_properties_of_characters} follow readily by definition and basic algebra. Fix an integer $a \in \Z$, we have
    $$
    \frac{1}{m_0} \sum_{j = 0}^{m_0 - 1} \mathrm{e}^{ \frac{2\pi \I a_0}{m_0} \cdot j } = \delta_{0, a \bmod{m_0}}.
    $$
    Since $\overline{\chi_{\mathbf{u}}} \chi_{\mathbf{w}} = \chi_{\mathbf{w} - \mathbf{u}}$, the group average factors into these one-coordinate averages with $a = w_j - u_j$, and combining the coordinate-wise matching the orthogonality follows immediately. The second clock-phase average is the same calculation in coordinate $0$.
\end{proof}

\paragraph{Phase operators and the child task.} We index the internal coordinates by $[s]$ as in \cref{prop:contracting_internal_block}, and define the frequency vector and the diagonal character unitary operator by
\begin{equation}
\label{eqn:frequency_vector_and_diagonal_characters}
\forall \, i \in [s], \quad\mathbf{v}_i = (1, i, i^2, \dots, i^{L}) \in G; \quad Z(\mathbf{z}) = \diag\left( \{\chi_{\mathbf{v}_i}(\mathbf{z}) \}_{i=1}^{s} \right),
\end{equation}
where the entries of $\mathbf{v}_i$ are reduced modulo $m_0$. For each $\mbf{z} \in G$, put $F(\mbf{z}), M(\mbf{z}), N(\mbf{z})$ be the maps of \cref{lemma:feedback_identity} for $Z = Z(\mbf{z})$, and set the controlled operation passed to the child task
$$
\mbf{F} = \sum_{\mathbf{z} \in G} \ps{\mbf{z}} \otimes F(\mbf{z}), \quad F(\mbf{z}) \coloneqq F(Z(\mbf{z})).
$$

\paragraph{Truncation and encoding over the phase group.} For practical implementability, we will use the Neumann series expansion of the matrix inverse appearing in $M(\mbf{z})$ and $N(\mbf{z})$, and track the error incurred by truncating it at a finite order.

\begin{lemma}[Neumann series and truncation error {\cite[Chapter VII, $\S$2, Lemma 2.1 and Corollary 2.3]{Conway2007}}]
\label{lemma:neumann_series}
Let $T$ be a square complex matrix satisfying $\|T\|_{\op}\leq r < 1$. Then $\id-T$ is invertible and
$$
\left(\id-T \right)^{-1}=\sum_{\ell=0}^{\infty}T^\ell,
$$
where the series converges in operator norm. Moreover, for every integer $L\geq 1$, the truncated sum $T_L=\sum_{\ell=0}^{L-1}T^\ell$ satisfies
$$
\left(\id-T \right)^{-1}-T_L=\sum_{\ell=L}^{\infty}T^\ell=T^L \left(\id-T \right)^{-1},\quad \left\|(\id-T)^{-1}-T_L \right\|_{\op}\leq\frac{r^L}{1-r}.
$$
\end{lemma}

\begin{proof}
The bound $\sum_{\ell=0}^{\infty}\|T^\ell\|_{\op}\leq\sum_{\ell=0}^{\infty}r^\ell=(1-r)^{-1}$ shows that the sequence $T_L$ converges in operator norm to $T_{\infty}:=\lim_{L\to\infty}T_L \coloneqq\sum_{\ell=0}^{\infty}T^\ell$. The finite geometric-sum identity gives
$$
(\id-T)T_L=T_L(\id-T)=\id-T^L.
$$
Since $\|T^L\|_{\op}\leq r^L$, taking $L\to\infty$ yields $(\id-T)T_{\infty}=T_{\infty}(\id-T)=\id$. Thus $T_{\infty}=(\id-T)^{-1}$. Subtracting $T_L$ gives the stated tail, and
$$
\left\|(\id-T)^{-1}-T_L \right\|_{\op} = \left\|\sum_{\ell=L}^{\infty} T^{\ell} \right\|_{\op}\leq\sum_{\ell=L}^{\infty} \left\|T^\ell \right\|_{\op}\leq\sum_{\ell=L}^{\infty}r^\ell=\frac{r^L}{1-r}.
$$
This completes the proof.
\end{proof}
For each $\mathbf{z}\in G$, we apply \cref{lemma:neumann_series} to $T=DZ(\mathbf{z})$ and $T=Z(\mathbf{z})^{\dagger}D^{\dagger}$, both of operator norm at most $r = 3/4 < 1$. Replacing the corresponding operator inverses in $M(\mathbf{z})$ and $N(\mathbf{z})$ by their first $L$ Neumann terms defines the truncated operators
\begin{equation}
\label{eqn:pointwise-truncations}
 M_L(\mathbf{z})=\begin{pmatrix}
 \id_k\\ \sum_{\ell=0}^{L-1}Z(\mathbf{z})(DZ(\mathbf{z}))^\ell C
 \end{pmatrix},\quad
 N_L(\mathbf{z})=\begin{pmatrix} \id_k\\ \sum_{\ell=0}^{L-1}
                   (Z(\mathbf{z})^{\dagger}D^{\dagger})^\ell Z(\mathbf{z})^{\dagger}B^{\dagger}
 \end{pmatrix}.
\end{equation}
Since $Z(\mbf{z})$ is unitary and the blocks $B, C$ of $V$ have operator norm at most $1$, \cref{lemma:neumann_series} with $r = 3/4$ thus ensures that
$$
\left\|  M_L(\mathbf{z}) - M(\mbf{z}) \right\|_{\op}, ~ \left\| N_L(\mathbf{z}) - N(\mbf{z}) \right\|_{\op} \leq \frac{r^L}{1 - r} = 4 r^L.
$$
Using the adjoints of these truncated maps, we define the input and output encoding maps $E_L,R_L \in \mathsf{L}(\C^d, \C^G\otimes\C^k)$, which relate the action of $V$ on the original space to that of the controlled child unitary $\mathbf{F}$ on the encoded space:
\begin{equation}
\label{eqn:encoding_map_definitions}
\forall \, \mathbf{x} \in \C^d, \quad  E_L  \mathbf{x} =\frac1{\sqrt{|G|}} \sum_{\mathbf{z}\in G}\ket{\mathbf{z}}\otimes M_L(\mathbf{z})^{\dagger} \mathbf{x} ,
\quad 
 R_L \mathbf{x} =\frac1{\sqrt{|G|}}\sum_{\mathbf{z}\in G}\ket{\mathbf{z}}\otimes N_L(\mathbf{z})^{\dagger} \mathbf{x} .
\end{equation}

\begin{proposition}[Near-isometry of the encoding maps]
\label{prop:near_isometry_of_encodings}
    The encoding maps $E_L$ and $R_L$ in \cref{eqn:encoding_map_definitions} satisfies
    $$
    \left( 1 - r^{2L} \right) \id_d \preceq T^\dagger T \preceq \id_d, \quad T = E_L, R_L.
    $$
\end{proposition}

\begin{proof}
    Plugging $T = E_L, R_L$ in $T^\dagger T$ gives
    $$
    \begin{aligned}
    E_L^\dagger E_L &= \frac{1}{|G|} \sum_{\mbf{z} \in G} M_L(\mbf{z}) M_L(\mbf{z})^\dagger = \E_{\mathbf{z}}\left[ M_L(\mbf{z}) M_L(\mbf{z})^\dagger\right], \\ R_L^\dagger R_L &= \frac{1}{|G|} \sum_{\mbf{z} \in G} N_L(\mbf{z}) N_L(\mbf{z})^\dagger = \E_{\mathbf{z}}\left[ N_L(\mbf{z}) N_L(\mbf{z})^\dagger\right].
    \end{aligned}
    $$
    For $\mbf{z} = (z_0, z_1, \dots, z_L)$ we fix $z_1, \dots, z_L$ and we average first over $z_0$. Recall $\omega = \mathrm{e}^{2\pi \I z_0 / m_0}$, then
    $$
    \begin{aligned}
   \forall \, i \in [s], \quad [Z(\mathbf{z})]_{i, i} &= \exp\left( \frac{2\pi \I}{m_0} \sum_{j=0}^{L} i^{j} z_j \right) = \exp\left( \frac{2\pi \I}{m_0} \left(\sum_{j=1}^{L} i^{j} z_j + z_0 \right) \right) = \omega [Z']_{i, i}, \\
   Z' &= \diag\left( \left\{ \exp\left(\frac{2\pi \I}{m_0} \sum_{j=1}^L i^j z_j \right)\right\}_{i=1}^s \right): \quad {Z'}^\dagger Z' = \id_s.
   \end{aligned}
    $$
    We then put 
    $$
    P(\omega) = \sum_{\ell = 0}^{L- 1} \omega^{\ell + 1} Z'(DZ')^{\ell} C, \quad Q(\omega) = \sum_{\ell = 0}^{L - 1} \omega^{-(\ell + 1)} ({Z'}^\dagger D^\dagger)^{\ell} {Z'}^\dagger B^\dagger.
    $$
    Then \cref{eqn:pointwise-truncations} tells
    $$
    M_{L}(\mbf{z}) M_{L}(\mbf{z})^\dagger = \begin{pmatrix}
        \id_k & P(\omega)^\dagger \\ P(\omega) & P(\omega) P(\omega)^\dagger
    \end{pmatrix}, \quad N_L(\mbf{z}) N_L(\mbf{z})^\dagger = \begin{pmatrix}
        \id_k & Q(\omega)^\dagger \\ Q(\omega) & Q(\omega) Q(\omega)^\dagger
    \end{pmatrix}.
    $$
    The exponents of $\omega$ that occur in $P(\omega)$ and $Q(\omega)$ are bounded from above by $L$, and by the group order setup in \cref{eqn:group_order_setup}, $m_0 = 2^{nL + \lceil \log_2 (L + 1) \rceil} \geq (L + 1) 2^{nL} > L$, so there is no wraparound and the exponents are all distinct and nonzero modulo $m_0$. Then by \cref{lemma:character_basic_properties}, and using the unitarity of $V$, we have
$$
\begin{aligned}
&\E_{z_0}\left[ P(\omega)\right] = \sum_{\ell = 0}^{L - 1} \E_{z_0} \left[\omega^{\ell + 1} \right] Z'(DZ')^{\ell} C = 0, \\
&\E_{z_0} \left[ P(\omega) P(\omega)^\dagger \right] = \sum_{\ell, \ell' = 0}^{L - 1} \E_{z_0} \left[ \omega^{\ell - \ell'}\right] Z'(DZ')^{\ell} C C^\dagger ({Z'}^\dagger D^\dagger)^{\ell'} {Z'}^{\dagger} \\
&\quad = Z' \sum_{\ell = 0}^{L - 1} (DZ')^{\ell} CC^\dagger ({Z'}^\dagger D^\dagger)^{\ell} {Z'}^\dagger = Z' \sum_{\ell = 0}^{L - 1} (DZ')^{\ell} \left( \id_s - DD^\dagger \right) ({Z'}^\dagger D^\dagger)^{\ell} {Z'}^\dagger \\
&\quad = Z' \left( \sum_{\ell = 0}^{L - 1} (DZ')^{\ell} ({Z'}^\dagger D^\dagger)^{\ell} - (DZ')^{\ell + 1} ({Z'}^\dagger D^\dagger)^{\ell + 1} \right){Z'}^\dagger = Z' \left(\id_s - (DZ')^{L} ({Z'}^\dagger D^\dagger)^{L} \right) {Z'}^\dagger \\
&\E_{z_0}\left[ Q(\omega)\right] = \sum_{\ell = 0}^{L - 1} \E_{z_0} \left[\omega^{-(\ell + 1)} \right] ({Z'}^\dagger D^\dagger)^{\ell} {Z'}^\dagger B^\dagger = 0, \\
&\E_{z_0}\left[ Q(\omega) Q(\omega)^\dagger \right] = \sum_{\ell, \ell' = 0}^{L - 1} \E_{z_0}\left[\omega^{\ell' - \ell} \right] ({Z'}^\dagger D^\dagger)^{\ell} {Z'}^\dagger B^\dagger B Z' (DZ')^{\ell'} \\
&\quad = \sum_{\ell = 0}^{L - 1} ({Z'}^\dagger D^\dagger)^{\ell} {Z'}^\dagger B^\dagger B Z' (DZ')^{\ell} = \sum_{\ell = 0}^{L - 1} ({Z'}^\dagger D^\dagger)^{\ell} {Z'}^\dagger \left(\id_s - D^\dagger D \right) Z' (DZ')^{\ell} \\
&\quad = \sum_{\ell=0}^{L-1} \left[(Z'^{\dagger}D^{\dagger})^\ell(DZ')^\ell -(Z'^{\dagger}D^{\dagger})^{\ell+1}(DZ')^{\ell+1}\right] = \id_s - ({Z'}^\dagger D^\dagger)^L (DZ')^{L}.
\end{aligned}
$$
Since 
$$
\begin{aligned}
&\left\| Z' (DZ')^{L} ({Z'}^\dagger D^\dagger)^{L} {Z'}^\dagger  \right\|_{\op} \leq \left\| (DZ')({Z'}^\dagger D^\dagger) \right\|_{\op}^L \leq r^{2L}, \\
&\left\|({Z'}^\dagger D^\dagger)^L (DZ')^{L} \right\|_{\op} \leq \left\|({Z'}^\dagger D^\dagger) (DZ')  \right\|_{\op} \leq r^{2L}.
\end{aligned}
$$
It follows readily that
$$
\begin{aligned}
\left(1 - r^{2L} \right) \id_d \preceq &\E_{z_0} \left[ M_L(\mbf{z}) M_L(\mbf{z})^\dagger\right] = \id_d - 0_{k \times k} \oplus Z'(DZ')^{L} ({Z'}^\dagger D^\dagger)^{L}  {Z'}^\dagger \preceq \id_d, \\
\left(1 - r^{2L} \right)  \id_{d} \preceq &\E_{z_0} \left[ N_L(\mbf{z}) N_L(\mbf{z})^\dagger\right] = \id - 0_{k \times k} \oplus ({Z'}^\dagger D^\dagger)^L (DZ')^{L} \preceq \id_{d}.
\end{aligned}
$$
Averaging both sides over the remaining indices $z_1,\dots,z_L$ then completes the proof.
\end{proof}

\paragraph{Approximate intertwining.} We also derive a truncation-robust version of the exact intertwining relation $F(\mbf{z})M(\mbf{z})^\dagger=N(\mbf{z})^\dagger V$ from \cref{lemma:feedback_identity}. Due to \cref{lemma:neumann_series}, 
$$
\begin{aligned}
    \|E_L - E\|_{\op}^2 &= \left\| \left(E_L - E\right)^\dagger \left(E_L - E\right) \right\|_{\op} \\
    &= \frac{1}{|G|} \left\| \sum_{\mbf{z} \in G} \left(M_L(\mathbf{z}) - M(\mbf{z})  \right) \left(M_L(\mathbf{z}) - M(\mbf{z})  \right)^\dagger \right\|_{\op} \\
    &\leq \frac{1}{|G|} \sum_{\mbf{z} \in G} \left\|M_L(\mathbf{z}) - M(\mbf{z}  \right\|_{\op}^2 \leq (4r^{L})^2,
\end{aligned}
$$
and similar for $\| R_L - R \|_{\op}$. It therefore follows that
\begin{equation}
\label{eqn:approximate_intertwining_of_child_task_passing}
\begin{aligned}
    \left\| \mathbf{F} E_L - R_L V \right\|_{\op} &= \left\| \mathbf{F}(E_L - E) + (R - R_L) V  \right\|_{\op} \\
    &\leq \left\| \mathbf{F}(E_L - E) \right\|_{\op} +  \left\|(R - R_L) V  \right\|_{\op} = \|E_L - E\|_{\op} + \|R - R_L\|_{\op} \leq 8r^{L}.
\end{aligned}
\end{equation}

\subsubsection{Row sparsity in the Fourier basis}
\label[appendix]{appendix:row_sparsity_in_Fourier_basis}
Recall from \cref{appendix:encoder_circuit_with_block_encoding} that column preparation leaves the input index entangled with the output register. If a row is dense, its output label can be compatible with exponentially many input indices, so the label alone does not yield the short list of candidates that our encoder needs to uncompute the input index efficiently. In this section, we overcome this obstacle by showing that, in the Fourier basis of the phase register, the encoding maps become sparse. 

We first recall Newton's identities, which allow us to recover the multiset of indices contributing to a given Fourier frequency from the power sums of these indices.

\begin{lemma}[Newton's identities \cite{symm_func_1995}]
\label{lemma:newton_identities}
    Let $m \geq 1$ and $x_1, x_2, \dots, x_m \in \C$. Define the following elementary symmetric functions and power sums
    $$
    e_0 = 1; \quad \forall \, \ell  \geq 1,~ e_\ell = \sum_{ 1 \leq i_1 < \cdots < i_\ell \leq m } x_{i_1} \cdots x_{i_\ell}, \quad p_\ell = \sum_{i=1}^m x_{i}^\ell ,
    $$
    where we set $e_\ell = 0$ for $\ell  > m$. Then for every degree $j \geq 1$, it holds that
    $$
    j \cdot e_j = \sum_{\ell =1}^{j} (-1)^{\ell  - 1} p_{\ell} e_{j - \ell }.
    $$
\end{lemma}
Newton’s identities guarantee that sufficiently many power sums uniquely determine a multiset. These identities allow us to show how the phase sums occurring in the matrix entries of the encoders determine the multiset of indices that produced it.

\begin{lemma}[From power sums to a short multiset, cf. {\cite[Lemma 9]{10.1063/1.4954983}}]
\label{lemma:power_sum_to_short_multiset}
    Let $G$ be the phase group and $\mathbf{v}_1,\dots,\mathbf{v}_s\in G$ the frequency vectors from \cref{appendix:nearly_isometric_encodings_over_phase_group}. For a multiset $S$ of elements of $[s]$, let $\mathbf{v}_S=\sum_{i\in S}\mathbf{v}_i\in G$, where both the sum and the size $|S|$ count elements with multiplicity. Then the map $S\mapsto\mathbf{v}_S$ is injective on multisets with $|S|\leq L$. In particular, $\mathbf{v}_S\neq\mathbf{0}$ for every nonempty multiset $S$ with $|S|\leq L$.
\end{lemma}

\begin{proof}
    Let $S$ be any multiset with $\ell=|S|\leq  L$. For $j=0,1, \dots,L$, we define the integer power sums
$$
 p_j(S)=\sum_{i\in S}i^j.
$$
Note that $p_0(S)=\ell$. Since every $i\in S$ satisfies $1\leq  i\leq  s<2^n$, we further have
$$
 0\leq  p_j(S)\leq  L \cdot 2^{nL}<(L+1)2^{nL}\leq 2^b=m_0.
$$
By \cref{eqn:frequency_vector_and_diagonal_characters}, the $j$-th coordinate of $\mathbf{v}_S$ has no wraparound and is thus $p_j(S)\bmod m_0$. Because $0\leq  p_j(S)<m_0$, the integer $p_j(S)$ is uniquely determined by this coordinate. So we can obtain all of $p_0(S), \dots, p_L(S)$ from $\mathbf{v}_S$, and in particular $\ell=p_0(S)$. If $\ell=0$, then $S$ must be the empty multiset.

Now we assume $\ell\ge1$, write the elements of $S$ as $i_1, \dots, i_\ell$ (with repetition), and define their elementary symmetric sums by
$$
 e_0(S)=1, \quad \forall\, j \in [\ell], ~
 e_j(S)=\sum_{1\leq a_1<\cdots<a_j\leq \ell}
 i_{a_1}\cdots i_{a_j}.
$$
Applying \cref{lemma:newton_identities} to $x_a=i_a$, we can obtain the elementary symmetric functions $e_1(S), \dots$, $e_\ell(S)$ from $p_1(S), \dots, p_\ell(S)$. This therefore gives us the polynomial
$$
 \prod_{a=1}^{\ell}(x-i_a)=x^\ell-e_1(S)x^{\ell-1}+\cdots+(-1)^{\ell}e_\ell(S),
$$
whose roots are exactly $i_1, \dots, i_\ell$. Hence $S$ is uniquely determined by $\mathbf{v}_S$, so that the map $S\mapsto\mathbf{v}_S$ is injective. Moreover, since $\mathbf{v}_{\varnothing}=\mathbf{0}$, by injectivity we have $\mathbf{v}_S\neq\mathbf{0}$ for any nonempty $S$ with $|S|\leq  L$, completing the proof.
\end{proof}

\paragraph{Encoding maps in the Fourier basis.} Let $\mathscr{F}_G$ be the Fourier transform on $G$ with positive phases \cite{Terras1999} obtained from $\mathscr{F}_b$ in \cref{lemma:fourier_transform_skeleton} via $\mathscr{F}_G = \mathscr{F}_b^{\otimes (L + 1)}$:
$$
\forall \, \mathbf{w} \in G, \quad \ket{\widehat {\mbf{w}}} = \mathscr{F}_G \ket{\mbf{w}} \coloneqq \frac{1}{\sqrt{|G|}} \sum_{\mathbf{z} \in G} \chi_{\mbf{w}}(\mbf{z}) \ket{\mbf{z}}.
$$
By \cref{lemma:character_basic_properties} the vectors $\{\ket{\widehat{\mathbf{w}}}\}_{\mbf{w} \in G}$ form an orthonormal basis, so $\mathscr F_G$ is unitary. Set the Fourier-transformed encodings on $\mathsf{L}(\C^d, \C^{G} \otimes \C^k)$:
\begin{equation}
    \label{eqn:Fourier_transformed_encodings}
     \widehat{E}_L=(\mathscr F_G^{\dagger}\otimes \id_k)E_L, \quad
 \widehat{R}_L=(\mathscr F_G^{\dagger}\otimes \id_k)R_L .
\end{equation}
And we index their rows by a tuple $(\mathbf{w}, a)$ with $\mathbf{w} \in G$ and $1 \leq a \leq k$. For instance,
\begin{equation}
\label{eqn:entries_of_encoding_E_L}
\left[\widehat{E}_L \right]_{(\mathbf{w}, a), j} = \frac{1}{|G|} \sum_{\mathbf{z} \in G} \overline{\chi_{\mbf{w}}(\mbf{z})} \left[ M_L(\mbf{z})^\dagger \right]_{a, j}
\end{equation}

\begin{proposition}[Row sparsity in the Fourier basis]
\label{prop:Fourier_sparsity_of_encodings}
    Every row of $\widehat{E}_L$ and of $\widehat{R}_L$ has at most $L$ nonzero entries.
\end{proposition}

\begin{proof}
We prove the claim for $\widehat{E}_L$ and extend it to $\widehat{R}_L$ with slight modifications. By \cref{eqn:pointwise-truncations}, each entry of $M_L(\mathbf{z})^{\dagger}$ is a polynomial in the diagonal entries $\chi_{-\mathbf{v}_1}(\mathbf{z}), \dots, \chi_{-\mathbf{v}_s}(\mathbf{z})$ of $Z(\mathbf{z})^{\dagger}$ with constant coefficients determined by $C$ and $D$. Every monomial in these entries is a single character under the composition $\chi_{\mathbf{u}}\chi_{\mathbf{u}'}=\chi_{\mathbf{u}+\mathbf{u}'}$. Therefore, for fixed indices $a \in [k]$ and $j \in [d]$,
$$
\exists \, c_{\mathbf{u}}\in\C, \quad \left[M_L(\mathbf{z})^{\dagger}\right]_{a,j}
 =\sum_{\mathbf{u}\in G}c_{\mathbf{u}}\,\chi_{\mathbf{u}}(\mathbf{z}).
$$
Plugging this into \cref{eqn:entries_of_encoding_E_L} and using the orthogonality relation of \cref{lemma:character_basic_properties}, we find
that only the term $\mathbf{u}=\mathbf{w}$ survives:
\begin{equation}
\label{eqn:only_one_coefficient_stays}
\begin{aligned}
\left[\widehat{E}_L \right]_{(\mathbf{w}, a), j} &= \frac{1}{|G|} \sum_{\mathbf{z} \in G} \overline{\chi_{\mbf{w}}(\mbf{z})} \sum_{\mathbf{u}\in G}c_{\mathbf{u}}\,\chi_{\mathbf{u}}(\mathbf{z}) =\sum_{\mathbf{u}\in G}c_{\mathbf{u}}\left(\frac1{|G|}\sum_{\mathbf z\in G} \overline{\chi_{\mbf{w}}(\mbf{z})}\chi_{\mathbf{u}}(\mathbf z) \right)=c_{\mathbf{w}}.
\end{aligned}
\end{equation}
So the question is which characters occur in each column of
$M_L(\mathbf{z})^{\dagger}$.

By \cref{eqn:pointwise-truncations}, the first $k$ columns of
$M_L(\mathbf{z})^{\dagger}$ form $\id_k$ for every $\mathbf{z}$, so they
contain only the constant character $\chi_{\mathbf{0}}$. Hence the port
columns of $\widehat{E}_L$ are zero except in the rows
$(\mathbf{0},1), \dots, (\mathbf{0},k)$, where they form $\id_k$. The last $s$ columns of $M_L(\mathbf{z})^{\dagger}$ form the adjoint of the lower block in \cref{eqn:pointwise-truncations}:
$$
\sum_{\ell=1}^{L}C^{\dagger}\left({Z(\mathbf{z})}^{\dagger}D^{\dagger} \right)^{\ell-1}{Z(\mathbf{z})}^{\dagger}.
$$
Multiplying out the $\ell$-th term, basic linear algebra tells its entry in row $a$ and internal column $i$ (equivalently, the column $k + i$ of the complete operator $M_L(\mathbf{z})$):
\begin{equation}\label{eqn:path-monomial}
 \left[M_L(\mathbf{z})^{\dagger}\right]_{a,k+i}
 = \sum_{\ell=1}^L \sum_{\substack{i_1, \dots, i_\ell\in[s]\\ i_\ell=i}}
 \left[C^{\dagger} \right]_{a,i_1} \left[D^{\dagger} \right]_{i_1,i_2}\cdots \left[ D^{\dagger} \right]_{i_{\ell-1},i_\ell}\,
 \chi_{-(\mathbf{v}_{i_1}+\cdots+\mathbf{v}_{i_\ell})}(\mathbf{z}).
\end{equation}
The terms are indexed by paths $(i_1, \dots, i_\ell)$ of internal coordinates that end at the input coordinate $i$, and we call $-(\mathbf{v}_{i_1}+\cdots+\mathbf{v}_{i_\ell})$ the frequency of the path. In the notation of \cref{lemma:power_sum_to_short_multiset} it is $-\mathbf{v}_S$ with $S=\{i_1, \dots ,i_\ell\}$ a multiset of elements of $[s]$, $|S|=\ell\leq L$ and $i\in S$.

Now fix a row $(\mathbf{w},a)$. Its port coordinate entries can be nonzero only if $\mathbf{w}=\mathbf{0}$, and its internal coordinate entries come from the paths of frequency $\mathbf{w}$. No path has frequency $\mathbf{0}$ since $\mathbf{v}_S\neq \mathbf{0}$ again by \cref{lemma:power_sum_to_short_multiset}. So if $\mathbf{w}=\mathbf{0}$, the only nonzero entry of the row is the $1$ in port column $a$. If $\mathbf{w}\neq \mathbf{0}$, every path of frequency $\mathbf{w}$ should satisfy $- \mathbf{v}_S=\mathbf{w}$ as we previously remarked in \cref{eqn:only_one_coefficient_stays}, and by \cref{lemma:power_sum_to_short_multiset} all these paths must be supported on the same multiset $S$. We list the elements of this specific $S$ with multiplicity as $j_1, \dots, j_\ell$, where $\ell=|S|\leq  L$. Each of these paths visits $j_1, \dots, j_\ell$ in some order, that is, sequentially along $(j_{\pi(1)}, \dots, j_{\pi(\ell)})$ for a permutation $\pi \in \mathfrak{S}_\ell$, and it finally contributes to column $j_{\pi(\ell)}$. Hence we have the implication $
[\widehat{E}_L]_{(\mathbf{w},a),\,k+i}\neq 0 \implies i\in\{j_1,\dots,j_\ell\}$, so the nonzero entries of the row lie in the columns $j_1, \dots, j_\ell$, and there are at most $\ell\leq  L$ of them.

For $\widehat{R}_L$, it suffices to replace $M_L$ by $N_L$ in the expression of \cref{eqn:entries_of_encoding_E_L} and argue analogously. The first $k$ columns of
$N_L(\mathbf{z})^{\dagger}$ again form $\id_k$, and the last $s$ columns form
$\sum_{\ell=1}^{L}BZ(DZ)^{\ell-1}$. Expanding the $\ell$-th term as before gives a similar form of \cref{eqn:path-monomial} with $B$ and $D$ in place of $C^{\dagger}$ and $D^{\dagger}$, and each path now has frequency $\mathbf{v}_S$ instead of $-\mathbf{v}_S$. The argument above then applies directly with $\mathbf{v}_S=\mathbf{w}$ in place of $\mathbf{v}_S=-\mathbf{w}$. This therefore proves the row sparsity of both $\widehat{E}_L$ and $\widehat{R}_L$, as desired.
\end{proof}

\subsubsection{Implementing the encoders}
\label[appendix]{appendix:implementing_encoders}
\paragraph{Extension to labeled families.} The results of \cref{appendix:closed_loop_identity,appendix:nearly_isometric_encodings_over_phase_group,appendix:row_sparsity_in_Fourier_basis} extend directly to a family $\{V_\lambda\}_\lambda$ controlled by a label $\lambda$. The group $G$, the dimensions, and the register layout are identical for all labels, so the per-label maps assemble into block-diagonal operators. Hence all bounds also hold for superpositions of labels and with the same constants.

This section constructs the quantum circuits for the encoders mentioned in \cref{appendix:encoder_circuit_with_block_encoding}, first for a general row-sparse near-isometry and then for the maps discussed in \cref{appendix:row_sparsity_in_Fourier_basis}.

\paragraph{Layout of registers.} Let $I\subseteq\{0,1\}^w$ be a nonempty set of valid rows. For maps in $\mathsf{L}(\C^d, \C^{I})$ with $1\leq d\leq  2^n$, the rows are given by the basis states $\ket u$ where $u\in I$ of a $w$-qubit register $\mathsf{Y}$, and the columns are the basis states $\ket i$, where $0\leq i<d$, of the $n$-qubit input register.
We number the basis states from $0$, so coordinate $j$ in the one-based matrix notation of the earlier sections is the basis state $j-1$. Then we fix $L\geq 1$ and put
$$
 h=\lceil\log_2(L+1)\rceil,\quad \nu=n+h .
$$
Extend the input register by padding $h$ high bits to a $\nu$-qubit register $\mathsf{X}$, and add two one-qubit flags $\mathsf{flag} = \mathsf{a}\mathsf{b}$ (cf. \cref{appendix:encoder_circuit_with_block_encoding}, one for normalization in state preparation, and one for tuning the rotation angle in exact amplitude amplification). Let the register ordered as $\mathsf{X},\mathsf{Y},\mathsf{a},\mathsf{b}$, the aforementioned input and output embeddings $\iota$ and $\gamma$ acts explicitly as
\begin{equation}
\label{eqn:embedding_actions_on_registers}
\forall\, 0 \leq i < d, \quad \iota\ket{i}=\ket{i}_{\mathsf{X}}\ket{0^w}_{\mathsf{Y}}\ket0_{\mathsf{a}}\ket0_{\mathsf{b}}; \quad \forall \, u \in I, \quad\gamma\ket{u}=\ket{0^\nu}_{\mathsf{X}}\ket{u}_{\mathsf{Y}}\ket1_{\mathsf{a}}\ket1_{\mathsf{b}}.
\end{equation}
The range of $\gamma$ is the \emph{good subspace}.

\begin{lemma}[Block encodings of a row-sparse near-isometry, \cite{10.1145/2591796.2591854, QSVT}, adapted]
\label{lemma:block_encoding_of_row_sparse_near_isometry}
    Let a linear operator $T \in \mathsf{L}(\C^d, \C^I)$, suppose every row of $T$ has at most $L$ non-zero entries and it satisfies
    $$
    \exists \, \eps \in [0, 1), \quad (1 - \eps) \id_d \preceq T^\dagger T \preceq \id_d,
    $$
    then the following statements hold:
    \begin{enumerate}
        \item[(1)] There exists a unitary $B_T$ acting on $\HH_{\mathsf{X}} \otimes \HH_{\mathsf{Y}} \otimes \HH_{\mathsf{a}} \otimes \HH_{\mathsf{b}}$ (cf. \cref{eqn:embedding_actions_on_registers}) that satisfies
        $$
        \gamma^\dagger B_T \iota = t_0 T, \quad t_0 = \sin\left( \frac{\pi}{4L + 2} \right).
        $$
        And synthesizing it uses at most $\nu + w + 4$ slots and is ancilla-free.
        \item[(2)] With the amplification operator $\mathscr{G}_T = (2 B_T \iota \iota^\dagger B_T^\dagger - \id)(\id - 2 \gamma \gamma^\dagger)$, the unitary $U_T = \mathscr{G}_T^L B_T$ satisfies
        $$
        \left\| U_T \cdot \iota - \gamma \cdot T \right\|_{\op} \leq 5 \eps.
        $$
        Synthesizing $U_T$ requires at most $(2 L + 1)(\nu + w + 4) + 2 L = \bigo(L(n + w + h))$ slots and $w + h + 2$ ancillae besides the $n$-qubit input. 
    \end{enumerate}
\end{lemma}

\begin{proof}
    (1) $B_T$ is built from a product of two controlled state preparations and a rotation on $\HH_{\mathsf{b}}$. The first preparation writes the columns of $T$ into register $\mathsf{Y}$, marked by $\ket{1}_{\mathsf{a}}$. For each $0\leq i<d$ we have $\|T\ket i\|^2 \leq 1$, and the vector
\begin{equation}\label{eqn:column-state}
 \ket{\alpha_i}
 =\sum_{u\in I}T_{ui}\ket u_{\mathsf{Y}}\ket1_{\mathsf{a}}
       +\sqrt{1-\|T\ket i\|^2}\,\ket{0^w}_{\mathsf{Y}}\ket0_{\mathsf{a}}
\end{equation}
is normalized. By \cref{lemma:diagonal_gates_controlled_state_prep_in_slot_formalism} with register $\mathsf{X}$ as the control and using $w+2$ slots, there exists a unitary $A$ that acts as $A \left(\ket i_{\mathsf{X}}\ket{0^w}_{\mathsf{Y}}\ket0_{\mathsf{a}} \right)=\ket i_{\mathsf{X}}\ket{\alpha_i}$ for every $0\leq i<d$, and with an arbitrary unit vector in place of $\ket{\alpha_i}$ for the other basis states on $\HH_{\mathsf{X}}$.

The second preparation produces the uniform superposition over a set $I_u$ of $L$ candidate columns for each row $u$. Moreover, its adjoint uncomputes the index register. For $u\in I$, let $I_u\subseteq\{0,1\}^\nu$ be a set of size $L$ that contains every $i$ such that $T_{ui}\neq 0$. Such a set exists because row $u$ has at most $L$ nonzero entries and $2^\nu \geq 2^h>L$. And for $u\notin I$, setting $I_u$ as any set of size $L$ would work. By \cref{lemma:diagonal_gates_controlled_state_prep_in_slot_formalism} again, with register $\mathsf{Y}$ as the control, with $\nu+1$ slots we synthesize a (controlled) unitary $W = \sum_{u} \ps{u} \otimes W_u$ (cf. \cref{appendix:encoder_circuit_with_block_encoding}) such that
$$
 W\ket{0^\nu}_{\mathsf{X}}\ket u_{\mathsf{Y}} = W_u \ket{0^{\nu}}_{\mathsf{X}} \ket{u}_{\mathsf{Y}} =\ket{s_u}_{\mathsf{X}}\ket u_{\mathsf{Y}},
\quad
 \ket{s_u}=\frac1{\sqrt L}\sum_{i\in I_u}\ket i .
$$
For $0\leq  i<d$ and $u\in I$, recall \cref{eqn:column-state}, we take the matrix element of $W^{\dagger}A$ between the states $\ket{i,0^w,0}_{\mathsf{XYa}}$ and $\ket{0^\nu,u,1}_{\mathsf{XYa}}$:
\begin{equation}
\label{eqn:entry_wise_extraction_of_T}
 \bra{0^\nu,u,1}W^{\dagger}A\ket{i,0^w,0} = \braket{ s_u , u, 1 | i, \alpha_i } = \frac{T_{ui}}{\sqrt L},
\end{equation}
where the last step uses that $T_{ui}\neq 0$ implies $i\in I_u$. This entrywise result gives a block encoding of $T / \sqrt{L}$ up to isometries. Finally, to turn the amplitude to $t_0$ from $1/\sqrt{L}$, we borrow the technique from \cite{Brassard2002, ntvs-c48s, grier2026mathsfqac0containsmathsftc0with} by applying a rotation on the second flag bit $\mathsf{b}$. Since $\sin(\frac{\pi L}{4L + 2}) \cdot \sqrt{L} \leq \frac{\pi \sqrt{L}}{4L + 2} < \frac{\pi}{4 \sqrt{L}} < 1$, there exists a real rotation $R_{\mathsf{b}}$ (as one slot) on $\HH_\mathsf{b}$ with $\bra1R_{\mathsf{b}}\ket0=t_0\sqrt L$. Since $A$ and $W$ do not act on $\mathsf{b}$, by \cref{eqn:entry_wise_extraction_of_T}, the unitary $B_T=R_{\mathsf{b}}W^{\dagger}A$ satisfies $\gamma^{\dagger}B_T\iota= \braket{1 | R_{\mathbf{b}} | 0} \cdot T/\sqrt L= t_0T$. The total number of slots used add up to $(\nu + 1) + (w + 2) + 1 = \nu + w + 4$, so (1) follows.

(2) Suppose the operator $T$ has a singular value decomposition $T = \sum_{j = 1}^d \sigma_j \mathbf{y}_j \mathbf{x}_j^\dagger$ with $\{ \mathbf{x}_j\}_{j=1}^d$ being an orthonormal basis for $\C^d$ and $\{\mathbf{y}_j\}_j \subseteq \C^I$ are orthonormal, and by assumption $\sqrt{1 - \eps} \leq \sigma_j \leq 1$ for each $j \in [d]$. Take $\beta_j \colon \arcsin(\sigma_j \cdot t_0) \in (0, \theta]$, statement (1) tells us that the component of the unit vector $B_T \iota \mathbf{x}_j$ in the good subspace is $t_0 \cdot \gamma T \mathbf{x}_j = t_0 \cdot \sigma_j \gamma \mathbf{y}_j$, so that
$$
B_T \iota \mathbf{x}_j = \sin \beta_j \cdot \gamma \mathbf{y}_j + \cos \beta_j \cdot \mathbf{y}_j^{\perp}
$$
for some unit vector $\mathbf{y}_j^{\perp}$ orthogonal to the good subspace. As in the Grover search, with the good subspace marker $\gamma$ and reflection about the prepared subspace $B_T \iota$, we compose multiple applications of the Grover-type iteration $\mathscr{G}_T = (2 B_T \iota \iota^\dagger B_T^\dagger - \id)(\id - 2 \gamma \gamma^\dagger)$ \cite{Brassard2002,QSVT}, such that after $\ell$ iterations it obliviously amplifies the good subspace to
\begin{equation}
\label{eqn:mapped_state_after_iteration}
\mathscr{G}_T^{\ell} B_T \iota \mathbf{x}_j = \sin \left((2 \ell + 1)\beta_j \right) \cdot \gamma \mathbf{y}_j + \cos \left( (2 \ell + 1)\beta_j \right) \cdot \mathbf{y}_j^{\perp}.
\end{equation}

\paragraph{Exact amplification.} As in \cite[Section 2.1]{Brassard2002}, if $\sigma_j = 1$ then correspondingly $\beta_j = \frac{\pi}{4L + 2}$, then with $\ell = L$ it gives $U_T \iota \mathbf{x}_j = \gamma \mathbf{y}_j = \gamma T \mathbf{x}_j$. In particular, if $T$ is an isometry, every singular value yields $1$ so iterating over all $j \in [d]$ yields $U_T \iota  = \gamma \mathbf{y}_j = \gamma T $ exactly.

\paragraph{Robustness from near-isometry.} For the near-isometry condition, the amplitude $t_0 \sigma_j$ depends on the singular direction, and the angles $\beta_j\leq \frac{\pi}{4L + 2}$ fall short of $\frac{\pi}{4L + 2}$. \cref{eqn:mapped_state_after_iteration} with $\ell=L$ stops short of the good subspace by the angle of difference $\Delta_j=(2L+1)(\frac{\pi}{4L + 2}-\beta_j)\ge 0$:
$$
 U_T\iota\mathbf{x}_j=\cos\Delta_j\,\gamma\mathbf{y}_j
 +\sin\Delta_j\,\mathbf{y}_j^{\perp}.
$$
Hence $\gamma^{\dagger}U_T\iota=\sum_{j = 1}^d \cos\Delta_j\,\mathbf{y}_j\mathbf{x}_j^{\dagger}$,
and since $U_T$ is unitary and both $\iota,\gamma$ are isometries,
\begin{equation}
\label{eqn:residual_of_approximation}
\begin{aligned}
 (U_T\iota-\gamma T)^{\dagger}(U_T\iota-\gamma T)
 &= \id_d -T^{\dagger}\gamma^{\dagger}U_T\iota
   -\left(T^{\dagger}\gamma^{\dagger}U_T\iota \right)^{\dagger}+ T^{\dagger}T\\
 &=\sum_{j=1}^{d} \left[(1-\sigma_j)^2+2\sigma_j(1-\cos\Delta_j) \right] \mathbf{x}_j\mathbf{x}_j^{\dagger}.
 \end{aligned}
\end{equation}
The shortfall is linear in $1-\sigma_j$: since $t_0\leq \sin\frac\pi6=\frac12$ and the derivative of $\frac{\mathrm{d}}{\mathrm{d} x} \arcsin(x) = \frac{1}{\sqrt{1 - x^2}}$ is bounded from above by $2/\sqrt3<2$ on $[0, 1/2]$, so the mean value theorem tells
$$
\begin{aligned}
\Delta_j = (2L + 1)( \theta-\beta_j) &= (2L + 1) \left(\arcsin t_0-\arcsin(t_0\sigma_j) \right) \\
&\leq (2 L + 1) t_0 (1 -\sigma_j) \cdot \max_{x \in [0, 1/2] } \left|\frac{\mathrm{d}}{\mathrm{d} x} \arcsin(x)  \right| \\
&\le2(2L + 1) \sin\left( \frac{\pi}{4L + 2} \right)(1-\sigma_j)
 \leq  \pi(1-\sigma_j).
 \end{aligned}
$$
With the handy inequality $1-\cos\Delta_j\leq \Delta_j^2/2$ and $\sigma_j\le1$, each coefficient in \cref{eqn:residual_of_approximation} is at most $(1 - \sigma_j)^2 + \pi^2(1 - \sigma_j)^2 = (1+\pi^2)(1-\sigma_j)^2$, and since $\sigma_j \in [1-\eps, 1]$ we have $1-\sigma_j\le1-\sqrt{1-\eps}\leq \eps$. Therefore
$\|U_T\iota-\gamma T \|_{\op}\leq \sqrt{1+\pi^2} \cdot \eps<5\eps$, proving the robustness statement in (2). It suffices to count the requirement for slots.

Using $2B_T\iota\iota^{\dagger}B_T^{\dagger}-\id=B_T(2\iota\iota^{\dagger}-\id)B_T^{\dagger}$, we can write $U_T$ as a product of $2L+1$ copies of $B_T$ or $B_T^{\dagger}$ and $2L$ reflections, each of the form $2\iota\iota^{\dagger}-\id$ or $\id-2\gamma\gamma^{\dagger}$. These reflections are diagonal in the computational basis, so by \cref{lemma:diagonal_gates_controlled_state_prep_in_slot_formalism} each of them takes exactly one slot. In total $U_T$ uses $(2L+1)(\nu+w+4)+2L$ slots. The only fresh qubits are the $h$ high bits of register $\mathsf{X}$, the $w$ qubits of register $\mathsf{Y}$, and the two flags that are cleanly uncomputed, completing the proof.
\end{proof}

\paragraph{Constructing the encoding circuits.} We now apply \cref{lemma:block_encoding_of_row_sparse_near_isometry} to the Fourier-transformed encoding operators $\widehat{E}_L$ and $\widehat{R}_L$ for their quantum circuit. Use the common layout consisting of the extended input index $\mathsf{X}$ on $\nu = n+h$ qubits, the phase register $\mathsf{Z}$ on $K= \log_2 |G| = b(L+1)$ qubits, the port register $\mathsf{P}$ on $n$ qubits\footnote{
So the register $\mathsf{Y} = \mathsf{ZP}$ as in the register arrangement in \cref{eqn:embedding_actions_on_registers}, with $w = K + n$.
}, and two flags, with embeddings
\begin{equation}
\label{eqn:register_embeddings}
 \iota\ket{i}=\ket i_{\mathsf{X}}\ket{0^K}_{\mathsf{Z}}\ket{0^n}_{\mathsf{P}}\ket{00}_{\mathsf{a}\mathsf{b}}; \quad \forall 0 \leq p < k, \quad \gamma(\ket{\mathbf{z}}\ket p)=\ket{0^{n+h}}_{\mathsf{X}}\ket{\mathbf{z}}_{\mathsf{Z}}\ket p_{\mathsf{P}}\ket{11}_{\mathsf{a}\mathsf{b}}.
\end{equation}
By \cref{prop:Fourier_sparsity_of_encodings}, every row of $\widehat{E}_L$ and of $\widehat{R}_L$ has at most $L$ nonzero entries, and by \cref{prop:near_isometry_of_encodings} and unitarity of the Fourier transform,
$$
 (1-r^{2L})\id_d\preceq
 \widehat{E}_L^{\dagger}\widehat{E}_L\preceq \id_d, \quad 
 (1-r^{2L})\id_d\preceq
 \widehat{R}_L^{\dagger}\widehat{R}_L\preceq \id_d.
$$
Using \cref{lemma:block_encoding_of_row_sparse_near_isometry} with $w=K+n$ and $\eps=r^{2L}$, we obtain the amplified circuits $\widehat U_E$ and $\widehat U_R$.

Let $\mathscr F_{\mathsf{Z}}$ denote the operation of applying $\mathscr F_G$ to register $\mathsf{Z}$. Since every frequency is a valid row, we have
$
 \mathscr F_{\mathsf{Z}}\gamma=\gamma(\mathscr F_G\otimes \id_k)
$. We thus set $U_E=\mathscr F_{\mathsf{Z}}\widehat U_E$ and $U_R=\mathscr F_{\mathsf{Z}}\widehat U_R$. Since $E_L=(\mathscr F_G\otimes \id_k)\widehat{E}_L$ and $R_L=(\mathscr F_G\otimes \id_k)\widehat{R}_L$, unitarity of $\mathscr F_{\mathsf{Z}}$ and \cref{lemma:block_encoding_of_row_sparse_near_isometry} give
\begin{equation}
\label{eqn:circuit_implementation_errors}
 \left\| U_E\iota-\gamma E_L \right\|_{\op}\leq 5r^{2L}, \quad \left\|U_R\iota-\gamma R_L\right\|_{\op} \leq 5r^{2L}.
\end{equation}
These $L+1$ Fourier transforms (recall that $\mathscr{F}_G = \mathscr{F}_{b}^{\otimes (L + 1)}$) cost $\bigo(K)$ slots and no workspace, so each encoder uses $\bigo(L(K+n+h)+K)$ slots and $K+n+h+2$ ancillae.

\subsubsection{Assembling the recursion: error and size}
\label[appendix]{recursion_with_error_and_size}
We provide no new construction in this section. Instead, we assemble the results of \cref{appendix:contracting_internal_block,appendix:closed_loop_identity,appendix:nearly_isometric_encodings_over_phase_group,appendix:row_sparsity_in_Fourier_basis,appendix:implementing_encoders} into one level and unroll the recursion to make rigorous the error and circuit-complexity estimates we previously briefed in \cref{appendix:proof_overview}.

For clarity, let us wrap up previous discussions and state the parameters again. Fix $n$ and $L$. For $d\geq 2$, \cref{prop:contracting_internal_block} gives $V=P_{\mathrm{out}}UP_{\mathrm{in}}$ with $\|D\|_{\op} < r=3/4$ and $k=d-s<d$, and recall (cf. \cref{eqn:dim_d_circuit_single_step}) the ideal circuit at dimension $d$ proceeds as 
\begin{equation}
\label{eqn:ideal_circuit}
C_d=P_{\mathrm{out}}^{\dagger}\,U_R^{\dagger}\,C_k\,U_E\,P_{\mathrm{in}}^{\dagger},
\end{equation}
where $C_k$ is the circuit at dimension $k$ for the child family $\mathbf F=\bigoplus_{\lambda \in \{0, 1\}^t, \mathbf{z}\in G}F_{\lambda}(\mathbf{z})$, with labels $(\lambda,\mathbf{z})$. Let $e(d)$ be the largest clean error of the circuit at dimension $d$, over all controlled families and all numbers of label qubits. We drop $\lambda$ as well for notational brevity.

We extend the embeddings $\iota,\gamma$ explicated in \cref{appendix:implementing_encoders} so that they also put the remaining workspace, including that required by $C_k$, into $\ket{0}$. By assumption $\|C_k\gamma-\gamma\mathbf F \|_{\op} \leq e(k)$ is guaranteed by the child task. Since the preconditioners are implemented exactly and map clean valid inputs to clean valid inputs, telescoping along the ideal maps of \cref{eqn:ideal_identities} and combining the bounds in \cref{eqn:approximate_intertwining_of_child_task_passing} and \cref{eqn:circuit_implementation_errors} then gives
\begin{equation}
\label{eqn:level_wise_error_recurrence}
\begin{aligned}
e(d) &= \left\| U_R^\dagger C_k U_E  \iota -   \iota  V \right\|_{\op} = \left\| C_k U_E  \iota -  U_R  \iota  V \right\|_{\op} \\
&= \| C_k\left(U_E  \iota - \gamma  E_L \right) + \left(C_k  \gamma -  \gamma  \mathbf{F} \right) E_L + \gamma\left( \mathbf{F} E_L - R_L V \right) + (\gamma  R_L - U_R  \iota) V  \|_{\op} \\
&\leq \| U_E  \iota - \gamma  E_L \|_{\op} + \|C_k  \gamma -  \gamma  \mathbf{F} \|_{\op}  + \|  \mathbf{F} E_L -  R_L V \|_{\op} + \|\gamma  R_L - U_R  \iota   \|_{\op} \\
&\leq  e(k) + 10 r^{2L} + 8r^{L} \leq e(k) + 18 r^{L}.
\end{aligned}
\end{equation}
We then unroll the recurrence to give the following bounds.

\begin{proposition}[Resource quantification for the recursive skeleton]
\label{prop:resource_quantification_of_skeleton}
   Let $n\ge1$, $1\leq d\le2^n$, $t\ge0$ and $L\ge1$, let $J = 32768(n + 1)^3$ (cf. \cref{appendix:contracting_internal_block}), and put $h=\lceil\log_2(L+1)\rceil$, $b=nL+h$ (cf. \cref{appendix:nearly_isometric_encodings_over_phase_group} and \cref{appendix:implementing_encoders}) and $K=b(L+1)$. There exists a deterministic Turing machine that writes down a skeleton that, for suitable tables, implements every controlled unitary family $\{U_\lambda\}_{\lambda\in\{0,1\}^t}$ on $\mathcal H_d$ with clean error at most $18Jr^L$ in $\poly(n, t, L)$ time, where $r=3/4$. Its number of qubits $Q$ and slots $S$ scale as 
\begin{equation}
\label{eqn:scaling_of_total_qubit_and_slot_count}
 Q =t+ \bigo \left((n+J(K+n+h) \right), \quad S= \bigo\left(1+J[L(K+n+h)+K+n] \right).
\end{equation}
\end{proposition}

\begin{proof}
    The error upper bound statement follows directly by applying the recurrence relation in \cref{eqn:level_wise_error_recurrence} plus a base case value. For $d = 1$, each $U_{\lambda}$ is a phase gate for some phase $\alpha_{\lambda}$ on the valid index $0$, and one slot on an index bit. That is, with table $\diag(\mathrm{e}^{\I\alpha_\lambda},1)$ controlled by $\lambda$, which can be implemented exactly. So $e(1) = 0$. Summing \cref{eqn:level_wise_error_recurrence} over the $\leq J$ levels therefore gives $e(d) \leq 18 J r^L + e(1) = 18 J r^L$.

    As remarked in \cref{appendix:implementing_encoders}, each level adds $K+n+h+2$ qubits of the encoders and $\bigo(n)$ clean ancillae for the preconditioners, which stay allocated during the inner call to the child. The phase register of a level becomes part of the label register of the next. It also adds $\bigo(L(K+n+h)+K)$ slots for the encoders and $\bigo(n)$ for the preconditioners. As the circuit $C_d$ contains the child $C_k$ only once, summing over fewer than $J$ levels gives the desired scaling of $Q$ and $S$ in \cref{eqn:scaling_of_total_qubit_and_slot_count}.

    Finally, $q = 2^{\lfloor \log_2 d \rfloor}$, $s$ (cf. \cref{eqn:block_size_parameters}), $k = d - s$ are computed from $d$ by arithmetic on only $\bigo(n)$ bits, and the skeletons of \cref{appendix:programmable_slots}, \cref{prop:contracting_internal_block} and \cref{lemma:block_encoding_of_row_sparse_near_isometry} depend only on $n,d,t,L$, so it suffices for the Turing machine to write down the $S$ slots and their control lists, each of length less than $Q$, in time $\poly(n, t, L)$. This completes the proof.
\end{proof}

\paragraph{Choosing the truncation length.} With the guarantee in \cref{eqn:level_wise_error_recurrence}, when setting $\beta \geq 1$ as the precision parameter in \cref{prob:USP} so that $\delta = 2^{-\beta - 1}$. We take the truncation length in the loop (cf. \cref{appendix:nearly_isometric_encodings_over_phase_group}) as 
\begin{equation}
\label{eqn:truncation_length}
L = 4 \left( \beta + \lceil \log_2(72 J) \rceil \right).
\end{equation}
Then the skeleton implemented in \cref{prop:resource_quantification_of_skeleton} has clean error $e(d) \leq 18 J r^{L} \leq 18 \cdot J \cdot 2^{-\beta} \cdot 1/(72 J) = \delta / 2$. Moreover, the parameters $J, L, h, b, K$ all scale as $\poly(n, t, \beta)$, and therefore so do $Q, S$ by \cref{eqn:scaling_of_total_qubit_and_slot_count}.

\subsubsection{The oracle circuit: tables in one Boolean oracle}
\label[appendix]{putting_tables_into_oracle}
As the final piece of the puzzle, this section proves the compilation step sketched at the end of \cref{appendix:wrap_up_the_proof}: we put the tables into the Boolean oracle and convert the skeleton into the oracle circuit, so they combine to approximately synthesize the target controlled unitary family. 

The main ingredient of the first step is the conventional Solovay-Kitaev theorem, where we also handle exact phases and explain how to read the gate tables when the control registers are in superposition. In our formulations, we interpret a \emph{word} as a finite product of operators from the gate set $\{\hdm, \tgate, \tgate^\dagger\}$, possibly padded with identity letters, and call the individual gates $\id, \hdm, \tgate, \tgate^\dagger$ of a word its \emph{letters}. The approximations are always measured in the operator-norm distance to track the effect of global phases.

\begin{lemma}[Solovay-Kitaev {\cite[Theorem 8.5]{KitaevShenVyalyi2002}}]
\label{lemma:solovay_kitaev_for_SU}
Let $\mathcal{S}\subseteq\SU(2)$ be a finite set that is closed under inverses and generates a dense subgroup of $\SU(2)$, and let $\alpha>3$. There exists a universal constant $\ell_0 > 0$ that depends only on $\mathcal{S}$ and $\alpha$, such that for every $0<\eps \leq 1/2$, every $U\in\SU(2)$ is within $\eps$ of a product of at most $\ell_0\log^{\alpha}(1/\eps)$ elements of $\mathcal{S}$.
\end{lemma}

Note that \cref{lemma:solovay_kitaev_for_SU} is the global phase-insensitive primitive of the universal approximation for qubit unitaries, and it does not apply to the letters directly: $\det(\hdm)=-1$ and $\det(\tgate)=\mathrm{e}^{\I\pi/4}$. The usual remedy is to rescale them by scalars into $\SU(2)$, but then any legitimate word approximates the target only up to a global phase. That is not enough here, since a slot applies a family of $W_y$ under control and a global phase of each $W_y$ becomes a relative phase among different control branches. To settle, we first construct a finite, inverse-closed set of determinant-one words that generates a dense subgroup of $\SU(2)$, and then apply \cref{lemma:solovay_kitaev_for_SU} to this set.

\begin{lemma}[Phase-aware Solovay-Kitaev, cf. \cite{shen2026quantumalgorithmsmultivariablepolynomial}]
\label{lemma:phase_aware_solovay_kitaev}
    There is a universal integer $\ell_0 \in \N$ such that for every integer $\kappa\ge1$ and every $U\in\SU(2)$, there exists some determinant-one word of length at most $\ell_0 \kappa^4$ approximates $U$ within $2^{-\kappa}$.
\end{lemma}

\begin{proof}
Let $G_0 = \langle \hdm, \tgate \rangle$ be the group of all words from our gate set, and let $\Gamma = G_0 \cap \SU(2)$. The determinant restricts to a homomorphism from $G_0$ onto the group $\mu_8 = \{\mathrm{e}^{\I\pi j/4} : j = 0, 1, \dots, 7\}$ of eighth roots of unity, with kernel $\Gamma$. By the first isomorphism theorem \cite[Chapter 3.3, Theorem 16]{DummitFoote2004}, $G_0/\Gamma \cong \mu_8$, so the index $[G_0 : \Gamma] = |\mu_8| = 8$.

Next, we are to find a finite generating set for $\Gamma$. For $k \in \mathbb{Z}/8\mathbb{Z}$, let $g_k = \tgate^k \hdm (\tgate^\dagger)^{k+4}$, where exponents of $\tgate$ are read modulo $8$ since $\tgate^8 = \id$. Each $g_k$ is a determinant-one word of length at most $11$, and $g_k^{-1} = g_{k+4}$. Because $\tgate^\dagger = \tgate^7$, every word can be written as $\tgate^{a_0} \hdm \tgate^{a_1} \hdm \cdots \hdm \tgate^{a_m}$. Setting $\mu_j = a_0 + \cdots + a_j + 4j$, we have
\begin{equation}
\label{eqn:det_one_word_reformulation}
  \tgate^{a_0} \hdm \tgate^{a_1} \hdm \cdots \hdm \tgate^{a_m} = g_{\mu_0} g_{\mu_1} \cdots g_{\mu_{m-1}} \tgate^{\mu_m},
\end{equation}
since adjacent factors on the right combine as $(\tgate^\dagger)^{\mu_{j-1}+4} \tgate^{\mu_j} = \tgate^{a_j}$ for $1 \leq j \leq m$. As $\det(g_k) = 1$ for all $g_k$, the word has determinant $\mathrm{e}^{\I\pi\mu_m/4}$. For a word in $\Gamma$, this determinant equals one, so $\mu_m \equiv 0 \bmod 8$ and $\tgate^{\mu_m} = \id$. Thus every element of $\Gamma$ is a product of the $g_k$, that is, $\mathcal{S} = \{g_0, \dots, g_7\}$ generates $\Gamma$, and clearly $\mathcal{S}$ is closed under inverses.

It remains to show that $\Gamma$ is dense in $\SU(2)$. Let $\pi\colon \U(2) \to \mathsf{PU}(2) = \U(2)/\U(1)$ be the quotient map. The image $\pi(G_0)$ is dense in $\mathsf{PU}(2)$ \cite[Section 4.5.3]{Nielsen2012}. Since $[G_0 : \Gamma] = 8$ with coset representatives $\id, \tgate, \dots, \tgate^7$, we have the left-coset decomposition $G_0 = \bigcup_{j=0}^{7} \tgate^j \Gamma$. Applying $\pi$ and taking closures on both sides gives
$$
  \mathsf{PU}(2) = \bigcup_{j=0}^{7} \pi(\tgate^j) \cdot \overline{\pi(\Gamma)},
$$
so $\overline{\pi(\Gamma)}$ is a closed subgroup of finite index in $\mathsf{PU}(2)$. A closed subgroup of finite index in a connected group is the whole group, so $\pi(\Gamma)$ is dense in $\mathsf{PU}(2)$. The closure $\overline{\Gamma}$ of $\Gamma$ in $\SU(2)$ is compact, so $\pi(\overline{\Gamma})$ is closed. It also contains the dense set $\pi(\Gamma)$, hence $\pi(\overline{\Gamma}) = \mathsf{PU}(2)$. As $\pm\id$ are the only elements of $\SU(2)$ with trivial image under $\pi$, the subgroup $\overline{\Gamma}$ has index at most $2$ in $\SU(2)$. Since $\SU(2)$ is connected, the same fact about closed subgroups of finite index therefore gives $\overline{\Gamma} = \SU(2)$. The set $\mathcal{S}$ then gives us a candidate that satisfies the requirements in the statement of \cref{lemma:solovay_kitaev_for_SU}, and we plug in $\alpha = 4$ and the precision $\eps = 2^{-\kappa}$ to obtain the word length $\ell_0 \log^\alpha(1/\eps) = \ell_0 \kappa^4$, as desired.
\end{proof}

\begin{lemma}[Code-controlled letters]
\label{lemma:code_controlled_letters}
    Define the code map $\mathrm{code}: \{\id, \hdm, \tgate, \tgate^\dagger\} \to \{0, 1\}^2$ explicitly via $\mathrm{code}(\id) = 00$, $\mathrm{code}(\hdm) = 01$, $\mathrm{code}(\tgate) = 10$, $\mathrm{code}(\tgate^\dagger) = 11$, and let $U_c$ be the letter with code $c \in \{0, 1\}^c$. Then the controlled operation $\sum_{c \in \{0, 1\}^c} \ps{c} \otimes U_c$ has an exactly synthesizable circuit of constant size over $\{\hdm, \tgate, \tgate^\dagger, \cnot\}$ using one clean ancilla qubit.
\end{lemma}

\begin{proof}
    Note that this gate is the diagonal gate $\diag(\id, \hdm, \tgate, \tgate^\dagger)$, with entries of $\hdm$ being $\pm 1/\sqrt{2}$, and the non-trivial diagonal entries of $\tgate, \tgate^\dagger$ are $\mathrm{e}^{\pm \I \pi / 4} = (1 \pm \I)/\sqrt{2}$. All these entries are in the ring $\Z[1/\sqrt{2}, \I]$, so are the matrix entries via matrix multiplication. The exact synthesis theorem of Giles and Selinger \cite[Theorem~1]{PhysRevA.87.032332} gives a result over the gate set $\{\hdm, \cnot, \tgate, \diag(1, \I),$ $\mathrm{e}^{\I \pi /4} \id\}$ for unitaries with entries in $\Z[1/\sqrt{2}, \I]$. The proof is therefore immediate by noting that $\diag(1, \I) = \tgate^2$ and $\mathrm{e}^{\I \pi / 4} \id = (\tgate^2 \hdm)^3$.
\end{proof}

\begin{proposition}[Compiling a skeleton into an oracle circuit]
\label{prop:skeleton_to_oracle_circuit_compilation}
Fix a skeleton on $Q$ qubits with $S \ge 1$ slots, and let $\kappa \ge 1$ be an integer.
Then there exists a deterministic Turing machine that, in $\poly(Q, S, \kappa)$ time, outputs an oracle circuit $\widetilde{C}$ over $\{\hdm, \tgate, \tgate^\dagger, \cnot\}$ with a single Boolean oracle, such that the width, size and address length of $\widetilde{C}$ also scale as $\poly(Q, S, \kappa)$. Moreover, for every assignment of tables to the
slots, there exists a Boolean oracle such that
$$
  \left\|\widetilde{C}\iota - \iota C\ \right\|_{\op} \leq S \cdot 2^{-\kappa},
$$
where $C$ denotes the skeleton instantiated with these tables, again with $\iota$ appending additional ancillae.
\end{proposition}

\begin{proof}
    We take steps to construct the circuit. 
    \paragraph{Compiling one slot.} For a table entry $W_y \in \U(2)$ we choose $\phi_y $ such that $\mathrm{e}^{2 \I \phi_y} = \det(W_y)$ and put $S_y = \mathrm{e}^{-\I \phi_y } W_y, P_y = \diag(\mathrm{e}^{\I \phi_y}, \mathrm{e}^{-\I \phi_y}) \in \SU(2)$. Note that $P_y \ket{0} = \mathrm{e}^{\I \phi_y} \ket{0}$, we thus have $\iota_0 W_y =S_y \otimes P_y \iota_0$ with $\iota_0$ introducing one ancilla qubit. We then apply \cref{lemma:phase_aware_solovay_kitaev} with precision parameter $\kappa + 1$ for determinant-one words $\widetilde{S}_y, \widetilde{P}_y$ that approximates $S_y, P_y$, respectively, one a common word length $\ell_{\kappa} = \ell_0 (\kappa + 1)^4$. Then the controlled action satisfies
    $$
    \left\| \sum_{y} \ps{y} \otimes \widetilde{S}_y \otimes \widetilde{P}_y - \sum_{y} \ps{y} S_y \otimes P_y\right\|_{\op} \leq \max_{y} \left\| \widetilde{S}_y \otimes \widetilde{P}_y - S_y \otimes P_y \right\|_{\op} \leq 2^{-\kappa - 1} \cdot 2 = 2^{-\kappa}.
    $$
    The error bound applies to every slot that occurs in composing the final oracle circuit.
    \paragraph{Coherent lookup of tables.} The oracle is specified by the Boolean function $f_{\mathbf{U}, \beta} \coloneqq f$ in \cref{appendix:wrap_up_the_proof}. For slot $j$ and control $y$ padded to the full $Q$-bit string $\widehat{y}$, and a position $0\leq i<\ell_\kappa$,
$$
 f(j,\widehat y,0,i,\cdot)=\mathrm{code}(g_{j,\widehat y,i}),\quad f(j,\widehat y,1,i,\cdot)=\mathrm{code}(g'_{j,\widehat y,i}),
$$
where $g_{j,\widehat y,i}$ and $g'_{j,\widehat y,i}$ are the $i$-th letters of the target word $\widetilde S_y$ and of the phase word $\widetilde P_y$ of slot $j$, and the last address bit selects one of the two code bits. The
address length is $m=\lceil\log_2(S+1)\rceil+Q+\lceil\log_2(\ell_\kappa+1)\rceil+2 = \poly(Q, S, \kappa)$. To apply one letter, (\textit{i}) copy the controls of the slot into an address register with $\cnot$s and set the constant fields with $X= \hdm \tgate^4 \hdm$, query the two code bits into two clean qubits, (\textit{ii}) apply \cref{lemma:code_controlled_letters} to the target if the letter belongs to the target word, and to the ancilla qubit for one slot if it belongs to the phase word, (\textit{iii}) then query the two bits again, and undo the address preparation to return back the ancillae.
The letters act neither on the controls nor on the address, so in every control branch the second queries erase the bits written by the first, and by linearity the slot is implemented exactly as $\sum_y \ps{y}\otimes\widetilde S_y\otimes\widetilde P_y$, with $8\ell_\kappa$ queries and $\bigo(\ell_\kappa(m+1))$ gates per slot and $\bigo(m+1)$ ancillae, and both scale as $\poly(Q, S, \kappa)$.

\paragraph{Extending to all slots.} Suppose the ideal circuit can be decomposed into sequential actions of ideal slots $C = V_S V_{S-1} \cdots V_1$, and we implement these slots as $\widetilde{V}_S, \widetilde{V}_{S-1}, \dots, \widetilde{V}_1$ accordingly as per our previous discussion. Without loss of generality, we extend the ideal $C$ and $V_j$s to the involved ancillae during the computation so that $\iota C = C \iota$. Then by a simple telescoping, the implementation error satisfies
$$
\begin{aligned}
\left \|\widetilde C\iota-\iota C \right\|_{\mathrm{op}} 
 = \left\| \left(\widetilde{C} - C\right) \iota \right\|_{\op} &=
 \left\|\sum_{j=1}^{S} \widetilde V_S\cdots\widetilde V_{j+1}
(\widetilde V_j-V_j)V_{j-1}\cdots V_1\iota \right\|_{\mathrm{op}}\\
 &\leq \sum_{j=1}^{S} \left\|(\widetilde V_j-V_j)V_{j-1}\cdots V_1\iota \right\|_{\mathrm{op}} \leq S \cdot 2^{-\kappa}.
\end{aligned}
$$
\paragraph{Uniformity.} The Turing machine writes the address computations, the lookups, and the fixed exact synthesis circuits of \cref{lemma:code_controlled_letters} from the skeleton and $\ell_\kappa$, and we simply hard-code the universal constant $\ell_0$. The resource requirement in time therefore scales as $\poly(Q, S, \kappa)$.

Combining the above arguments completes the proof.
\end{proof}

\begin{proof}[Proof of \cref{thm:usp}]
    Fix $n,d,t,\beta$ as in the input, and we put the clean error $\delta=2^{-\beta-1}$. Take $L$ as in \cref{eqn:truncation_length}, \cref{prop:resource_quantification_of_skeleton} gives a skeleton with clean error at most $\delta/2$. Let $C_d$ denote this exact-slot circuit with its tables filled in, and let $S$ be the number of slots in $C_d$. We then use \cref{prop:skeleton_to_oracle_circuit_compilation} to compile $C_d$ and take
$$
 \kappa=\beta+2+\lceil\log_2(S+1)\rceil .
$$
Then $2^{-\kappa}\leq\delta/(2(S+1))$, so the compilation adds at most $S \cdot 2^{-\kappa} \leq \delta \cdot S/(2(S+1)) < \delta/2$ approximation error. The final circuit produced by the compiled oracle circuit is $C = \widetilde C_d$. Suppose $\iota_1$ extends the workspace of $C_d$ and $\iota_2$ introduces the ancillae for the compiler, and we set $\iota=\iota_2\iota_1$. It follows that
$$
\begin{aligned}
 \left\| C\iota-\iota\mathbf{U} \right\|_{\op}
 &= \left\| \left(\widetilde C_d\iota_2-\iota_2C_d \right)\iota_1
 +\iota_2 \left(C_d\iota_1-\iota_1\mathbf{U} \right) \right\|_{\op} \\
 &\leq \left\| \left(\widetilde C_d\iota_2-\iota_2C_d \right)\iota_1 \right\|_{\op} + \left\| \iota_2 \left(C_d\iota_1-\iota_1\mathbf{U} \right) \right\|_{\op} \\
 &\leq \left\| \widetilde C_d\iota_2-\iota_2C_d  \right\|_{\op} + \left\| C_d\iota_1-\iota_1\mathbf{U} \right\|_{\op} \leq \frac{\delta}{2} + \frac{\delta}{2} = \delta.
 \end{aligned}
$$

All the required resources are $\poly(n, t, \beta)$: we have
$\kappa=\bigo(\beta+\log(S+1))$, and the compiler is polynomial in $Q,S,\kappa$. The single Boolean function stores the words required to synthesize the tables for all slots of all levels. We thus obtain, in time $\poly(n, t, \beta)$, an oracle circuit whose size, width, and query length scale as $\poly(n, t, \beta)$. This completes the proof.
\end{proof}

\section{\texorpdfstring
  {Proof of \cref{lemma:usynthesis_with_pspace_oracle}}
  {Proof of Lemma 4.7}}
\label[appendix]{appendix:proof_of_pspace_USP} 
This appendix is dedicated to the proof of \cref{lemma:usynthesis_with_pspace_oracle}, the $\pspace$-oracle version of the USP theorem stated in \cref{thm:usp}, which serves as a key ingredient in the proof of \cref{thm:upspace-in-uqip}. Specifically, we use the $\pspace$ arithmetic results from \cref{sec:pspace_arthm} to show that, for $\upspace$ families, the building blocks of the USP algorithm are $\pspace$-computable.

Throughout this section, fix the $\upspace$ unitary family $W = \{W_n\}_{n \in \mathbb{N}}$ of \cref{lemma:usynthesis_with_pspace_oracle} with $\pspace$-computable entries (see \cref{sec:pspace_arthm}). Its entries are $\pspace$-computable by hypothesis, so \cref{def:pspace_computable_complex_num} guarantees that there exists a deterministic Turing machine that, on input $(1^{n}, a, b, 1^{g})$, outputs a complex dyadic number $\widetilde{w}$ satisfying
$$
  \left|\widetilde{w} - [W_n]_{a,b}\right| \leq 2^{-g}
$$
using $\poly(n, g)$ space. We fix one such machine and denote it by $M_W$, so the same machine would serve all dimensions, entry indices and precisions.

Recall from the proof of \cref{thm:usp} in \cref{appendix:answer_of_USP}, that the oracle circuit is produced from the sizes alone: the circuit generator writes the skeleton via \cref{prop:resource_quantification_of_skeleton} and compiles it by \cref{prop:skeleton_to_oracle_circuit_compilation}, while everything that depends on the target sits in the tables of the slots and in the words that the Boolean oracle stores for them. Several proofs in \cref{appendix:answer_of_USP} are conducted non-constructively. To prove \cref{lemma:usynthesis_with_pspace_oracle}, it therefore suffices to (\textit{i}) fix every table by an explicit rule such that the error analysis of \cref{appendix:answer_of_USP} still applies up to a small additional error, (\textit{ii}) show that every table entry, and hence every stored word, can be computed in polynomial space from the answers of $M_W$, and finally (\textit{iii}) collect the words for all $n$ and $\beta$ into the corresponding $\pspace$ language $A_W$. 

\paragraph{Notational conventions.} We keep the notation of \cref{appendix:answer_of_USP} for the target $U=W_n$, so that $d=2^n$ and $t=0$ in the setup \cref{prop:resource_quantification_of_skeleton} and the only target is $U^{(0)}=W_n$ in \cref{appendix:poly_size_recursion_by_one_call_per_step}. We write $V^{(j)}_{\lambda_j}=P_{\mathrm{out}}U^{(j)}_{\lambda_j}P_{\mathrm{in}}$ with the preconditioners of level $j$ and label $\lambda_j$. From now on, the target family always carries its index $n$, so the symbols $W$ and $W_y$ retain their meanings from \cref{appendix:answer_of_USP}. The tilde $\widetilde{(\cdot)}$ marks a numerical or circuit approximation of scalars or operators on space-uniform circuits.

\subsection{Admissible tables: reusing the error analysis}
Recall that \cref{appendix:answer_of_USP} proves the error bound $18Jr^L$ of \cref{prop:resource_quantification_of_skeleton} for a particular choice of tables, some of which are only shown to exist. Under $\pspace$-computable restriction, we are to replace them by tables that can be computed in polynomial space. To this end, we first single out the properties of the tables on which this bound depends, so that every such choice inherits the bound without replicating the error analysis. In this section, we fix $n$ and $L$ and isolate these properties.

\begin{lemma}[Admissible tables]
\label{lemma:admissible_tables}
    Suppose we fill the skeleton given by \cref{prop:resource_quantification_of_skeleton} with $d = 2^n$ and $t = 0$ level by level as in \cref{appendix:answer_of_USP}, except that at every level $j < m \leq J$ and for every label, the preconditioners and the two Fourier-transformed encoders obtained by applying \cref{lemma:block_encoding_of_row_sparse_near_isometry} to $T = \widehat{E}_L, \widehat{R}_L$ satisfy the following conditions:
    \begin{enumerate}
        \item \label{item:admissible_condition_1} The preconditioners $P_{\mathrm{in}}, P_{\mathrm{out}}$ have the form used in \cref{prop:contracting_internal_block}, and the internal block $D$ of the preconditioned operator $V^{(j)}_{\lambda_j}$ (cf. \cref{eqn:block_decomposition_of_unitary,fig:closing_internal_coorinate_loop}) satisfies $\|D\|_{\op}<r$.
        \item \label{item:admissible_condition_2} For every row $u$ of $T$, the set $I_u$ consists of exactly $L$ basis indices of the register $\mathsf{X}$ and includes every $i$ with $T_{ui} \neq 0$ (cf. proof of \cref{lemma:block_encoding_of_row_sparse_near_isometry}).
    \end{enumerate}
    Then the operator $C$ implemented by the skeleton with these tables filled in satisfies $\| C \cdot \iota - \iota \cdot W_n \|_{\op} \leq 18 Jr^L$.
\end{lemma}

\begin{proof}
    The proof of \cref{prop:resource_quantification_of_skeleton} uses the preconditioners only through the bound $\|D\|_{\op} < r$, under which \cref{lemma:feedback_identity} through \cref{prop:Fourier_sparsity_of_encodings} hold for every preconditioned unitary $V$, and the candidate sets only via condition 2, which yields $\gamma^\dagger B_T\iota=t_0T$ for the block encoding $B_T=R_{\mathsf b}W^\dagger A$ via \cref{eqn:entry_wise_extraction_of_T} in the proof of \cref{lemma:block_encoding_of_row_sparse_near_isometry}. All other tables are as in \cref{appendix:answer_of_USP}, so the proof can be reused verbatim. This completes the proof.
\end{proof}

\subsection{\textsf{PSPACE}-computable preconditioners}
\label[appendix]{appendix:pspace_computable_preconditioners}
In this section, we choose the preconditioners of \cref{prop:contracting_internal_block} canonically. The choices of signs $S_1$ and $S_2$ in the discussion of \cref{appendix:contracting_internal_block} have been shown to exist by a probabilistic argument based on tail bounds for independent signs (cf. \cref{lemma:random_sign_entry_and_row_bounds,lemma:matrix_rademacher}). A machine with polynomial space, however, cannot even store a sign pattern of length $q = 2^{\lfloor \log_2 d \rfloor}$. To keep the construction $\pspace$-computable, we follow the small-sample-space approach of \cite{ALON1986567, doi:10.1137/0215074} and generate $\bigo(n)$-wise independent pseudorandom signs from polynomial-length seeds. These signs preserve the moments required by the tail bounds, so some pair of seeds still can make the internal block contractive. We then find such a pair by exhaustive search in polynomial space.

\begin{lemma}[Canonical space-uniform preconditioners]
\label{lemma:canonical_space_uniform_preconditioners}
    Put $N = n + 1$ and $g_0 = 8N^{2}$. There exists a deterministic algorithm that, given approximations to within $2^{-g_0}$ of the entries of a unitary $U \in \U(d)$ with $2 \leq d \leq 2^{n}$, selects the preconditioners that satisfy condition 1 in \cref{lemma:admissible_tables}. The selection depends only on the supplied approximations, has an $\bigo(n^{2})$-bit description, and is computable with $\poly(n)$ space apart from the requests for the entries of $U$.
\end{lemma}

\begin{proof}
    To start with, we reuse the notation $q, s, W, M_0, M_1, S_1, S_2, \mathbf{h}_i$ and $\mathbf{m}_i$ from \cref{appendix:contracting_internal_block}.

    \paragraph{The simple case $s = 1$.} In this case, $D \in \C$. As in the proof of \cref{prop:contracting_internal_block}, there exists some entry of the last column of $U$ that has modulus $\leq 1/\sqrt{2}$. Therefore, we estimate the squared moduli of these entries to within $1/64$, take the first row whose estimate is at most $33/64$ and swap it into the last position, so that the corresponding entry becomes $D$. Then $\|D\|_{\op}^{2} = |D|^{2} \leq 33/64 + 1/64 < (3/4)^{2} = r^{2}$. We proceed to recover the result for $s \geq 2$.

    \paragraph{Seeded signs.} We use the finite-field construction of \cite[Proposition 6.3]{ALON1986567}. For $s \geq 1$, with the choice of $s$ in \cref{eqn:block_size_parameters}, we have $q \geq 4096s N^2 \geq 8192 N^2$. We will identify $[q]$ through binary digits with the field $\F_q$, represented by the first irreducible binary polynomial of degree $\log_2 q$. That is, we fix a degree-$\log_2 q$ irreducible polynomial $f(t)$ over $\F_2[t]$, and implement $\F_q$ via $\F_q = \F_2[t] / f(t)$. The binary representation of $\F_q$ elements $(y_0, y_1, \dots, y_{\log_2 q - 1})$ is then identified via the translation
    $$
    (y_0, y_1, \dots, y_{\log_2 q - 1}) \iff  \quad y_0+y_1t+\cdots+y_{\log_2 q-1}t^{\log_2 q-1} \bmod{f(t)}.
    $$
    The sample space therefore consists of the $q^{2N}$ polynomials of degree less than $2N$ over $\F_q$, and we write them as
    $$
     p_{\mathbf{a}}=\sum_{v=0}^{2N-1}a_v\zeta^v\in\F_q[\zeta],\quad \mathbf{a}=(a_0, \dots ,a_{2N-1})\in\F_q^{2N},
    $$
    and we call each $\mathbf{a}$ a \emph{seed}. Let $c(y) \coloneqq y_0$ denote the constant coefficient of $y\in\F_q$, and set
$$
 \forall \, x\in\F_q,\quad
 \varepsilon_x(\mathbf{a})=(-1)^{c(p_{\mathbf{a}}(x))}.
$$
For distinct $x_1, \dots ,x_{2N} \in \F_q$, an invertible Vandermonde matrix can recover $\mathbf{a}$ from the values of $p_{\mathbf{a}}(x_1), \dots ,p_{\mathbf{a}}(x_{2N}) \in \F_q$, so for a uniformly distributed seed $\mathbf{a}$, they are independent and uniform. Since the map $c\colon\F_q\to\F_2$ is linear and onto, exactly $q/2$ elements of $\F_q$ are mapped to $0$, and the other $q/2$ are mapped to $1$. Hence, any $2N$ of the signs are independent and uniform. We take $S_2=\diag(\varepsilon_x(\mathbf{a}))_{x\in\F_q}$ and $S_1=\diag(\varepsilon_x(\mathbf b))_{x\in\F_q}$ for two independent uniform seeds $\mathbf{a}$ and $\mathbf b$. Given the seed and index, the sign matrices take $\bigo(N)$ operations in $\F_q$ to evaluate via the conventional Horner’s method \cite{Horner1819}.

\paragraph{Moments.} We use this $2N$-wise independence to transfer the following moment bounds from independent uniform signs to the seeded signs. For $x_1,\dots,x_k\in\F_q$ with $k\leq 2N$, let $T$ be the set of distinct points among $\{x_j\}_{j=1}^k$ and let $m_x$ be the number of occurrences of each $x \in T$. Since $|T|\leq k\leq 2N$, the signs indexed by $T$ are independent and uniform by the previous discussion. Therefore, with a uniformly distributed seed,
\begin{equation}
\label{eqn:expectation_identity_unchanged}
\begin{aligned}
    \E_{\mathbf{a}} \left[\varepsilon_{x_1}(\mathbf{a})\cdots\varepsilon_{x_k}(\mathbf{a}) \right] &=\prod_{x\in T}\E_{\mathbf a}\left[\varepsilon_x(\mathbf a)^{m_x}\right] = \prod_{x\in T} \frac{1+(-1)^{m_x}}{2} = \mathbf{1}\left\{\forall\, x \in T,  m_x \in 2 \Z \right\} \\
    &= \E_{\substack{x \in T \\ \eps_{x} \overset{\mathrm{i.i.d.}}{\sim} \mathrm{Unif}(\{-1, 1\}) }} \left[ \eps_1 \eps_2 \cdots \eps_k \right].
\end{aligned}
\end{equation}
Since every sign squares to one, each monomial of degree at most $2N$ reduces to a product of at most $2N$ distinct signs. The identity in \cref{eqn:expectation_identity_unchanged} shows that the expectation of each such monomial coincides with that under independent uniform signs. By linearity of expectation, the same holds for every polynomial of degree at most $2N$ in the signs. After taking $S_1, S_2$, in the proof of \cref{lemma:random_sign_entry_and_row_bounds}, every entry of $M_1=M_0S_2W$ is a sign sum
$$
 \xi=[M_1]_{i,a}=\sum_{j\in[q]}\varepsilon_j\alpha_j,\quad 
 \alpha_j=[M_0]_{i,j}W_{j,a},\quad \sum_{j\in[q]}|\alpha_j|^2\leq \frac1q,
$$
so $|\xi|^{2N}=(\xi\bar\xi)^N$ is a polynomial of degree $2N$ in the signs $\eps_j$ of $S_2$. For a fixed $S_2$, \cref{eqn:internal_block_via_rank-one-sum} gives $D=\sum_{i\in[q]}\tau_i\mathbf h_i\mathbf m_i^\dagger$, so $\Tr[(DD^\dagger)]^N$ has degree $2N$ in the signs $\tau_i$ of $S_1$. 

For fixed $S_2$, recall that in \cref{prop:contracting_internal_block} we set $B_i=\mathbf h_i\mathbf m_i^\dagger$, and $\sum_i\mathbf h_i\mathbf h_i^\dagger=W^\dagger W= \id_s$, $\norm{\mathbf h_i}^2=s/q$ and $\|M_1\|_{\op}\le1$, then 
$$
 \begin{aligned}
 \sum_iB_iB_i^\dagger
 &=\sum_i \left\|\mathbf m_i\right\|^2\mathbf h_i\mathbf h_i^\dagger
 \preceq s\max_{i\in[q],a\in[s]}\bigl|[M_1]_{i,a}\bigr|^2\id_s, \quad \sum_iB_i^\dagger B_i =\frac{s}{q}M_1^\dagger M_1\preceq\frac{s}{q}\id_s.
 \end{aligned}
$$
Thus the variance parameter $\sigma$ in the Rademacher inequality of \cref{lemma:matrix_rademacher} satisfies
\begin{equation}
\label{eqn:variance_param_upper_bound}
 \sigma^2\leq  s\cdot\max \left\{\max_{i\in[q],a\in[s]}\bigl|[M_1]_{i,a}\bigr|^2,\frac1q \right\}.
\end{equation}

Recall that for fully independent uniform signs $\{\eps_j\}_{j=1}^q$ and $\{\tau_i\}_{i=1}^q$, \cref{lemma:random_sign_entry_and_row_bounds,lemma:matrix_rademacher} show that $\Pr[|\xi| \geq x] \leq 4 \mathrm{e}^{-q x^2 /4}$ and $\Pr[\|D\|_{\op} \geq x] \leq 2s^2 \mathrm{e}^{-x^2 / (2 \sigma^2)}$, respectively. Together with the tail-integral formula \cite[Exercise 1.7.2]{Durrett2019} and the identity $x! = \int^{\infty}_{0} \mathrm{e}^{-t} t^{x} \, \mathrm{d} t$ for $x \in \Z_{\geq 0}$ \cite[Example 2.1.17]{Durrett2019}, these bounds give
\begin{equation}
\label{eqn:entrywise_magnitude_bound}
 \begin{aligned}
 \E\left[|\xi|^{2N}\right]
 &=\int_0^\infty2Nx^{2N-1}\Pr[|\xi|\ge x]\,\mathrm dx\\
 &\le4\int_0^\infty2Nx^{2N-1}\mathrm{e}^{-qx^2/4}\,\mathrm dx =\left(\frac4q\right)^N N
   \int_0^\infty t^{N-1}\mathrm{e}^{-t}\,\mathrm dt =4N!\left(\frac4q\right)^N.
 \end{aligned}
\end{equation}
And for each fixed $S_2$, with norm inequality $\Tr[(XX^\dagger)^N] \leq \rank(X) \|XX^\dagger\|^N_{\op} = \rank(X)\|X\|_{\op}^{2N}$,
\begin{equation}
\label{eqn:internal_block_inner_product_upper_bound}
 \begin{aligned}
\E_{S_1}\left[\Tr\left[(DD^\dagger)^N\right]\right]
 \leq  s\E_{S_1}\left[\|D\|_{\op}^{2N}\right] & =s\int_0^\infty2Nx^{2N-1}\Pr_{S_1}[\|D\|_{\op}\ge x]\,\mathrm dx\\
 &\le2s^2 \int_0^\infty2Nx^{2N-1}\mathrm{e}^{-x^2/(2\sigma^2)}\,\mathrm dx \\
 &= 2s^2 N(2\sigma^2)^N
   \int_0^\infty t^{N-1}\mathrm{e}^{-t}\,\mathrm dt = 2s^2N!(2\sigma^2)^N.
 \end{aligned}
\end{equation}
Replacing the signs of $S_1, S_2$ with previous construction via $\eps_j = \eps_j(\mathbf{a})$ and $\tau_i = \eps_i (\mathbf{b})$. By the $2N$-wise independence properties, the expectations of $|\xi|^{2N}$ and, for fixed $S_2$, $\Tr[(DD^\dagger)^N]$ are unchanged for seeded signs due to \cref{eqn:expectation_identity_unchanged}. The corresponding upper bounds for their expectation under the fully independent uniform assumption also hold for our seeds. Averaging first over $\mathbf b$ and then over $\mathbf a$ and using $\sigma^{2N}\leq  s^N \cdot \max\{\sum_{i,a}\bigl|[M_1]_{i,a}\bigr|^{2N}, q^{-N}\} \leq s^N(\sum_{i,a}\bigl|[M_1]_{i,a}\bigr|^{2N}+q^{-N})$ from \cref{eqn:variance_param_upper_bound}, plugging in the moment upper bounds in \cref{eqn:entrywise_magnitude_bound,eqn:internal_block_inner_product_upper_bound} and using $\max_{i \in [q], a \in [s]} |[M_1]_{i,a}|^2$ $\leq \sum_{i \in [q], a \in [s]}|[M_1]_{i,a}|^2$ then give
$$
 \begin{aligned}
 \E_{\mathbf{a}, \mathbf{b}}\left[\Tr\left[(DD^\dagger)^N\right]\right] &\leq 2s^2N!(2s)^N\left( \sum_{i \in [q], a \in [s]} \left|[M_1]_{i,a} \right|^2+q^{-N}\right) \\
 &\leq 2s^2N!(2s)^N\left(qs\cdot4N!\left(\frac4q\right)^N+q^{-N}\right)\\
 &\leq 10qs\cdot s^2(N!)^2\left(\frac{8s}q\right)^N
 \le4s^2\left(\frac{32N^2s}{q}\right)^N  \le2^{-5N},
 \end{aligned}
$$
where the third inequality follows from $qs\le4^n=4^{N-1}$ and $N!\leq  N^N$, and the last one uses $4s^2\le4^N$ and $s/q\leq  1/(4096N^2)$. So some pair of seeds $(\mathbf{a}, \mathbf{b})$ gives
$\Tr[(DD^\dagger)^N]\le2^{-5N}$.

\paragraph{Choosing the seeds.} To find a suitable pair of seeds, we enumerate the pairs of seeds in lexicographic order and estimate the corresponding $\Tr[(DD^\dagger)^{N}]$ to precision $2^{-5N}$. Then we select the first pair for which the estimate is at most $2^{1-5N}$.
Any pair with $\Tr[(DD^\dagger)^{N}] \leq 2^{-5N}$ that we have shown to exist will pass this test, so that the selected pair satisfies $\|D\|_{\op}^{2N} \leq \Tr[(DD^\dagger)^{N}] \leq 2^{1 - 5N} + 2^{-5N} = 3 \cdot 2^{-5N}$, and therefore $\|D\|_{\op} \leq 3^{1/(2N)} \cdot 2^{-5/2} < 1/2 < r$.
To compute the estimate, note that every entry of $D = W^\dagger S_1 M_0 S_2 W$ is $q^{-1}$ (cf. \cref{eqn:internal_block_D_expression}) times a signed sum of $q^{2}$ entries of $M_0$, so $\Tr[(DD^\dagger)^{N}]$ is a sum of $s^{2N} q^{4N} \leq 2^{6nN}$ terms, each $\pm q^{-2N}$ times a product of $2N$ entries of either $M_0$ or their conjugates.
If we evaluate this sum exactly on the supplied approximations, the error is bounded from above by 
$$ 3 \cdot (2N) \cdot 2^{6nN - g_0} = 6N \cdot 2^{6nN-8N^2} = 6N \cdot 2^{-2N^2 - 6N} \leq 2^{-5N}
$$ by the error bound provided in statement 2 of \cref{lemma:more_computation_within_pspace}. For $s=1$, a squared modulus is a product of two entries, so the same lemma applies with $v = 2$ and bounds its error by $3\cdot2\cdot2^{-g_0} \leq 1/64$.

\paragraph{Space bound.} We find the field polynomial in $\bigo(\log_2 q) = \bigo(n)$ space by enumerating the binary polynomials of degree $\log_2{q}$ in lexicographic order and taking the first irreducible one, and irreducibility is tested via trial division by all polynomials of degree at most $(\log_2{q})/2$ \cite[Lemma 3.21]{cook2025structureinplacespaceboundedcomputation}. The search over pairs of seeds stores the current pair, of $4N\log_2q=\bigo(n^2)$ bits, and for each pair, it enumerates the terms of $\Tr[(DD^\dagger)^N]$ one at a time through their $2N$ indices in $[s]$ and $4N$ indices in $[q]$, which take at most $2N\lceil\log_2s\rceil+4N\log_2q\le6nN$ bits. For each term, it recomputes the needed signs, requests the entries one at a time, and keeps only the current product and the running sum. Since the supplied approximations have polynomial length $b = \poly(n)$, the product, including the factor $q^{-2N}$, has $\bigo(2Nb+2N\log_2q)=\bigo(N(b+n))$ bits, and summing at most $2^{6nN}$ terms adds $\lceil\log_2(s^{2N}q^{4N})\rceil+1\le6nN+1=\bigo(nN)$ bits. This space is reused across terms and pairs, so the search eventually runs in space $\poly(n)$ apart from the requests to the target unitary. For $s=1$, a row counter and one estimate would suffice. The output is simply the selected pair of seeds (or index of the selected row when $s=1$). Therefore, a space-uniform procedure can construct the seeds for preconditioners that satisfy the contractive condition in \cref{lemma:admissible_tables}.
\end{proof}

\subsection{\textsf{PSPACE}-computable target at all levels}

In this section, we fix the preconditioners for all recursion levels constructed via the polynomial-space procedure in \cref{appendix:pspace_computable_preconditioners} and show that the targets $U^{(j)}_{\lambda_j}$ of the inner levels that \cref{eqn:recursion_of_parameters_setup,eqn:feedback-definitions} define through exact matrix inverses have $\pspace$-computable entries. Both are done level by level. At level $0$, the target is $W_n$, and the Turing machine $M_W$ approximates its entries directly\footnote{
Note that this will be the only place where the hypothesis of the computability of $W$ in \cref{lemma:usynthesis_with_pspace_oracle} enters
}. Once the entries of $U^{(j)}_{\lambda_j}$ can be sufficiently approximated, the preconditioners of level $j$ and label $\lambda_j$ are those that \cref{lemma:canonical_space_uniform_preconditioners} selects from the approximations of these entries within $2^{-g_0}$ for some bit precision $g_0$. This fixes the preconditioned operator $V^{(j)}_{\lambda_j}$ and hence the targets of level $j+1$. Thus $\pspace$-computability propagates through the recursion in a top-down manner, while the required precision grows only polynomially across all levels, so the gate words can be computed in polynomial space and their bits collected can be into a single $\pspace$ oracle.

\begin{lemma}[Approximating the targets at all levels]
\label{lemma:approximating_targets_at_all_levels}
    The entries of $U^{(j)}_{\lambda_j}$ and of $V^{(j)}_{\lambda_j}$, indexed by $(1^n,1^L,j,\lambda_j,a,b)$, are $\pspace$-computable. Moreover, approximating an entry of $U^{(j)}_{\lambda_j}$ within $2^{-g}$ calls $M_W$ only at precisions at most $\max\{g,g_0\}+(3n+13)j$ bits. Here $g_0 = 8(n + 1)^2$ from \cref{lemma:canonical_space_uniform_preconditioners}.
\end{lemma}
\begin{proof}
    We take steps to approximate the entries of the targets of level $j+1$ from those of level $j$.

    \paragraph{A perturbation bound.} Fix a level $j<m$ and a label with $m \leq J$ being the actual number of recusion levels, and let $V,A,B,C,D$ be  $V^{(j)}_{\lambda_j}$ and its corresponding blocks as in \cref{eqn:block_decomposition_of_unitary}. For a diagonal unitary $Z$, let $F(Z)$ be as in \cref{eqn:feedback-definitions}, and denote $T=DZ$, so that $\|T\|_{\op} <3/4$. Let $\widetilde V,\widetilde Z$ be approximations that $\|\widetilde V-V\|_{\op}, \|\widetilde Z-Z\|_{\op}\leq \vartheta$ for some precision $\vartheta \le1/40$, and put $\widetilde T=\widetilde D\widetilde Z$. Let $L_{\mathrm{num}} \geq 1$ be the numerical truncation length used to approximate the exact feedback target to the requested precision, and let $\widetilde{T}_{L_{\mathrm{num}}} = \sum_{\ell = 1}^{L_{\mathrm{num}} - 1} \widetilde{T}^{\ell}$ denote the truncated sum of \cref{lemma:neumann_series} when applied to $\widetilde T$, and we set $\widetilde F_{L_{\mathrm{num}}}=\widetilde{A}+\widetilde B\widetilde Z\widetilde{T}_{L_{\mathrm{num}}}\widetilde C$. Then $\|\widetilde{T}-T\|_{\op} \le \|\widetilde{D}-D\|_{\op}\|\widetilde{Z}\|_{\op} + \|D\|_{\op}\|\widetilde{Z}-Z\|_{\op} \le \vartheta(1+\vartheta)+ 3\vartheta /4 \le 2\vartheta$ and thus $\|\widetilde T\|_{\op} \leq \| T\|_{\op} + \|\widetilde T - T\|_{\op} < 3/4 + 2 \vartheta \leq 4/5$.
Using a simple geometric-series bound, 
$$
 \begin{aligned}
 &\left\|(\mathbf I_s-T)^{-1}\right\|_{\op} \le \sum_{\ell\ge0}\left(\frac{3}{4}\right)^\ell = 4, \\
 &\max\left\{\left\|\left(\mathbf I_s-\widetilde T \right)^{-1}\right\|_{\op}, \left\|\widetilde{T}_{L_{\mathrm{num}}}\right\|_{\op}\right\} \leq \sum_{\ell \geq 0}  \left\| \widetilde{T}_{L_{\mathrm{num}}} \right\|_{\op}^{\ell} \leq \sum_{\ell\ge0}\left(\frac{4}{5}\right)^\ell = 5.
 \end{aligned}
$$
The resolvent identity \cite[Proposition 1.9]{Hislop1996}
$$
\left(\mathbf I_s-\widetilde T\right)^{-1} -\left(\mathbf I_s-T\right)^{-1} = \left(\mathbf I_s-\widetilde T\right)^{-1} \left(\widetilde T-T\right) \left(\mathbf I_s-T\right)^{-1}
$$
and \cref{lemma:neumann_series} applied to $\widetilde T$ give 
$$
\begin{aligned}
\left\|\widetilde{T}_{L_{\mathrm{num}}}-(\mathbf I_s-T)^{-1} \right\|_{\op} & \leq \left\|  \left(\mathbf I_s-\widetilde T\right)^{-1} \left(\widetilde T-T\right) \left(\mathbf I_s-T\right)^{-1} \right\|_{\op} + \left\| \sum_{\ell \geq L_{\mathrm{num}}} \widetilde{T}^{\ell} \right\|_{\op}
\\
&\leq \left\|  \left(\mathbf I_s-\widetilde T\right)^{-1} \right\|_{\op} \left\| \widetilde T-T\right\|_{\op} \left\| \left(\mathbf I_s-T\right)^{-1}\right\|_{\op} + 5 \cdot \left(\frac{4}{5} \right)^{L_{\mathrm{num}}} 
\\
&\leq 5 \cdot 2 \vartheta \cdot 4 + \left(\frac{4}{5} \right)^{L_{\mathrm{num}}}  =40\vartheta+5 \cdot \left(\frac{4}{5} \right)^{L_{\mathrm{num}}}.
\end{aligned}
$$
We plug this into the telescoped difference to have
\begin{equation}
\label{eqn:perturbed_target_expression}
\begin{aligned}
\widetilde F_{L_{\mathrm{num}}}-F(Z)
=&\left(\widetilde{A}-A\right) +\left(\widetilde B-B\right)\widetilde Z\widetilde{T}_{L_{\mathrm{num}}}\widetilde C +B\left(\widetilde Z-Z\right)\widetilde{T}_{L_{\mathrm{num}}}\widetilde C\\ &+BZ\left(\widetilde{T}_{L_{\mathrm{num}}} -\left(\mathbf I_s-T\right)^{-1}\right)\widetilde C +BZ\left(\mathbf I_s-T\right)^{-1}\left(\widetilde C-C\right).
\end{aligned}
\end{equation}
Then summing up the accumulated error by applying the triangle inequality to \cref{eqn:perturbed_target_expression} gives
\begin{equation}
\label{eqn:feedback_perturbation_bound}
 \left\|\widetilde F_{L_{\mathrm{num}}}-F(Z) \right\|_{\op} \le60\vartheta+6 \cdot \left(\frac{4}{5} \right)^{L_{\mathrm{num}}}.
\end{equation}
\paragraph{Approximating the child target's entries.}
To approximate an entry of $F(\mathbf z)=F(Z(\mathbf z))$ within $2^{-g}$, we approximate all entries of $V$ within $2^{-g-n-8}$, and the diagonal entries of $Z(\mathbf z)$ defined in \cref{eqn:character_of_phase_group,eqn:frequency_vector_and_diagonal_characters}, which are values $ \mathrm{e}^{2\pi\I \theta}$ at dyadic and rational $\theta$, to the same precision stated in statement 1 of \cref{lemma:more_computation_within_pspace}, and keeping all remaining entries $0$. Since these approximations are computed deterministically, they form fixed matrices $\widetilde V,\widetilde Z$, and since $\|X\|_{\op} \leq \|X\|_F \leq 2^n \max_{a, b} |X_{a, b}|$ for a matrix of dimensions $\leq 2^n$. Taking approximation precision $2^{-g - n -8}$ suffices to guarantee that $\widetilde{V}, \widetilde{Z}$ approximates $V, Z$ with $\vartheta\le2^{-g-8} < 1/40$ in operator norm distance. We then put  $L_{\mathrm{num}}=4g+16$, evaluate each entry of $\widetilde F_{L_{\mathrm{num}}}$ exactly as a sum over] paths of products of these approximations, and round it within $2^{-g-2}$ that loses us at most $2^{-g-2}$.
By the perturbation bound \cref{eqn:feedback_perturbation_bound} and that $(4/5)^4<1/2$, so $(4/5)^{L_{\mathrm{num}}} < 2^{-g - 4}$, the error accumulates to at most $60\vartheta+6 \cdot \left(4/5 \right)^{L_{\mathrm{num}}} + 2^{-g-2} < (60\cdot2^{-8}+6\cdot2^{-4}+2^{-2})\cdot2^{-g}<2^{-g}$ by triangle inequality.

\paragraph{The choice of precisions and required space.} An entry of $V = P_{\mathrm{out}}U^{(j)}_{\lambda_j} P_{\mathrm{in}}$ is a sum of at most $4^n$ terms, and each is the product of an entry of $U^{(j)}_{\lambda_j}$ and at most two entries of the preconditioners, which are $0$, $\pm1$ or $\pm q^{-1/2}$ from the sign gates and Hadamard gates. By \cref{lemma:more_computation_within_pspace}, entries of $U^{(j)}_{\lambda_j}$ within $2^{-g'-2n-5}$ give the entries of $V$ within $2^{-g'}$. Unwrapping $g'$ from our previous discussion gives the total precision bits requirement $g' + 2n + 5 = (n + 8) + 2n + 5 = 3n + 13$. Approximating an entry of level $j+1$ within $2^{-g}$ therefore uses entries of level $j$ only at precisions at most $\max\{g+3n+13,g_0\} $, where $g_0$ comes from the selection of the preconditioners in \cref{lemma:canonical_space_uniform_preconditioners}. Induction on $j$ gives rise to the upper bound of precision requirement $\max\{g, g_0\} + (3n + 13) j$, since we have
$$
 \max\{g+3n+13,g_0\}+(3n+13)j \leq \max\{g,g_0\}+(3n+13)(j+1).
$$
There are fewer than $J$ levels, and for each level a depth-first evaluation stores a label of at most $jK \leq JK$ bits, the selected preconditioner data, a path counter using $\bigo(nL_{\mathrm{num}}) = \poly(n, g)$ bits, and accumulators that store the sums as in the proof of \cref{lemma:arthm_of_PSPACE_num} (cf. \cite{SAVITCH1970177}). Here we recall $K=b(L+1)=\bigo(nL^2)$, and every precision in use is at most $\max\{g,g_0\}+(3n+13)J+\bigo(n)$. As $J = \mathcal{O}(n^{3})$, all requested precisions and truncation lengths are polynomial in $n$ and $g$. Evaluating each path term contains $\mathcal{O}(L_{\mathrm{num}})$ dyadic factors of polynomial length, so its exact product also has polynomial length. Since $\widetilde Z$ is diagonal, an entry of $\widetilde B\widetilde Z(\widetilde D\widetilde Z)^\ell\widetilde C$ is a sum over $\ell+1$ indices in $[s]$. Thus there are at most
$$
 1+\sum_{\ell=0}^{L_{\mathrm{num}}-1}s^{\ell+1}
 \le(L_{\mathrm{num}}+1) \cdot 2^{nL_{\mathrm{num}}}
 =2^{\bigo(nL_{\mathrm{num}})}
$$
path terms. We sum them one at a time with only $\mathcal{O}(nL_{\mathrm{num}})$ additional bits. The entrywise evaluation of sums of products in \cref{lemma:arthm_of_PSPACE_num,lemma:more_computation_within_pspace}, the preconditioner search in \cref{lemma:canonical_space_uniform_preconditioners} and the calls to $M_W$ also use polynomial space for these input lengths and precisions, since all of them are space-uniform. The evaluation would keep these data for at most $\mathcal{O}(J) = \bigo(n^3)$ active levels and reuse the workspace after each call, so the total space stays within $\poly(n, L, g)$.
\end{proof}

The proof of \cref{lemma:approximating_targets_at_all_levels} evaluates the approximations used in the selection above at every level, deterministically from the outputs of $M_W$. The preconditioners of all levels are therefore well defined, and they and the targets of the deeper levels depend on these outputs.

\subsection{\textsf{PSPACE}-computable encoders and tables}
In this section, we show that the encoders of every level, and then all tables of the skeleton, are $\pspace$-computable. For consistency, we inherit the register layout of \cref{eqn:register_embeddings} and number the basis states from $0$, same as the convention of \cref{appendix:implementing_encoders}.

\begin{lemma}[Computing entries of the encodings and candidate sets]
\label{lemma:computing_entries_and_candidate_sets}
    Let $T\in\{\widehat E_L,\widehat R_L\}$ at some level and label. Then the following statements hold:
    \begin{enumerate}
        \item The entries of $T$ and the numbers $\| T \ket{i} \|^2$ are $\pspace$-computable.
        \item For every row $u$ of $T$, a set $I_u$ satisfying condition 2 in \cref{lemma:admissible_tables} is computed in space $\poly(n,L)$, using only the row index and the dimension parameters, without querying any entries of the target unitaries.
    \end{enumerate}
\end{lemma}

\begin{proof}
    1. By the entry expression in \cref{eqn:path-monomial} and the proof of \cref{prop:Fourier_sparsity_of_encodings}, an entry of $T$ reads $0$ or $1$ in a port column, and otherwise a sum over paths selected by exact comparisons in $G$ of at most $\sum_{\ell = 1}^L s^{\ell  -1} \leq Ls^L \leq 2^{nL+h}$ products of at most $L$ entries of $V$ or their conjugates. Note also that $\|T\ket i\|^2$ is a sum of at most $|I| = k \cdot |G| \leq 2^n \cdot 2^K =2^{K+n}$ squared moduli of entries. Applying the exponential-length sum closure property in \cref{lemma:arthm_of_PSPACE_num} and the $\pspace$-computability result in \cref{lemma:approximating_targets_at_all_levels} to these sums ensures the $\pspace$-computability.

    2. Let $u=(\mathbf{w},a)$ as in the entry-wise expression of \cref{eqn:entries_of_encoding_E_L}. Recall the proof of \cref{prop:Fourier_sparsity_of_encodings}, the nonzero entries of the row only lie in the port column $a$ if $\mathbf{w}=\mathbf 0$, and otherwise in the internal columns $k+i$ with $i\in S$, where $S$ is the multiset of at most $L$ elements of $[s]$ subject to $\mathbf v_S=-\mathbf{w}$ for $\widehat E_L$, or $\mathbf v_S=\mathbf{w}$ for $\widehat R_L$, which must be unique due to \cref{lemma:newton_identities} if it exists. We find such an $S$ by enumerating all nondecreasing tuples of at most $L$ elements of $[s]$. Let $I_u$ initially consist of the basis indices of these columns, namely $a-1$ if $\mathbf{w}=\mathbf 0$ and $k+i-1$ for the distinct $i\in S$ otherwise. We then pad $I_u$ to $L$ indices using the smallest basis indices at least $2^n$ of the register $\mathsf X$. Since $2^h\ge L+1$, the number of such padding indices is $2^{n+h}-2^n=2^n(2^h-1)\ge2^nL\ge L$. If no tuple matches, and for control values that are not rows of $T$, $I_u$ would consist of these padding indices only. In the enumeration, we store the current tuple, which has at most $L$ indices of at most $n$ bits each. To test the frequency match, we read $\mathbf{w}$ from the read-only input one coordinate at a time and compute $\sum_{i \in S} i^{v} \bmod m_0$ for $v = 0, \dots, L$. For the modular powers and the running sum, we need a constant number of $b$-bit registers with $b = nL + \lceil \log_2(L + 1) \rceil = \mathcal{O}(nL)$ (cf. \cref{eqn:phase_group_setup}). We output the distinct column indices and then the padding indices one at a time. All counters and the current output index use $\mathcal{O}(n + \log(L + 1))$ bits, so collecting everything, the procedure uses $\bigo(nL + b + n + \log(L + 1)) = \poly(n, L)$ auxiliary bits in total.
\end{proof}

\begin{lemma}[Approximating the column preparations]
\label{lemma:pspace_computability_of_column_prep}
Let the bit precision $g_{\mathrm{prep}}\ge 0$. For each encoder, there exists a Grover-Rudolph type state preparation unitary $\widetilde{A}$ on the skeleton of the column preparation in \cref{lemma:block_encoding_of_row_sparse_near_isometry}, whose tables are $\pspace$-computable uniformly also in $g_{\mathrm{prep}}$, and a column preparation $A$ as in the proof of \cref{lemma:block_encoding_of_row_sparse_near_isometry}, such that $\|\widetilde{A}-A\|_{\op}\leq  2^{-g_{\mathrm{prep}}}$.
\end{lemma}

\begin{proof}
    We take steps to construct the state preparation unitary $\widetilde{A}$ and compare it with the desired $A$. This is in a similar vein to the polynomial-space approximation of pure states \cite[Section 6.1]{RY22}, but now we have many vectors to track. We take steps to control the precision on a space-uniform machine.

    \paragraph{Rounding the column vectors.} Recall from the proof of \cref{lemma:block_encoding_of_row_sparse_near_isometry} and \cref{eqn:column-state} that the column preparation acts as follows:
    \begin{equation}
    \label{eqn:column_preparation_unitary_recalled}
     \begin{aligned}
 A\bigl(\ket i_{\mathsf X}\ket{0^w}_{\mathsf Y}\ket0_{\mathsf a}\bigr) =\ket i_{\mathsf X}\ket{\alpha_i}, \quad 
 \ket{\alpha_i}
 &=\sum_{u\in I}T_{ui}\ket u_{\mathsf Y}\ket1_{\mathsf a}
   +\sqrt{1-\|T\ket i\|^2}\,
    \ket{0^w}_{\mathsf Y}\ket0_{\mathsf a},
 \end{aligned}
    \end{equation}
    here $I\subseteq\{0,1\}^w$ is the set of valid rows as $T \in \mathsf{L}(\C^d, \C^I)$. We put the column preparation bit precision $g_{\mathrm{col}}=g_{\mathrm{prep}}+w+3$ and let $\ket{\alpha_i^{\mathrm{dy}}}$ be the $(w + 1)$-qubit vector obtained by rounding these coordinates to the grid $2^{-g_{\mathrm{col}}-1}(\mathbb Z+\I\mathbb Z)$ as complex numbers within $2^{-g_{\mathrm{col}}}$, which the computability results in \cref{lemma:arthm_of_PSPACE_num} and \cref{lemma:computing_entries_and_candidate_sets} allow. Then it satisfies
    $$
    \begin{aligned}
    \left\| \ket{\alpha_i^{\mathrm{dy}}} - \ket{\alpha_i} \right\| &= \sqrt{ \sum_{x \in \{0, 1\}^{w + 1}} \left| \braket{x | \alpha_i^{\mathrm{dy}}} - \braket{x | \alpha_i} \right|^2 } \\
    &\leq \sqrt{2^{w + 1}} \cdot \sqrt{2^{-2 g_{\mathrm{col}}}} = 2^{(w + 1)/2 - g_{\mathrm{prep}} - w - 3} \leq 2^{-g_{\mathrm{prep}} - 2}.
    \end{aligned}
    $$
    In particular, we adjust to ensure $\ket{\alpha_i^{\mathrm{dy}}} \neq 0$. For the normalized unit vector $\ket{\widetilde\alpha_i}=\ket{\alpha_i^{\mathrm{dy}}}/\norm{\alpha_i^{\mathrm{dy}}}$ the triangle inequality and its reverse guarantees that
$$
\begin{aligned}
 \norm{\ket{\widetilde{\alpha}_i}-\ket{\alpha_i}}
 &\le \norm{\ket{\widetilde{\alpha}_i}-\ket{\alpha_i^{\mathrm{dy}}}}
       +\norm{\ket{\alpha_i^{\mathrm{dy}}}-\ket{\alpha_i}}\\
 &= \left|1-\norm{\ket{\alpha_i^{\mathrm{dy}}}} \right|
       +\norm{\ket{\alpha_i^{\mathrm{dy}}}-\ket{\alpha_i}}
 \le 2\norm{\ket{\alpha_i^{\mathrm{dy}}}-\ket{\alpha_i}}
 \le 2^{-g_{\mathrm{prep}}-1}.
\end{aligned}
$$
\paragraph{The rounded preparation unitary.} Let $\widetilde{A}$ be the preparation as in \cref{eqn:column_preparation_unitary_recalled}, but with $\ket{\alpha_i}$ replaced by the rounded vectors $\ket{\widetilde\alpha_i}$, and prepares $\ket{0^w}_{\mathsf Y}\ket0_{\mathsf a}$ when $i$ is not valid. As in the proof of \cref{lemma:diagonal_gates_controlled_state_prep_in_slot_formalism}, its rotations are determined by the ratios of the prefix probabilities $\Pr[p]$, which we take for the unnormalized vector $\ket{\alpha_i^{\mathrm{dy}}}$ as the normalization cancels in the ratios, and its phases by the phases of the nonzero coordinates of $\ket{\alpha_i^{\mathrm{dy}}}$. For a prefix $p$ of the label $x$ on $\mathsf Y\mathsf a$ carrying $\ket{\alpha_i}$, and the coordinate grid gives
$$
 \Pr[p]=\sum_{y\in\{0,1\}^{w+1-|p|}}
    \bigl|\braket{py | \alpha_i^{\mathrm{dy}}}\bigr|^2
 \in2^{-2g_{\mathrm{col}}-2}\mathbb Z_{\ge0}.
$$
We compute this sum exactly, so the tests for zero are exact, and a nonzero prefix probability is at least $2^{-2g_{\mathrm{col}}-2}$. The same grid gives modulus at least $2^{-g_{\mathrm{col}}-1}$ for every nonzero coordinate. Thus the rotations and phases are $\pspace$-computable by \cref{lemma:arthm_of_PSPACE_num}, and a zero prefix probability allows arbitrary operation, so we do nothing for simplicity.

\paragraph{Bounding the preparation error.} For unit vectors $\ket{x}$ and $\ket{y}$, let $\mu = \ket{x}^\dagger \ket{y}$ and write $\ket{y} = \mu\ket{x} + c\ket{x}^\perp$, where $c \geq 0$ and $\ket{x}^\perp$ is a unit vector orthogonal to $\ket{x}$. Let $R$ be the unitary that acts on $\operatorname{span}\{\ket{x}, \ket{x}^\perp\}$ by $\left(\begin{smallmatrix} \mu & -c \\ c & \bar{\mu} \end{smallmatrix}\right)$ in the basis $(\ket{x}, \ket{x}^\perp)$ and as the identity on its orthogonal complement.
Then $R\ket{x} = \mu\ket{x} + c\ket{x}^\perp = \ket{y}$. On this span, $R + R^\dagger = 2\Re(\mu) \, \id$, so
$$
  (R - \id)^\dagger (R - \id) = 2\id - R - R^\dagger = (2 - 2\Re(\mu))\id.
$$
Since $R - \id$ maps this span into itself and vanishes on its orthogonal complement, we get $\|R - \id\|_{\op} = \sqrt{2 - 2\Re(\mu)}$. Finally, $\norm{\ket{x} - \ket{y}}^{2} = 2 - 2\Re(\ket{x}^\dagger \ket{y}) = 2 - 2\Re(\mu)$, so $\|R - \id\|_{\op} = \norm{\ket{x} - \ket{y}}$. By this identity we can connect $\widetilde{A}$ and $A$ by such a controlled rotation $A = R_{\mathrm{ctrl}} \widetilde{A}$ with $R_{\mathrm{ctrl}} = \sum_{i} \ps{i} \otimes R_i$ and $R_i$ is from taking $\ket{x} = \ket{\widetilde{\alpha}_i}$ and $\ket{y} = \ket{\alpha_i}$ above. Then we have 
$$
\left\| A - \widetilde{A} \right\|_{\op} = \max_{i} \left\| R_i - \id \right\|_{\op} = \max_{i} \left\| \ket{\widetilde{\alpha}_i} - \ket{\alpha_i} \right\|_{\op} \leq 2^{-g_{\mathrm{prep}} - 1} < 2^{-g_{\mathrm{prep}}},
$$
as desired. This completes the proof.
\end{proof}

\paragraph{Filling the skeleton.} We apply \cref{lemma:pspace_computability_of_column_prep} with a common precision parameter $g_{\mathrm{prep}}$ to every encoder at every level. Let $\widetilde C_{n,L,g_{\mathrm{prep}}}$ denote the skeleton of \cref{prop:resource_quantification_of_skeleton} filled with the preconditioners we fixed in \cref{lemma:approximating_targets_at_all_levels}, the candidate sets $I_u$ computed via \cref{lemma:computing_entries_and_candidate_sets}, the resulting rounded column preparations $\widetilde{A}$ from \cref{lemma:pspace_computability_of_column_prep}, and in all other slots, the tables that \cref{appendix:answer_of_USP} gives
by explicit formulas. The next lemma shows that all of its tables can be computed in polynomial space with a bounded error.

\begin{lemma}[Approximating the target by the filled skeleton]
\label{lemma:filled_skeleton_error_bound}
The following statements hold:
\begin{enumerate}
\item There exists a deterministic machine that, on the input $(1^n,1^L,1^{g_{\mathrm{prep}}})$, a slot of $\widetilde C_{n,L,g_{\mathrm{prep}}}$, a control value $y$ of the slot, and the precision $1^g$, outputs the entries of the table members $W_y$ within $2^{-g}$ in space $\poly(n,L,g_{\mathrm{prep}},g)$. It accesses the target only through queries to the Turing machine $M_W$.
\item The operator $\widetilde{C}_{n,L,g_{\mathrm{prep}}}$ satisfies the inequality
\begin{equation}
\label{eqn:slot-error}
\left\|\widetilde C_{n,L,g_{\mathrm{prep}}}\cdot\iota-\iota\cdot W_n \right\|_{\op} \le18Jr^L+2J(2L+1)\cdot 2^{-g_{\mathrm{prep}}}.
\end{equation}
\end{enumerate}
\end{lemma}

\begin{proof}
    1. To prove the first statement, we illustrate the workflow of the machine: it runs the skeleton generator of \cref{prop:resource_quantification_of_skeleton} up to the requested slot, which gives the level $j$ of the slot and the rule that defines its table, and it extracts the label $\lambda_j$ from $y$. For $s>1$, the preconditioners at level $j$ consist of Hadamard slots and index permutations with tables that depend only on the sizes, and of sign slots. For $s=1$, they consist of a row permutation. The sign slots and the row permutation depend on the target only through the seeds or the row that \cref{lemma:canonical_space_uniform_preconditioners} selects from the approximations within $2^{-g_0}$ of the entries of $U^{(j)}_{\lambda_j}$, so the machine computes these approximations by \cref{lemma:approximating_targets_at_all_levels}, reruns the selection and evaluates the signs or the permutation that moves the selected row to the last position. The single slot of the bottom level $m$ has the table entry $\diag(\mathrm{e}^{\I\alpha_{\lambda_m}},1)$, where $\mathrm{e}^{\I\alpha_{\lambda_m}}$ is the only entry of the $1\times1$ unitary $U^{(m)}_{\lambda_m}$ which can be approximated analogously. The column preparation operator $\widetilde{A}$ depends on the target through the rounded vector $\ket{\alpha_i^{\mathrm{dy}}}$ as in \cref{lemma:pspace_computability_of_column_prep} for the input $i$ in $y$: the machine computes its coordinates from the entries of $T$ and from $\norm{T\ket i}^2$ that \cref{lemma:computing_entries_and_candidate_sets} obtains from the entries of $V^{(j)}_{\lambda_j}$, and the conditional probabilities $\Pr[p0]$ and $\Pr[p1]$ exactly for the prefix $p$ in $y$, and finally the rotation or the phase within $2^{-g}$ by \cref{lemma:arthm_of_PSPACE_num}. The other tables are independent of the target unitary. Those of the candidate preparations $W$ arise as for $\widetilde{A}$ from the vector $\ket{s_u}=L^{-1/2}\sum_{i\in I_u}\ket i_{\mathsf X}$ of the proof of \cref{lemma:block_encoding_of_row_sparse_near_isometry} for the row $u$ in $y$. Its coordinates are either $L^{-1/2}$ or $0$, so not relative phase adjustment is required, and its prefix probability $\Pr[p]$, defined as in \cref{lemma:pspace_computability_of_column_prep}, is $1/L$ times the number of $i\in I_u$ whose computational-basis labels begin with $p$; the machine counts these $i$ exactly while it enumerates the candidate set $I_u$ by \cref{lemma:computing_entries_and_candidate_sets}. The tables of the Hadamard slots and index permutations, the flag rotation $R_{\mathsf b}$, the reflections and the Fourier transforms $\mathscr F_{\mathsf Z}$ follow from their explicit formulas by \cref{lemma:arthm_of_PSPACE_num} and \cref{lemma:more_computation_within_pspace}. These analysis applies readily to adjoint of tables.

    It suffices to analyze the space complexity of the machine. In this way, it accesses the level-wise targets only through the approximations of \cref{lemma:approximating_targets_at_all_levels}, and hence only through $M_W$. The column vectors are rounded at precision $g_{\mathrm{col}}=g_{\mathrm{prep}}+w+3$ with $w=K+n$ (cf. \cref{eqn:register_embeddings}), and the amplitude $\sqrt{1-\|T\ket i\|^2}$ of $\ket{\alpha_i}$ is $\pspace$-computable by \cref{lemma:computing_entries_and_candidate_sets} combined with the basic arithmetic in \cref{lemma:arthm_of_PSPACE_num}. For a sum of at most $2^a$ products of at most $v$ factors, \cref{lemma:more_computation_within_pspace} tells us that adding $a+\lceil\log_2v\rceil+3$ extra precision bits suffices. Applying this to the term counts in the statement of \cref{lemma:computing_entries_and_candidate_sets} adds $nL+h+\lceil\log_2L\rceil+K+n+\bigo(1)=\bigo(nL+K)$ bits. Whenever this requires an entry of $V^{(j)}_{\lambda_j}$ within $2^{-g'}$, the bit precision requirement of \cref{lemma:approximating_targets_at_all_levels} bounds
the precision of the resulting calls to $M_W$ by
$$
 \max\{g'+2n+5,g_0\}+(3n+13)J
 \le g'+g_0+(3n+13)J+2n+5.
$$
Therefore the arithmetic uses space polynomial in $n$, $L$, $g_{\mathrm{prep}}$ and $g$. Apart from its calls, which run one at a time in the space bounds shown in proof of \cref{lemma:approximating_targets_at_all_levels}, the machine needs to store the slot, $y$, at most two seeds for generating the signs (cf. \cref{lemma:canonical_space_uniform_preconditioners}), and a constant number of indices, counters and number accumulators of polynomial length, so overall it runs in space $\poly(n,L,g_{\mathrm{prep}},g)$.

2. Let $C$ be the original operator obtained by exact column preparation. As the contractivity and sparsity conditions in \cref{lemma:admissible_tables} are still satisfied with our polynomial-space alternatives \cref{lemma:canonical_space_uniform_preconditioners,lemma:computing_entries_and_candidate_sets}, the error analysis is still valid so $\|C \iota - \iota C \|_{\op} \leq 18 J r^L$. For both encoders $T \in \{ \widehat{E}_L, \widehat{R}_L \}$, the ideal block-encoded operator $B_T = R_{\mathsf{b}} W^\dagger A$ (cf. (1) of \cref{lemma:block_encoding_of_row_sparse_near_isometry}) we work out its approximate version $\widetilde{B}_T = R_{\mathsf{b}} W^\dagger \widetilde{A}$. Then 
$$
\begin{aligned}
\left\|\widetilde{B}_T-B_T\right\|_{\op} &= \left\|R_{\mathsf{b}}W^\dagger(\widetilde{A}-A)\right\|_{\op} = \left\|\widetilde{A}-A\right\|_{\op} \le 2^{-g_{\mathrm{prep}}}, \\
\left\|\widetilde{B}_T^\dagger-B_T^\dagger\right\|_{\op} &= \left\|(\widetilde{B}_T-B_T)^\dagger\right\|_{\op} = \left\|\widetilde{B}_T-B_T\right\|_{\op} \le 2^{-g_{\mathrm{prep}}}.
\end{aligned}
$$
Therefore, the approximate amplified unitary $\widetilde{U}_T=(\widetilde{B}_T(2\iota\iota^\dagger-\mathbf I)\widetilde{B}_T^\dagger(\mathbf I-2\gamma\gamma^\dagger))^L \widetilde{B}_T$ satisfies
\begin{equation}
\label{eqn:encoding_circuit_perturbation_bound}
\begin{aligned}
 \left\|\widetilde{U}_T-U_T\right\|_{\op} &\leq  \sum_{\ell=1}^{L}\left(\left\|\widetilde{B}_T-B_T\right\|_{\op} +\left\|\widetilde{B}_T^\dagger-B_T^\dagger\right\|_{\op}\right) + \left\|\widetilde{B}_T-B_T\right\|_{\op}\\
 &\leq  L\left(2^{-g_{\mathrm{prep}}}+2^{-g_{\mathrm{prep}}}\right)
       +2^{-g_{\mathrm{prep}}}
 =(2L+1)\cdot2^{-g_{\mathrm{prep}}}.
\end{aligned}
\end{equation}
As in \cref{appendix:implementing_encoders}, the encoders of level $j$ are $U_E^{(j)}=\mathscr F_{\mathsf Z}U_T$ for $T=\widehat E_L$ and $U_R^{(j)}=\mathscr F_{\mathsf Z}U_T$ for $T=\widehat R_L$, and define their approximate counterparts in $\widetilde C_{n,L,g_{\mathrm{prep}}}$ as $\widetilde U_E^{(j)}=\mathscr F_{\mathsf Z}\widetilde U_T$ and $\widetilde U_R^{(j)}=\mathscr F_{\mathsf Z}\widetilde U_T$, respectively.
By unitary invariance and the invariance under adjoints, the factors $U_E^{(j)}$ and $(U_R^{(j)})^\dagger$ therefore change by at most $(2L+1)\cdot2^{-g_{\mathrm{prep}}}$ in operator norm as in \cref{eqn:encoding_circuit_perturbation_bound}, uniformly in the label. Each level $j < m \leq J$ contains one $U_E^{(j)}$, one $(U_R^{(j)})^\dagger$, the preconditioners and the call to level $j+1$, with the preconditioners and the slot of the bottom level unaffected by the replacement. Telescoping over all the levels therefore gives
\begin{equation}
\label{eqn:filled_circuit_perturbation_bound}
\begin{aligned}
 \left\|\widetilde{C}_{n,L,g_{\mathrm{prep}}}-C\right\|_{\op}
 &\leq  \sum_{j=0}^{m-1}
    \left(\left\|\widetilde{U}_E^{(j)}-U_E^{(j)}\right\|_{\op} +\left\|(\widetilde{U}_R^{(j)})^\dagger -(U_R^{(j)})^\dagger\right\|_{\op}\right)\\
 &\leq  \sum_{j=0}^{m-1}2(2L+1)\cdot2^{-g_{\mathrm{prep}}} =2m(2L+1)\cdot2^{-g_{\mathrm{prep}}}
 \leq  2J(2L+1)\cdot2^{-g_{\mathrm{prep}}}.
\end{aligned}
\end{equation}
Finally, by the triangle inequality with the performance guarantee of $C$ and \cref{eqn:filled_circuit_perturbation_bound}, we have
$$
\begin{aligned}
&\left\| \widetilde{C}_{n, L, g_{\mathrm{prep}}} \cdot \iota - \iota \cdot W_n \right\|_{\op} \leq \left\|\left(\widetilde{C}_{n, L, g_{\mathrm{prep}}} - C \right) \iota  \right\|_{\op} + \left\| C \iota - \iota W_n \right\|_{\op} \\
&\qquad \leq \left\|\widetilde{C}_{n, L, g_{\mathrm{prep}}} - C  \right\|_{\op} \left\|\iota  \right\|_{\op} + \left\| C \iota - \iota W_n \right\|_{\op} \leq 2J(2L + 1) \cdot 2^{-g_{\mathrm{prep}}} + 18 J r^L,
\end{aligned}
$$
which gives us the desired inequality in \cref{eqn:slot-error}, concluding the proof.
\end{proof}

\subsection{A \textsf{PSPACE} oracle for all tables}
\label[appendix]{appendix:pspace_oracle_for_all_tables}
    To complete the oracle construction, we show how the stored words can be chosen in polynomial space and then collected into the target $\pspace$ language $A_W$. We inherit the subscript notation $\ell_{\kappa} = \ell_0 (\kappa + 1)^4$ as in the proof of \cref{prop:skeleton_to_oracle_circuit_compilation} and $\iota_0$ as the operation of introducing one fresh ancilla.

    \begin{lemma}[Selecting gate words in polynomial space]
    \label{lemma:selecting_gate_words_in_polynomial_space}
        There exists a deterministic procedure that, given as input $1^\kappa$ with $\kappa\geq 1$ and approximations within $2^{-\kappa-6}$ of the matrix entries of a table entry $W_y\in\mathsf U(2)$, outputs two words $\widetilde S_y$ and $\widetilde P_y$ in $\left<\hdm, \tgate \right>$ of length $\ell_{\kappa+2}$ satisfying
        \begin{equation}
        \label{eqn:word_wise_error}
    \left\|(\widetilde S_y\otimes\widetilde P_y)\iota_0-\iota_0W_y \right\|_{\op} < 2^{-\kappa}. 
    \end{equation}
This procedure uses $\poly(\kappa)$ space.
    \end{lemma}

\begin{proof}
    We construct such a procedure explicitly: it enumerates all the pairs of words of length $\ell_{\kappa+2}$ in lexicographic order and outputs the first pair accepted by the following test. For a enumerated pair $(\widetilde S_y,\widetilde P_y)$, consider the residual matrix $\Delta=(\widetilde S_y\otimes\widetilde P_y)\iota_0-\iota_0W_y \in \C^{4 \times 2}$. The test approximates $\Delta$ by a matrix $\widetilde{\Delta}$ and accepts the pair if $\|\widetilde{\Delta}\|_{F}\leq 2^{-\kappa-1}$. Recall the proof of \cref{prop:skeleton_to_oracle_circuit_compilation}, the word we enumerate has entries in $\Z[1/\sqrt{2}, \I]$ so we invoke \cite{PhysRevA.87.032332} to compute the entries of $(\widetilde S_y\otimes\widetilde P_y)\iota_0$ exactly in this ring and rounds them within $2^{-\kappa-6}$. The entries of $\iota_0W_y$ are either entries of $W_y$ or $0$, so the supplied approximations give them within $2^{-\kappa-6}$ as well. Subtracting the two approximations yields
    $$
    \forall \, a \in [4], b \in [2], \quad \left| \widetilde{\Delta}_{a, b} - \Delta_{a, b} \right| \leq 2^{-\kappa - 6} + 2^{- \kappa - 6} = 2^{-\kappa - 5},
    $$
    so using this bound and Cauchy-Schwarz gives
    $$
    \left\| \widetilde{\Delta} - \Delta \right\|_{F} = \sqrt{ \sum_{a=1}^4 \sum_{b = 1}^2 \left| \widetilde{\Delta}_{a, b} - \Delta_{a, b} \right|^2 } \leq \sqrt{8} \cdot 2^{-\kappa - 5} < 2^{-\kappa - 3}.
    $$
    We first show that some such pair is accepted. Let $S_y,P_y\in\SU(2)$ be the factors of $W_y$ in the proof of \cref{prop:skeleton_to_oracle_circuit_compilation}, so that $(S_y\otimes P_y) \iota_0=\iota_0 W_y$. Applying the approximation of \cref{lemma:phase_aware_solovay_kitaev} with precision parameter $\kappa+3$ then gives determinant-one words within $2^{-\kappa-3}$ of both $S_y$ and $P_y$, of length at most $\ell_0(\kappa+3)^4=\ell_{\kappa+2}$ possibly with padding. For this pair, expanding
the difference of the tensor products then gives
$$
 \begin{aligned}
\left\|\Delta\right\|_{\op} &= \left\|\left(\left(\widetilde{S}_y-S_y\right)\otimes\widetilde{P}_y + S_y\otimes\left(\widetilde{P}_y-P_y\right)\right)\iota_0 \right\|_{\op} \\
&\leq \left\|\widetilde{S}_y-S_y\right\|_{\op}\left\|\widetilde{P}_y\right\|_{\op} + \left\|S_y\right\|_{\op}\left\|\widetilde{P}_y-P_y\right\|_{\op} \leq 2\cdot2^{-\kappa-3}=2^{-\kappa-2}.
 \end{aligned}
$$
A simple norm inequality gives
$$
\|\Delta\|_{F}
  \le\sqrt{\rank(\Delta)} \cdot \norm{\Delta}_{\op}
 \le\sqrt2\cdot2^{-\kappa-2}.
$$
This pair is therefore accepted, since
$$
 \norm{\widetilde{\Delta}}_{F}
 \le\norm\Delta_{F}+\norm{\widetilde{\Delta}-\Delta}_{F}
 <\sqrt2\cdot2^{-\kappa-2}+2^{-\kappa-3}<2^{-\kappa-1}.
$$
Thus the procedure always output a pair. Conversely, every accepted pair $(\widetilde S_y, \widetilde P_y)$ should satisfy
$$
\left\|(\widetilde S_y\otimes\widetilde P_y)\iota_0-\iota_0W_y \right\|_{\op} = \norm{\Delta}_{\op} \le\norm\Delta_{F}
 \le\norm{\widetilde{\Delta}}_{F}+\norm{\widetilde{\Delta}-\Delta}_{F} <2^{-\kappa-1}+2^{-\kappa-3}<2^{-\kappa},
$$
so the output pair satisfies \cref{eqn:word_wise_error}. Note that none of the bounds on $\Delta$ and $\widetilde{\Delta}$ above depends on the slot or on $y$, so these bounds hold for every table entry $W_y$ with the same setup of precision $\kappa$. Finally, the procedure stores two words of $\bigo(\ell_{\kappa+2})$ bits alone and a constant number of $\Z[1/\sqrt{2}, \I]$ ring elements and numbers of length $\bigo(\kappa+\ell_{\kappa+2})$ for addition and comparison, so the space complexity reads $\bigo(\ell_{\kappa+2})+\bigo(\kappa+\ell_{\kappa+2}) =\poly(\kappa)$. This concludes the proof.
\end{proof}

\paragraph{Setting the parameters.} Given $(1^n,1^\beta)$, and we still put $\delta=2^{-\beta-1}$ and, in place of \cref{eqn:truncation_length} and of the choice of $\kappa$ in the proof of \cref{thm:usp}, we update the parameter settings:
\begin{equation}
\label{eqn:parameters_for_pspace_oracle}
L =4\left(\beta+ \left\lceil\log_2\left(144J\right) \right\rceil\right), ~ g_{\mathrm{prep}} =\beta+3+ \left\lceil\log_2\left(2J\left(2L+1\right)\right)\right\rceil, ~ \kappa =\beta+3+ \left\lceil\log_2\left(S+1\right) \right\rceil,
\end{equation}
\cref{prop:resource_quantification_of_skeleton} for this $L$. These numbers all scale as $\poly(n, \beta)$ as both $J$ and $S$ do, and are computed in polynomial time.

\paragraph{The language.}
Let $f_{W_n,\beta}$ be the Boolean function of \cref{eqn:boolean_function_definition} (see also \cref{prop:skeleton_to_oracle_circuit_compilation}) for the skeleton $\widetilde C_{n,L,g_{\mathrm{prep}}}$ and the word length $\ell_{\kappa+2}$, in which the target and phase words of every slot and control value $y$ are those that \cref{lemma:selecting_gate_words_in_polynomial_space} selects for the table entry $W_y$ from the approximations of \cref{lemma:filled_skeleton_error_bound} with precision $g=\kappa+6$. For unused addresses setting $f_{W_n,\beta} = 0$ suffices. The language then follows readily from $f_{W_n, \beta}$:
\begin{equation}
\label{eqn:the_pspace_language}
 A_W=\{ 1^n0 1^\beta0 x: \ n,\beta\ge1,\ f_{W_n,\beta}(x)=1\}.
\end{equation}
A query $1^n0\,1^\beta0\,x$ to $A_W$ returns $f_{W_n,\beta}(x)$, and this single language will serve all inputs $n, \beta\ge1$.

\paragraph{The language is in $ \pspace$.}
On input $1^n0\,1^\beta0\,x$, a decider computes the parameters in \cref{eqn:parameters_for_pspace_oracle}, parses $x$ in the address format of \cref{eqn:boolean_function_definition}, and rejects all malformed inputs. It then runs the machine of \cref{lemma:filled_skeleton_error_bound} with $g=\kappa+6$ to obtain the requested slot and control value, runs the search for words as per \cref{lemma:selecting_gate_words_in_polynomial_space} on the output, and returns the requested code bit. Since $L$, $g_{\mathrm{prep}}$ and $\kappa$ are $\poly(n, \beta)$, the statements of \cref{lemma:filled_skeleton_error_bound,lemma:selecting_gate_words_in_polynomial_space} bound the space of the decider by a polynomial in the length of its input. Therefore, the language $A_W\in \pspace$.

\subsection{Completing the proof}
\begin{proof}[Proof of \cref{lemma:usynthesis_with_pspace_oracle}]
Given inputs $(1^n, 1^\beta)$, the algorithm first computes parameters in \cref{eqn:parameters_for_pspace_oracle} and outputs the oracle circuit $C_{n, \beta}^{(\cdot)}$ of the proof of \cref{thm:usp} for $d = 2^n$ and $t = 0$ with word length $\ell_{\kappa + 2}$ and prefix $1^n 0 1^\beta 0$ attached to every oracle address. This takes time $\poly(n, \beta)$ and $C_{n, \beta}^{(\cdot)}$ has the inequality required in \cref{lemma:usynthesis_with_pspace_oracle}, still with $\iota = \iota_2 \iota_1$ (cf. \cref{putting_tables_into_oracle}). Let $V_j$ denote the $j$-th slot of the circuit $\widetilde C_{n,L,g_{\mathrm{prep}}}$ and $\widetilde{V}_j$ its corresponding compiled implementation, as in the proof of \cref{prop:skeleton_to_oracle_circuit_compilation}. With the oracle $A_W\in \pspace$ of \cref{appendix:pspace_oracle_for_all_tables}, \cref{lemma:selecting_gate_words_in_polynomial_space} bounds the error of every $\widetilde V_j$ by $2^{-\kappa}$ by extending to all controlled actions and extending to absorb all clean ancillae used, then we have
$$
\left\|C^{A_W}_{n,\beta}\cdot\iota_2-\iota_2\cdot
          \widetilde C_{n,L,g_{\mathrm{prep}}} \right\|_{\op} \le\sum_{j=1}^{S} \left\|\widetilde V_j\iota_2-\iota_2V_j \right\|_{\op} \le S\cdot2^{-\kappa}.
$$
Since $r^4<1/2$, plugging \cref{eqn:parameters_for_pspace_oracle} yields
$$
\begin{aligned}
 &18Jr^L =18J(r^4)^{\beta+\lceil\log_2(144J)\rceil} \le18J \cdot 2^{-\beta-\lceil\log_2(144J)\rceil}
 \le\frac{18J}{144J} \cdot 2^{-\beta}=\frac{\delta}{4},\\
 &2J(2L+1)\cdot2^{-g_{\mathrm{prep}}} =2^{-\beta-3} \cdot 
   \frac{2J(2L+1)}{2^{\lceil\log_2(2J(2L+1))\rceil}}
 \le2^{-\beta-3}=\frac{\delta}{4},\\
 &S\cdot2^{-\kappa} =2^{-\beta-3} \cdot \frac{S}{2^{\lceil\log_2(S+1)\rceil}}
 \le2^{-\beta-3} \cdot \frac{S}{S+1}<\frac{\delta}{4}.
\end{aligned}
$$
Finally, as in the proof of \cref{thm:usp}, using triangle inequality and \cref{lemma:filled_skeleton_error_bound} gives
$$
\begin{aligned}
    \left\| C_{n, \beta}^{A_W} \cdot \iota - \iota \cdot W_n \right\|_{\op} &\leq \left\| C_{n, \beta}^{A_W} \cdot \iota_2 - \iota_2 \cdot \widetilde{C}_{n, L, g_{\mathrm{prep}}} \right\|_{\op} + \left\|\widetilde{C}_{n, L, g_{\mathrm{prep}}} \cdot \iota_1 - \iota_1 \cdot W_n  \right\|_{\op} \\
    &\leq S \cdot 2^{-\kappa} + 18 J r^L + 2J(2 L + 1) \cdot 2^{-g_{\mathrm{prep}}} < \frac{3}{4} \delta < 2^{-\beta - 1}.
\end{aligned}
$$
This satisfies the precision requirement and thus completes the proof.
\end{proof}

\endgroup

\end{document}